\documentclass[twocolumn,showpacs,preprintnumbers,amsmath,amssymb]{revtex4}

\newcommand{\bea}{\begin{eqnarray}}
\newcommand{\eea}{\end{eqnarray}}
\usepackage{graphicx}
\usepackage{dcolumn}
\usepackage{bm}
\usepackage{here}

\begin{document}
\title{Evolution of linear perturbations through a bouncing world model: \\
       Is the near Harrison-Zel'dovich spectrum possible via a bounce?}
\author{Han Seek Kim}
\email{hansik@knu.ac.kr}
\author{Jai-chan Hwang}%
\affiliation{Department of Astronomy and Atmospheric Sciences,
Kyungpook National University, Daegu, Korea}

\date{\today}

\begin{abstract}

We present a detailed numerical study of the evolutions of
cosmological linear perturbations through a simple bouncing world
model based on two scalar fields. We properly {\it identify} the
relatively growing and decaying solutions in expanding and
collapsing phases. Using a decomposition based on the large-scale
limit exact solution of curvature (adiabatic) perturbations with two
independent modes, we assign the relatively growing/decaying one in
an expanding phase as the $C/d$-mode. In the collapsing phase, the
roles are reversed, and the $C/d$-mode is relatively
decaying/growing. The analytic solution shows that, as long as the
large scale and the adiabatic conditions are met, the $C$- and
$d$-modes {\it preserve} their nature throughout the bounce. Here,
by using a concrete nonsingular bouncing world model based on two
scalar fields, we numerically follow the evolutions of the correctly
identified $C$- and $d$-modes which preserve their nature through
the bounce, thus confirming our previous anticipation based on the
analytic solution. Since we are currently in an expanding phase the
observationally relevant one in the expanding phase is the
relatively growing $C$-mode whose nature is preserved throughout the
bounce. The spectrum of $C$-mode generated from quantum fluctuations
in the collapsing phase has a quite blue spectrum compared with the
near Harrison-Zel'dovich scale-invariant one. Thus, while the
large-scale condition is satisfied and the adiabatic condition is
met during the bounce, we conclude that it is {\it not possible} to
obtain the near Harrison-Zel'dovich scale-invariant density spectrum
through a bouncing world model as long as the seed fluctuations were
generated from quantum fluctuations of the curvature perturbation in
the collapsing phase. We also study the tensor-type perturbation.
For the tensor-type perturbation, however, both $C$- and $d$-modes
(of the tensor-type perturbations) in the collapsing phase survive
as the relatively growing $C$-mode in the expanding phase. Due to
its growing nature in the collapsing phase, the initial $d$-mode
dominates the surviving $C$-mode spectrum after the bounce.

\end{abstract}
\maketitle

\tableofcontents

\section{Introduction}
                                               \label{sec:Introduction}

In order to explain the large-scale structures and motions of the
galaxy distribution and the temperature anisotropies of cosmic
microwave background we need near Harrison-Zel'dovich
scale-invariant spectrum for the initial density fluctuations which
was suggested in the early 70's \cite{HZ-spectrum}. Currently,
inflation (early acceleration phase) scenario provides the only
known mechanism which can successfully generate seed fluctuations
for such cosmological structures. Inflation provides near
Harrison-Zel'dovich spectrum rather naturally by amplifying the ever
present quantum fluctuations in the matter and space-time in
microscopic scales during an early acceleration phase.

Recently, several bouncing cosmological world models were suggested
as alternatives to the inflation scenario. The pre-big bang scenario
\cite{pbb} is based on the low-energy tree-level string theory
effective action. However, pre-big bang scenario fails to explain
the current large-scale structure because the generated spectrum is
quite bluer than near Harrison-Zel'dovich spectrum, see
\cite{pbb-pert}. More recently, ekpyrotic and cyclic scenarios
\cite{Ek} were suggested based on the collision of two branes in a
five dimensional bulk. According to the authors of these scenarios,
near Harrison-Zel'dovich spectrum follows from the $d$-mode of
curvature perturbation in the zero-shear gauge ($\varphi_\chi$;
terms will be explained shortly). Such a claim has been disputed by
several authors \cite{Lyth,BF,hw1,hw2,HN-bounce}. We pointed out
that although the $d$-mode is the dominating mode in the collapsing
phase it decays away as the background model switches to expanding
phase, thus irrelevant observationally \cite{hw1,hw2,HN-bounce}.
Also the $d$-mode of $\varphi_\chi$ is known to exaggerate the
perturbations due to overly distorted hypersurface (temporal gauge)
condition compared with the curvature perturbation in the comoving
gauge ($\varphi_v$), see \cite{Bardeen-1980}. In fact, the authors
of cyclic scenario in \cite{Ek} claimed that in effective four
dimensional space-time their world model goes through a singularity.
In such a case not only the background world model, but also the
perturbations inevitably become singular, thus we have no tool to
handle the situation \cite{Lyth}. In order to handle the situation
theoretically in \cite{hw1,hw2,HN-bounce} we have considered several
smooth and nonsingular bouncing world model assuming both the
background world model and linear perturbations remain valid. The
present work is a continuation of our previous studies using a
specific bouncing world model and concrete numerical study of
perturbations under such a bounce.

Here we summarize our previous reason, for not considering the
$d$-mode as seed fluctuations, used in \cite{hw1,hw2,HN-bounce}. A
scalar-type cosmological perturbation of a single component medium
is described by a second-order differential equation. As long as the
large-scale condition is met we have an analytic solution with two
independent modes which we term $C$- and $d$-modes, see Sec.\
\ref{lsccd}. We assign the relatively growing/decaying one in an
expanding phase as the $C/d$-mode; in the collapsing phase, the
roles are reversed, and the $C/d$-mode is relatively
decaying/growing. In fact, for $\varphi_v$ the $C$-mode simply
remains constant in time and the $d$-mode decays in an expanding
phase, thus suggesting their names. Since such a large-scale
condition is well met near the bounce we are considering, we
anticipate the nature of the $C$- and $d$-modes to be preserved even
after the bounce. Because the observationally relevant one in the
expanding phase is the relatively growing $C$-mode, even in the
collapsing phase, we have to consider the seed perturbation in the
$C$-mode generated from quantum fluctuations. This shows quite blue
spectra both for the pre-big bang and the ekpyrotic (or cyclic)
scenarios. Although not successful as alternative to the inflation,
these models have spurred interests in the evolution of
perturbations through nonsingular bounces in the early universe
\cite{bounce-pert-others,FB,AW}. In this work we will show the same
conclusion, now using a numerical study of perturbation through a
specific nonsingular bouncing world model.

In order to have a realistic bounce in practice, we have to consider
an additional field or fluid component which drives the nonsingular
bounce in additional to the field or fluid which gives rise to the
seed fluctuations for the large-scale structures
\cite{HN-bounce,bounce-pert-others}. In such cases we have to
consider perturbations of the additional field introduced to have
bounce, in addition to the field for fluctuation seeds. Even in such
cases, however, we expect our above conclusion based on the analytic
solutions for the single component is relevant for the analysis; for
our previous analytic argument, see Sec.\ V.C of \cite{HN-bounce}.
Recently, however, authors of \cite{AW} have reported that by
numerical analysis, for a specific collapse phase, the $d$-mode of
$\varphi_v$ in the collapsing phase survives as the $C$-mode in the
expanding phase, thus allowing a scale-invariant spectrum available
for that specific collapse, see also \cite{Wands-1999,FB,BF}. The
numerical results in \cite{AW} show the growing mode ($d$-mode) in a
collapsing phase survives as the growing mode ($C$-mode) in an
expanding phase. In this work we will show that this is a result
based on mixed initial conditions of our precise $C$- and $d$-modes;
the latter initial conditions produce evolutions where $C$- and
$d$-mode natures are preserved even after the bounce.

In this work, we would like to address and clarify whether switching
of $d$-mode to the $C$-mode is possible as the perturbation goes
through a nonsingular bounce. We will resolve such an issue by
numerically following the evolution of diverse $C$- and $d$-mode
initial conditions through a simple toy bounce model based on two
scalar fields. In order to have the $C$- and $d$-mode decomposition,
the the curvature (adiabatic) perturbations should evolve
independently (we will call it the adiabatic condition) from the
isocurvature perturbations, and the large-scale condition should be
met. Our results show that, as long as the large-scale condition and
adiabatic condition are met, we can find our $C$- and $d$-mode
initial conditions where the nature of $C$- and $d$-mode are
preserved through the bounce; as some background variables diverge
near the bounce we will encounter some sharp breakdowns of the above
conditions caused by the divergence of the background variables; we
will argue that such sharp breakdowns do not have physical impacts
on interpreting the $C$- and $d$-mode natures.

In order to get these results which are supported by previous
analytic methods in \cite{HN-bounce}, it is {\it important} to find
the precise $C$- and $d$-mode initial conditions which allow the
proper evolutions of these two modes. Since our realistic bounce
model is based on two scalar fields, the naive analytic form $C$-
and $d$-mode initial conditions based on analytic solutions of a
single component medium are not the precise ones. We will show that
the previous different reports were based on taking these naive
initial conditions which in fact correspond to mixed initial
conditions of our (precise) $C$- and $d$-modes. Under such mixed
initial conditions the adiabatic condition breaks down near the
bounce, thus one {\it cannot} trace the resulting $C$-mode in the
expanding phase to the $d$-mode in the collapsing phase. We will
show that under such mixed initial conditions the isocurvature
perturbations are excited near the bounce. From our study we will
conclude that, as long as the large-scale conditions and the
adiabatic conditions are met, it is not possible to have near
Harrison-Zel'dovich spectrum from the $C$-mode generated from
quantum fluctuations during the collapsing phase before bounce. We
emphasize the conditions we use to get such a strong conclusion: (i)
Einstein's gravity and linear perturbation theory work throughout
the evolution, (ii) quantum fluctuations of a minimally coupled
scalar field in a collapsing phase provide seed curvature
(adiabatic) mode fluctuations for the large-scale structure, (iii)
the large-scale conditions are met during the bounce, (iv) the
adiabatic conditions are met during the bounce. We will show that
the latter two conditions are often violated subtly near the bounce
without affecting curvature perturbation natures in the large-scale
limit.

In the case of the tensor-type perturbation (gravitational waves) we
will present the $C$-mode which remains constant throughout the
bounce. However, the $d$-mode in the collapsing phase {\it
generates} an additional $C$-mode after the bounce. Since the
$d$-mode grows very rapidly during the collapsing phase, the
$C$-mode generated by the initial $d$-mode is likely to be dominant
over the original $C$-mode initially generated during the collapsing
phase. In such a case, it should be the $d$-mode initial condition
in the collapsing phase which gives rise to the dominating $C$-mode
tensor-type perturbation in an expanding phase after the bounce. We
will show that for a collapsing phase with matter-dominated-like
equation of state we could have near Harrison-Zel'dovich spectrum
for the tensor-type perturbation. Its amplitude, however, should
depend on the duration of the collapsing phase after its generation
from quantum fluctuations.

In Sec.\ \ref{sec:Background} we introduce our nonsingular bouncing
world model based on an ordinary minimally coupled scalar field
together with a ghost scalar field without potential; the latter
field causes a nonsingular bounce. In Sec.\ \ref{sec:Perturbations}
we present basic perturbation equations, large-scale analytic
solutions and exact solutions under special situations in three
different temporal gauge conditions. We include the tensor-type
perturbations. In Sec.\ \ref{sec:Numerical-results} we present our
numerical integrations based on our precise initial conditions which
give correct behaviors of the $C$- and $d$-modes in both collapsing
and expanding phases. We will check the conditions required to have
such a decomposition. We compare our results with the ones based on
the naive initial conditions which are motivated by the analytic
initial conditions in a single component situation. We will show
that these latter initial conditions in fact correspond to the mixed
initial conditions, and show that as the evolution proceeds certain
conditions needed for the mode decomposition are violated near the
bounce. We also analyze the tensor-type perturbations. Since
perturbation equations and variables often show singular behaviors
near the bounce, in this section we carefully examine and explain
the evolutions and the required conditions. We also explain the
method of finding our precise $C$- and $d$-mode initial conditions.
In Sec.\ \ref{sec:Implication} we consider the quantum fluctuations
as the initial conditions and present the generated power spectra
for the scalar- and tensor-type perturbations. We also present
implications of our results. Section \ref{sec:Discussion} presents
discussion.

In this work, we {\it assume} Einstein's gravity and spatially
homogeneous-isotropic background world model with linear
perturbations valid throughout the bounce. We set $8 \pi G = 1 = c$
and $\hbar = 1$.

\section{Background}
                                               \label{sec:Background}

In order to have a collapsing phase followed by an expanding phase
which is smoothly connected by a nonsingular bounce we consider a
model based on two scalar fields introduced by Allen and Wands in
\cite{AW}. One field $\phi$ is an ordinary minimally coupled scalar
field with a positive kinetic energy and a self-interaction
potential. The other field $\sigma$ has a negative kinetic energy (a
ghost field) without a potential. The ghost field could violates
null-energy condition \cite{ghost-field}, $\mu + p < 0$, which is
necessary to have a smooth and nonsingular bounce. The ghost scalar
field dominates near the singularity and causes a non-singular
bounce. The action and the energy-momentum tensor are \bea
   & & S=\int
       \sqrt{-\tilde g}\left[ { 1 \over 2}\tilde R
       -{1 \over 2} \tilde \phi^{,c}\tilde \phi_{,c}
       -\tilde V (\tilde \phi)
       +{1 \over 2}\tilde \sigma^{,c}\tilde \sigma_{,c} \right] {d^4}x,
   \nonumber \\
   \label{eq-action} \\
   & & \tilde T_{ab} = \left[-{1\over 2} {\tilde \phi^{,c}}{\tilde\phi_{,c}}
       +{1 \over 2}{\tilde\sigma^{,c}}{\tilde\sigma_{,c}}
       - \tilde V(\tilde \phi) \right]\tilde g_{ab}
   \nonumber \\
   & & \qquad
       +
       {\tilde \phi_{,a}}{\tilde \phi_{,b}} - {\tilde \sigma_{,a}}{\tilde
       \sigma_{,b}},
   \label{eq-energymomoentum}
\eea where tildes indicate covariant variables; the Latin indices
indicate the space-time.

As the background world model we consider a spatially {\it flat},
homogeneous and isotropic, Friedmann model. To the background order
the Robertson-Walker metric can be written as \bea
   && ds^2 = a^2 \left( - d \eta^2 + g^{(3)}_{\alpha\beta} d x^\alpha
       d x^\beta \right),
\eea where $a(\eta)$ is the cosmic scale factor, $\eta$ is a
conformal time, and $g^{(3)}_{\alpha\beta}$ could become
$\delta_{\alpha\beta}$ in our flat background; the Greek indices
indicate the space. The effective fluid quantities are\bea
    && \mu={1 \over 2}{\dot\phi}^2-{1 \over
       2}{\dot\sigma}^2 + V, \quad
       p={1 \over 2}{\dot\phi}^2-{1 \over 2}{\dot\sigma}^2 - V,
    \label{BG-fluid}
\eea where $\mu$ and $p$ are the energy density and pressure,
respectively; a dot indicates a time derivative based on the time
$t$ with $dt \equiv a d \eta$. The basic equations for background
are \bea
    && H^2={1 \over 3}\left({1 \over 2}{\dot\phi}^2-{1 \over
       2}{\dot\sigma}^2+V\right),
    \label{eq-friedmann}\\
    && \dot H=-{1 \over
       2}({\dot\phi}^2-{\dot\sigma}^2),
    \label{eq-scaledd}\\
    && \ddot \phi + 3 H \dot \phi + V_{,\phi}=0,
    \label{eq-ordinary}\\
    && \ddot\sigma + 3H\dot\sigma=0,
    \label{eq-ghostscalarfield}
\eea where $H \equiv {\dot a / a}$. Equation (\ref{eq-friedmann})
and (\ref{eq-scaledd}) are the Friedmann equations, and Eqs.\
(\ref{eq-ordinary}) and (\ref{eq-ghostscalarfield}) are Klein-Gordon
equations for each of the scalar fields. We {\it take} the potential
of the ordinary scalar field to be a simple exponential form \bea
    &V=V_{0}\exp(-\lambda\phi).\label{eq-potential}
\eea

\begin{figure*}
\begin{center}
\includegraphics[width=8cm]{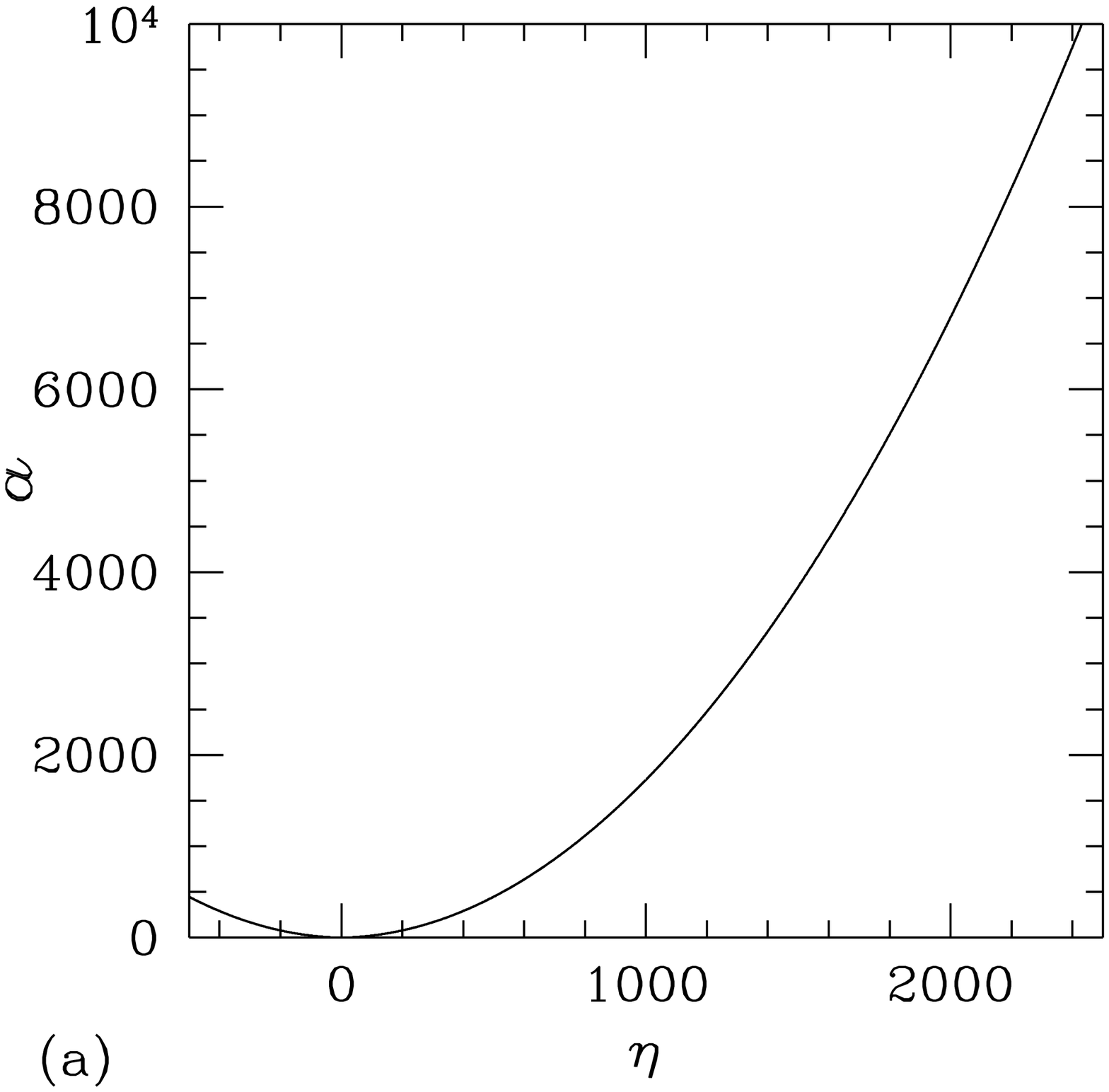}\hfill
\includegraphics[width=8cm]{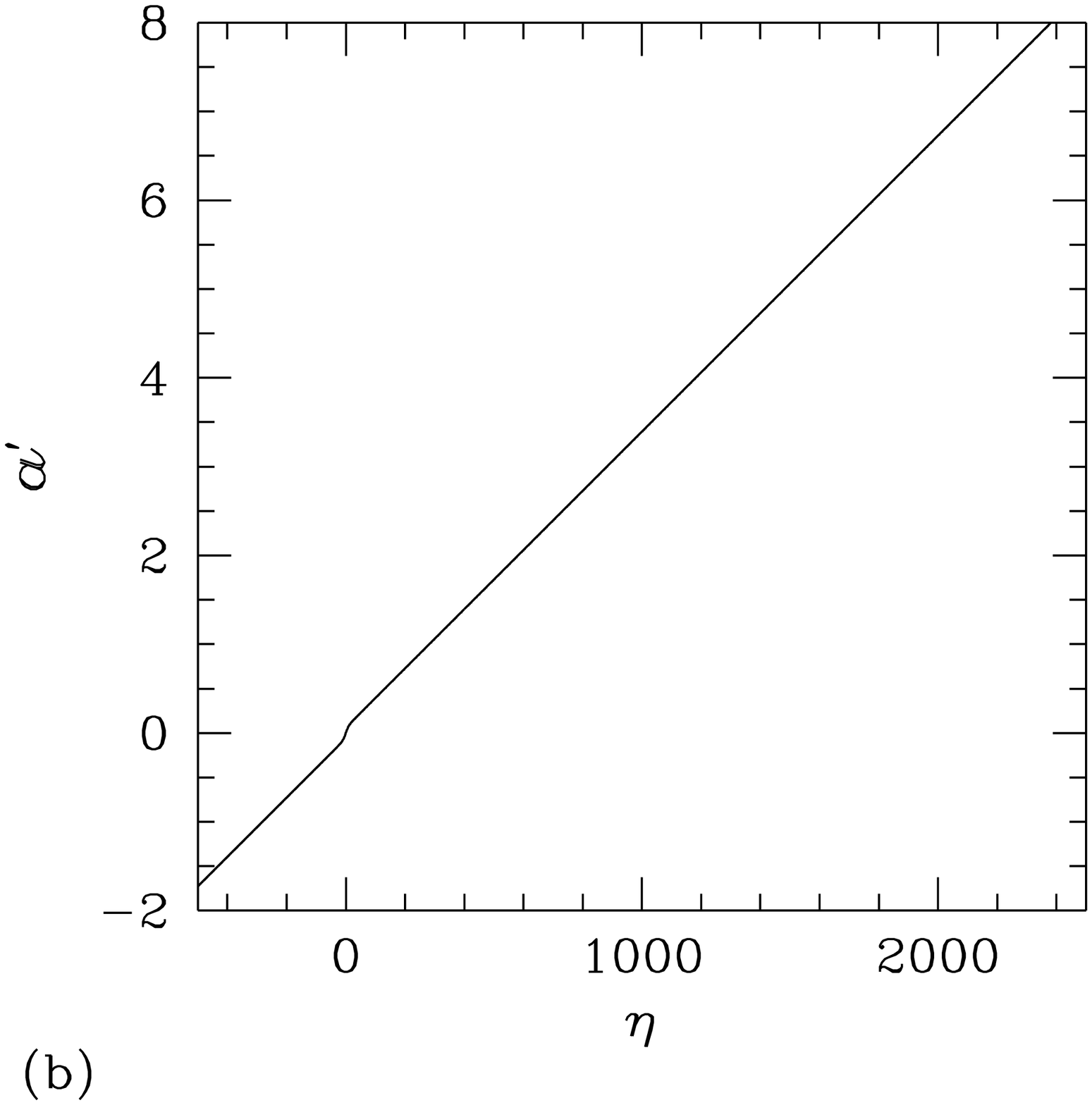}
\caption{Evolution of the scale factor and its time derivative.
         In all figures in this work
         we take $\lambda=\sqrt{3}$. This leads to matter-dominated-era-like
         behaviors ($a \propto \eta^2 \propto t^{2/3}$) away from the bounce.
         Panel (a) shows evolution of the scale factor.
         The bounce is nonsingular, and we set $a_{min} \equiv 1$ at the bounce.
         Panel (b) shows time derivative of the scale factor.
         At the bounce, the derivative of the scale factor becomes zero.
         }
         \label{fig-scaledsf}
\end{center}
\end{figure*}

\begin{figure*}
\begin{center}
\includegraphics[width=8cm]{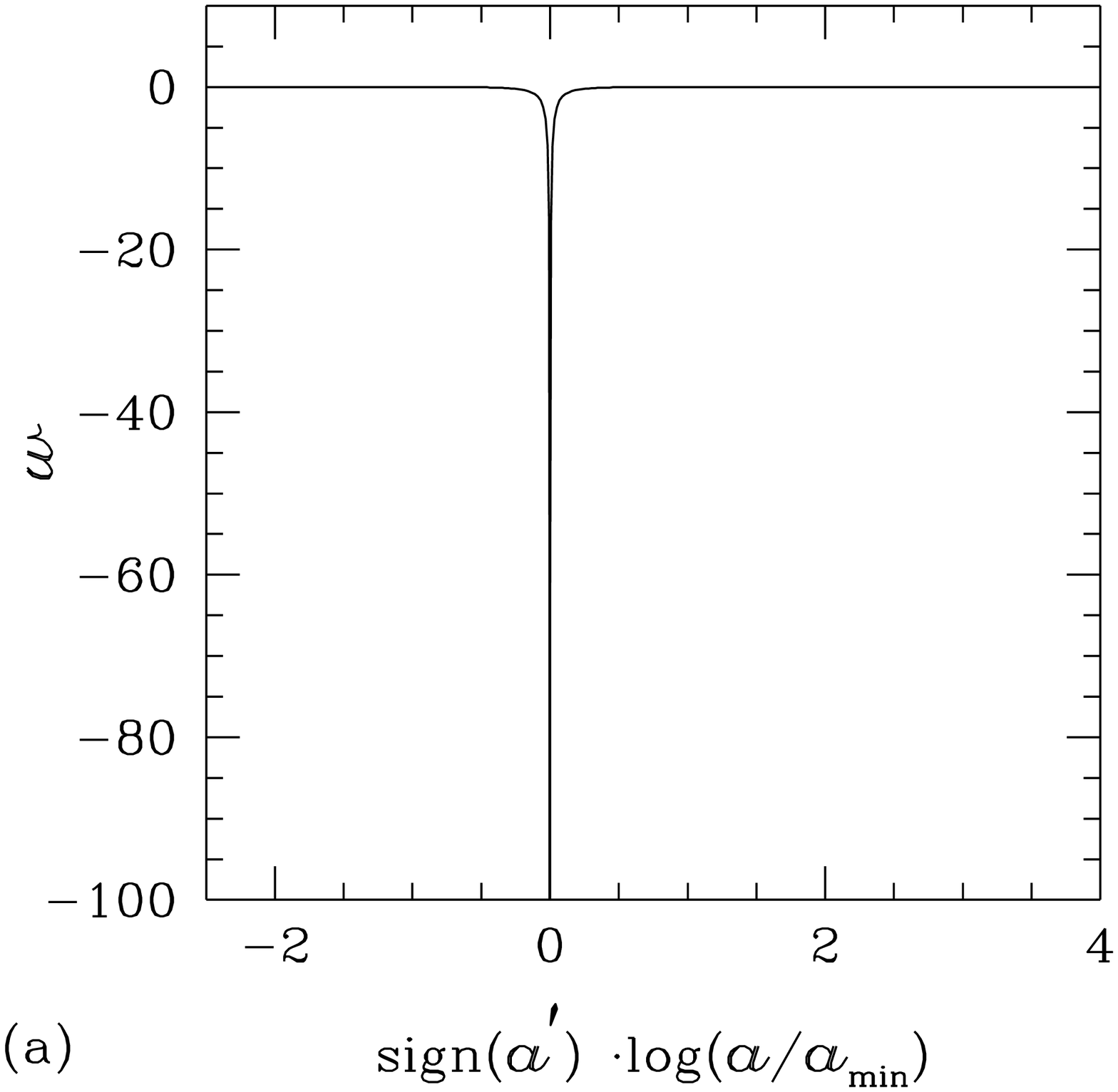}\hfill
\includegraphics[width=8cm]{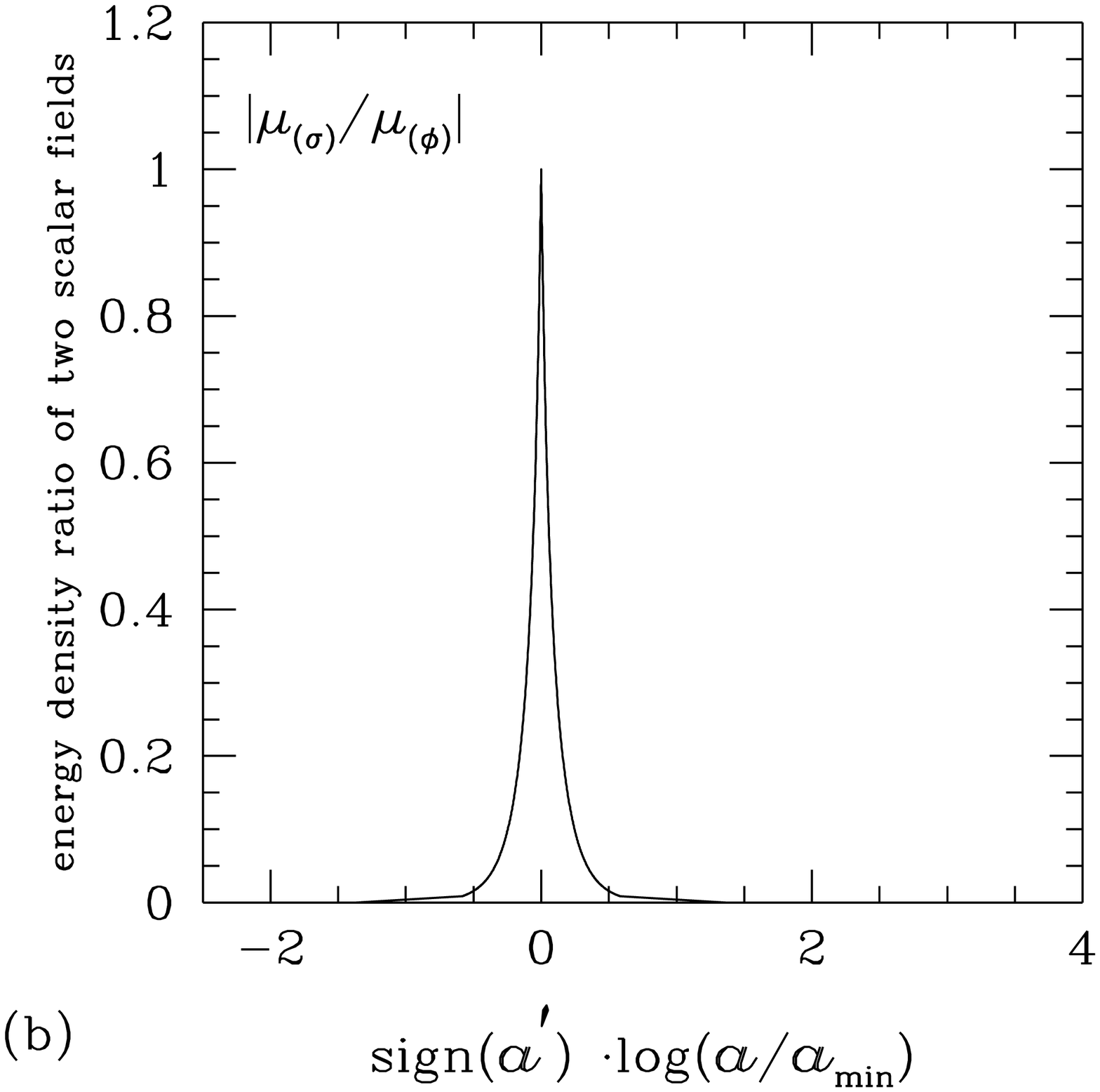}
\caption{Evolutions of the equation of state and the energy density
         ratio of the two fields. Panel (a) shows evolution of the
         equation of state $w \equiv p / \mu$. The equation of state $w$ is
         zero at early and late times which behaves like the matter
         dominated eras. Panel (b) shows the energy density ratio of the
         two scalar fields: $\mu_{(\phi)} \equiv {1
         \over 2} \dot \phi^2 + V$ and $\mu_{(\sigma)} \equiv - {1 \over 2}
         \dot \sigma^2$. The energy density of ordinary scalar field $\phi$
         dominates except near the bounce where the ghost field $\sigma$
         becomes significant. The bounce is caused by the presence of $\sigma$. The
         horizontal axis indicates the time axis using the scale factor; time
         increases from left to right, and $-$ (+) sign denotes the
         collapsing (expanding) phase.
         }
         \label{fig-eosedr}
\end{center}
\end{figure*}

We solve Eqs.\ (\ref{eq-friedmann})-(\ref{eq-ghostscalarfield})
numerically to follow the background evolution. We can integrate
Eq.\ (\ref{eq-ghostscalarfield}) to obtain \bea
    &&\dot\sigma \propto {1 \over a^3}.
    \label{eq-sigmaprime}
\eea Thus, the energy density and pressure of the ghost field
dominate near the bounce. Figure \ref{fig-scaledsf} shows the
evolution of scale factor and its derivative near the bounce. Figure
\ref{fig-eosedr} shows that the energy density of ghost scalar field
dominates only near the bounce. Away from the bounce, where the
ghost scalar field is negligible, assuming the exponential potential
in Eq.\ (\ref{eq-potential}) we have the power-law expansion or
collapse \cite{LM} \bea
    &&a \propto t^q \propto \eta^{q \over 1-q}, \quad
    q\equiv{2\over 3(1+w_{\phi})}
    = {2 \over \lambda^2}; \nonumber\\
     &&V \propto e^{-\sqrt{3(1+w_{\phi})}\phi}, \quad
    \dot\phi =\sqrt{3(1+w_{\phi})}H,
     \label{eq-repo}
\eea where $w_{\phi}={ p_{(\phi)} / \mu_{(\phi)}}$ with
$\mu_{(\phi)} \equiv {1 \over 2} \dot \phi^2 + V$ and $p_{(\phi)}
\equiv {1 \over 2} \dot \phi^2 - V$. Thus, we have
$\lambda=\sqrt{3(1+w_{\phi})}$. The value of $\lambda$ determines
the evolution of scale factor far away from the bounce. For
$\lambda=2$ we have $w_{\phi}= 1/3$ and the scale factor evolves
like the radiation dominate era, and for $\lambda=\sqrt{3}$ we have
$w_{\phi}=0$ which is a matter dominate era.  In this work we will
{\it take} $\lambda=\sqrt{3}$, thus $w_{\phi}=0$.

In Figs.\ \ref{fig-scaledsf} and \ref{fig-eosedr}, we show the
background evolution through the bounce. In order to help numerical
reproduction, we present the precise numerical value of our initial
conditions in Table \ref{Table-binitial}.  A prime indicates a time
derivative based on the conformal time $\eta$.

\begin{table}[H]
\caption{The initial conditions we used for the background
         evolution. We set $a_{min} \equiv 1$, $V_0 \equiv 0.01$,
         and
         $\lambda = \sqrt{3}$.
         }
         \label{Table-binitial}
\begin{center}
\begin{tabular}{ c  r } \hline\hline
       variable  & initial value  \\  \hline
       $ \eta $ & {$-$5.0E+02}\\
       $\phi $ &{1.05727052142872E+01} \\
       $\phi^\prime$ & {$-$6.6836661819519E$-$03}\\
       $\sigma$ & {$-$1.8136774745271E+00}\\
       $\sigma^\prime$ & {7.0552762997745E$-$07}\\
       $a$ & {4.4771380493724E+02}\\
       $a^\prime$ & { $-$1.7276454036056E+00}\\ \hline\hline
\end{tabular}
\end{center}
\end{table}

\section{Perturbations}
                                               \label{sec:Perturbations}

We introduce completely general perturbed order variables in the
Robertson-Walker background. The metric becomes \bea
    && \tilde g_{00} = -a^2 \left( 1 + 2 \alpha \right), \quad
       \tilde g_{0\alpha}
       = -a^2 \left( \beta_{,\alpha} + B^{(v)}_{\alpha} \right),
    \nonumber \\
    && \tilde g_{\alpha\beta}
        = a^2 \left( g^{(3)}_{\alpha\beta}
        + 2 g^{(3)}_{\alpha\beta} \varphi
        + 2\gamma_{,\alpha\vert\beta}
        + 2 C^{(v)}_{(\alpha|\beta)}
        + 2C^{(t)}_{\alpha\beta} \right),
    \nonumber\\
    \label{eq9-metric}
\eea where an index $0$ indicates $\eta$; a vertical bar $\mid$
indicates a covariant derivative based on ${g^{(3)}_{\alpha\beta
}}$, and $2 C^{(v)}_{(\alpha|\beta)} \equiv
C^{(v)}_{\alpha\vert\beta} + C^{(v)}_{\beta\vert\alpha}$. The
perturbed order variables $\alpha$, $\beta$, $\gamma$ and $\varphi$
indicate the scalar-type perturbations. The transverse
$B^{(v)}_\alpha$ and $C^{(v)}_\alpha$ indicate the vector-type
perturbations. The transverse-tracefree $C^{(t)}_{\alpha\beta}$
indicates the tensor-type perturbation. Indices $(v)$ and $(t)$
indicate the vector- and tensor-type perturbations, respectively. In
our background world model the three types of perturbations evolve
independently to the linear order. We follow conventions in
\cite{Bardeen-1988,hw4,hw11}. The energy-momentum tensor can be
decomposed into the effective fluid quantities as \bea
    & & T^0_{0}=-(\mu + \delta\mu), \quad
      T^0_{\alpha} = (\mu +p)(-v_{,\alpha}+v^{(v)}_{\alpha}),
      \nonumber\\
     && T^\alpha _{\beta}
      = (p + \delta p) \delta^{\alpha}_{\beta} + \Pi^\alpha_\beta,
    \label{eq-talphabeta}
\eea where tracefree $\Pi^\alpha_\beta$ indicates anisotropic
stress. For the scalar fields we set \bea
   & & \tilde \phi = \phi + \delta \phi, \quad
       \tilde \sigma = \sigma + \delta \sigma.
\eea The two scalar fields contribute to the effective fluid
quantities as \bea
    && \delta\mu
        = {\dot
        {\phi}}{\delta\dot \phi} -{\dot {\sigma}}{\delta \dot\sigma}-
        ({\dot \phi}^2-{\dot \sigma}^2)\alpha + V_{,\phi} \delta \phi
        , \nonumber \\
     &&\delta p
        = {\dot \phi}{ \delta\dot \phi} -{\dot \sigma}{ \delta\dot \sigma}-
       ({\dot \phi}^2-{\dot \sigma}^2)\alpha - V_{,\phi}\delta\phi,
    \nonumber \\
    && (\mu+p)v
       ={1 \over a}(\dot \phi \delta \phi - \dot \sigma
       \delta \sigma) , \quad
       (\mu+p)v^{(v)}_\alpha = 0,
    \nonumber\\
    & &
       \Pi^\alpha_\beta = 0.
    \label{perturbed-fluid}
\eea

When we solve the Einstein equations, we have freedom to choose the
spatial and temporal gauge conditions. Practically, it is important
to take a gauge which suits the problem, but generally we do not
know the suitable gauge condition, \textit{a priori}. The
tensor-type perturbation is gauge-invariant and the vector-type
perturbation can be written in a uniquely gauge-invariant form. By
using a spatially gauge-invariant combination \bea
    && \chi \equiv a(\beta +\alpha\dot\gamma),
    \label{eq-metricdecompose}
\eea instead of $\beta$ and $\gamma$ individually, all scalar-type
perturbation variables are spatially gauge-invariant. For the
scalar-type perturbation, a proposal made in \cite{hw4,Bardeen-1988}
is that we write the set of equation without fixing the temporal
gauge (hypersurface) condition and arrange the equations so that we
can implement easily various fundamental temporal gauge conditions.
The temporal gauge conditions should be chosen to achieve either
mathematical simplicity or physical interpretation.

The complete set of equations describing the scalar-type
perturbation without fixing the temporal gauge condition, i.e., in a
gauge-ready form, can be found in Eqs.\ (10)-(15) of
\cite{HN-multi-CQG}. Using Eqs.\ (\ref{BG-fluid}) and
(\ref{perturbed-fluid}) for the fluid quantities we have
\begin{widetext}
\bea
    && \kappa \equiv 3 \left( - \dot {\varphi} + H\alpha \right) -
    {\Delta \over a^2}\chi,
     \label{eq-kappa}\\
    &&  {\Delta \over a^2}\varphi + H \kappa = -{1 \over 2} \left[ \dot \phi \delta\dot \phi-\dot \sigma \delta \dot\sigma
    - ({\dot\phi}^2-{\dot\sigma}^2)\alpha+V_{,\phi}\delta \phi
    \right],
    \label{eq-kavarphi}\\
    && \kappa + {\Delta \over a^2} \chi = {3 \over 2} \left( \dot \phi \delta \phi
    -\dot \sigma \delta \sigma \right),
    \label{eq-kappachi}\\
    && \dot {\chi} + H \chi -\alpha -\varphi = 0,
     \label{eq-dotchi}\\
    && \dot {\kappa} +2H\kappa= -\left({\Delta \over a^2} + 3 \dot {H}
    \right)\alpha +\left\{ 2\left[\dot \phi {\delta\dot\phi}
    -\dot \sigma {\delta\dot \sigma} -({\dot\phi}^2-{\dot \sigma}^2)\alpha
    \right]
    -V_{,\phi}\delta \phi \right\}, \label{eq-kappadot} \\
    &&  \delta\ddot \phi +3H\delta\dot  \phi +V_{,\phi\phi}\delta\phi - {\Delta \over a^2}
    \delta \phi = \dot
    {\phi}(\kappa + \dot {\alpha} + 3H\alpha ) + 2 \ddot{\phi}\alpha,
     \label{eq-dpdd}\\
    &&  \delta\ddot \sigma +3H \delta \dot\sigma - {\Delta \over a^2}
    \delta \sigma = \dot
    {\sigma}(\kappa + \dot {\alpha} + 3H\alpha ) + 2 \ddot{\sigma}\alpha.
     \label{eq-dsdd}
\eea
\end{widetext}
The scalar fields do not support the vector-type perturbations. The
tensor-type perturbation is described by \cite{gw} \bea
    && \ddot{C}_{\alpha\beta}^{(t)}
       +3 H \dot{C}_{\alpha\beta}^{(t)}
       - {\Delta \over a^2} C_{\alpha\beta}^{(t)} = 0.
    \label{eq-tensor}
\eea The presence of scalar fields affects the tensor-type
perturbation only through the background evolution.

The gauge transformation property can be found in
\cite{hw4,Bardeen-1988}. Under the gauge transformation ${\hat
{x}}^a \equiv x^a+\tilde {\xi}^a$ with ${\xi}^t \equiv a\tilde
{\xi}^0$ $(0=\eta)$ and ${\xi}^\alpha \equiv
{\xi}^{,\alpha}+{\xi}^{{(v)}\alpha}$ we have \bea
   & & \hat {\alpha}=\alpha -{\dot \xi}^t, \quad
       \hat {\varphi}=\varphi-H{\xi}^t,
   \nonumber\\
   & &
       \hat {\kappa}=\kappa+\left(3\dot H+{\Delta \over a^2}\right){\xi}^t, \quad
       \hat {\chi}=\chi -{\xi}^t,
   \nonumber\\
   & &
       \hat{\delta {\mu}}=\delta \mu -\dot \mu {\xi}^t, \quad
       \hat{\delta{p}}=\delta p-\dot{p} {\xi}^t, \quad
       \hat {v}=v-{1 \over a}{\xi}^t,
   \nonumber \\
   & &
       \hat{\delta{\phi}}=\delta \phi - \dot \phi {\xi}^t, \quad
       \hat{\delta{\sigma}}=\delta \sigma - \dot \sigma {\xi}^t.
    \label{eq-trans}
\eea Thus, all variables in our Eqs.\
(\ref{eq-kappa})-(\ref{eq-dsdd}) are spatially gauge-invariant. It
will be convenient to introduce several gauge invariant combinations
\bea
    & & \varphi_v \equiv \varphi - a H v, \quad
        \varphi_\chi \equiv \varphi-H \chi, \quad
        \delta \phi_\chi \equiv \delta \phi - \dot \phi \chi,
        \nonumber\\
    & & \delta \sigma_\chi \equiv \delta \sigma - \dot \sigma \chi,
        \quad
        \delta \phi_{\delta \sigma} \equiv \delta \phi
        - {\dot \phi \over \dot \sigma} \delta \sigma.
    \label{GI-combinations}
\eea Our notation of the gauge-invariant variables explicitly shows
the variable and the gauge condition: for example, $\varphi_v$ is
the same as $\varphi$ in the comoving gauge which sets $v \equiv 0$,
etc. It also allows algebraic manipulations like \bea
   & & \varphi_{\delta \phi}
       \equiv \varphi - {H \over \dot \phi} \delta \phi
       \equiv - {H \over \dot \phi} \delta \phi_\varphi,
\eea etc, which is convenient in practice.

\subsection{Large-scale solutions with $C$- and $d$-modes}
                                                      \label{lsccd}

Here, we introduce the large-scale asymptotic solutions with the
$C$- and $d$-mode decomposition. The evolution of a single component
medium is described by a second-order differential equation which
has two independent solutions. In the large-scale asymptotic limit,
we decompose these two solutions into the $C$-mode and the $d$-mode.
Since we are considering two fields, we have to check whether and
when such a decomposition is possible.

\begin{figure*}
\begin{center}
\includegraphics[width=8cm]{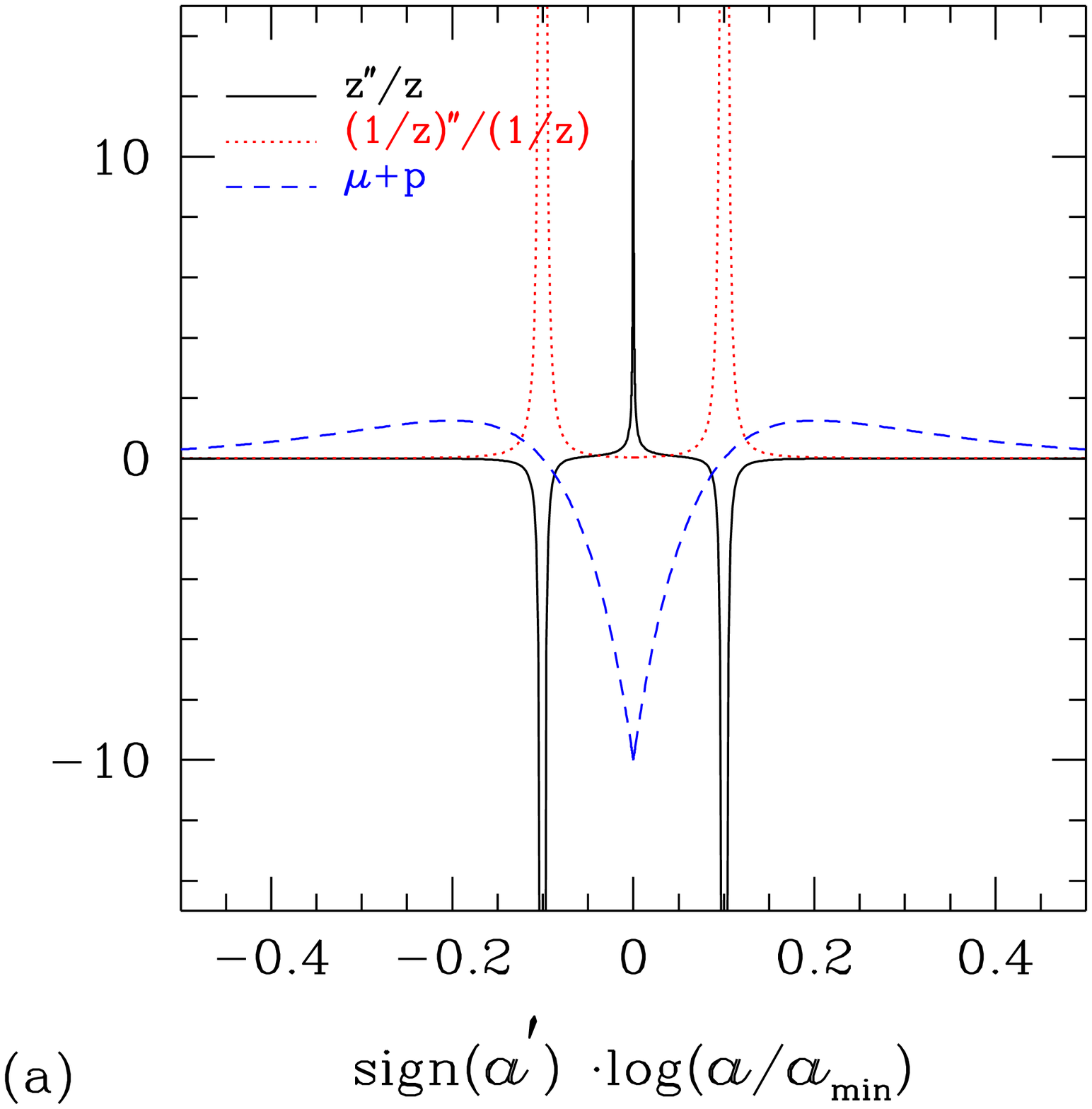}\hfill
\includegraphics[width=8cm]{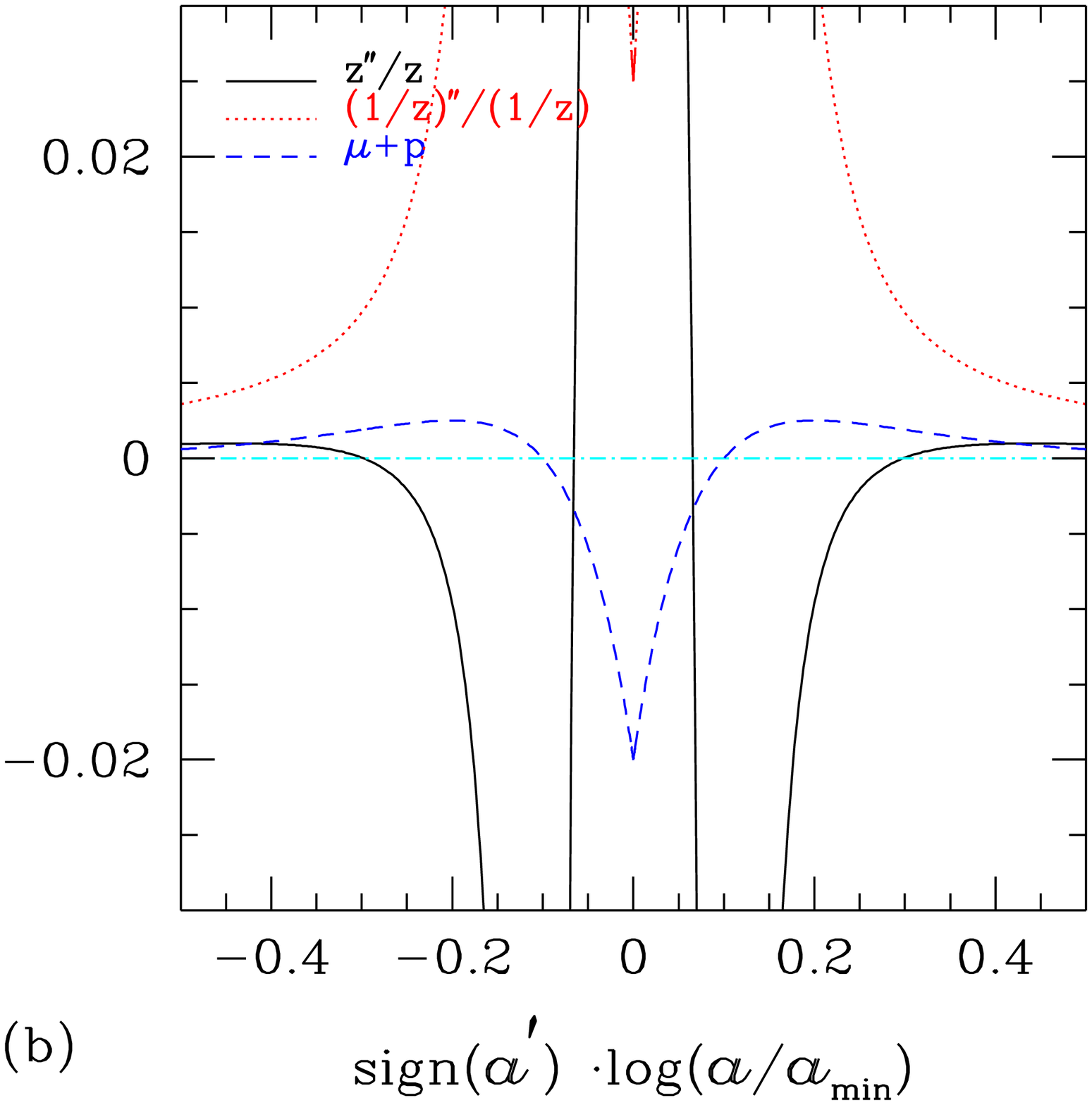}\\
\includegraphics[width=8cm]{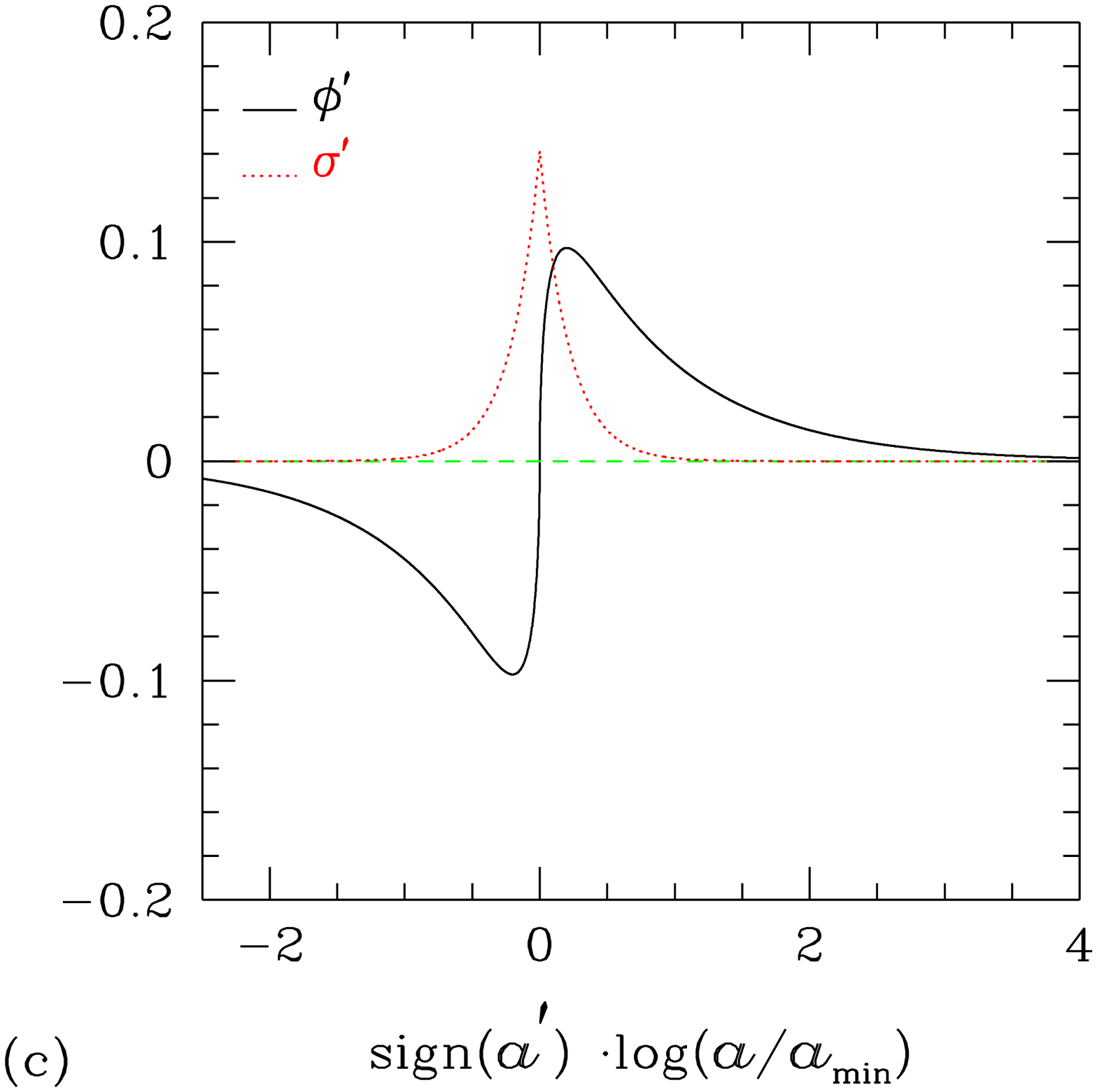}
\caption{Evolutions of $z^{\prime\prime}/z$,
         $(1/z)^{\prime\prime}/(1/z)$, $\mu+p$ near the bounce
         [Panel
         (a)], and $\phi^\prime$, $\sigma^\prime$ [Panel (c)].
         Panel (b) magnifies Panel (a) in vertical axis.
         In Panel (a), $(\mu + p)$ is plotted with $500$ times
         magnification.
         Panel (b) shows that $z^{\prime\prime}/z=0$ and
         $\mu+p=0$ occur at different epochs;
         $z^{\prime\prime}/z=0$ diverges at $\mu + p = 0$ and at the bounce.
         These combinations appear in Eqs.\
         (\ref{eq-varphiz})-(\ref{eq-isocurvaturechi}) which
         we use to distinguish the adiabatic (curvature) and isocurvature
         modes, and the large and small scales.
         As these variables (except for $\sigma^\prime$)
         diverge or vanish some of our
         large-scale conditions or adiabatic conditions
         often become singular for very brief periods,
         see Figs.\ \ref{Fig-c}-\ref{Fig-d} and \ref{Fig-anlc}-\ref{Fig-c-atzero}.
         }
         \label{fig-zvzu}
\end{center}
\end{figure*}

Using Eqs.\ (\ref{eq-kappa})-(\ref{eq-dotchi}) we have \bea
    \varphi_v=-{H^2 \over \dot H a}\left({a \over H} \varphi_\chi\right)^\cdot.
    \label{eq-new6}
\eea Using Eqs.\ (\ref{eq-kappa}), (\ref{eq-kavarphi}),
(\ref{eq-dotchi}), (\ref{eq-trans}), and (\ref{eq-new6}) we can show
\bea
    \dot\varphi_v={k^2 \over a^2}{H \over \dot H}\varphi_\chi
       -{ 2H \over (\mu + p)^2}
       V_{,\phi}\dot\phi{\dot\sigma}^2\left({\delta\phi \over \dot\phi} -{\delta\sigma \over
       \dot\sigma} \right),
    \label{eq-new8}
\eea where $k$ is the wave number in Fourier space with $\Delta
\equiv - k^2$. By combining Eqs.\ (\ref{eq-new6}) and
(\ref{eq-new8}) we have
\begin{widetext}\bea
    {H^2 \over (\mu + p)a^3} \left[{(\mu +p)a^3 \over H^2}\dot
        \varphi_{v}\right]^\cdot
        +{k^2 \over a^2}\varphi_{v}
    &=& {H \over a^3 \sqrt{\mu +
        p}}\left[\hat{v}^{\prime\prime}-\left({z^{\prime\prime} \over
        z}-k^2\right)\hat{v}\right]\nonumber\\
    &=& -{2H^2 \over (\mu+p)a^3}\left\{{a^3 \over
        (\mu+p)H}\left[V_{,\phi}\dot\phi {\dot\sigma}^2\left({\delta\phi
        \over \dot\phi}-{\delta\sigma \over
        \dot\sigma}\right)\right]\right\}^\cdot,
    \label{eq-varphiz}\\
    {\mu + p \over H}\left[{H^2 \over a(\mu +p)}
        \left({a \over H}\varphi_{\chi}\right)^\cdot\right]^\cdot
        +{k^2 \over a^2} \varphi_{\chi}
    &=& {\sqrt{\mu +p} \over a^2}\left\{ u^{\prime\prime}
        -\left[{({1 / {z}})^{\prime\prime} \over ({1 / {z}})}-k^2\right]u \right\}\nonumber\\
    &=& -{1 \over \mu+p}V_{,\phi}\dot\phi{\dot \sigma}^2\left( {\delta\phi \over \dot\phi}- {\delta\sigma
        \over \dot\sigma}\right),
    \label{eq-varphizb}
\eea \end{widetext} where \bea
    && \hat{v}=z\varphi_v, \quad u={a \over {z} H}\varphi_{\chi};\quad z^2={a^2 (\mu +p) \over
    H^2}.
    \label{u-v-def}
\eea We note that, despite the notation, $z^2$ defined in Eq.\
(\ref{u-v-def}) could have the negative value depending on the sign
of $(\mu + p)$. Indeed, in order to have a smooth nonsingular bounce
it is necessary to change sign of $(\mu + p)$ near the bounce, see
Fig.\ \ref{fig-zvzu}.

In order to make Eqs.\ (\ref{eq-varphiz}) and (\ref{eq-varphizb})
complete we need equations for $({\delta\phi /
\dot\phi}-{\delta\sigma / \dot\sigma})$. From Eqs.\
(\ref{perturbed-fluid}),
 (\ref{eq-kavarphi}), (\ref{eq-kappachi}), (\ref{eq-dpdd}), and
(\ref{eq-dsdd}) we can derive
\begin{widetext}\bea
    &&{\mu +p \over a^3{\dot\phi}^2{\dot\sigma}^2}
    \left[{a^3{\dot\phi}^2{\dot\sigma}^2 \over \mu +p}
    \left({\delta\phi \over \dot\phi}-{\delta\sigma \over \dot\sigma}\right)^\cdot\right]^\cdot
    +\left[-3\dot H+4\left({V_{,\phi}{\dot\sigma} \over \mu +p}\right)^2+{k^2 \over a^2} \right]
    \left({\delta\phi \over \dot\phi}-{\delta\sigma \over \dot\sigma}\right)
    =-{2 \over H}{V_{,\phi} \over
    \dot\phi}\dot\varphi_v,\label{eq-isocurvature}\\
    &&{\mu +p \over a^3{\dot\phi}^2{\dot\sigma}^2}
    \left[{a^3{\dot\phi}^2{\dot\sigma}^2
    \over \mu +p}\left({\delta\phi \over \dot\phi}-{\delta\sigma
    \over \dot\sigma}\right)^\cdot\right]^\cdot+
    \left[-3\dot H+{k^2 \over a^2} \right]
    \left({\delta\phi \over \dot\phi}-{\delta\sigma \over \dot\sigma}\right)\
    =-{2 \over \dot H}{V_{,\phi} \over \dot\phi}
    {k^2 \over a^2}\varphi_\chi,
    \label{eq-isocurvaturechi}
\eea\end{widetext} where we used Eq.\ (\ref{eq-new8}) to derive Eq.\
(\ref{eq-isocurvaturechi}). Equations (\ref{eq-varphiz}),
(\ref{eq-varphizb}), (\ref{eq-isocurvature}), and
(\ref{eq-isocurvaturechi}) provide a complete set of equations in
terms of the curvature (adiabatic) perturbation variable $\varphi_v$
or $\varphi_\chi$, and the isocurvature perturbation variable
$({\delta\phi /\dot\phi}-{\delta\sigma / \dot\sigma})$. We introduce
a notation for the isocurvature perturbation variable \bea
    & & S_{\phi\sigma} \equiv {\delta\phi \over \dot\phi}-{\delta\sigma
        \over \dot\sigma}={\delta\phi_{\delta\sigma} \over
        \dot\phi}.
\eea

The above equations are known in the literature in a more general
context with multiple minimally coupled scalar fields, see
\cite{HN-multi-CQG}. Here we showed that the general equations in
\cite{HN-multi-CQG} are satisfied when one of the scalar field has
negative kinetic term. The gauge-invariant variables $\varphi_v$ and
$\varphi_\chi$ are the spatial curvature perturbation variable
$\varphi$ in the comoving gauge ($v \equiv 0$) and the zero-shear
gauge ($\chi \equiv 0$), respectively. Thus, both of these two
variables represent the curvature (adiabatic) perturbation. The
second-order differential equation for $(\delta \sigma / \dot\sigma
- \delta \phi / \dot\phi)$ represents the isocurvature perturbation,
see \cite{HN-multi-CQG}. In this work, we are mainly interested in
the evolution of the curvature-type perturbation (we call it the
curvature mode) where isocurvature perturbation is suppressed; we
will show that even in such a case, the isocurvature perturbation
does not necessarily vanish, and needs to be monitored.

The large-scale conditions are satisfied if $k^2$ terms are
negligible in Eqs.\ (\ref{eq-varphiz}) and(\ref{eq-varphizb}), i.e.,
\bea
   & & k^2 \ll {z^{\prime\prime} \over z}, \quad
       k^2 \ll {({1 / z})^{\prime\prime} \over ({1 / z})},
   \label{LS-conditions}
\eea for $\varphi_v$ and $\varphi_\chi$, respectively. The adiabatic
conditions are satisfied if terms in the RHSs of Eqs.\
(\ref{eq-varphiz}) and (\ref{eq-varphizb}) are negligible compared
with terms in the LHSs involving $\hat v$ and $u$, respectively.
That is, our adiabatic conditions are satisfied if \bea
    && \left| {2H^2\left\{{a^3 \over
        (\mu+p)H}\left[V_{,\phi}\dot\phi {\dot\sigma}^2\left({\delta\phi
        \over \dot\phi}-{\delta\sigma \over
        \dot\sigma}\right)\right]\right\}^\cdot \over
        {a\left(\mu+p\right)\left({z^{\prime\prime} \over z
        }-k^2\right)\varphi_v}} \right| < 1 ,
    \nonumber \\
    && \left| {a^2V_{,\phi}\dot\phi{\dot \sigma}^2
       \left( {\delta\phi \over \dot\phi}- {\delta\sigma
       \over \dot\sigma}\right) \over (\mu+p)
       \left[{({1 / {z}})^{\prime\prime} \over ({1 /
       {z}})}-k^2\right]\varphi_\chi} \right| < 1,
    \label{adiabatic-conditions}
\eea for $\varphi_v$ and $\varphi_\chi$, respectively. In order to
monitor these conditions it is convenient to have
 \bea
   {z^{\prime\prime} \over z}&=&
    -5a^2V-{{a^\prime}^2 \over a^2}
    +2{a^6 \over {a^\prime}^2}V^2
    -{a^3 \over{a^\prime}}V_{,\phi}\phi^\prime
    \nonumber\\
    & & + {1 \over {\phi^\prime}^2-{\sigma^\prime}^2}
    \Bigg[
    -a^2V_{,\phi\phi}{\phi^\prime}^2
    -6{a{a^\prime}}V_{,\phi}\phi^\prime
    \nonumber\\
    & &
    +a^4 \left( V_{,\phi} \right)^2
    +2{a^5 \over {a^\prime}}VV_{,\phi}\phi^\prime
    -a^4(V_{,\phi})^2{{\phi^\prime}^2 \over {\phi^\prime}^2-{\sigma^\prime}^2}
    \Bigg],
    \nonumber \\
    {(1/z)^{\prime\prime} \over (1/z)}&=&
    a^2V+3{{a^\prime}^2 \over a^2}
    +{a^3 \over {a^\prime}}V_{,\phi}\phi^\prime
    \nonumber\\
    &&+{1 \over {\phi^\prime}^2-{\sigma^\prime}^2}
    \Bigg[a^2V_{,\phi\phi}{\phi^\prime}^2
    +2{a{a^\prime}}V_{,\phi}\phi^\prime
    -a^4(V_{,\phi})^2
    \nonumber\\
    &&+2{a^5 \over {a^\prime}}VV_{,\phi}\phi^\prime
    +3a^4(V_{,\phi})^2{{\phi^\prime}^2 \over
    {\phi^\prime}^2-{\sigma^\prime}^2}\Bigg].
    \eea
 Later we will show that
our large-scale conditions and adiabatic conditions are often
sharply broken near the bounce. Such sharp divergences occur because
some background variables like $\mu + p$, $z$, $H$, and $\dot \phi$
often vanish or diverge near the bounce. Later we will argue that
such a very brief timescale breaking of our conditions do not have
physical impact on interpreting the evolution of curvature
(adiabatic) perturbations in the large-scale limit. In Fig.\
\ref{fig-zvzu} we present evolutions of ${z^{\prime\prime} / z}$,
${(1/z)^{\prime\prime} / (1/z)}$, $\sigma^\prime$, $\phi^\prime$,
and $(\mu + p)$, for later use.

If both the large scale condition and adiabatic condition are
satisfied, even in our two component situation, we have the
following general solutions \bea
    && \varphi_v( k, \eta)=C( k)-2d(k)k^2 \int^\eta {d\eta \over z^2},
    \label{eq-varphiv}\\
    && \varphi_{\chi}(k, \eta)={1 \over 2}C(k){H \over a}\int^\eta {z}^2d\eta
    + d(k){H \over a}.
    \label{eq-varphichi}
\eea Notice that the $d$-mode of $\varphi_v$ is $(k\eta)^2$-higher
order in the large-scale expansion compared with the $d$-mode of
$\varphi_\chi$. To the next order in the large-scale expansion we
have \cite{HN-multi-CQG}
\begin{widetext}\bea
    && \varphi_v( k, \eta)
    =C(k) \left\{1+k^2\left[\int^\eta { z}^2 \left(\int^\eta {d\eta \over
    z^2} \right) d\eta -\int^\eta { z}^2 d\eta \int^\eta {d\eta \over
    z^2} \right] \right\}-2d(k)k^2\int^\eta {d\eta \over z^2},
    \label{eq-varphive}\\
    && \varphi_\chi(k,\eta)
    ={1 \over 2} C(k){H\over a}\int^\eta {{z}}^2 d\eta+{H \over a}d(k) \left\{
    1+ k^2\left[\int^\eta {1 \over z^2} \left(\int^\eta {{z}}^2 d\eta \right)d\eta
    - \int^\eta {{z}}^2 d\eta \int^\eta {d\eta \over z^2}\right] \right\}.
    \label{eq-varphichie}
\eea \end{widetext} We note that the integrations in Eqs.\
(\ref{eq-varphiv})-(\ref{eq-varphichie}) do not consider
contributions from the lower bounds of integrations; this, in fact,
becomes a subtle point in the case of the bounce.

In an expanding universe, $C$-mode is relatively growing and
$d$-mode is decaying. In a collapsing phase, however, the $d$-mode
grows rapidly and the $C$-mode relatively decays. Actually, the
$C$-mode usually remains constant in either phase. Since we are
considering two fields, in order to use the concept of $C$- and
$d$-mode decomposition, we have to check whether both (i) the
adiabatic conditions and (ii) the large-scale conditions are
satisfied throughout the evolution.

\subsection{Equations in three different gauges}
                                           \label{three-gauges}

We will investigate the behaviors of $C$- and $d$-modes by following
their evolutions from the collapsing phase to the expanding phase
through our simple non-singular bounce model. In order to obtain
numerical results, it is practically important to take a suitable
temporal gauge condition. If we have the behavior of any one
perturbation variable in any one gauge condition the rest of the
variables even in the other gauge conditions can be algebraically
derived using our equations in the gauge-ready form in Eqs.\
(\ref{eq-kappa})-(\ref{eq-dsdd}) and the gauge-invariant
combinations; the latters are made from the gauge-transformation
properties in Eq.\ (\ref{eq-trans}), as in Eq.\
(\ref{GI-combinations}). We have chosen two gauge-invariant
combinations $\varphi_v$ and $\varphi_\chi$ because of their
conserved properties in the large-scale limit and popularity in the
literature. We could not directly solve $\varphi_v$, however,
because its equations show singular behavior near the bounce. We
prefer a gauge condition that has no singular behavior through the
bounce. We find several suitable gauge conditions and numerically
solve them independently. The solution of any gauge-invariant
variable should be the same independently of the gauge conditions.
This naturally provides a way to check the numerical accuracy. For
instance, $\varphi_v$ and $\varphi_\chi$ can be obtained from the
solutions known in the uniform-$\sigma$ gauge ($\delta \sigma \equiv
0$) and in the zero-shear gauge ($\chi \equiv 0$) as \bea
   && \varphi_v =
      \varphi_{\delta\sigma}
      - {a^\prime \over a} {\phi^\prime\delta\phi_{\delta\sigma}
      \over
      {\phi^\prime}^2
      - {\sigma^\prime}^2}
      = \varphi_\chi-{a^\prime \over a}{\phi^\prime
      \delta\phi_{\chi}-\sigma^\prime \delta\sigma_{\chi} \over
      {\phi^\prime}^2-{\sigma^\prime}^2}, \nonumber\\
   && \varphi_\chi=\varphi_{\delta\sigma}-{a^\prime \over a^2}\chi_{\delta\sigma}.
   \label{varphi_v-reconstruct}
\eea In the following we consider three different temporal gauge
conditions. In each of these temporal gauge conditions, the temporal
gauge degree of freedom is fixed completely, thus each of the
variables in these gauge conditions has unique gauge-invariant
counterparts, see \cite{hw4}. Thus, we can regard all the variables
as gauge-invariant ones.

\subsubsection{Uniform-curvature gauge}
                                            \label{uniform_curvature-gauge}

The uniform-curvature gauge takes $\varphi \equiv 0$ as the temporal
gauge condition. This gauge condition is particularly suitable to
handle quantum fluctuations generated during collapsing phase which
will provide the initial seed fluctuations for the large-scale
structures. This provides our initial conditions for numerical
integration. From Eqs.\
(\ref{eq-kappa})-(\ref{eq-kappachi}),(\ref{eq-dpdd}), and
(\ref{eq-dsdd}) we can derive
\begin{widetext}
\bea
    & & \delta\phi_{\varphi}^{\prime\prime}
        + 2{a^\prime \over a}\delta\phi_{\varphi}^{\prime}
        + k^2\delta\phi_{\varphi}
    +a^2\left(V_{,\phi\phi}+
    {a^2 \over {a^\prime}^2}{\phi^\prime}^2V+2{a \over
    a^\prime}V_{,\phi}\phi^\prime\right)
    \delta\phi_{\varphi}-a^2{ a^2\over
    {a^\prime}^2}\sigma^\prime\left(V\phi^\prime
    +{a^\prime \over a}V_{,\phi}\right)\delta \sigma_{\varphi}\nonumber \\
    & & \quad
    ={1 \over a}\left[u^{\prime\prime}_{\delta\phi}
    +\left(k^2-{a^{\prime\prime} \over a}\right)u_{\delta\phi}\right]
    +\left[aV\left({a^2 \over {a^\prime}^2}{\phi^\prime}^2
    +2{a \over a^\prime}{V_{,\phi} \over V}\phi^\prime
    +{V_{,\phi\phi} \over V}\right)u_{\delta\phi}
    -a{ a\over {a^\prime}}\sigma^\prime V\left({a \over a^\prime}\phi^\prime
    +{V_{,\phi} \over V}\right) u_{\delta\sigma}\right]=0,
     \label{eq-ucdp}\\
    & & \delta \sigma_{\varphi}^{\prime\prime}
        + 2{a^\prime \over a}\delta \sigma_{\varphi}^{\prime}
        + k^2\delta \sigma_{\varphi}
        - a^2{a^2 \over {a^\prime}^2} {\sigma^{\prime}}^2 V\delta \sigma_{\varphi}
        +a^2{ a^2\over {a^\prime}^2}\sigma^\prime\left(V\phi^\prime
        +{a^\prime \over a}V_{,\phi}\right)\delta\phi_{\varphi}\nonumber \\
    && \quad
    ={1 \over a}\left[u^{\prime\prime}_{\delta\sigma}
    +\left(k^2-{a^{\prime\prime} \over a}\right)u_{\delta\sigma}\right]
    +\left[- aV{a^2 \over {a^\prime}^2} {\sigma^{\prime}}^2u_{\delta\sigma}
    +a{ a\over {a^\prime}}\sigma^\prime V\left({a \over a^\prime}\phi^\prime
    +{V_{,\phi}\over V}\right)u_{\delta\phi}\right]=0,
    \label{eq-ucds}
\eea \end{widetext} where \bea
   & & u_{\delta\sigma}\equiv a\delta\sigma_{\varphi}, \quad
       u_{\delta \phi}\equiv a \delta\phi_{\varphi}.
\eea Away from the bounce, the effect of $\sigma$ is negligible, and
we have the power law regime. From Eq.\ (\ref{eq-repo}) we have
$\phi^\prime \propto {a^\prime / a}$. Thus, we have \bea
    && u^{\prime \prime}_{\delta\sigma}+\left(k^2-{a^{\prime\prime} \over
       a}\right)u_{\delta \sigma}=0,
    \label{eq-u1.}\\
    && u^{\prime \prime}_{\delta\phi}+\left(k^2-{a^{\prime\prime} \over
       a}\right)u_{\delta \phi}=0.
    \label{eq-u2.}
\eea For a general $\lambda$ we have the following exact solutions
\bea
    & & u_{(i)}=\sqrt{k|\eta|}[c_1(k)H_{\nu}^{(1)}(k|\eta|)
        +c_2(k)H_{\nu}^{(2)}(k|\eta|)], \nonumber \\
     && \qquad
        \nu \equiv {3(w-1) \over 2(3w+1)}
        = {3q-1 \over 2(q-1)},
    \label{eq-Hankel.}
\eea where $(i) = \delta \phi$ or $\delta \sigma$; $\lambda$ is
related to $q$ and $w$ as in Eq.\ (\ref{eq-repo}). If we take
$\lambda=\sqrt{3}$, thus $w=0$ and $q={2/3}$, we have \bea
    u_{(i)}&&=\sqrt{2 \over \pi}\Bigg\{-\left[\sin(\vert k\eta
      \vert)+{\cos(\vert k\eta \vert) \over \vert k\eta
      \vert}\right]\left[c_1(k)+c_2(k)\right]\nonumber\\
      &&+i\left[\cos(\vert k \eta \vert)-{\sin(\vert k\eta \vert)
      \over \vert k \eta
      \vert}\right]\left[c_1(k)-c_2(k)\right]\Bigg\}.
    \label{eq-usolutions}
\eea These will provide the initial seed fluctuations.
Identification of these classical solutions with the initial
fluctuations generated from the quantum fluctuations will be
considered in Sec.\ \ref{sec:Implication}. The equations in this
gauge, however, show the singular behavior as the background passes
through the bounce. Thus, in order to follow the evolution
numerically we need to use other gauge conditions which are
well-defined throughout the bounce.

\subsubsection{Uniform-$\sigma$ gauge}
                                           \label{uniform-sigma-gauge}

The uniform-$\sigma$ gauge takes $\delta \sigma \equiv 0$ as the
temporal gauge condition. Authors of \cite{AW} used this gauge
condition for their numerical calculation. From Eqs.\
(\ref{eq-kappa}), (\ref{eq-kappachi}), (\ref{eq-dotchi}),
(\ref{eq-dpdd}), and (\ref{eq-dsdd}) we have
\begin{widetext} \bea
     &&\delta \phi_{\delta\sigma}^{\prime\prime}
    +2{a^\prime \over a}\delta \phi_{\delta\sigma}^{\prime}
    +a^2V_{,\phi\phi}\delta\phi_{\delta\sigma}
    +k^2\delta\phi_{\delta\sigma}
    +2a^2V_{,\phi}\alpha_{\delta\sigma}=0,
    \label{eq-dsdp}\\
    && \varphi_{\delta\sigma}^{\prime\prime}
    +2{a^\prime \over a}\varphi_{\delta\sigma}^{\prime}
    +{1 \over 2}\left({\phi}^{\prime}\delta\phi_{\delta\sigma}^{\prime}
    -a^2V_{,\phi}\delta\phi_{\delta\sigma} \right)
    -a^2V\alpha_{\delta\sigma}-{a^\prime \over a}\alpha_{\delta\sigma}^{\prime}=0,
    \label{eq-dsvp}\\
     &&\alpha_{\delta\sigma}^{\prime}
    -3\varphi_{\delta\sigma}^{\prime}+{k^2 \over a}\chi_{\delta\sigma}=0,
    \label{eq-dsal}\\
    &&\chi_{\delta\sigma}^{\prime}+{a^\prime \over a}\chi_{\delta\sigma}
    -a(\alpha_{\delta\sigma}+\varphi_{\delta\sigma})=0.
    \label{eq-dschi}
\eea Thus, in this gauge we solve sixth-order differential
equations. Although, this set of sixth-order differential equations
can be easily reduced to a fourth-order differential equation, in
order to follow numerically the evolution of perturbations through
the bounce without singular behavior these sixth-order differential
equations are preferred. In order to provide initial conditions for
$\alpha_{\delta\sigma}$ and $\chi_{\delta\sigma}$ we use the
following constraint equations \bea
    & & \alpha_{\delta\sigma}
        ={a \over a^\prime}\left(\varphi^\prime_{\delta\sigma}
        +{1 \over 2} \phi^\prime\delta\phi_{\delta\sigma}\right),
    \label{eq-dsinal}\\
    & & \chi_{\delta\sigma}
        =a\Bigg\{{a \over a^\prime}\varphi_{\delta\sigma}-{ 3 \over 2}
        {1 \over k^2}\phi^\prime\delta\phi_{\delta\sigma}
        -{1 \over 2k^2}{a \over a^\prime}
    \left[\phi^\prime\delta \phi_{\delta\sigma}^\prime
    -({\phi^\prime}^2-{\sigma^\prime}^2)\alpha_{\delta\sigma}
    +a^2V_{,\phi}\delta\phi_{\delta\sigma}\right]\Bigg\},
    \label{eq-dsinchi}
\eea which follow from Eqs.\ (\ref{eq-kappa}) and
(\ref{eq-kappachi}) and Eqs.\ (\ref{eq-kavarphi}) and
(\ref{eq-kappachi}), respectively. Since the initial conditions are
provided in the uniform-curvature gauge, in order to construct
initial conditions for the variables in the uniform-$\sigma$ gauge
it is convenient to have \bea
    &&\delta\phi_{\delta\sigma}=\delta \phi_{\varphi}
    -{\phi^\prime \over \sigma^\prime}\delta\sigma_{\varphi}, \quad
    {\delta\phi}^\prime_{\delta\sigma}
    = {\delta \phi}^\prime_{\varphi}
    +a^2V_{,\phi}{\delta \sigma_{\varphi} \over \sigma ^\prime}
    -{\phi^\prime \over \sigma^\prime}\delta\sigma^\prime_{\varphi},
    \nonumber\\
    &&\varphi_{\delta\sigma}=-{a^\prime \over a}
    {\delta \sigma_{\varphi} \over \sigma^\prime},\quad
    \varphi^\prime_{\delta \sigma}
    =- a^2V{\delta \sigma_{\varphi} \over \sigma^\prime}
    -{a^\prime \over a}{{\delta \sigma}^\prime_{\varphi} \over \sigma ^\prime}.
    \label{eq-connectin}
\eea

From Eq.\ (\ref{eq-varphiv}) we can read the initial conditions for
the $C$- and $d$-modes of $\varphi_v$. In order to have the $C$- and
$d$-mode decomposition, it is necessary to satisfy the large-scale
and the adiabatic conditions. From Eq. (\ref{eq-varphiz}) the
adiabatic condition implies ${\delta\sigma / \dot\sigma}={\delta\phi
/ \dot\phi}$. From this, we construct $C$- and $d$-modes of
$\varphi_v$ and the related variables in the uniform-$\sigma$ gauge.
Thus, the initial conditions for the $C$-mode and $d$-mode are,
respectively \bea
    && C{\rm -mode}: \delta\phi_{\delta\sigma}=0, \quad
       {\delta\phi^\prime_{\delta\sigma}}=0, \quad
       \varphi_{\delta\sigma}=C, \quad
       {\varphi^\prime_{\delta\sigma}}=0,
    \nonumber \\
    && d{\rm -mode}: \delta\phi_{\delta\sigma}=0, \quad
       {\delta\phi^\prime_{\delta\sigma}}=0, \quad
       \varphi_{\delta\sigma}= d, \quad
       {\varphi^\prime_{\delta\sigma}}=-{3 \over \eta} d.
    \label{eq-connectioncd}
\eea \end{widetext} It is important to notice, however, that these
are initial conditions for the $C$- and $d$-modes in an {\it ideal}
situation when the large-scale and the adiabatic conditions are
satisfied {\it exactly}. Since both of these conditions are not
satisfied exactly in the two-component medium, we have to {\it find}
precise $C$- and $d$-mode initial conditions which have correct
behaviors in the initial phase and preferably throughout the bounce.

In Sec.\ \ref{sec:Numerical-results} we will present evolution of
perturbations under such precise initial conditions (we call these
`our initial $C$- and $d$-mode' initial conditions) which preserve
the $C$- and $d$-mode natures throughout the bounce; we will show
that, under this evolution, the isocurvature perturbation is not
excited near the bounce. We will also present the perturbation
behaviors under the naive initial conditions based on the analytic
solutions in an ideal situation in Eq.\ (\ref{eq-connectioncd}) (we
call these `analytic $C$- and $d$-mode' initial conditions) which
fail to produce the desired results. In short, the initial
conditions in Eq.\ (\ref{eq-connectioncd}), in fact, correspond to
mixtures of the $C$- and $d$-modes in the initial phase, and lead to
loss of these mode identities as the evolution proceeds through the
bounce; under this evolution, the isocurvature perturbation is also
excited near the bounce, see Sec.\ \ref{sec:Numerical-results}.

\subsubsection{Zero-shear gauge}
                                                  \label{zero-shear-gauge}

The zero-shear gauge takes $\chi \equiv 0$ as the temporal gauge
condition. From Eqs.\ (\ref{eq-kappa}), (\ref{eq-kappachi}),
(\ref{eq-dotchi}), (\ref{eq-dpdd}), and (\ref{eq-dsdd}) we have
\begin{widetext} \bea
     &&\varphi^\prime_{\chi}
     =-{a^\prime \over a}\varphi_{\chi}
     -{1 \over 2}(\phi^\prime\delta\phi_{\chi}
     -\sigma^\prime\delta\sigma_{\chi}),
    \label{eq-zsvp}\\
     &&\delta\phi_{\chi}^{\prime\prime}
     +2{a^\prime \over a}\delta\phi_{\chi}^{\prime}
    +a^2V_{,\phi\phi}\delta\phi_{\chi}
    +k^2\delta\phi_{\chi}
    =4\phi^{\prime}\left[{a^\prime \over a}\varphi_{\chi}
       +{1 \over 2}(\phi^\prime\delta\phi_{\chi}
       -\sigma^\prime\delta\sigma_{\chi})\right]
    +2a^2V_{,\phi}\varphi_{\chi},
     \label{eq-zsdpdd}\\
    && \delta\sigma_{\chi}^{\prime\prime}
    +2{a^\prime \over a}\delta\sigma_{\chi}^{\prime}
    +k^2\delta\sigma_{\chi}
    =4\sigma^\prime\left[{a^\prime \over a}\varphi_{\chi}
    +{1 \over 2}(\phi^\prime\delta\phi_{\chi}
    -\sigma^\prime\delta\sigma_{\chi})\right].
     \label{eq-zsdsdd}
\eea In order to construct the initial conditions in this gauge we
need the following relations \bea
    && \hspace{-.3cm}\delta\phi_{\chi}
    =\delta \phi_{\delta\sigma}
    -{{\phi^\prime} \over a}\chi_{\delta\sigma},\quad \delta\sigma_{\chi}
    =-{{\sigma^\prime} \over
    a}\chi_{\delta\sigma}, \quad
    \delta\phi_{\chi}^\prime = \delta \phi_{\delta\sigma}^\prime
    +3{a^\prime \over a^2}\phi^{\prime}\chi_{\delta\sigma}
    +aV_{,\phi}\chi_{\delta\sigma}
    -{\phi^\prime \over a}\chi_{\delta\sigma}^\prime,
    \nonumber\\
    &&\hspace{-.3cm}\delta\sigma_{\chi}^\prime
    =3{a^\prime \over a^2}\sigma^{\prime}\chi_{\delta\sigma}
    -{\sigma^\prime \over a}\chi_{\delta\sigma}^\prime, \quad
    \varphi_\chi=\varphi_{\delta\sigma}-{a^\prime \over
    a^2}\chi_{\delta\sigma},
    \label{eq-connectindszs}\eea
where \bea
    \chi_{\delta\sigma}
       &=&
       {a^2 \over a^\prime}
       \Bigg\{\varphi_{\delta\sigma}
       +{1\over 2k^2}\Bigg[
       {a \over a^\prime}\left({\phi^\prime}^2-{\sigma^\prime}^2\right)
       {\varphi}^\prime_{\delta\sigma}
       -\phi^\prime\delta\phi^\prime_{\delta\sigma}
       -a^2\left(V_{,\phi}
       +{a\over a^\prime}\phi^\prime V\right)\delta\phi_{\delta\sigma}\Bigg]\Bigg\},
       \nonumber\\
       \chi^\prime_{\delta\sigma}
       &=&
       - {a^\prime \over a} \chi_{\delta\sigma}
       +{a^2 \over a^\prime}\varphi^\prime_{\delta\sigma}
       +a\varphi_{\delta\sigma}
       +{a^2 \over 2a^\prime}\phi^\prime\delta\phi_{\delta\sigma}.
    \label{zsg-chi}
\eea \end{widetext} Relations in Eq.\ (\ref{zsg-chi}) follow from
Eqs.\ (\ref{eq-dschi})-(\ref{eq-dsinchi}). Equations in this gauge
also behave well throughout the bounce. Numerical results in this
gauge condition should coincide with the ones from the
uniform-$\sigma$ gauge for any gauge-invariant combination which
provides a self-consistent numerical check.

\subsection{Tensor-type perturbations}
                                                  \label{sec:GW}

The tensor-type perturbations correspond to the gravitational waves.
It is described by Eq.\ (\ref{eq-tensor}). The scalar fields do not
directly contribute to the tensor-type perturbations, i.e.,
$\Pi_{\alpha\beta} \equiv 0$. Equation (\ref{eq-tensor}) becomes
\bea
    && {1 \over a^3} \left( a^3 \dot{C}_{\alpha\beta}^{(t)}
       \right)^\cdot
        + {k^2 \over a^2}C_{\alpha\beta}^{(t)}
    \nonumber \\
    & & \qquad
        ={1 \over a^3}\left[\hat v_t^{\prime\prime}
        -\left({a^{\prime\prime} \over a}
        -k^2\right)\hat v_t\right]=0,
    \label{eq-tensoru}
\eea where $\hat v_t \equiv aC^{(t)}_{\alpha\beta}$. Notice the
similarity of this equation compared with Eq.\ (\ref{eq-varphiz})
for $\varphi_v$. In the large-scale limit we have \bea
   & & C_{\alpha\beta}^{(t)}(k,\eta)
       = \bar{C}_{\alpha\beta}(k)+\bar{d}_{\alpha\beta}(k)\int^\eta {1 \over
       a^2}d\eta,
   \label{eq-analten}
\eea  where $\bar{C}_{\alpha\beta}$ and $\bar{d}_{\alpha\beta}$ are
integration constants. To the next order in the large-scale
expansion we have\bea
    C_{\alpha\beta}^{(t)}(k,\eta)
        &=& \bar{C}_{\alpha\beta}(k)
        \Bigg\{1+k^2\Bigg[\int^\eta
        \left({a^2}\int^\eta{1\over a^2}d\eta\right)d\eta
        \nonumber\\
        &&-\int^\eta{1 \over a^2}d\eta\int^\eta{a^2} d\eta
        \Bigg]\Bigg\}
        + \bar{d}_{\alpha\beta}(k)\Bigg\{\int^\eta{1 \over a^2}d\eta
        \nonumber\\
        &&+k^2\Bigg[\int^\eta {a^2} \left(\int^\eta{1 \over
        a^2}d\eta \right) \left(
        \int^\eta{1 \over a^2}d\eta\right)d\eta
        \nonumber\\
        &&-\left(\int^\eta{1 \over
        a^2}d\eta\right)\int^\eta  \left({a^2}\int^\eta{1 \over
        a^2}d\eta\right)d\eta\Bigg]\Bigg\}.
    \label{eq-tensore}
\eea The above solutions can be compared with Eqs.\
(\ref{eq-varphiv}),(\ref{eq-varphive}) for $\varphi_v$.

For \bea
   & & {a^{\prime\prime} \over a} = {n_g \over \eta^2}, \quad
       n_g \equiv { q(2q-1) \over (1-q)^2},
\eea Eq.\ (\ref{eq-tensoru}) has an exact solution \bea
        C_{\alpha\beta}^{(t)}(k,\eta)
        &=& {\sqrt{|\eta|} \over a}
        \Big[\tilde c_{1\alpha\beta}({
        k})H_{\nu_g}^{(1)}(k|\eta|)\nonumber\\
       &&+\tilde c_{2\alpha\beta}({
       k})H_{\nu_g}^{(2)}(k|\eta|)\Big], \nonumber\\
      &&\nu_g \equiv {1 \over 2}{3q-1 \over 1-q}.
    \label{eq-tensorhankel}
\eea For $q=2/3$, we have \bea
    C_{\alpha\beta}^{(t)}(k,\eta)
        &=& {1 \over a}\sqrt{2 \over \pi k}
        \Bigg\{\Bigg[-\cos(\vert k\eta \vert)
        +{\sin(\vert k\eta \vert) \over \vert k\eta \vert}\Bigg]
        \Big[\tilde c_{1\alpha\beta}({k})\nonumber\\
        &&+\tilde c_{2\alpha\beta}({k})\Big]
        -i\Big[\sin(\vert k \eta \vert)
        +{\cos(\vert k\eta \vert) \over \vert k \eta
        \vert}\Big]\nonumber\\
        &&\times\Big[\tilde c_{1\alpha\beta}({k})
        -\tilde c_{2\alpha\beta}({k})\Big]\Bigg\}.
    \label{tensoranals}
\eea

\section{Numerical results}
                                            \label{sec:Numerical-results}

\subsection{Scalar-type perturbations}
                                            \label{scalar-numerical}

\subsubsection{Our $C$- and $d$-modes}
                                            \label{scalar-numerical-our}

\begin{figure*}
\centering%
\includegraphics[width=8cm]{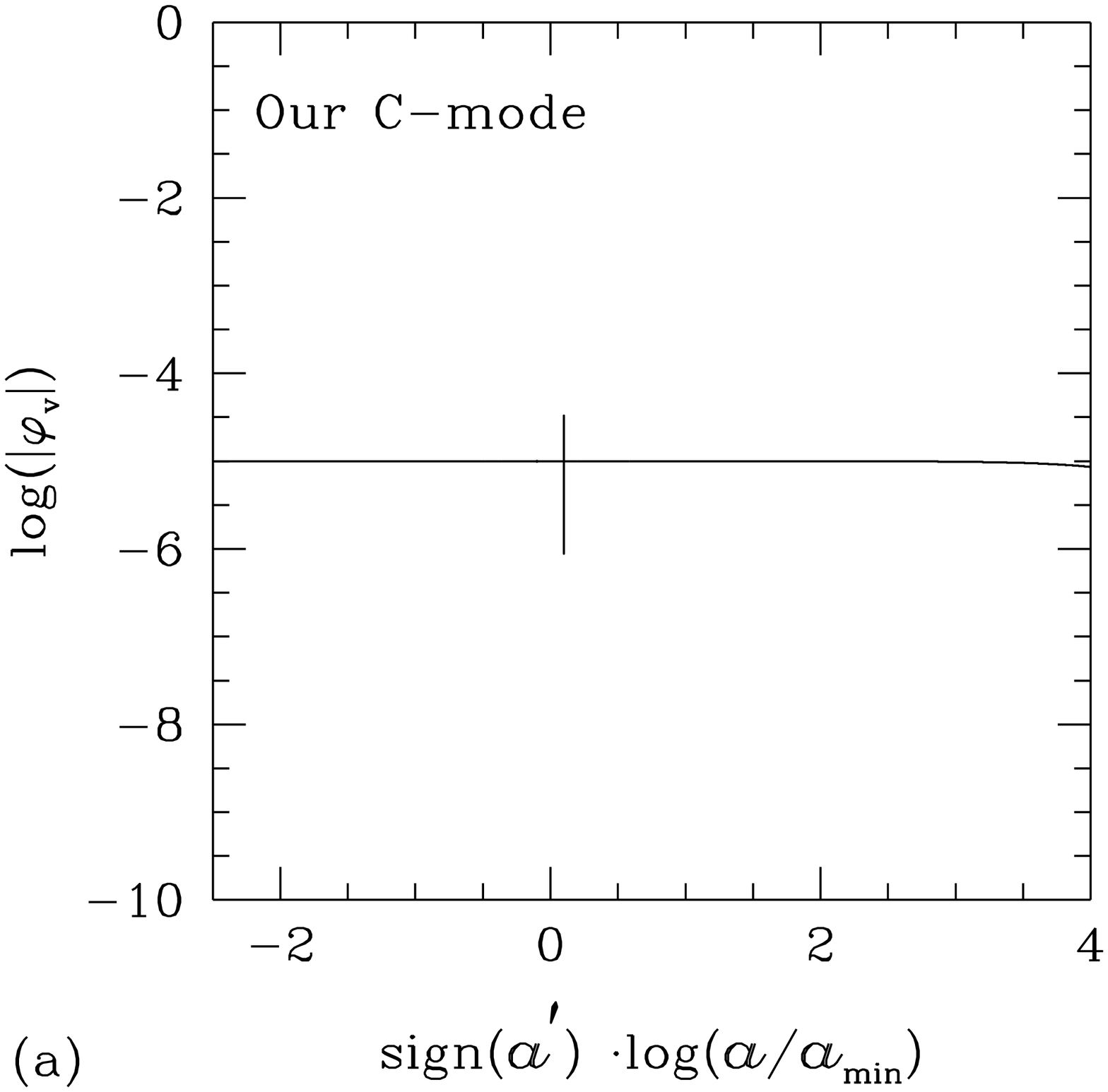}\hfill
\includegraphics[width=8cm]{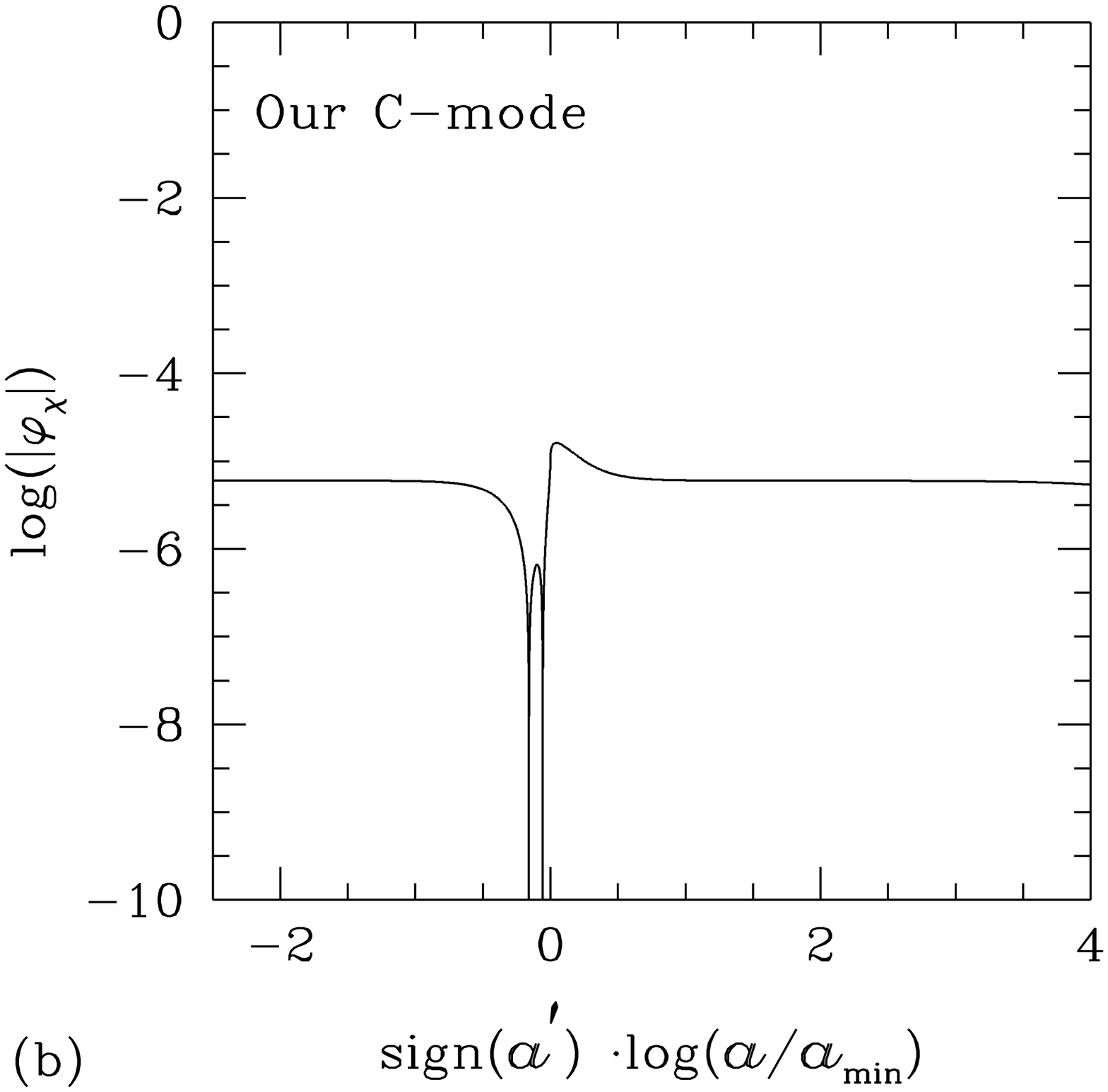}\\
\includegraphics[width=8cm]{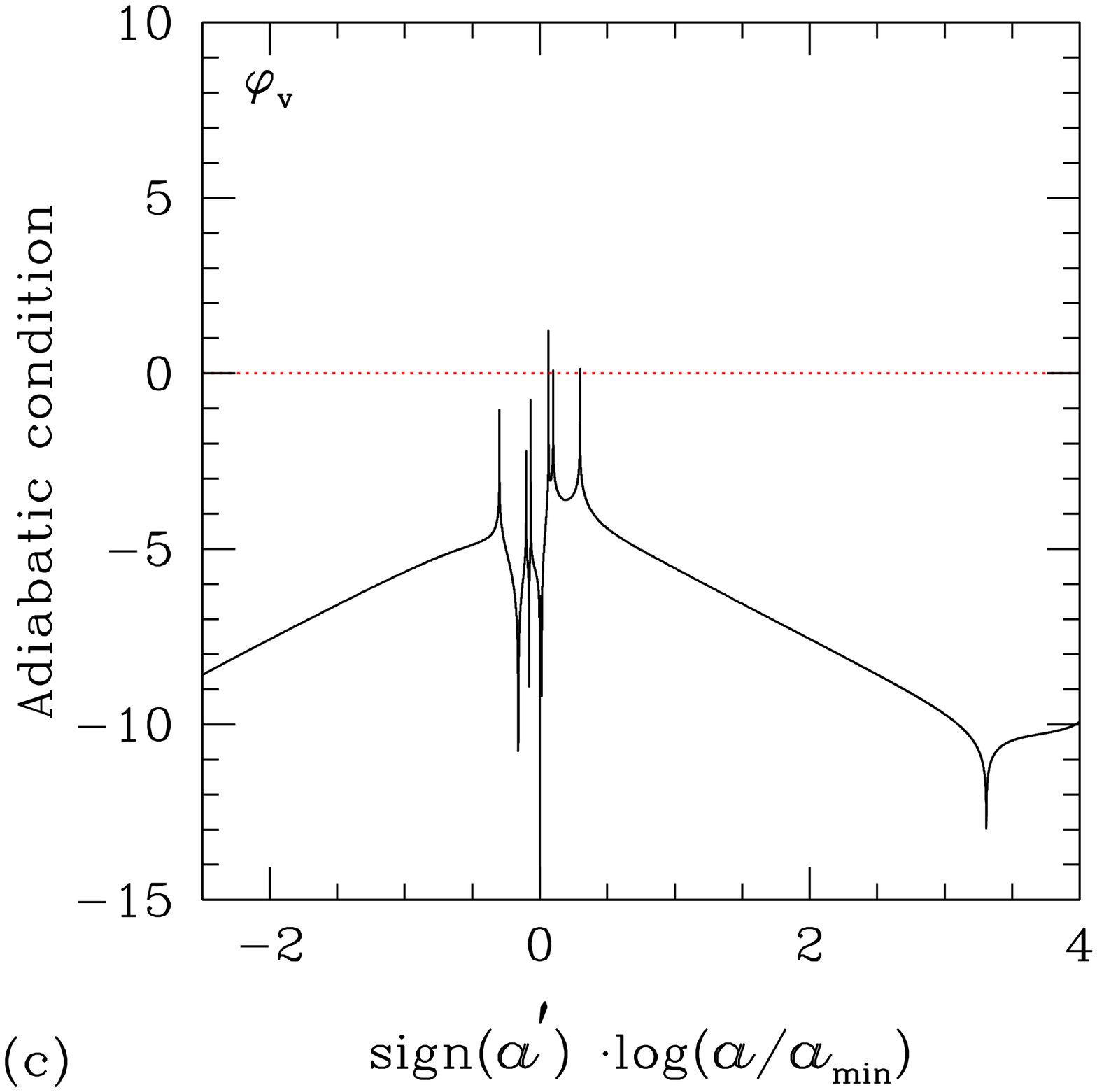}\hfill
\includegraphics[width=8cm]{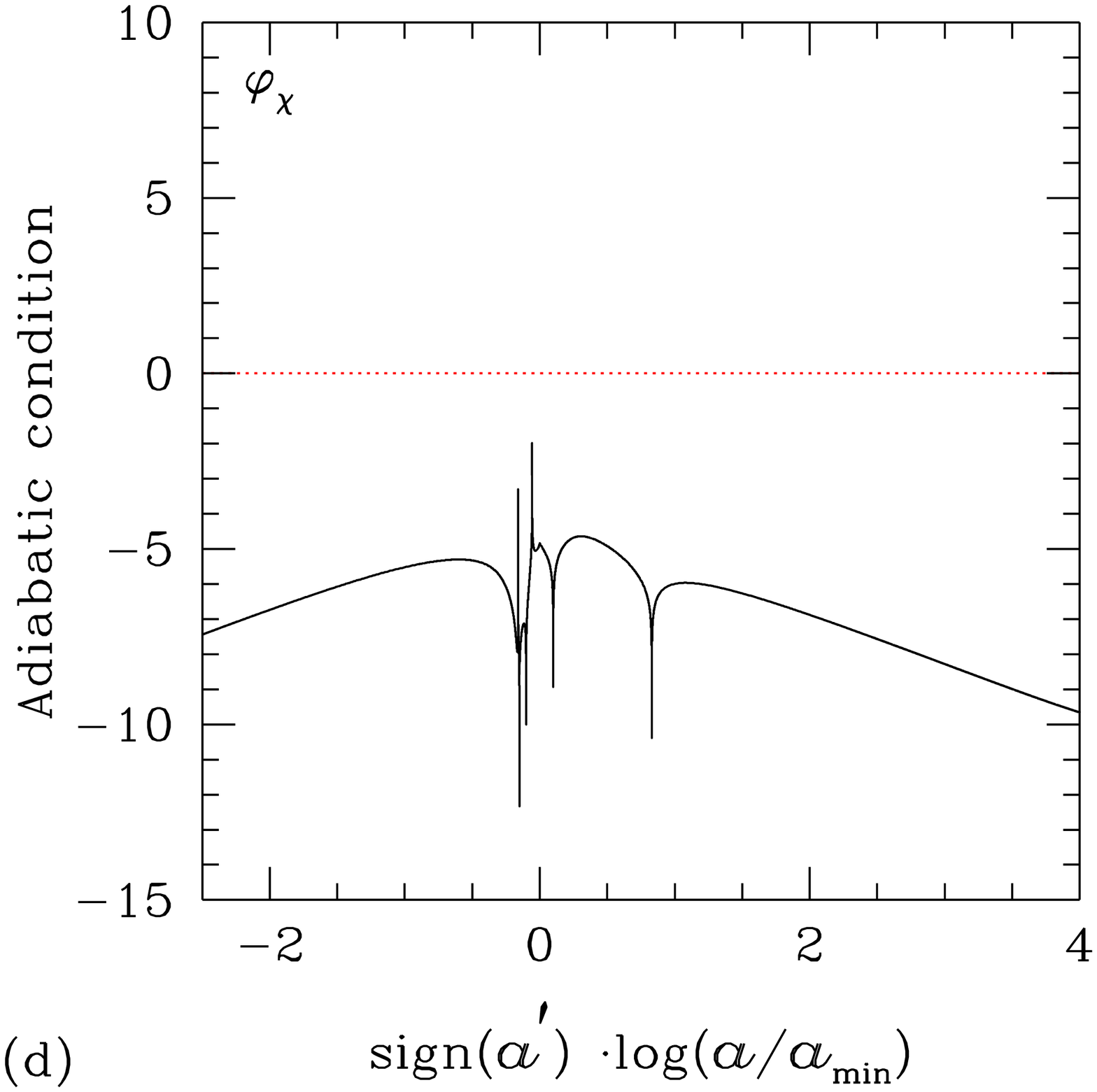}\\
\caption{Evolutions of our $C$-modes for $\varphi_v$ and
         $\varphi_\chi$, and the adiabatic conditions;
         the large-scale conditions are presented in Fig.\ \ref{Fig-LS}.
         Panels (a)
         and (b) show the behaviors of our $C$-modes for $\varphi_v$ and
         $\varphi_\chi$.
         We use the initial conditions in
         Table.\ \ref{Table-initiall2}.
         Panels (c) and (d) show the adiabatic conditions for $\varphi_v$ and
         $\varphi_\chi$, respectively;
         for $\varphi_v$ and $\varphi_\chi$ we plot
         the LHSs of Eq.\ (\ref{adiabatic-conditions}).
         Panel (c) shows that the adiabatic
         condition for $\varphi_v$ is broken by sharp peaks (divergences)
         six times near the
         bounce; we can show that the four sharp peaks at the outer edge
         and the inner edge surrounding the bounce are caused by
         $z^{\prime\prime}/z=0$, and the other two sharp peaks
         in the middle are caused by $\mu +p=0$.
         The adiabatic condition for $\varphi_\chi$ in Panel (d)
         also shows two sharp peaks
         just before the bounce
         because $\varphi_\chi$ crosses zero at these two points;
         see Eq.\ (\ref{adiabatic-conditions}) and Fig.\
         \ref{Fig-usecalc}(f) which shows
         evolution of $\varphi_\chi$ in a linear scale.
         Later, in Fig.\ \ref{Fig-isocurvature-C-mode} we will show
         that the isocurvature perturbation is not affected by the
         presence of these sharp peaks in the adiabatic condition
         and in the large-scale condition; the latter will be
         presented in Fig.\ \ref{Fig-LS}.
         Based on this observation and the physical behaviors of the
         curvature variables throughout the bounce, we suggest that
         such peaks do not have physical impact on
         the $C$- and $d$-mode natures of the perturbation.
         Panel (a) shows a spike in the evolution of
         $\varphi_v$ just after the bounce; it actually have two spikes
         at $\mu + p = 0$, see Eq.\  (\ref{varphi_v-reconstruct}) and Fig.\ \ref{Fig-usecalc}.
         Away from the bounce we have $\varphi_\chi = (3/5)
         \varphi_v$ which is the well known ratio for $w = 0$
         background.
         }
         \label{Fig-c}
\end{figure*}

In Figs.\ \ref{Fig-c}(a), \ref{Fig-c}(b), \ref{Fig-d}(a), and
\ref{Fig-d} (b), we present evolutions of `our $C$- and $d$-modes'.
These are based on our $C$- and $d$-mode initial conditions
presented in Table \ref{Table-initiall2}. These figures apparently
show that evolutions away from the bounce {\it coincide} with the
known behaviors of the $C$- and $d$-modes based on the large-scale
analytic solutions in Eqs.\ (\ref{eq-varphiv}) and
(\ref{eq-varphichi}). These analytic solutions are valid when both
the large-scale conditions and the adiabatic conditions are
satisfied.

\begin{figure*}
\centering%
\includegraphics[width=8cm]{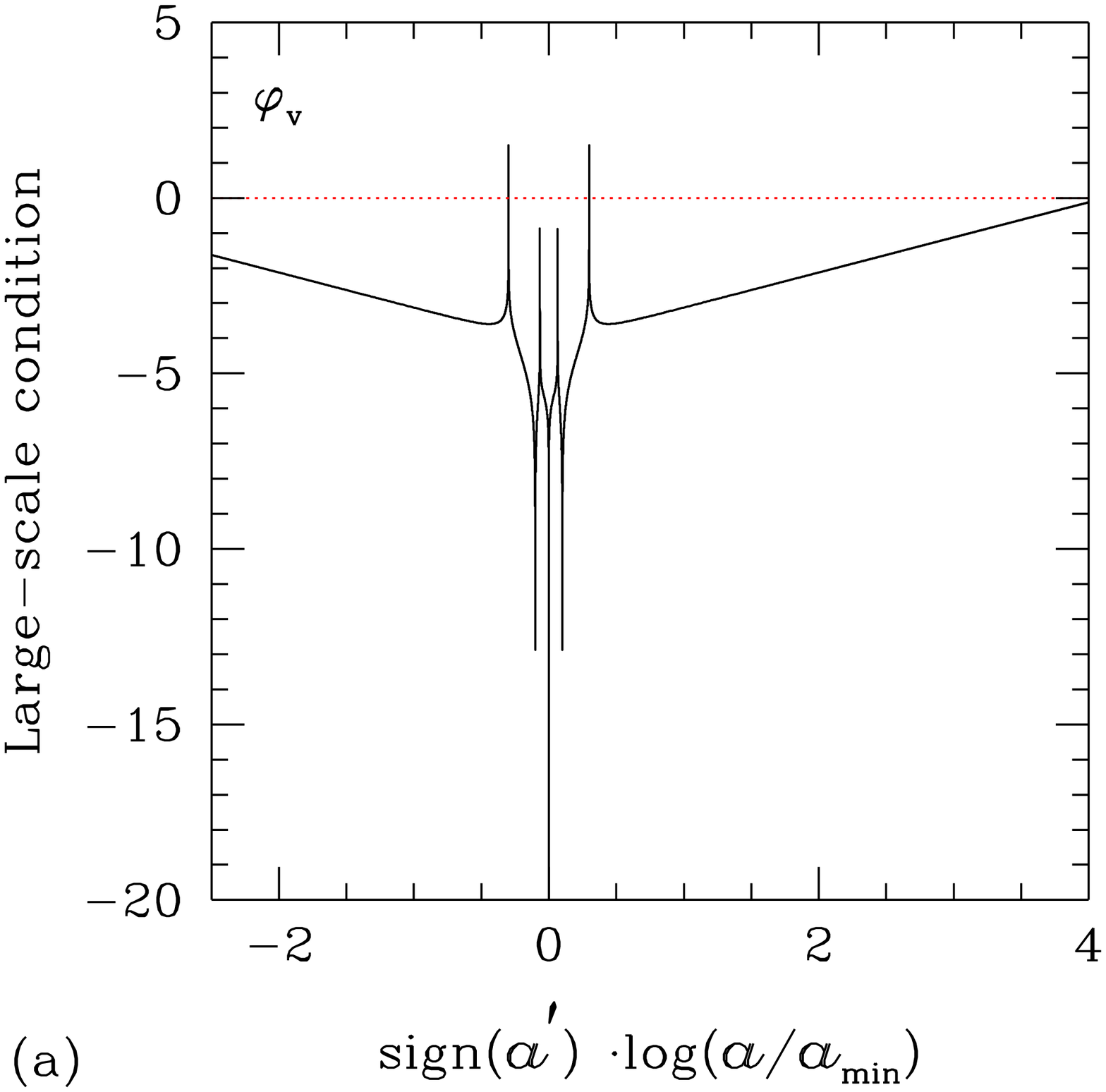}\hfill
\includegraphics[width=8cm]{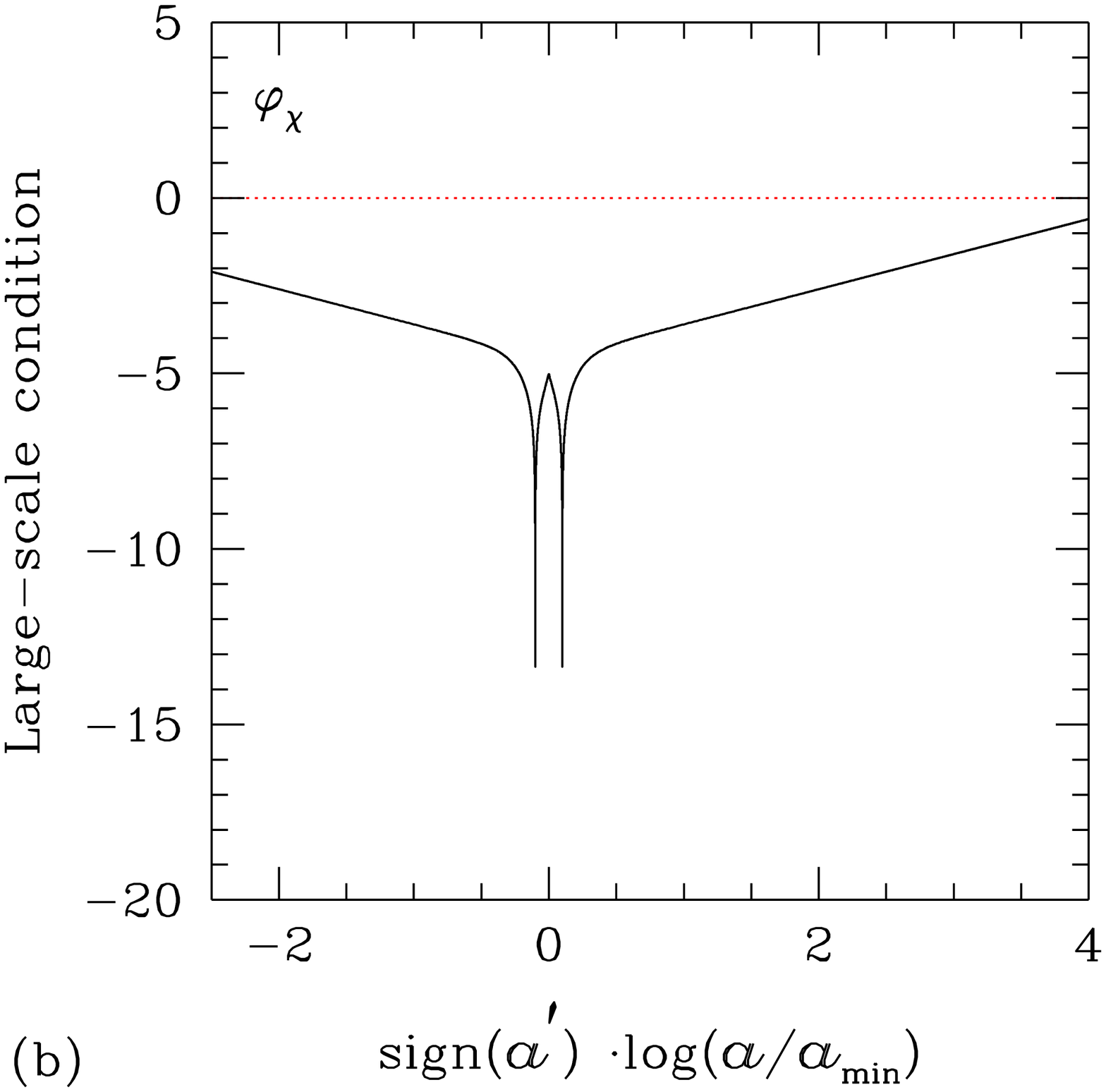}\\
\caption{Large-scale conditions for $\varphi_v$ and
         $\varphi_\chi$.
         Panels (a) and (b) show the large-scale conditions
         for
         $\varphi_v$ and $\varphi_\chi$ equations;
         for $\varphi_v$ and $\varphi_\chi$ we plot
         $k^2/(z^{\prime\prime}/z)$, and
         $k^2/[(1/z)^{\prime\prime}/(1/z)]$, respectively.
         As the large-scale conditions involve only the background
         evolution and the wavenumber, the same large-scale conditions
         apply to both $C$- and $d$-modes independently of the initial condition.
         Panel (a) shows that the
         large-scale condition for $\varphi_v$ is broken four times by sharp peaks
         (divergences) near the bounce.
         These divergences occur at $z^{\prime\prime}/z=0$.
         As explained in Fig.\ \ref{Fig-c}, as
         the isocurvature perturbation is not affected by the
         presence of these sharp peaks in the large-scale condition,
         such peaks are not supposed to have physical impact on
         interpreting the $C$- and $d$-mode natures.
         }
         \label{Fig-LS}
\end{figure*}

\begin{figure*}
\centering%
\includegraphics[width=8cm]{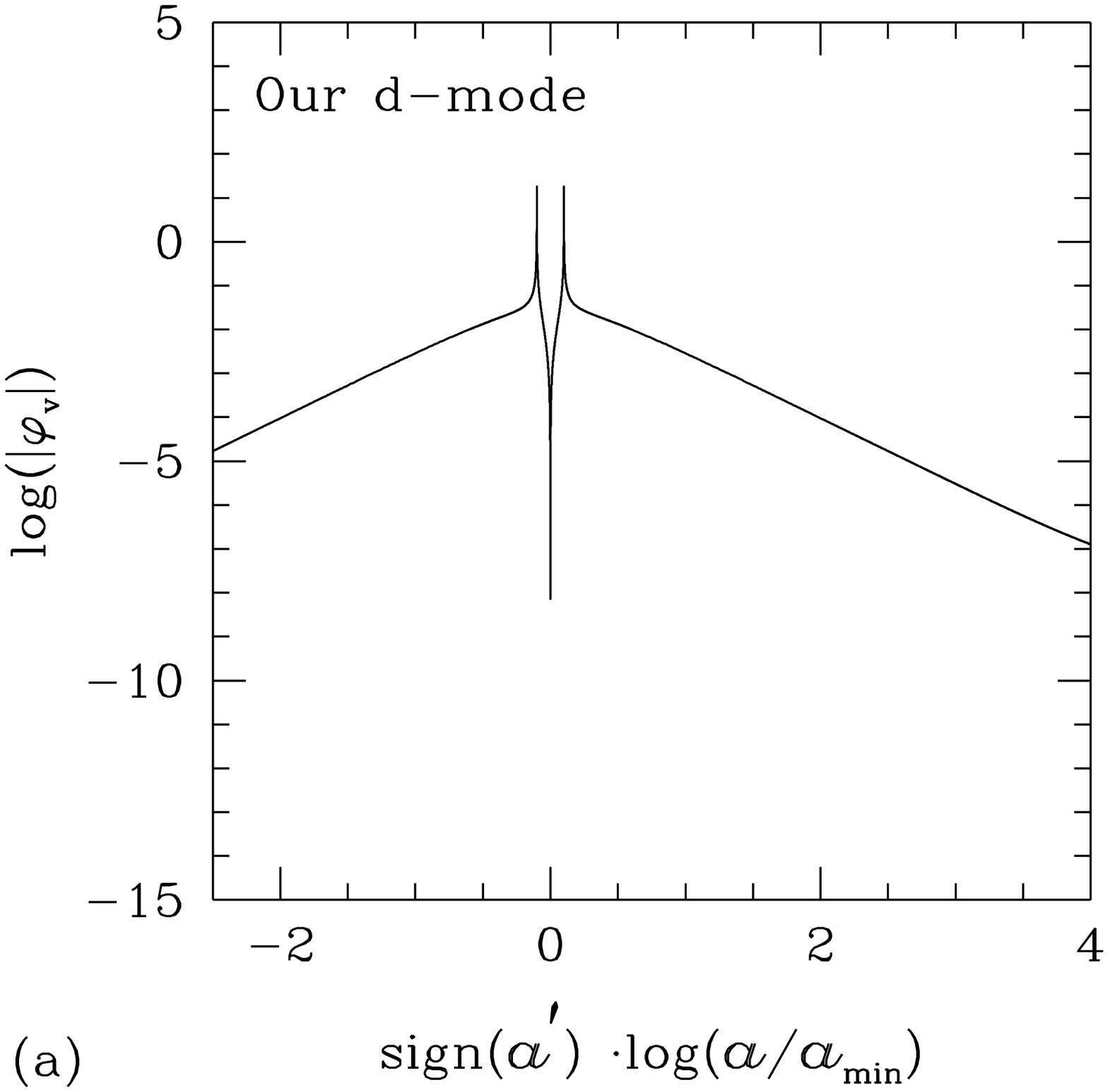}\hfill
\includegraphics[width=8cm]{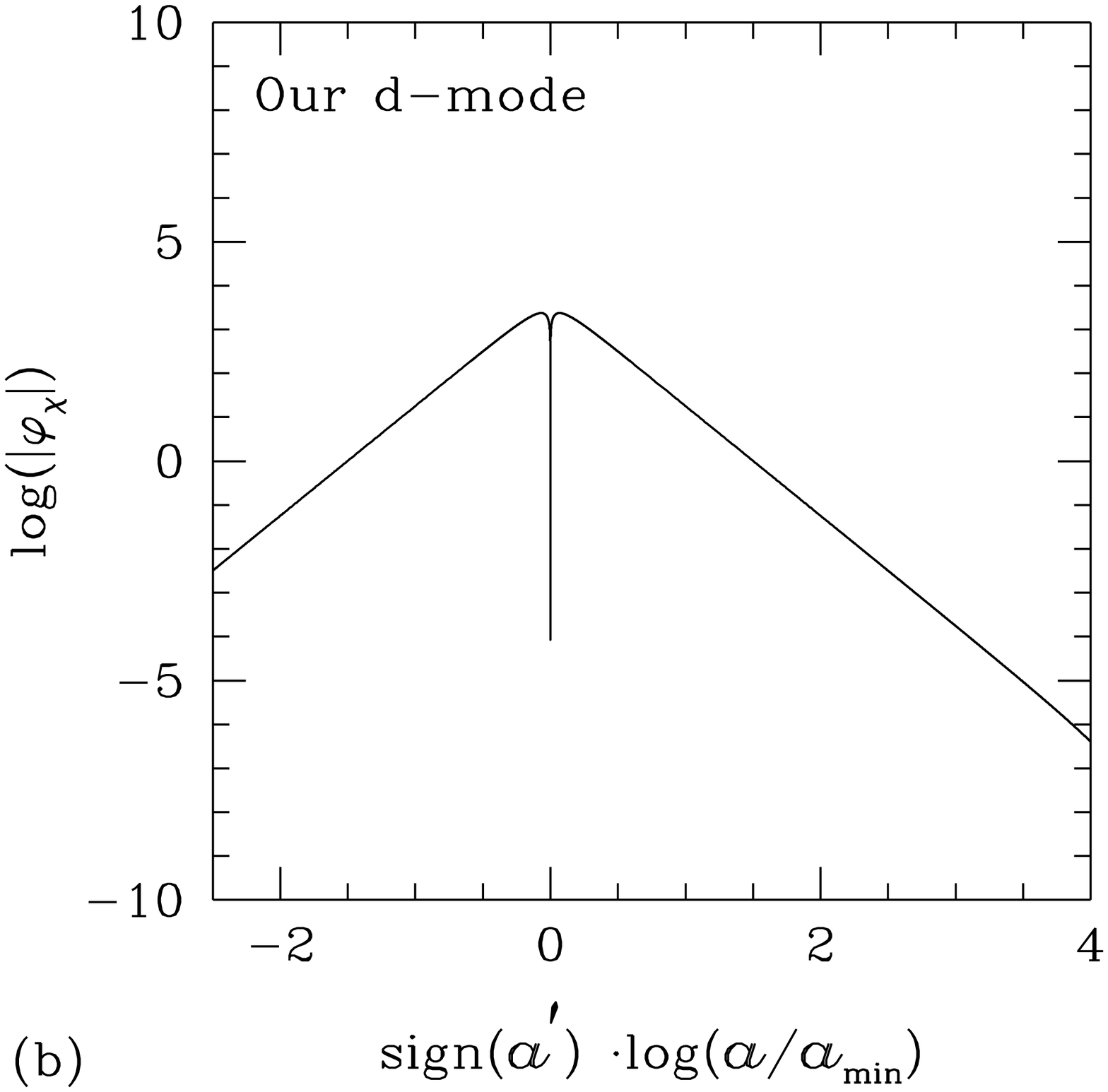}\\
\includegraphics[width=8cm]{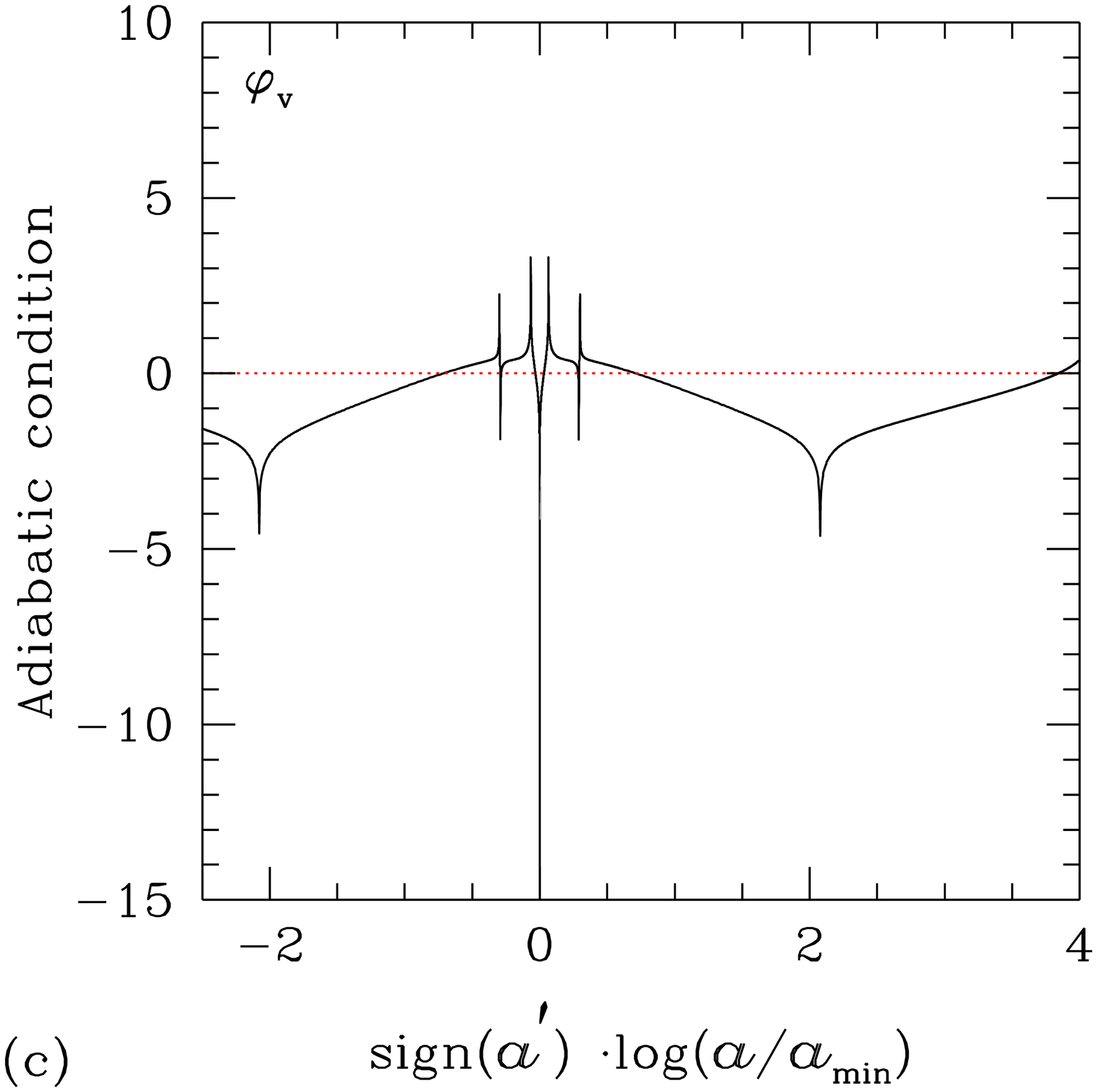}\hfill
\includegraphics[width=8cm]{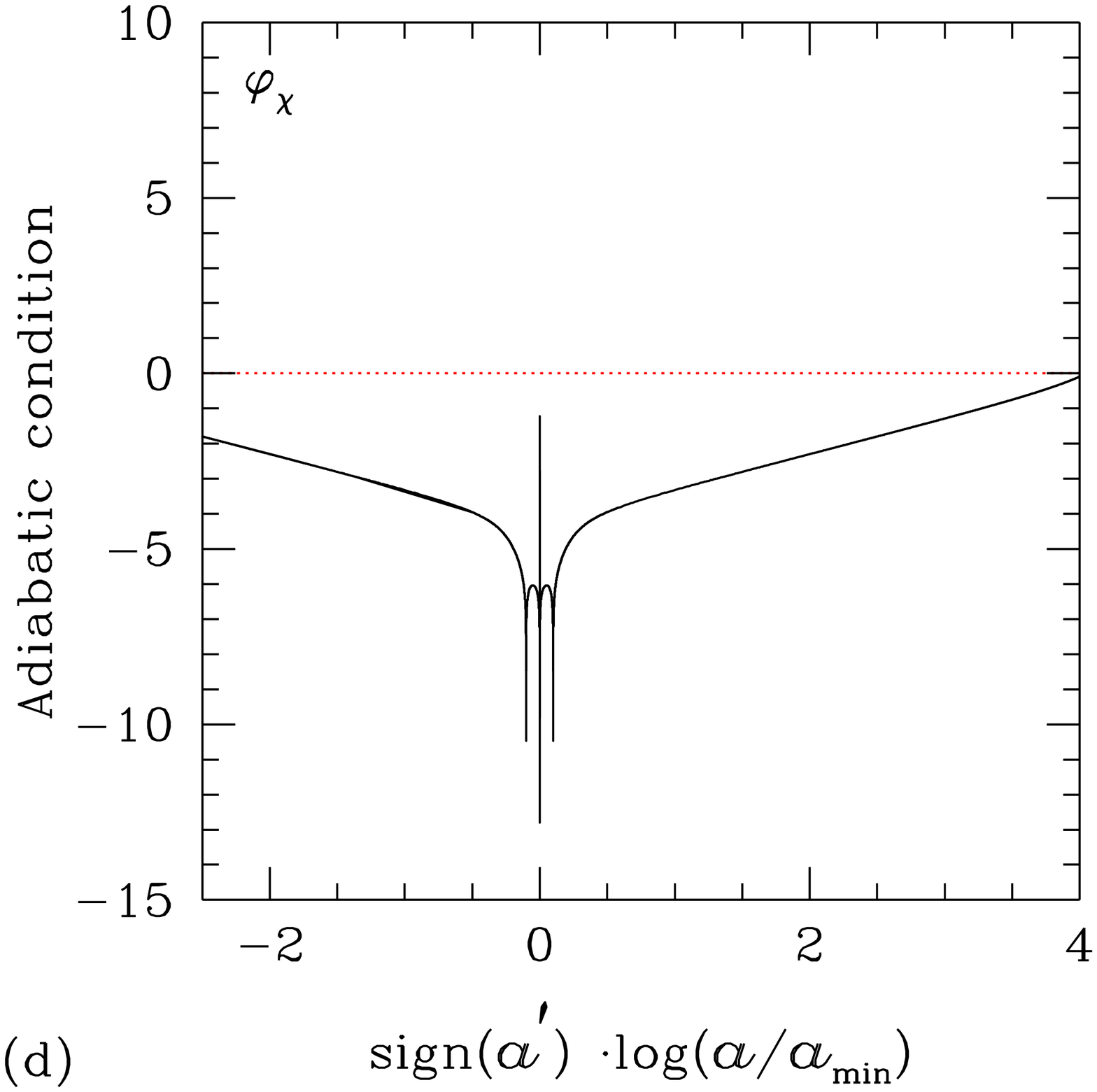}
\caption{Evolutions of our $d$-modes for $\varphi_v$ and
         $\varphi_\chi$, and the adiabatic conditions;
         the large-scale conditions are presented in Fig.\
         \ref{Fig-LS}.
         Panels (a) and (b) show the behaviors of $d$-modes for
         $\varphi_v$ and $\varphi_\chi$, respectively.
         We use the initial conditions in Table
         \ref{Table-initiall2}.
         Both $\varphi_v$ and $\varphi_\chi$ cross zero at the bounce,
         see Figs. \ref{Fig-usecald}(d) and \ref{Fig-usecald}(e)
         which show the evolution in linear scale.
         $\varphi_v$ diverges twice near the bounce at $\mu + p =
         0$, see Eq.\ (\ref{varphi_v-reconstruct}) and Fig.\ \ref{Fig-usecald}(d).
         Panels (c) and (d) show the adiabatic conditions for
         $\varphi_v$ and $\varphi_\chi$.
         Although, the adiabatic condition for $\varphi_v$ equation is severely broken
         near the bounce, we interpret this happens because the $d$-mode
         of $\varphi_v$ is $(k/aH)^2$-order higher in the large-scale expansion compared
         with the same mode of $\varphi_\chi$, i.e, the leading
         order $d$-mode of $\varphi_v$ has been canceled out;
         see Eqs.\ (\ref{eq-varphiv}) and (\ref{eq-varphichi}),
         or Eqs. (\ref{eq-varphive}) and (\ref{eq-varphichie}).
         Panel (c) also shows four sharp peaks (divergences)
         near the bounce which are caused by $z^{\prime\prime}/z=0$;
         compared with Fig.\ \ref{Fig-c}(c),
         the two sharp peaks potentially caused by $\mu + p = 0$
         disappeared because the $d$-mode of $\varphi_v$ diverges
         in proportion to $1/(\mu + p)$, e.g., see Eq.\
         (\ref{eq-varphiv}) and Panel (a).
         The adiabatic condition for $\varphi_\chi$ has one sharp peak
         at the bounce
         because $\varphi_\chi$ crosses zero at this points;
         see Eq.\ (\ref{adiabatic-conditions}) and Fig.\ \ref{Fig-usecald}(e).
         As explained in Fig.\ \ref{Fig-c},
         the isocurvature perturbation is not affected by the
         presence of these sharp peaks in the adiabatic condition,
         and the $d$-mode before the bounce remains as the $d$-mode
         after the bounce.
         Thus, such peaks do not have physical impact on
         the $d$-mode natures of the perturbations.
         }
         \label{Fig-d}
\end{figure*}

In Fig.\ \ref{Fig-LS} we examine the large-scale conditions using
Eq.\ (\ref{LS-conditions}). Figure \ref{Fig-LS}(a) apparently shows
that the large-scale condition for $\varphi_v$ is sharply broken
four times near the bounce at $z^{\prime\prime}/z = 0$, see Fig.\
\ref{fig-zvzu}. Despite these sharp divergences in our large-scale
condition, we argue that these do not have physical impact on
interpreting the $C$- and $d$-mode natures of $\varphi_v$ in Figs.\
\ref{Fig-c}(a) and \ref{Fig-d}(a). This is because our large-scale
condition is sharply broken as the $z^{\prime\prime}/z$ term in Eq.\
(\ref{eq-varphiz}) crosses zero, thus becomes smaller than
$k^2$-term, for very brief time intervals, see Fig.\ \ref{fig-zvzu}.
Figure \ref{Fig-LS}(b) shows that the large-scale condition is well
satisfied for $\varphi_\chi$. The large-scale conditions involve
only the background evolution and the wavenumber. Thus, the same
large-scale condition applies to both $C$- and $d$-modes
independently of the initial conditions for perturbations.

In Figs.\ \ref{Fig-c}(c), \ref{Fig-c}(d), \ref{Fig-d}(c), and
\ref{Fig-d}(d) we examine adiabatic conditions using Eq.\
(\ref{adiabatic-conditions}). All these figures show that the
adiabatic conditions are sharply broken near the bounce. These occur
at $z^{\prime\prime}/z = 0$, $\mu+p = 0$, or $\varphi_\chi = 0$.
Despite these sharp divergences in our adiabatic conditions, we
argue that these do not have physical impact on interpreting the
$C$- and $d$-mode natures of $\varphi_v$ in Figs.\ \ref{Fig-c}(a)
and \ref{Fig-d}(a). We have two reasons. First, $\varphi_v$ in
Figs.\ \ref{Fig-c}(a) and \ref{Fig-d}(a) shows that the $C$- and
$d$-mode natures are preserved away from the bounce. Second, later
in Figs.\ \ref{Fig-isocurvature-C-mode} and
\ref{Fig-isocurvature-d-mode} we will show that for our $C$- and
$d$-modes the isocurvature perturbation is not excited near the
bounce.

Compared with the above sharp spikes in the adiabatic conditions,
Fig.\ \ref{Fig-d}(c) shows that the adiabatic condition for
$\varphi_v$ is severely broken near the bounce region. Remember that
the $d$-mode of $\varphi_v$ in Eq.\ (\ref{eq-varphiv}) is higher
order in the large-scale expansion compared with the same mode of
$\varphi_\chi$ in Eq.\ (\ref{eq-varphichi}); this happens because
the leading order large-scale solution of the $d$-mode of
$\varphi_v$ has canceled out, see Eqs.\ (\ref{eq-new6}) and
(\ref{eq-varphichie}). Thus, as the $d$-mode of $\varphi_v$ is
$(k/aH)^2$-factor smaller than the $d$-mode of $\varphi_\chi$ we
could have such a breakdown in Fig.\ \ref{Fig-d}(c) without
affecting the $d$-mode nature of $\varphi_v$ before and after the
bounce.

Based on the above arguments, we suggest that `our $C$- and $d$-mode
initial conditions' in Table \ref{Table-initiall2} produce the
proper $C$- and $d$-modes which are valid throughout the bounce, as
presented in Figs.\ \ref{Fig-c}(a), \ref{Fig-c}(b), \ref{Fig-d}(a),
and \ref{Fig-d}(b). In order to obtain these numerical results
producing proper $C$- and $d$-mode decomposition, we have to find
the precise initial condition; the method will be explained in Sec.\
\ref{scalar-numerical-IC}.

In an ideal situation we may use the initial conditions of
$\varphi_v$ and $\varphi_\chi$ in Eq.\ (\ref{eq-connectioncd}) which
are valid when the large-scale and the adiabatic conditions are
satisfied exactly; these initial conditions are summarized in Table
\ref{Table-initiall1}, and numerical results will be presented in
Sec.\ \ref{sec:analytic-IC}. Since these two conditions are not
exactly satisfied in reality, we have to find more accurate initial
conditions which allow the correct $C$- and $d$-mode behavior at the
initial phase, and preferably throughout the bounce. The correct
initial conditions we found are presented in Table
\ref{Table-initiall2}, and the numerical results are presented in
Figs.\ \ref{Fig-c} and \ref{Fig-d}.

\begin{figure*}
\centering%
\includegraphics[width=8cm]{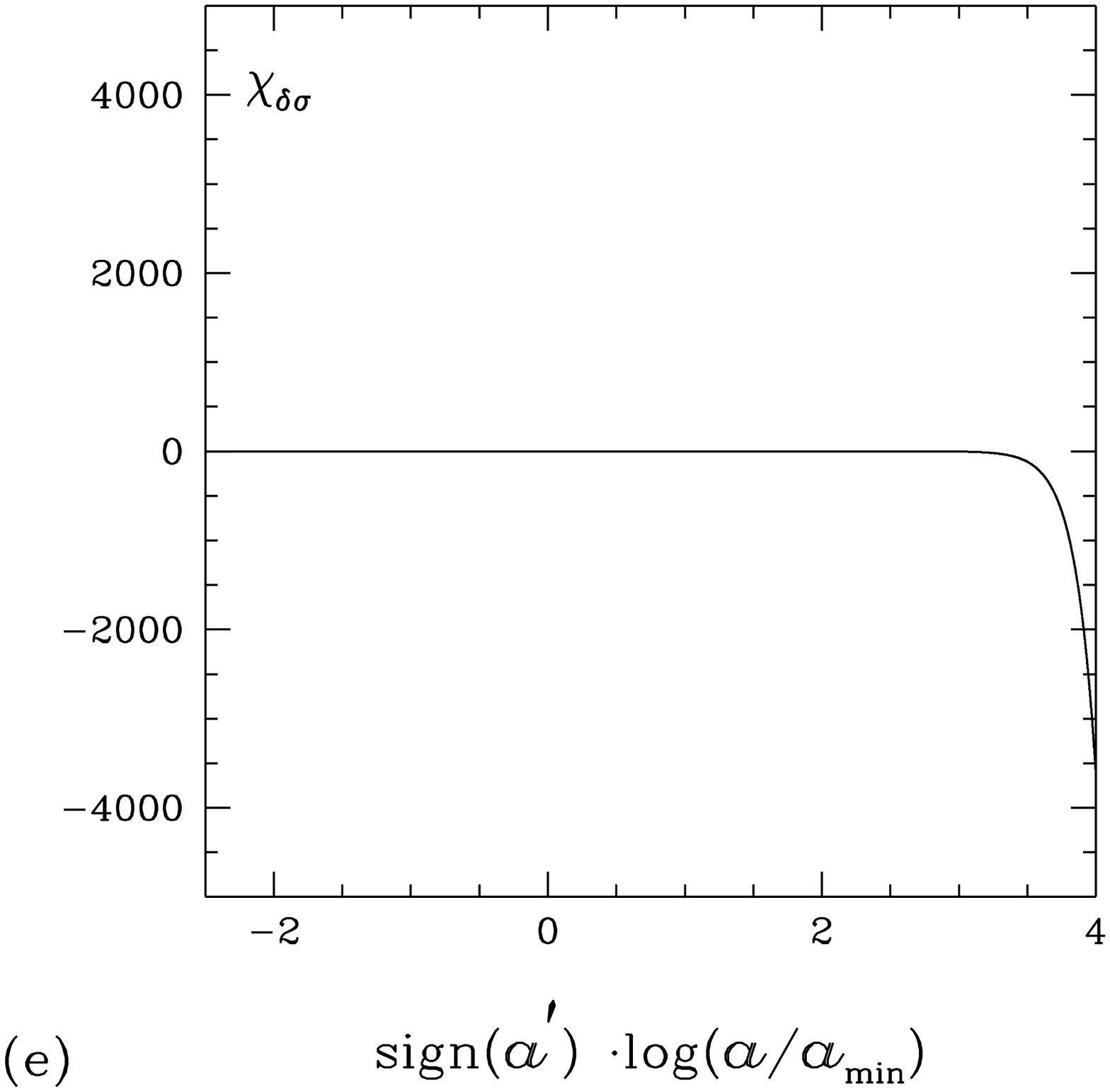}\hfill
\includegraphics[width=8cm]{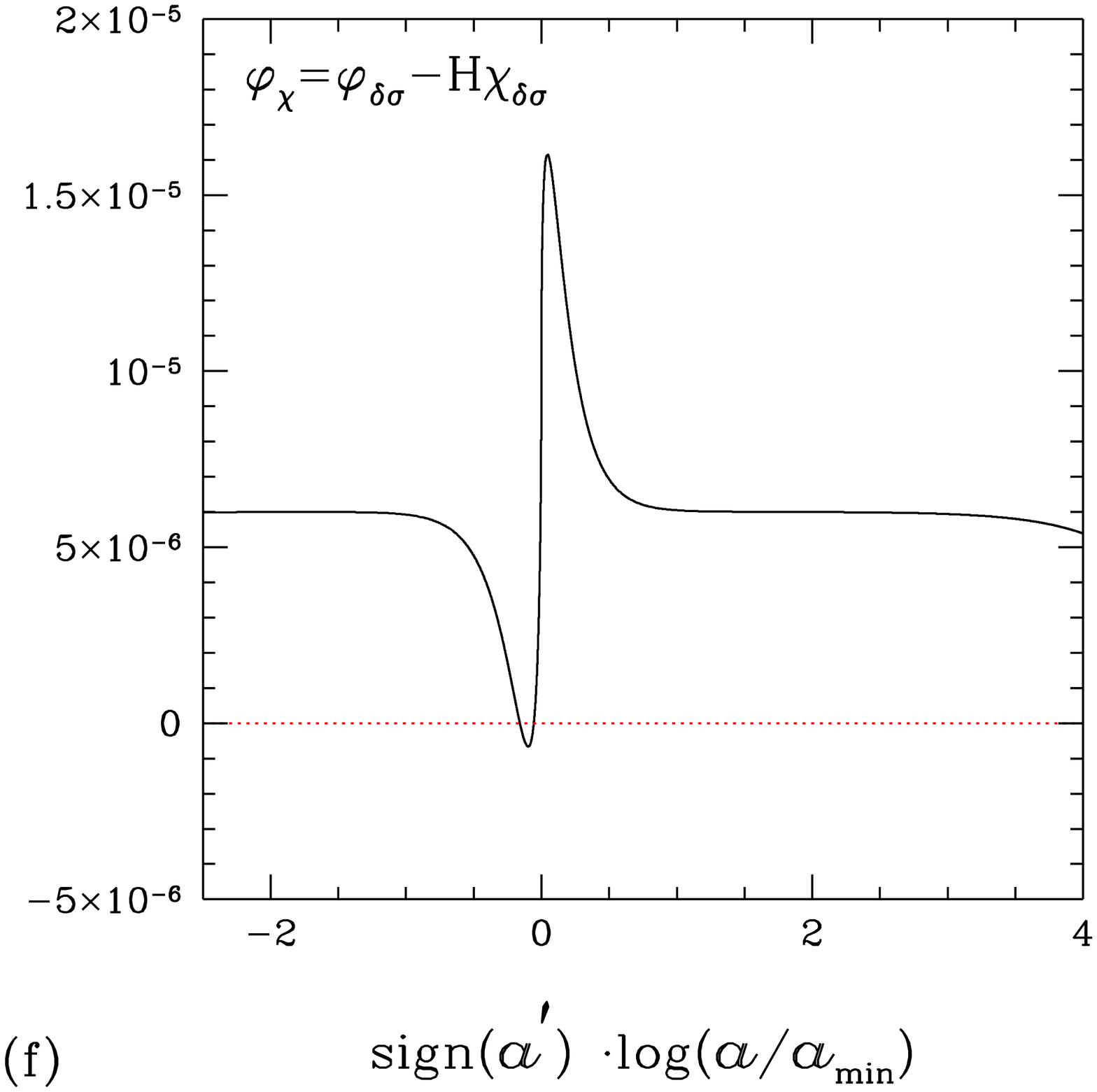}\\
\caption{The raw behaviors of gauge-invariant variables numerically
         solved using Eqs.\ (\ref{eq-dsdp})-(\ref{eq-dschi}) for $C$-modes of
         $\varphi_v$ and $\varphi_\chi$.
         Panels (a)-(c),(e) show that
         the variables we used in numerical integration behave smoothly
         throughout the evolution.
         We obtain the behaviors of $\varphi_v$ and $\varphi_\chi$
         in Panels (d) and (f) using
         Eq.\ (\ref{varphi_v-reconstruct}).
         Panel (c) magnifies Panel (b) near the bounce region.
         This figure explains why we have only one sharp spike
         after the bounce visible in Panel (d) and Fig.\ \ref{Fig-c}(a);
         the spikes are caused by vanishing $\mu + p$ in Eq.\
         (\ref{varphi_v-reconstruct}).
         The arrows indicate the epochs where $\mu + p = 0$.
         }
         \label{Fig-usecalc}
\end{figure*}

\begin{figure*}
\centering%
\includegraphics[width=8cm]{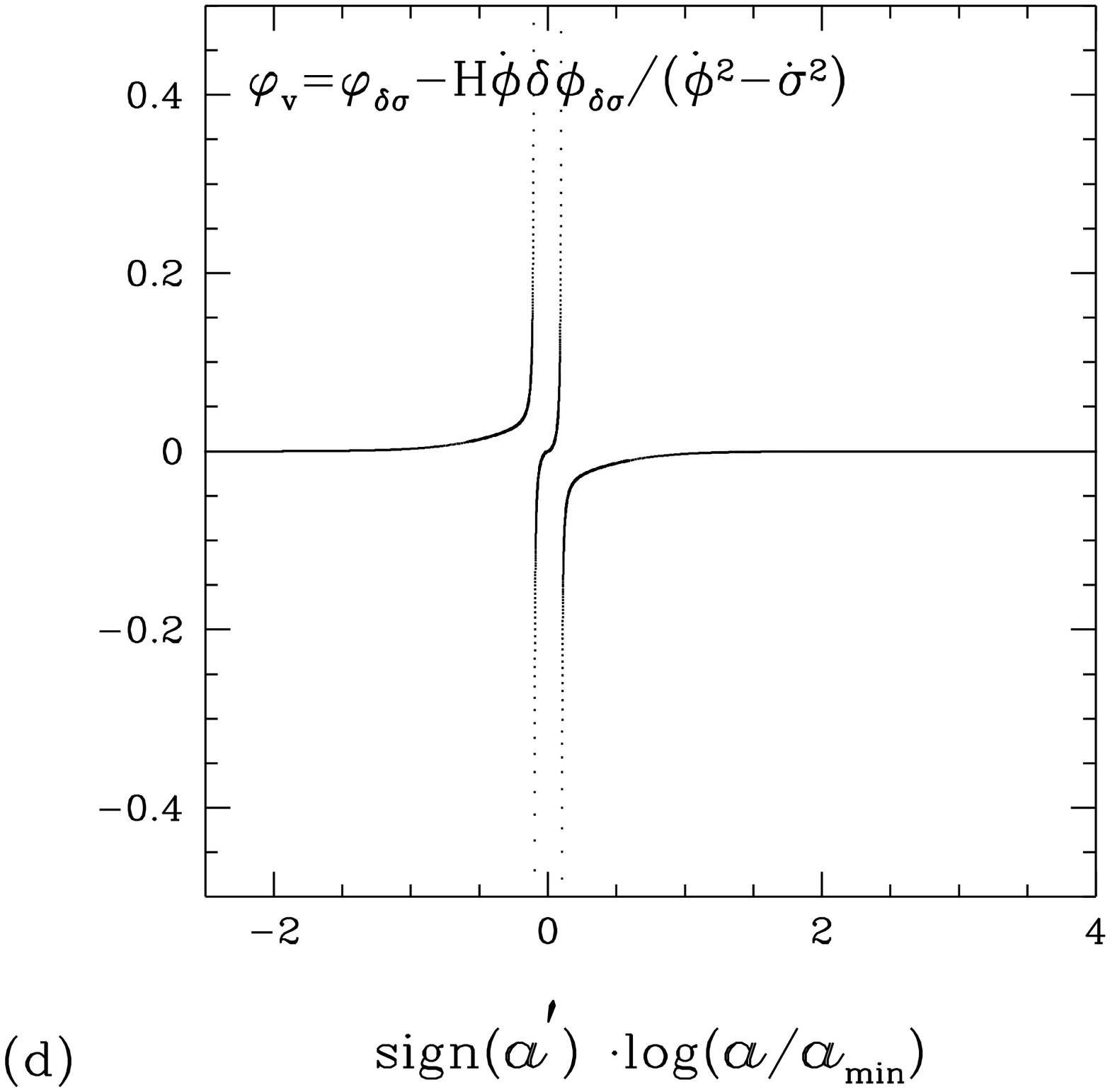}\hfill
\includegraphics[width=8cm]{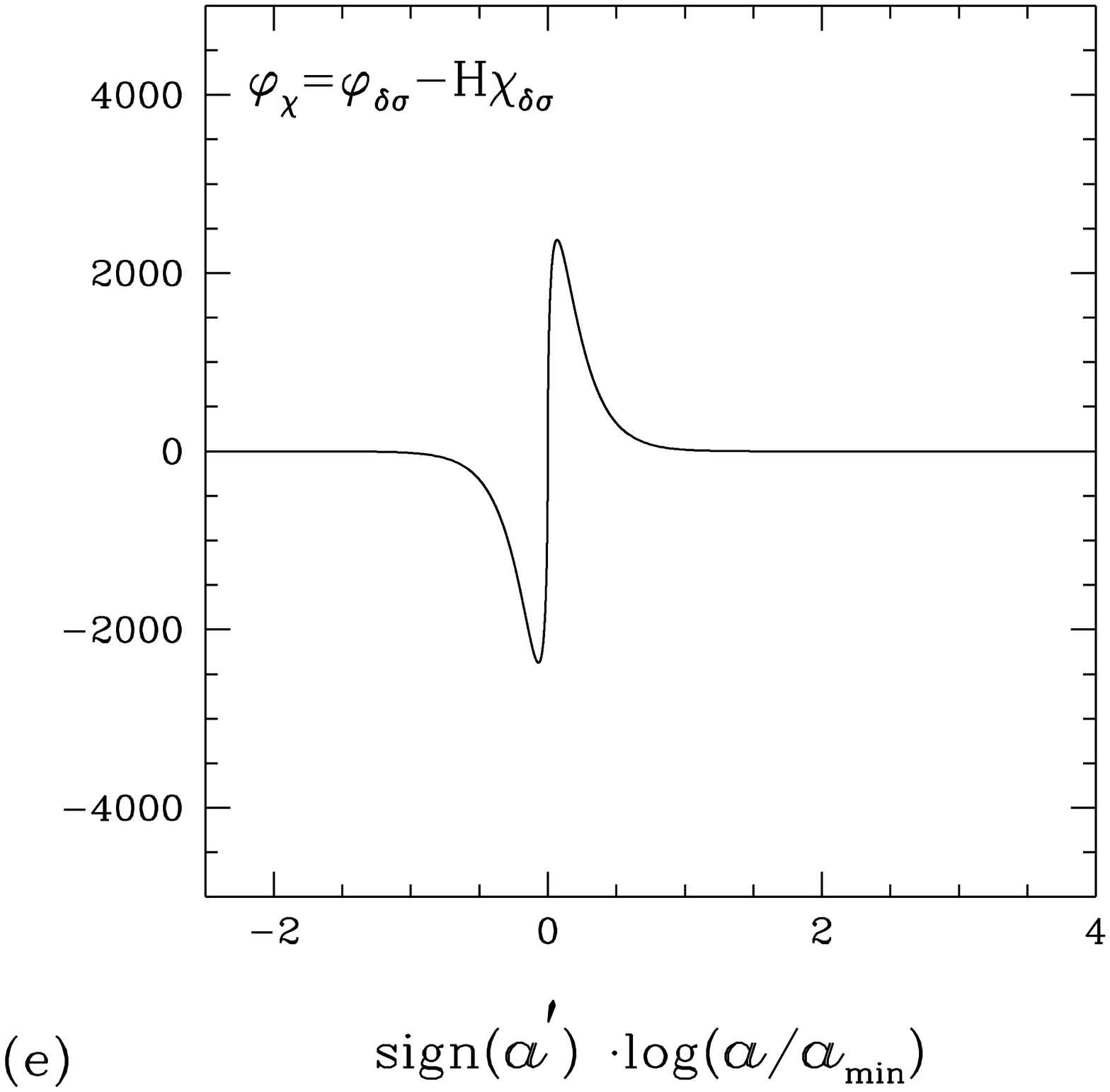}
\caption{The same as Fig.\ \ref{Fig-usecalc},
         now, for the $d$-mode.
         Panels (a)-(c) show that
         the variables we used in numerical integration behave smoothly
         throughout the evolution.
         }
         \label{Fig-usecald}
\end{figure*}

As noted in Sec.\ \ref{three-gauges}, we solve numerically two
independent sets of equations based on two different temporal gauge
conditions: these are Eqs.\ (\ref{eq-dsdp})-(\ref{eq-dschi}) in the
uniform-$\sigma$ gauge, and Eqs.\ (\ref{eq-zsvp})-(\ref{eq-zsdsdd})
in the zero-shear gauge. We derive initial conditions for variables
in the zero-shear gauge using relations between the uniform-$\sigma$
gauge and the zero-shear gauge in Eq.\ (\ref{eq-connectindszs}). Any
gauge-invariant combination of variables should show the same
behavior independently of the gauge conditions we took. This
provides a numerical check of the accuracy of our integration.
Although evolutions of $\varphi_v$ and $\varphi_\chi$ in Figs.\
\ref{Fig-c} and \ref{Fig-d} sometimes show singular behaviors near
the bounce, these are due to the singular behavior of background
variables multiplied to form these gauge-invariant combinations,
e.g., see Eq.\ (\ref{varphi_v-reconstruct}). In Figs.\
\ref{Fig-usecalc} and \ref{Fig-usecald} we show that during our
numerical integration the original variables used in the integration
behave smoothly.

\begin{figure}
\centering%
\includegraphics[width=8cm]{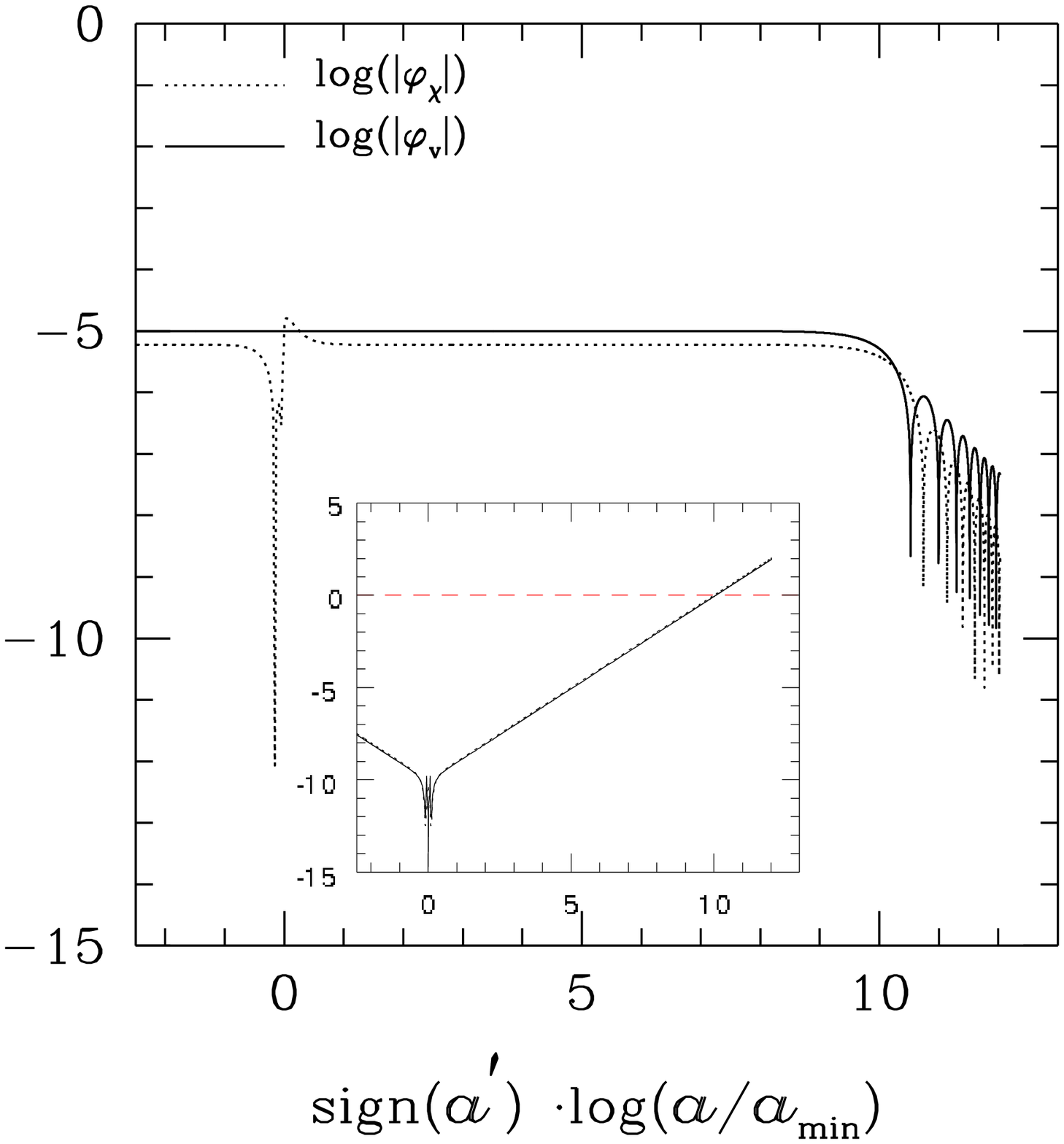}
\caption{Long term evolution of the $C$-mode.
         This figure shows the evolutions of
         $C$-modes using $k=10^{-6}$.
         The inset figure shows that the scale becomes subhorizon
         near $\log{(a/a_{min})}=10$.
         Both $\varphi_v$ and $\varphi_\chi$ begin oscillations
         as the scale comes inside the (visual) horizon.
         In a realistic situation, the horizon crossing of
         relevant cosmic scales should occur long after the bounce.
         }
         \label{numericalc}
\end{figure}

In Figure \ref{numericalc}, we show the $C$-mode evolution of
$\varphi_v$ and $\varphi_\chi$ for a different scale which enter the
horizon at later epoch; here we set $k=10^{-6}$ whereas in other
figures we used $k = 0.0005$, and the initial conditions differ from
the other figures.

Our results in Figs.\ \ref{Fig-c} and \ref{Fig-d} show that by
taking the correct initial conditions representing the $C$- and
$d$-modes for $\varphi_v$ and $\varphi_\chi$, the identities of $C$-
and $d$-modes remain valid throughout the bounce. That is, the
$C$-mode remains as the $C$-mode throughout the bounce, and
similarly for the $d$-mode. Although in the collapsing phase it is
the $d$-mode which is relatively growing, it decays away in the
expanding phase. Thus, if we are interested in the relatively
growing mode in the final expanding phase we have to consider the
$C$-mode even in the collapsing phase. This is one of the major
points we have emphasized in our previous works based on analytic
arguments in \cite{hw1,hw2,HN-bounce}, and here we confirm it by
using the numerical integration based on a specific bounce model.

\begin{table}
\caption{The initial conditions which give the
         precise (thus, correct) $C$- and $d$-mode behaviors
         in both collapsing and expanding phases away
         from the bounce.
         We call these `our $C$- and $d$mode initial conditions,
         or simply `our $C$- and $d$-modes.
         We take $k=0.0005$ and the initial $\eta=-500$. The
         numerical results based on these initial conditions are presented in
         Figs.\ \ref{Fig-c}-\ref{Fig-usecald}.
         }
         \label{Table-initiall2}
\begin{center}
\begin{tabular}{ r r r} \hline\hline
       variable   & C-mode  & d-mode \\  \hline
       $\delta\phi_{\delta\sigma}$ & $-$7.8058229814554E$-$06
       & $-$1.5568456514629E$+$03\\
       $\delta\phi_{\delta\sigma}^\prime$ &4.6325084877449E$-$08
       & 8.9724639867256E+00\\
       $\varphi_{\delta\sigma}$ &5.4370235008682E$-$06
       & $-$8.9884525573535E$+$02\\
       $\varphi_{\delta\sigma}^\prime$ & 2.7003984444483E$-$08
       & 5.1802544967425E$+$00\\
       $\varphi_v$  &9.9439752296909E$-$06
       & 1.0015746283898E$-$05\\
       $\varphi_v^\prime$  &2.5769163366172E$-$10
       & 5.8225396948330E$-$08\\
                                \hline\hline
\end{tabular}
\end{center}
\end{table}
\begin{figure*}
\centering%
\includegraphics[width=8cm]{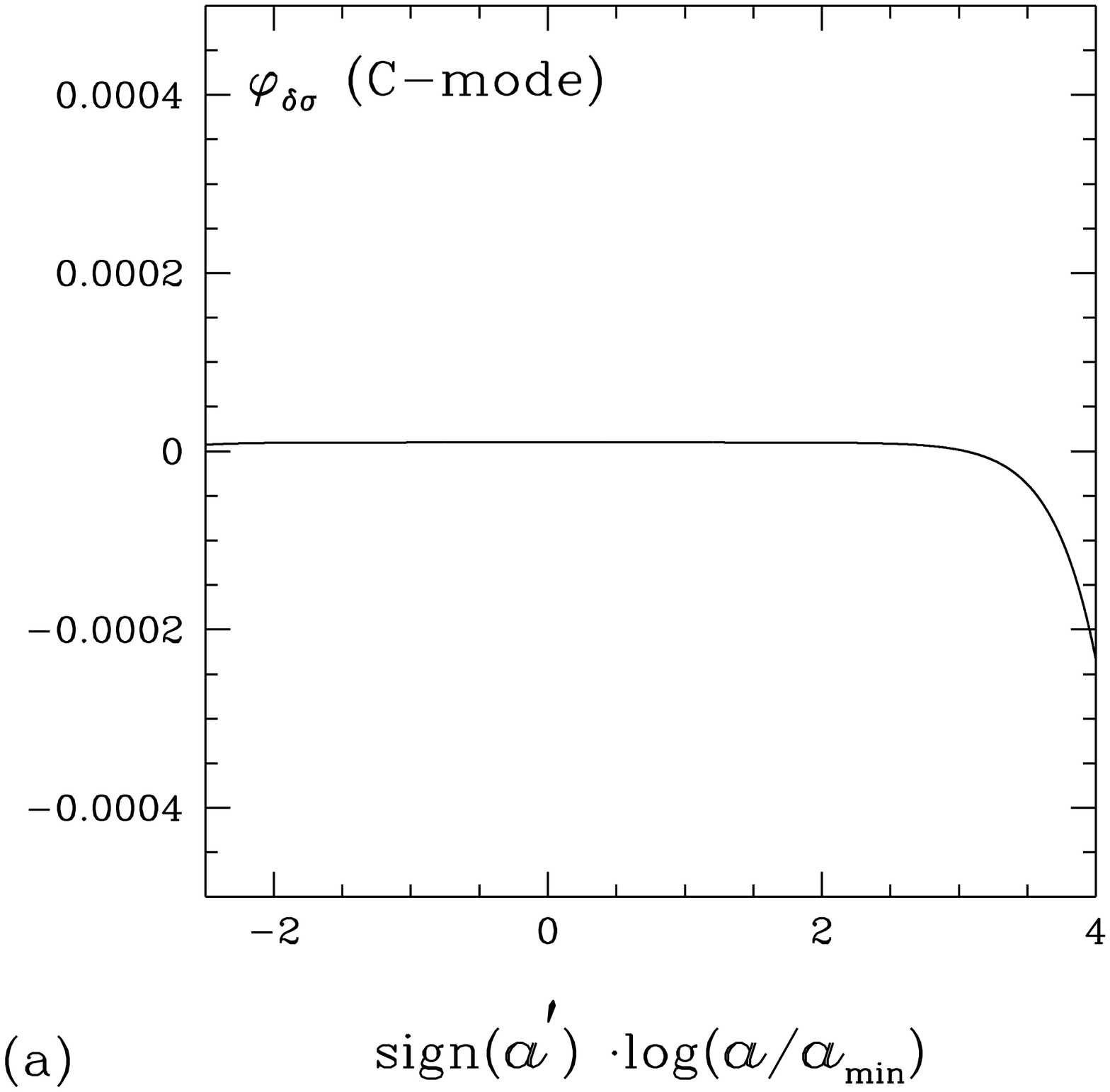}\hfill
\includegraphics[width=8cm]{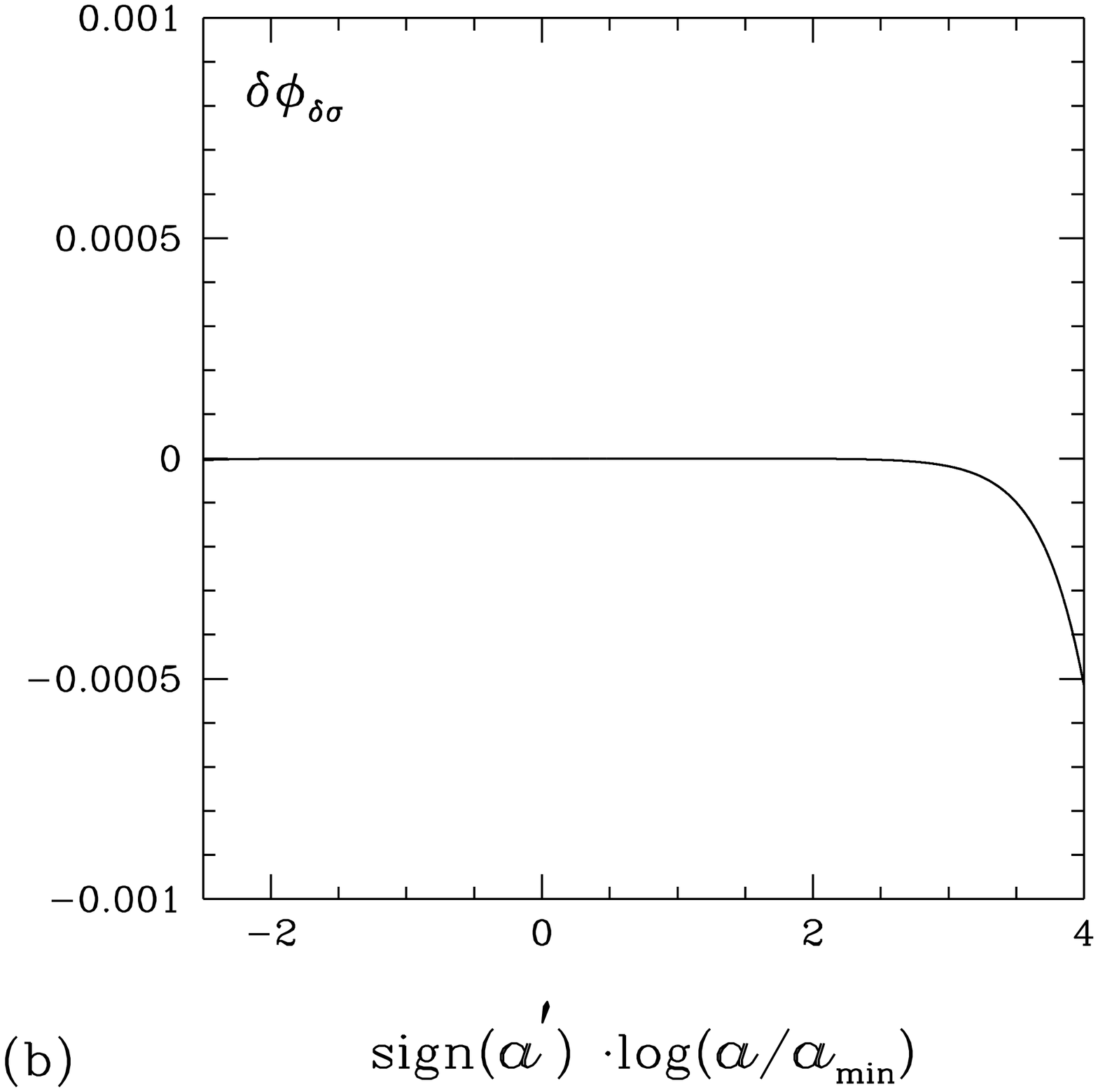}\\
\includegraphics[width=8cm]{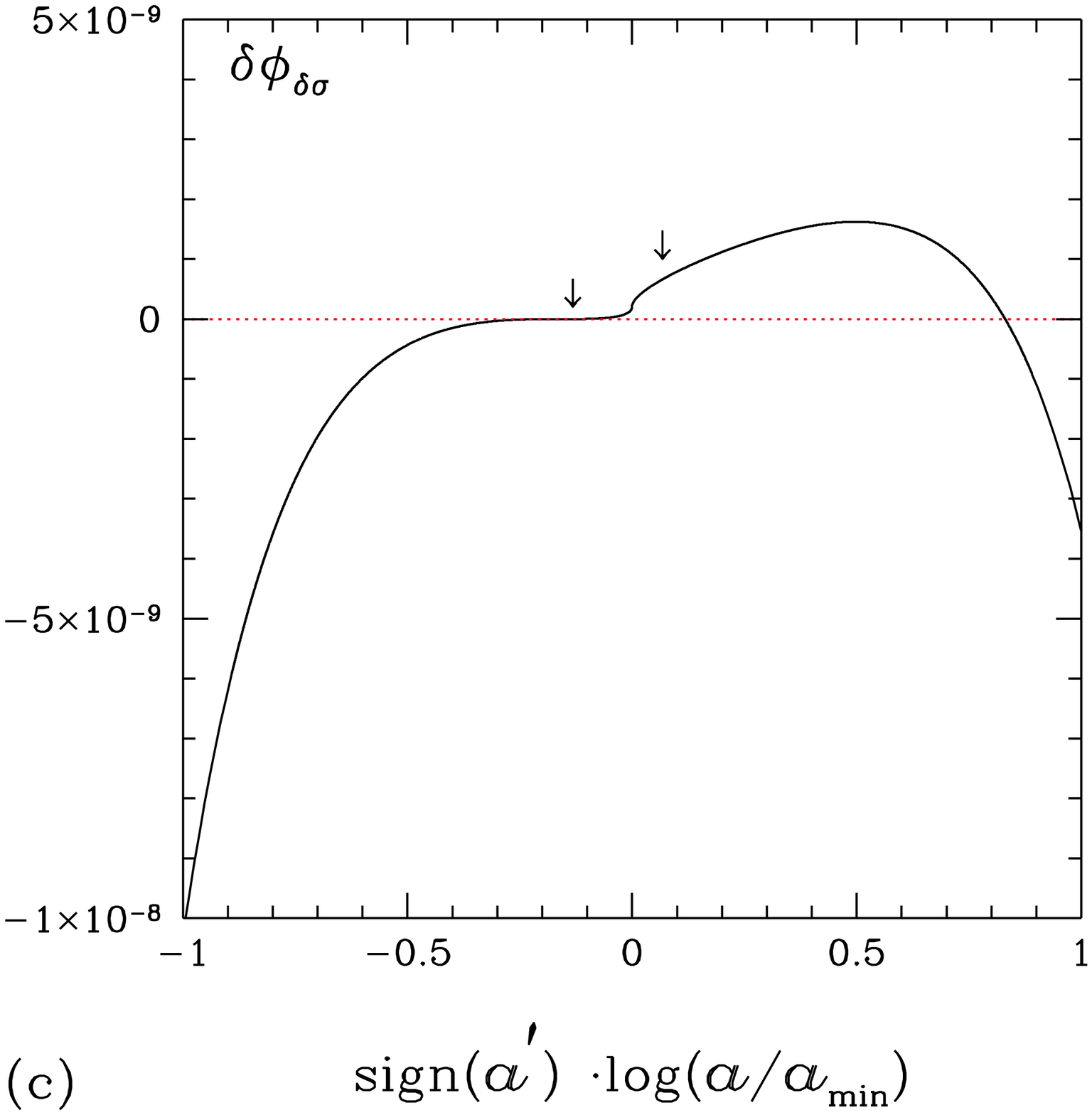}\hfill
\includegraphics[width=8cm]{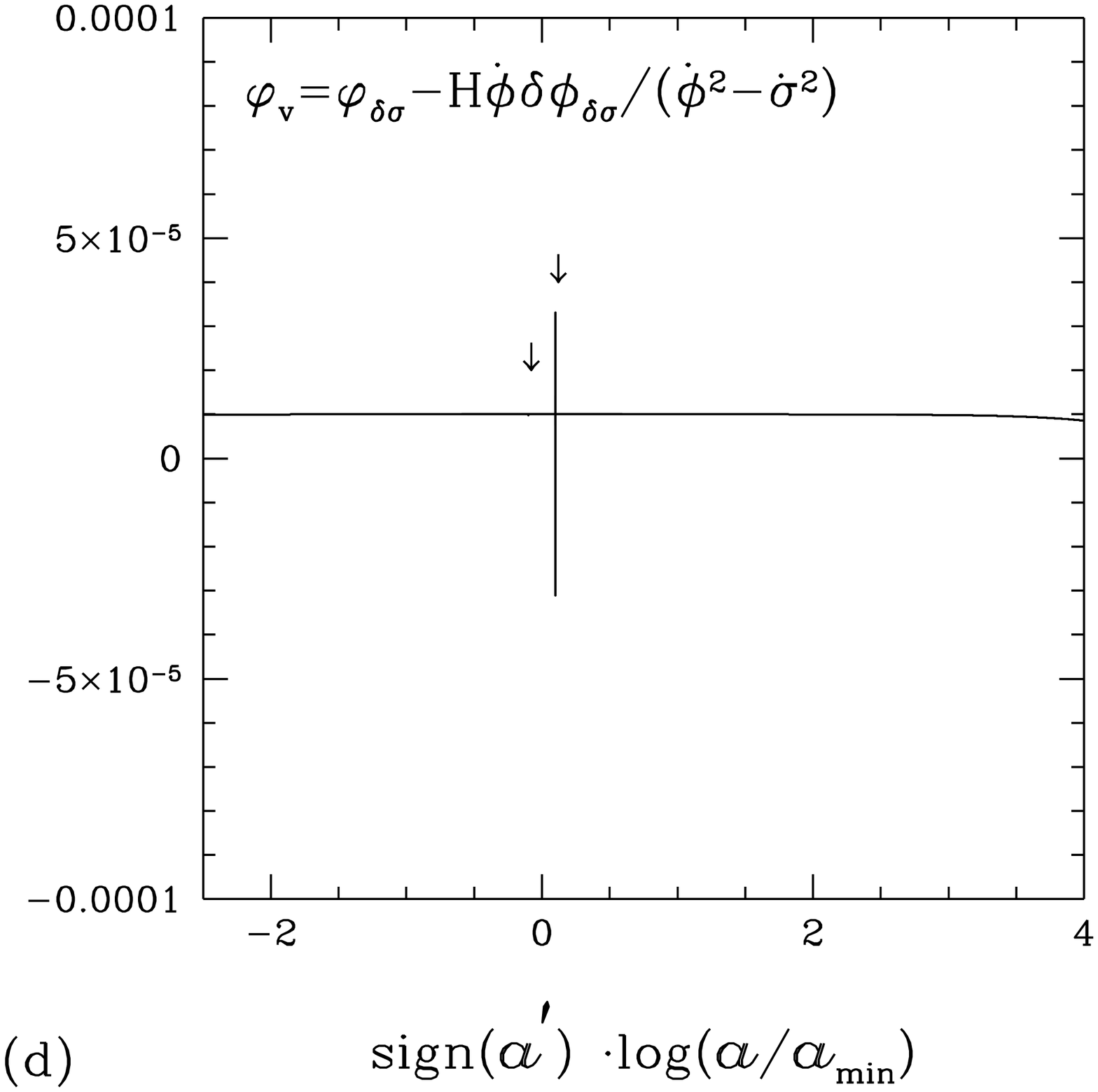}\\
\end{figure*}
\begin{figure*}
\centering%
\includegraphics[width=8cm]{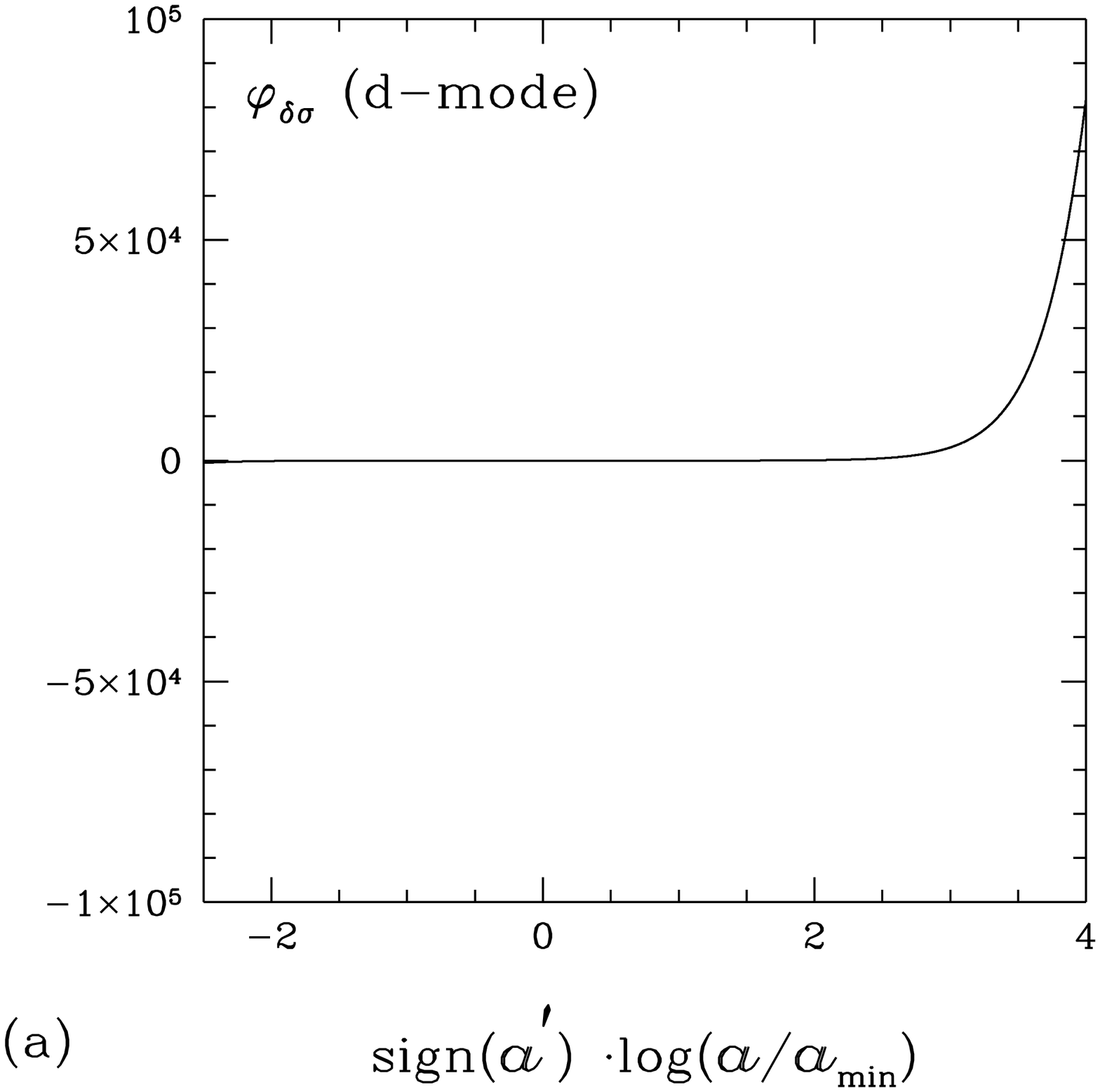}\\
\includegraphics[width=8cm]{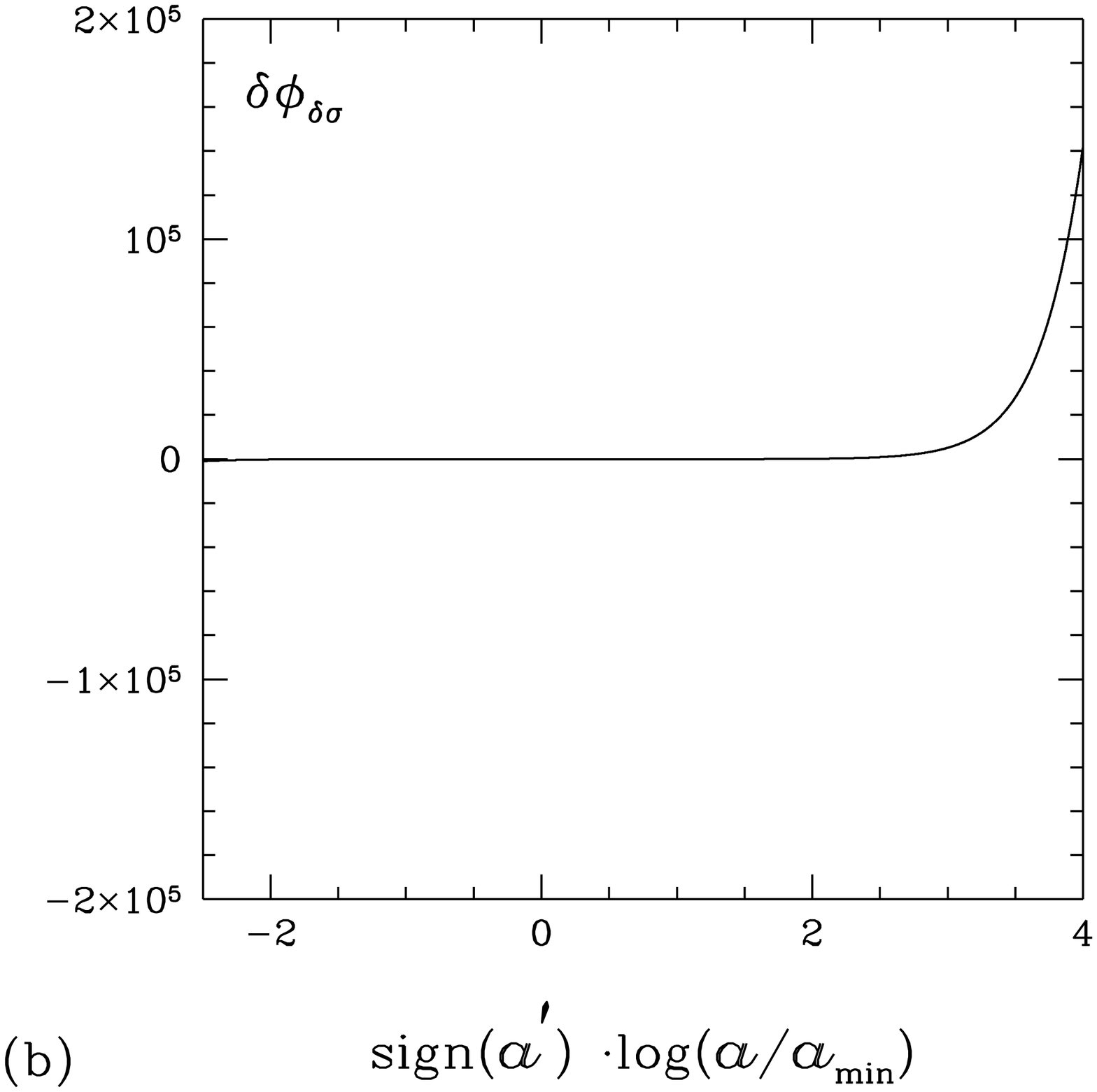}\hfill
\includegraphics[width=8cm]{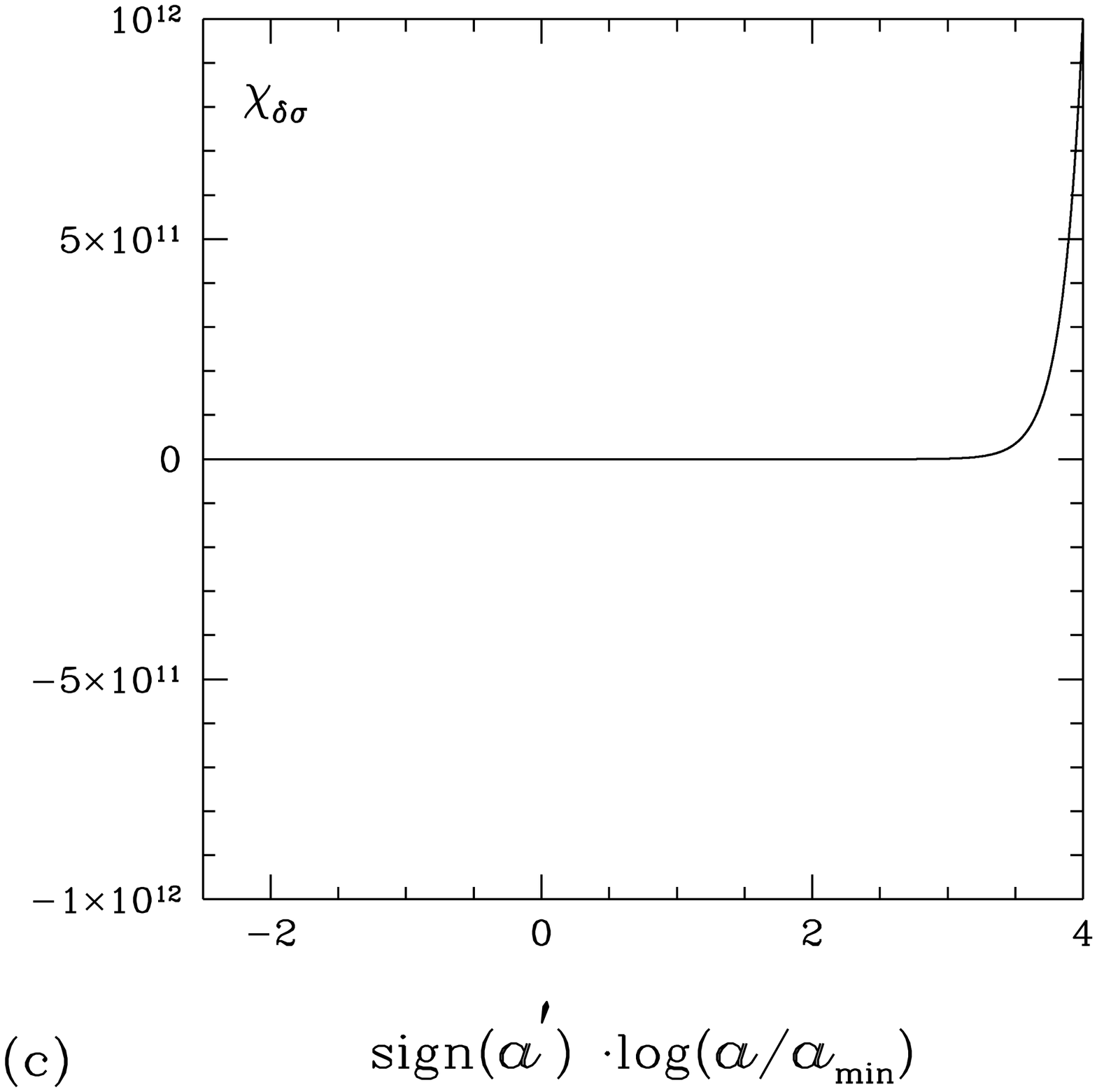}\\
\end{figure*}

\subsubsection{Analytic $C$- and $d$-modes}
                                              \label{sec:analytic-IC}

\begin{figure*}
\centering%
\includegraphics[width=8cm]{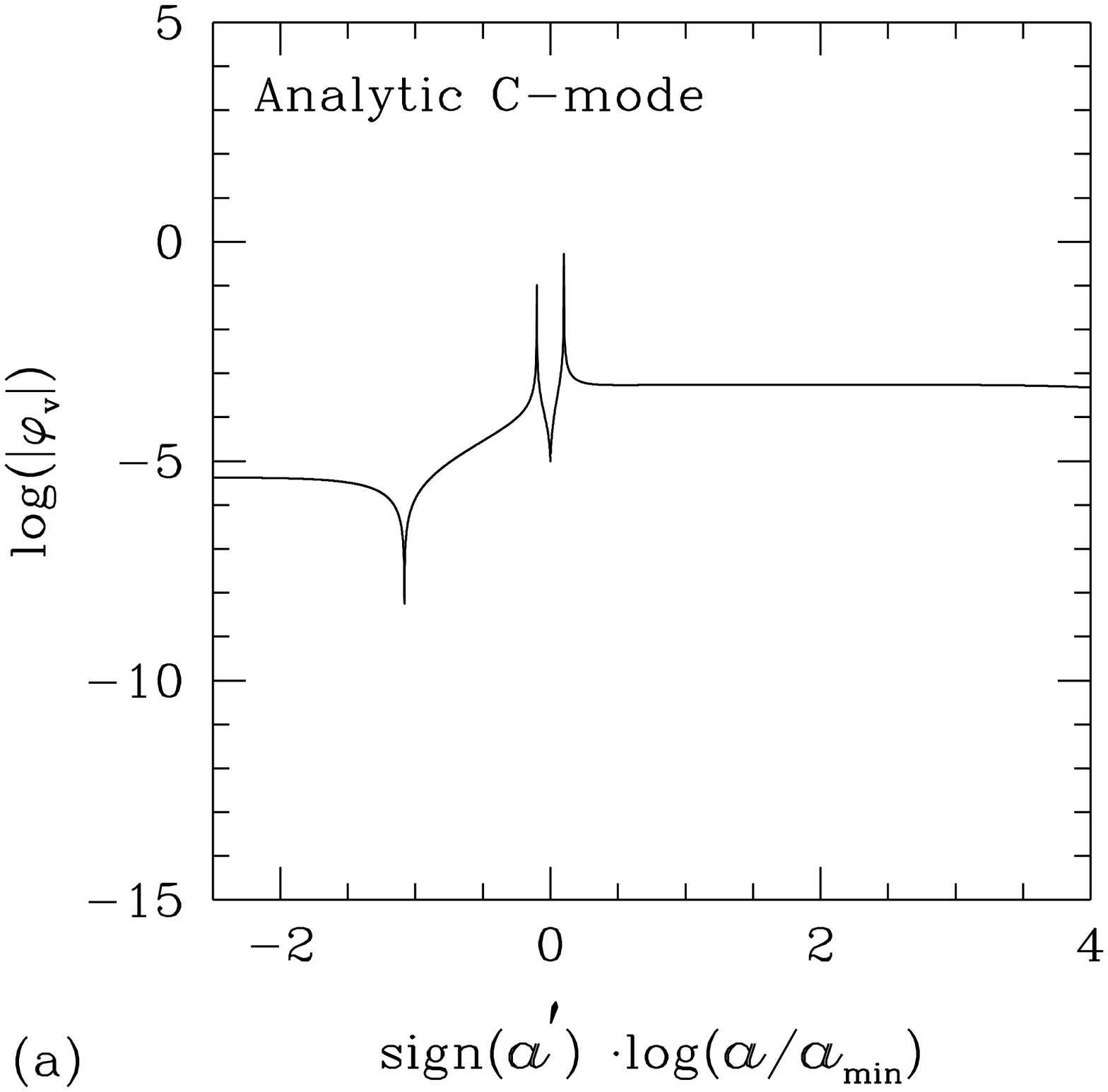}\hfill
\includegraphics[width=8cm]{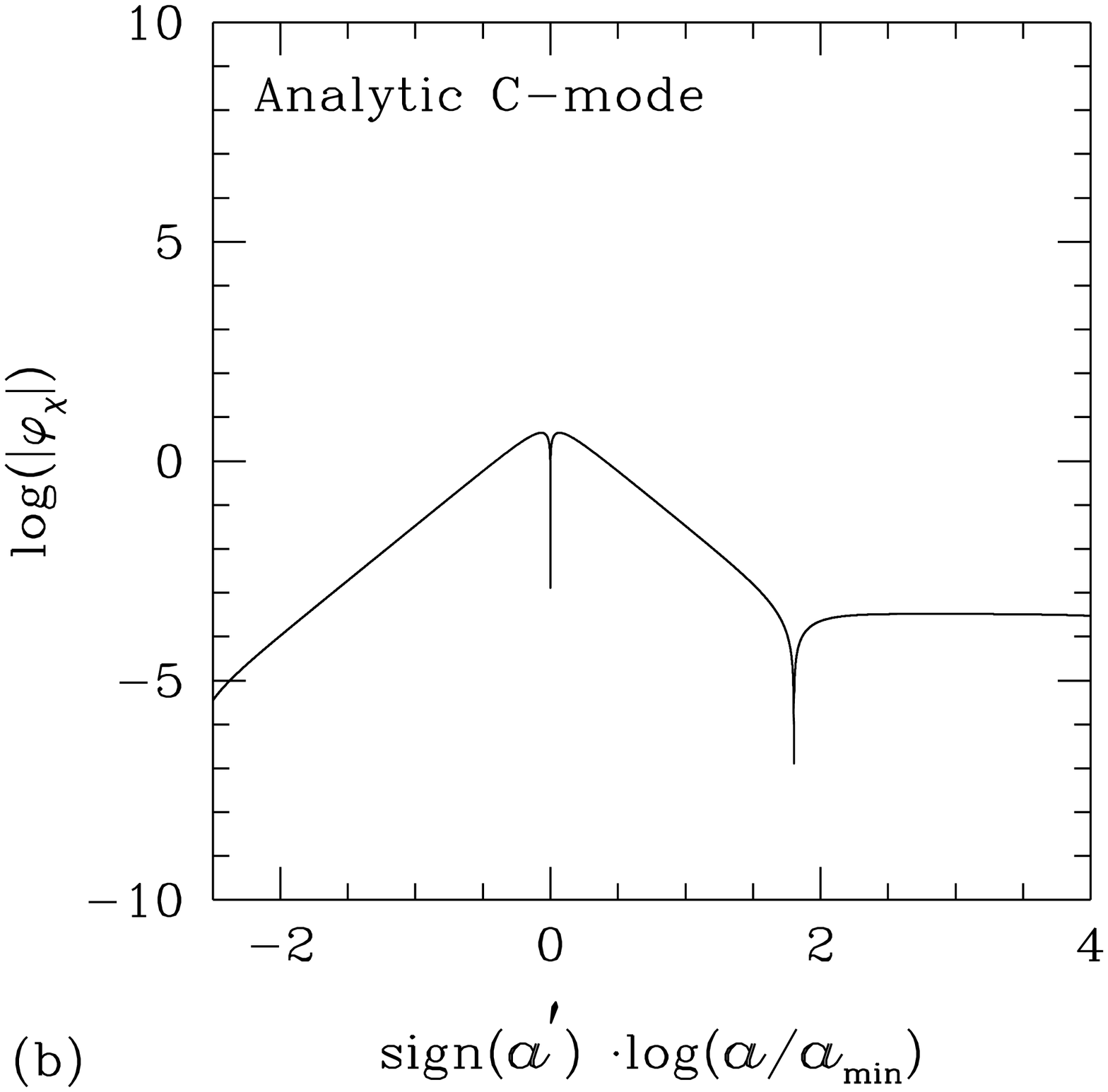}\\
\includegraphics[width=8cm]{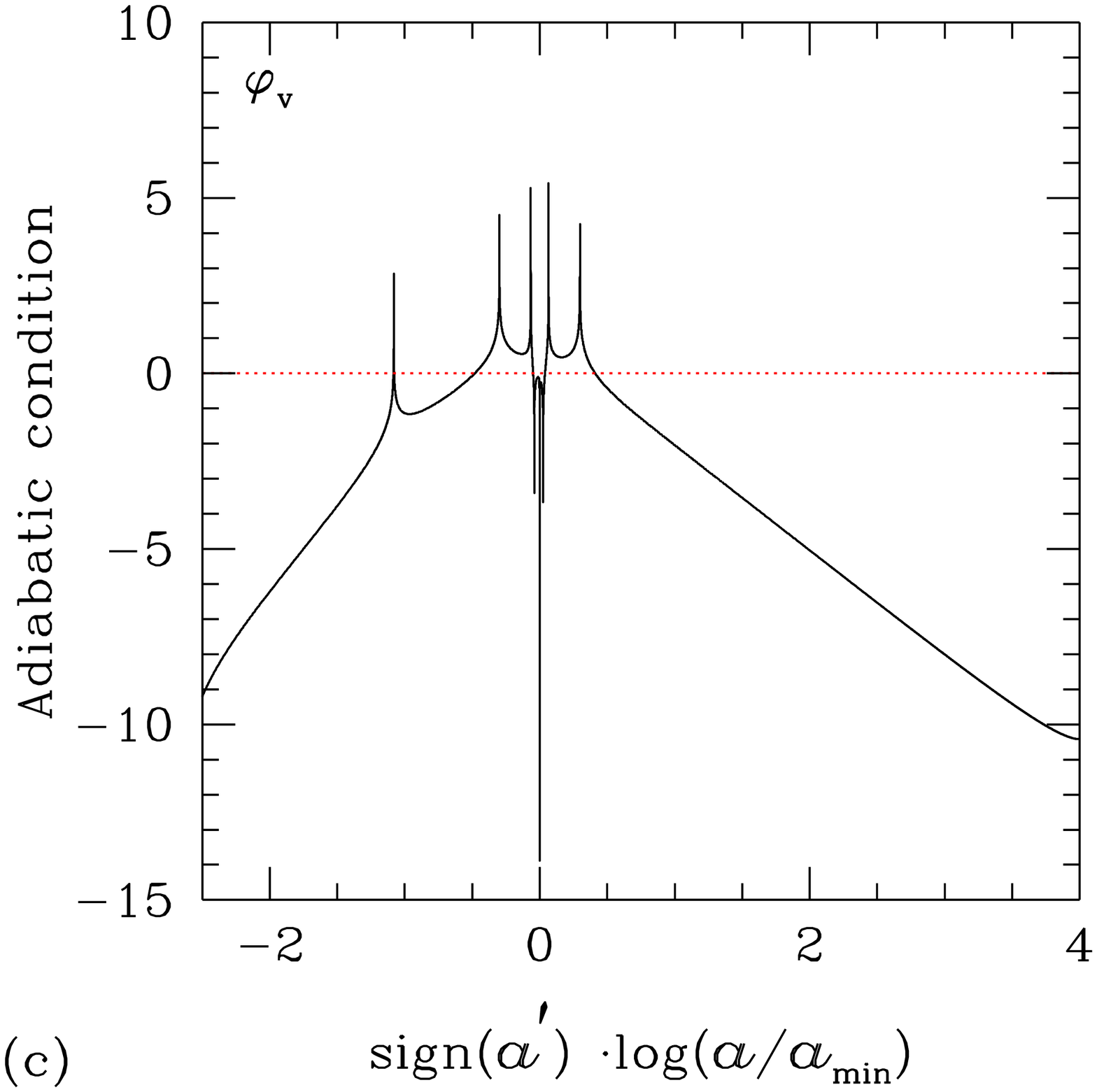}\hfill
\includegraphics[width=8cm]{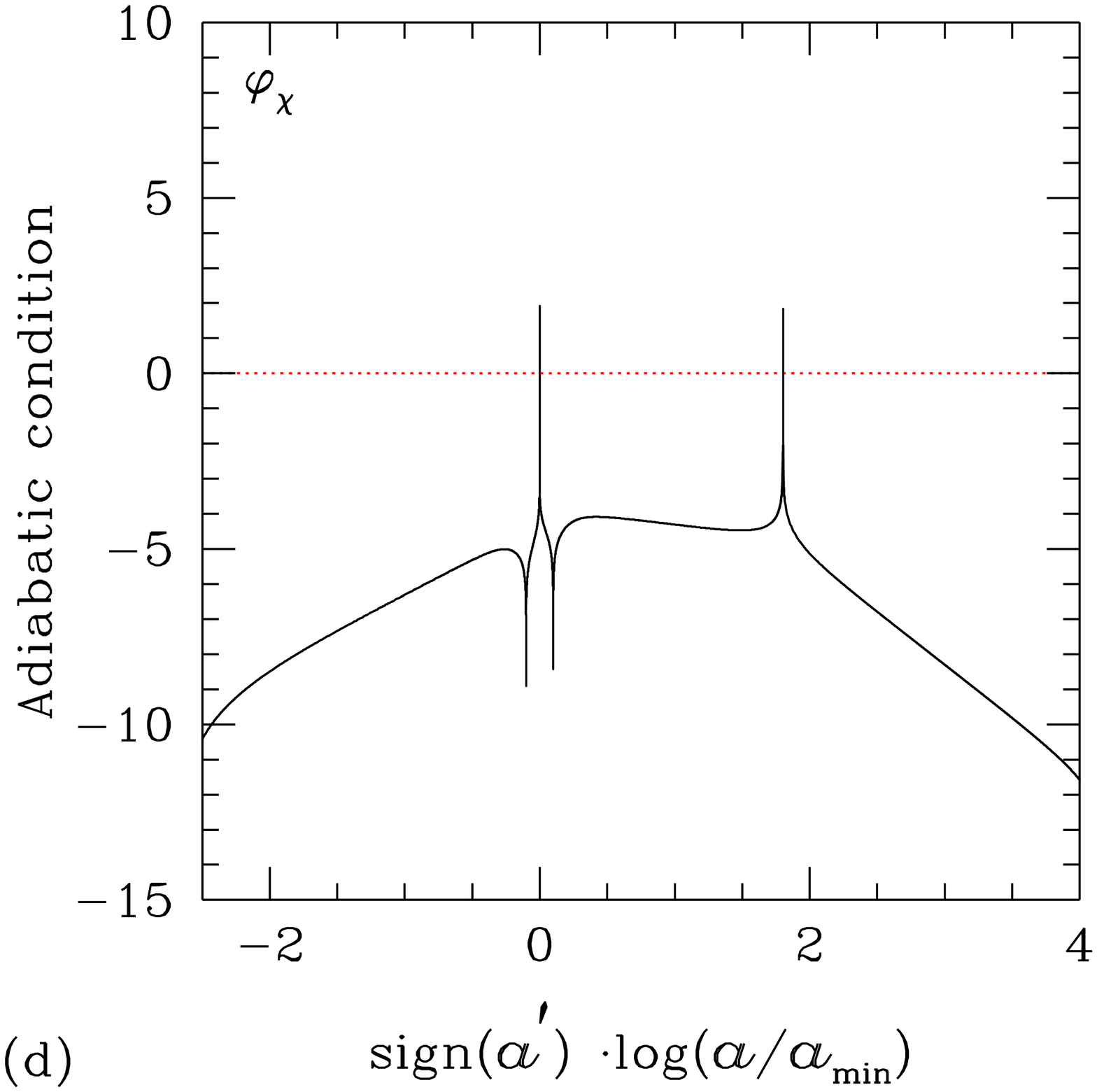}\\
\caption{Evolutions under the `analytic $C$-mode'
         initial condition in Table
         \ref{Table-initiall1} which
         is valid for an ideal situation based on an analytic
         solution.
         This figures is the same as Fig.\ \ref{Fig-c}, now for an
         analytic $C$-mode initial condition.
         Panels (a) and (b) show evolutions of $\varphi_v$ and
         $\varphi_\chi$.
         Panels (c) and (d) show adiabatic conditions for $\varphi_v$ and
         $\varphi_\chi$.
         The large-scale conditions are the same as in Fig.\
         \ref{Fig-LS}.
         Compared with `our $C$-mode' presented in Fig.\
         \ref{Fig-c}, which shows that the $C$-mode nature is
         maintained before and after the bounce,
         the present `analytic $C$-mode' leads to
         mixed behaviors of the $C$- and $d$-modes.
         In the present case, the initial condition in
         Table \ref{Table-initiall1} corresponds to a mixture of
         our precise $C$- and $d$-modes presented in
         Figs.\ \ref{Fig-c} and \ref{Fig-d}, respectively.
         Panels (c) and (d) show that the present initial conditions lead to
         failures of the adiabatic conditions both for
         $\varphi_v$ and $\varphi_\chi$ during the evolution.
         Two sharp spikes away from the bounce in Panels (c) and (d)
         are caused because the curvature variables $\varphi_v$ and
         $\varphi_\chi$ vanish as the dominating
         mode switches between the $C$-mode and the $d$-mode,
         see Panels (a) and (b).
         The sharp spike at the bounce in Panel (d)
         is caused because $\varphi_\chi$ vanishes at the bounce;
         apparently, the behavior of $\varphi_\chi$ shows its
         $d$-mode nature before and after the bounce, thus vanishes
         at $H = 0$, see Eq.\ (\ref{eq-varphichi}).
         As in our $C$-mode in Fig.\ \ref{Fig-c}, such a sharp peak
         does not
         affect the $d$-mode nature of $\varphi_\chi$ at the bounce.
         The adiabatic condition for $\varphi_v$ in Panel (c),
         however, is severely broken near the bounce.
         The evolution of $\varphi_v$ in Panel (a) also shows
         switching of the $d$-mode before the bounce into
         the $C$-mode after the bounce.
         Furthermore, later in Fig.\ \ref{Fig-isocurvature-C-mode}
         we will show that, for the analytic $C$-mode,
         the isocurvature perturbation is significantly
         excited at the bounce.
         Based on these observations we conclude that
         we {\it cannot} trace the $C$-mode after the bounce to the
         one provided initially.
         }
         \label{Fig-anlc}
\end{figure*}

\begin{figure*}
\centering%
\includegraphics[width=8cm]{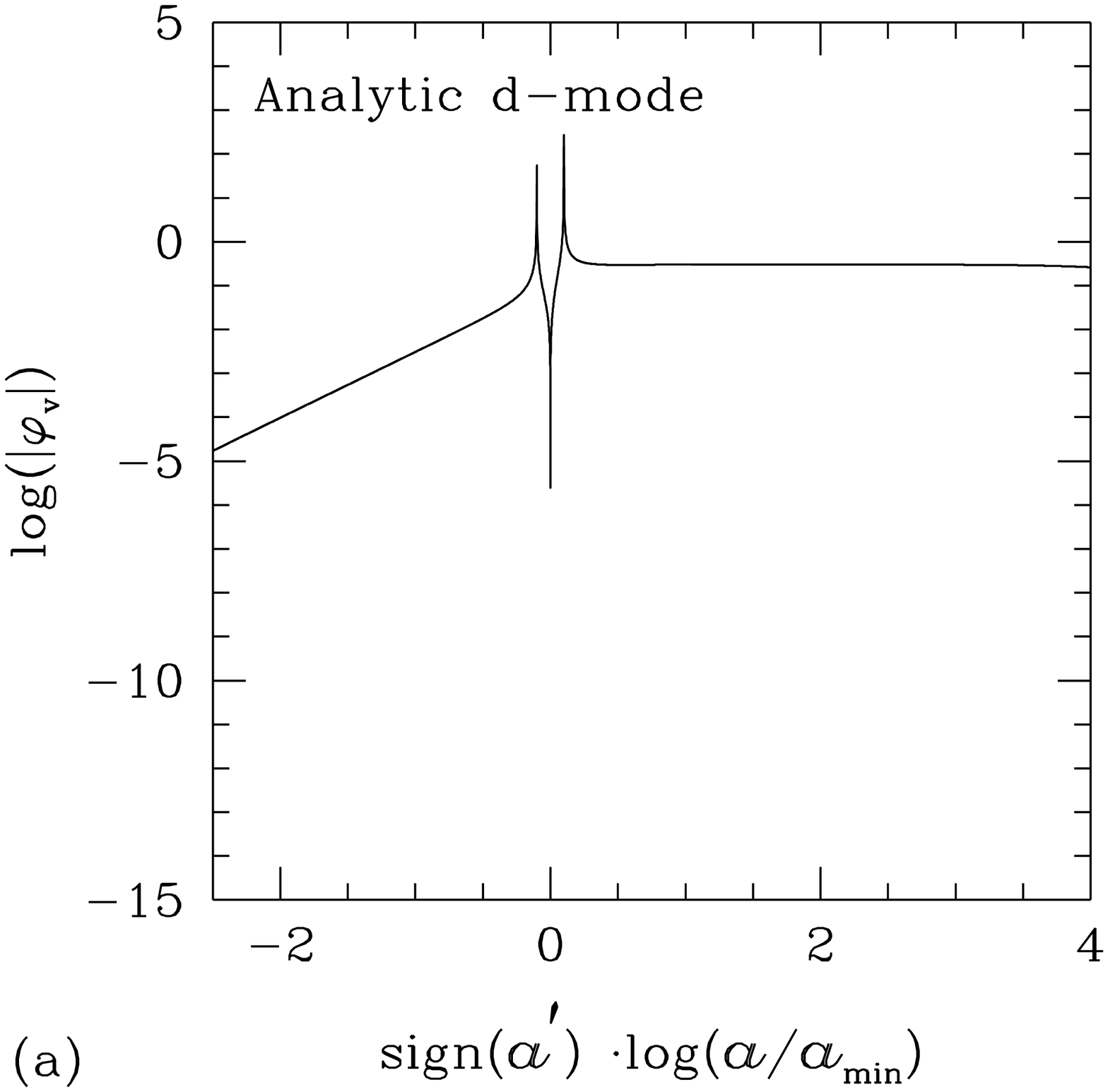}\hfill
\includegraphics[width=8cm]{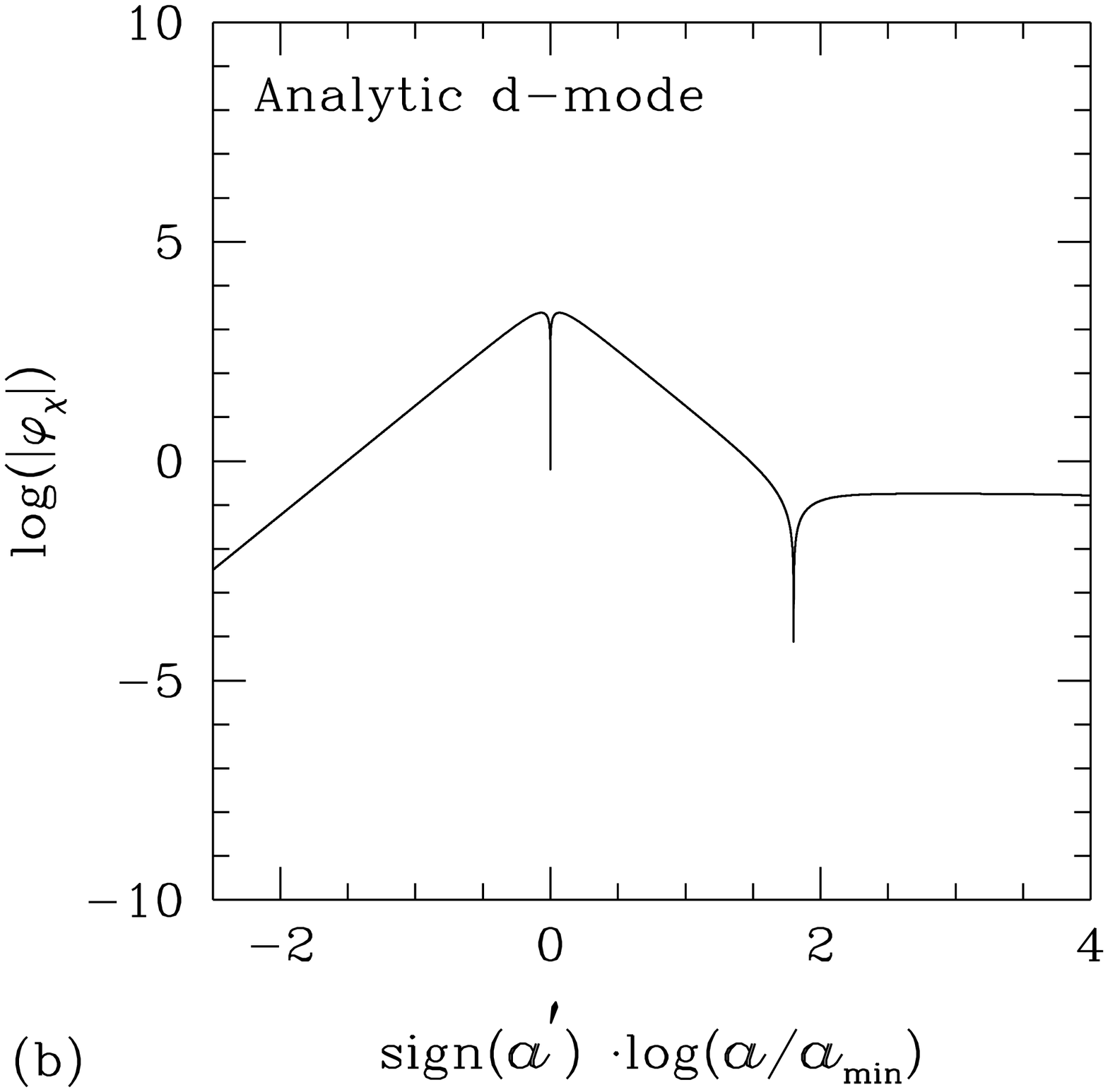}\\
\includegraphics[width=8cm]{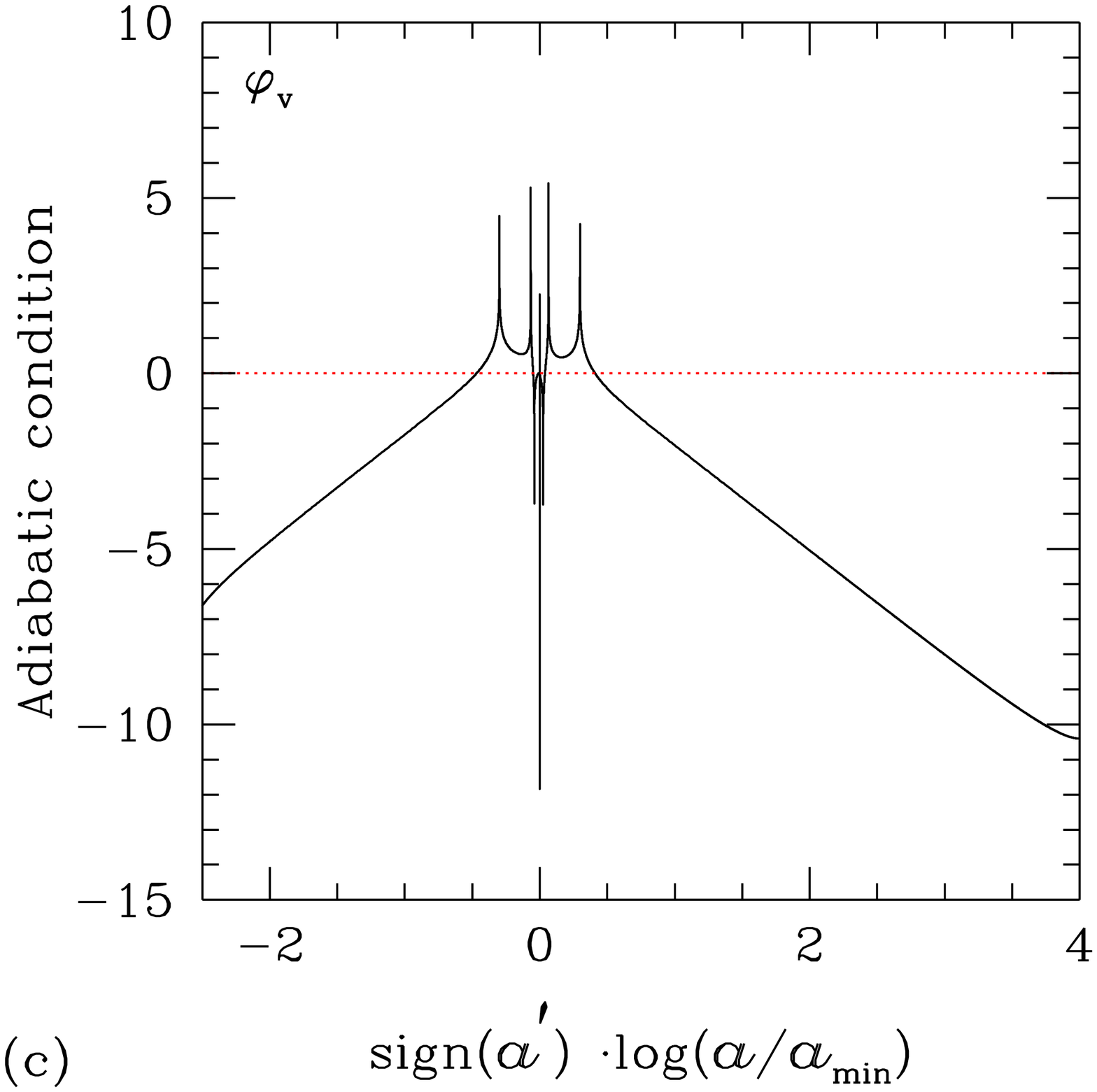}\hfill
\includegraphics[width=8cm]{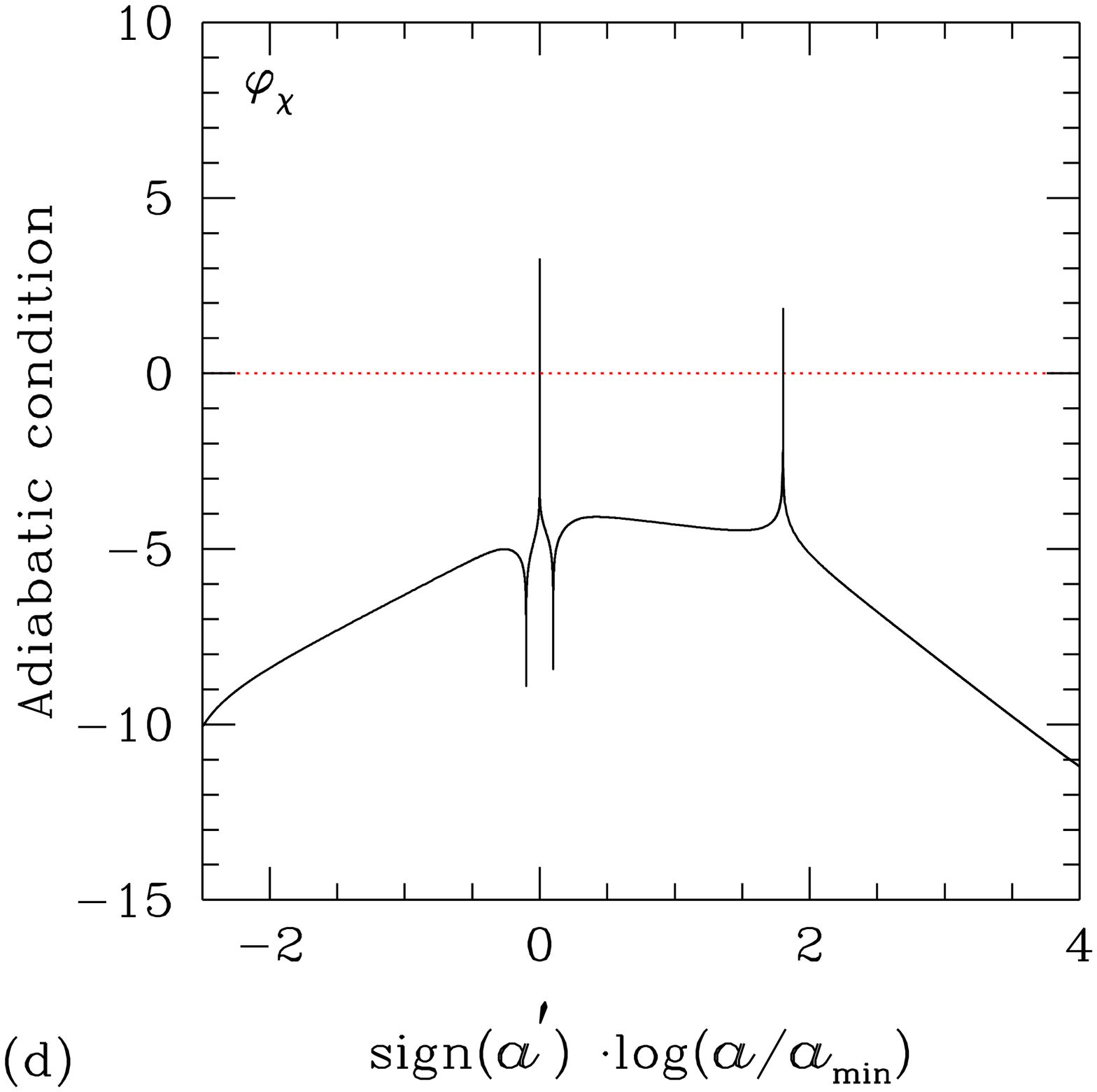}
\caption{The same figures as in Fig.\ \ref{Fig-anlc}, now for the
         analytic
         $d$-mode initial condition in Table \ref{Table-initiall1}.
         Compared with `our $d$-mode' presented in Fig.\
         \ref{Fig-d}, which show that the $d$-mode nature is
         maintained before and after the bounce,
         the present `analytic $d$-mode' leads to
         mixed behaviors of the $C$- and $d$-modes.
         In the present case, the initial condition in
         Table \ref{Table-initiall1} corresponds to a mixture of
         our precise $C$- and $d$-modes presented in
         Figs.\ \ref{Fig-c} and \ref{Fig-d}, respectively.
         Panels (c) and (d) show that the present initial conditions lead to
         failures of the adiabatic conditions both for
         $\varphi_v$ and $\varphi_\chi$ during the evolution.
         One sharp spike away from the bounce in Panel (d)
         is caused because
         $\varphi_\chi$ vanishes as the dominating
         mode switches from $d$-mode to $C$-mode, see Panel (b).
         As explained in Fig.\ \ref{Fig-anlc},
         the sharp spike at the bounce in Panel (d), caused by vanishing
         $\varphi_\chi$ at $H = 0$, does not
         affect the $d$-mode nature of $\varphi_\chi$ at the bounce.
         As in Fig.\ \ref{Fig-anlc}, the adiabatic condition for $\varphi_v$ in Panel (c)
         is severely broken near the bounce.
         The evolution of $\varphi_v$ in Panel (a) also shows
         switching of the $d$-mode before the bounce into
         the $C$-mode after the bounce.
         Later in Fig.\ \ref{Fig-isocurvature-d-mode}
         we will show that, for the analytic $d$-mode,
         the isocurvature perturbation is significantly
         excited at the bounce.
         Based on these observations we conclude that
         we {\it cannot} trace the $d$-mode after the bounce to the
         one provided initially.
         The adiabatic condition for $\varphi_v$ is also
         severely broken for `our $d$-mode' in Fig.\ \ref{Fig-d}.
         In the case of our $d$-mode, in Fig.\ \ref{Fig-d} we have argued that
         this happens because the $d$-mode of $\varphi_v$ is $(k/aH)^2$-order
         smaller than the $d$-mode of $\varphi_\chi$, thus quite small,
         see Eq.\ (\ref{eq-varphiv}).
         Furthermore, in the case of our $d$-mode the isocurvature
         perturbation is not excited near the bounce, despite the
         apparent breakdown of the adiabatic condition for
         $\varphi_v$, see Fig.\ \ref{Fig-isocurvature-d-mode}.
         }
         \label{Fig-anld}
\end{figure*}

In Figs.\ \ref{Fig-anlc}(a), \ref{Fig-anlc}(b), \ref{Fig-anld}(a),
and \ref{Fig-anld}(b) we present some generic behaviors of
imprecisely imposed $C$- and $d$-mode initial conditions. Unless we
set up initial conditions precisely as in `our $C$- and $d$-modes'
presented in Table \ref{Table-initiall2}, it is very likely that the
curvature perturbations behave similar to the ones in Figs.\
\ref{Fig-anlc}(a), \ref{Fig-anlc}(b), \ref{Fig-anld}(a), and
\ref{Fig-anld}(b). What is actually presented in these figures is
based on `analytic $C$- and $d$-mode' initial conditions in Table
\ref{Table-initiall1}. These initial conditions are based on the
large-scale limit solutions of adiabatic perturbation in Eqs.\
(\ref{eq-varphiv}) and (\ref{eq-varphichi}). Thus, these are valid
for an ideal situation assuming exact validity of the adiabatic
condition and the large-scale limit.

In Figs.\ \ref{Fig-anlc}(c), \ref{Fig-anlc}(d), \ref{Fig-anld}(c),
and \ref{Fig-anld}(d) we present the adiabatic conditions. The
large-scale conditions are the same as in Fig.\ \ref{Fig-LS}.
Detailed descriptions of the evolutions and the adiabatic conditions
can be found in the figure captions. Sharp spikes away from the
bounce in the adiabatic conditions in Figs.\ \ref{Fig-anlc}(c),
\ref{Fig-anlc}(d), and \ref{Fig-anld}(d) occur at $\varphi_v = 0$ or
$\varphi_\chi = 0$ which can be identified in Figs.\
\ref{Fig-anlc}(a), \ref{Fig-anlc}(b), and \ref{Fig-anld}(b). The
curvature perturbations vanish at these points as the dominating
mode changes between the $C$-mode and the $d$-mode. Sharp spikes at
the bounce in Figs.\ \ref{Fig-anlc}(d) and \ref{Fig-anld}(d) occur
because $\varphi_\chi = 0$ at the bounce. Even in our $d$-mode case
we have $\varphi_\chi = 0$ at the bounce, see Fig.\ \ref{Fig-d}.
Figures \ref{Fig-anlc}(b) and \ref{Fig-anld}(b) are effectively the
same; although we call Fig.\ \ref{Fig-anlc}(b) an analytic $C$-mode,
apparently, the initial condition is dominated by the $d$-mode from
the beginning.

\begin{figure*}
\centering%
\includegraphics[width=8cm]{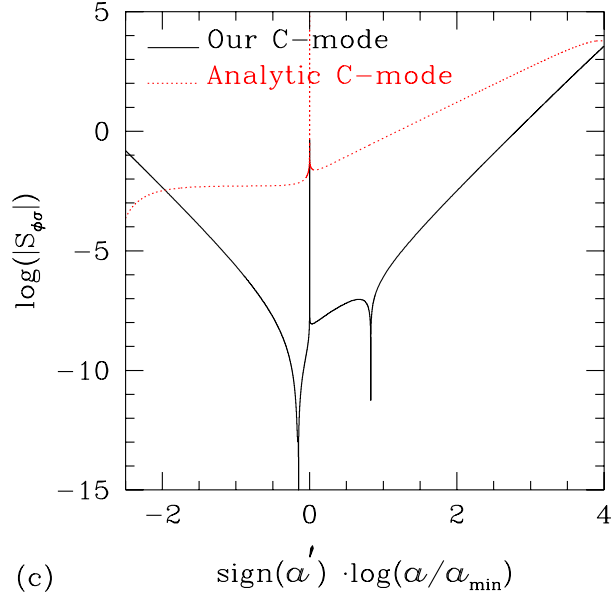}
\caption{Evolutions of the isocurvature perturbation variable
         $S_{\phi\sigma}$
         for `our $C$-mode' (solid line) and the `analytic $C$-mode' (dotted line).
         Panel (b) magnifies Panel (a) near the bounce.
         Panel (c) shows Panel (a) in logarithmic scale.
         The sharp divergences at the bounce occur because we have $\dot \phi =
         0$ at the bounce.
         For our $C$-mode, $S_{\phi\sigma}$ nearly vanishes near
         $\log{(a/a_{min})} \simeq -.2$; this is because we have found
         our $C$-mode initial condition beginning at this epoch, see
         the text for explanation.
         Although Figs.\ \ref{Fig-c}(c) and (d) show that the adiabatic conditions for
         our $C$-modes of $\varphi_v$ and $\varphi_\chi$ are broken by sharp
         peaks near the bounce, the above figures show that the isocurvature
         variable $S_{\phi\sigma}$ show smooth behavior near the bounce.
         As explained in Fig.\ \ref{Fig-c}, such sharp divergences
         occur because we have $\mu + p = 0$ near the bounce and
         $\varphi_\chi$ crosses zero twice near the bounce, see Fig.\
         \ref{Fig-c}(b).
         In case of the analytic $C$-mode, Figs.\ \ref{Fig-anlc}(c) and (d)
         show that the adiabatic conditions are severely broken near
         the bounce, and are sharply broken in two places away from the bounce.
         As explained in Fig.\ \ref{Fig-anlc}, the two sharp spikes away from
         the bounce occur because of the change in dominating mode between
         $C$-mode
         and $d$-mode, where the curvature variables $\varphi_v$ or
         $\varphi_\chi$ vanish.
         The above figures show that for the analytic $C$-mode the
         isocurvature variable significantly changes its behavior at the
         bounce.
         Based on these behaviors, we argue that for
         our $C$-mode initial condition although the adiabatic conditions
         are broken by sharp peaks near the bounce, because the isocurvature
         perturbation shows the smooth behavior, the isocurvature
         perturbation does not give physical impact on the adiabatic nature
         of our $C$-mode throughout the bounce.
         Thus, although our $C$-mode initial condition shows
         apparently non-vanishing isocurvature mode (see Table
         \ref{Table-initiall2}), we argue that our $C$-mode initial
         condition keeps the adiabatic nature throughout the bounce.
         On the other hand, for the analytic $C$-mode,
         the adiabatic conditions are severely broken near the
         bounce,
         and the isocurvature perturbation shows sharp change at the
         bounce. Thus, we argue that the isocurvature perturbation
         gives physical impact on the adiabatic nature
         of the analytic $C$-mode;
         this analytic $C$-mode is based on an initial condition with
         `precise'
         adiabatic condition (see Table \ref{Table-initiall1}).
         That is, for the analytic $C$-mode, although the initial
         condition is imposed assuming the adiabatic condition is exactly met,
         as the evolution proceeds the isocurvature perturbation
         is also excited, especially near the bounce.
         Thus, the numerical evolution shows that after the bounce the analytic
         initial $C$-mode in fact becomes a substantial mixture of the adiabatic and
         isocurvature modes.
         }
         \label{Fig-isocurvature-C-mode}
\end{figure*}

\begin{figure*}
\centering%
\includegraphics[width=8cm]{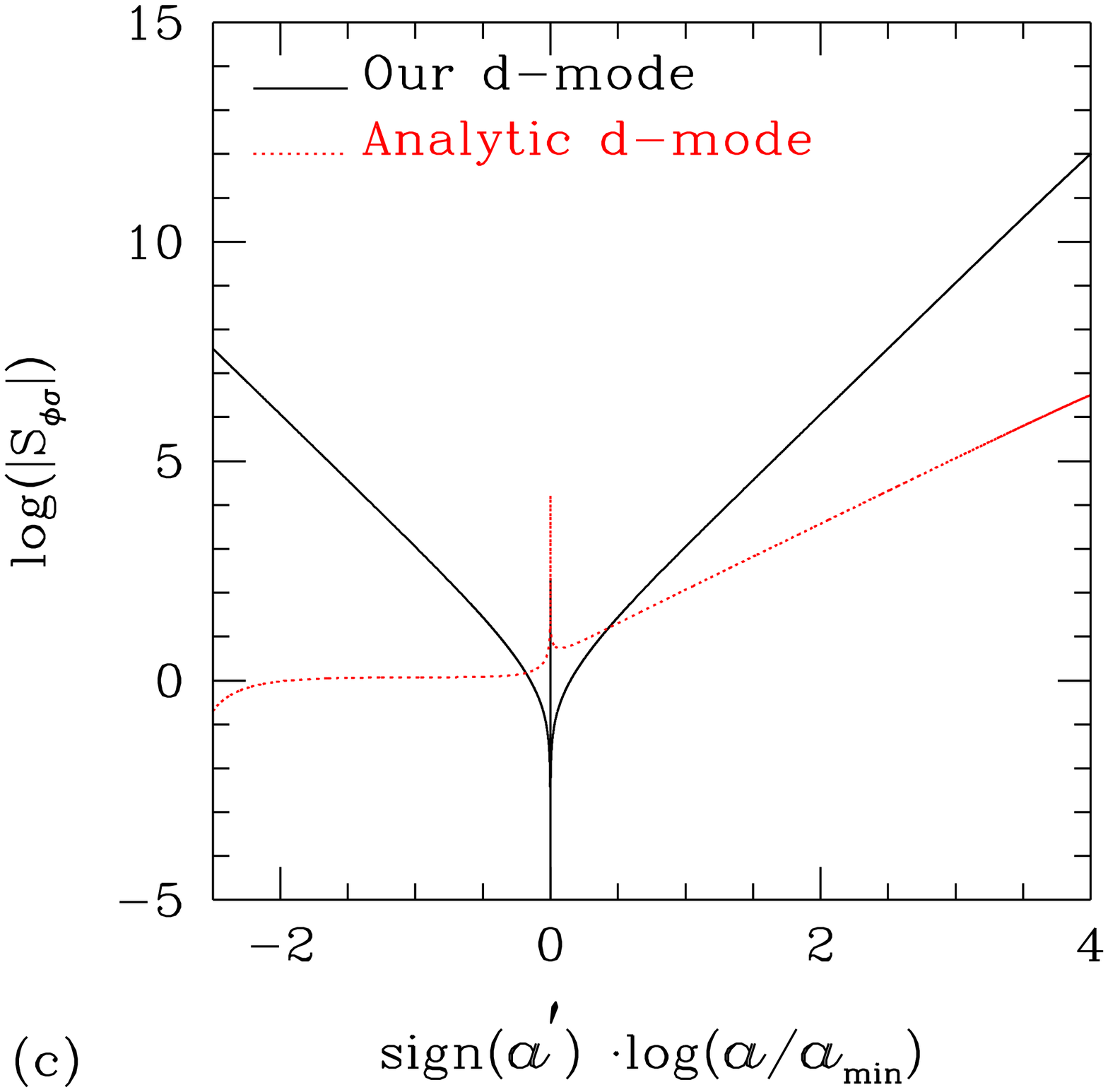}
\caption{Evolutions of the isocurvature perturbation variable
         $S_{\phi\sigma}$
         for our $d$-mode (solid line) and the analytic $d$-mode (dotted line).
         Panel (b) magnifies Panel (a) near the bounce.
         Panel (c) shows Panel (a) in logarithmic scale.
         The sharp
         divergences at the bounce occur because we have $\dot \phi =
         0$ at the bounce.
         For our $d$-mode $S_{\phi\sigma}$ nearly vanishes just
         before bounce near
         $\log{(a/a_{min})} \simeq 0.$; this is because we have found
         our $d$-mode initial condition beginning at this epoch, see
         the text for explanation.
         Figure \ref{Fig-d} (c) shows that the adiabatic conditions for
         our $d$-modes of $\varphi_v$ are severely broken
         near the bounce.
         However, in Fig.\ \ref{Fig-d} we argued that
         the $d$ mode of $\varphi_v$ is $(k/aH)^2$-order higher
         in the large-scale expansion compared with the
         $d$-mode of $\varphi_\chi$; i.e.,
         the leading order $d$-mode of $\varphi_v$ has been canceled out,
         see Eqs.\ (\ref{eq-varphiv})-(\ref{eq-varphichie}).
         Notice that in Fig.\ \ref{Fig-d}(d) the adiabatic condition for
         $d$-mode of $\varphi_\chi$ is well satisfied.
         Based on this observation, in Fig.\ \ref{Fig-d} we have argued that
         despite the apparent severe breaking of the adiabatic condition
         for $\varphi_v$ the adiabatic nature of our $d$-mode
         initial condition is preserved throughout the bounce.
         In order to reinforce our such an argument in this figure
         we show the evolution of isocurvature perturbation variable $S_{\phi\sigma}$.
         The above figures show smooth behavior of our $d$-mode
         near the bounce.
         In case of the analytic $d$-mode, Figs.\ \ref{Fig-anld}
         (c) and (d) show that the adiabatic condition for $\varphi_v$
         is severely broken near the bounce, and the adiabatic condition
         for $\varphi_\chi$ is sharply broken at the bounce and away from the bounce.
         As explained in Fig.\ \ref{Fig-anld}, one sharp spike for $\varphi_\chi$
         away from
         the bounce occurs because of the change of dominating mode between $d$
         and $C$ mode, where the curvature variable $\varphi_\chi$ vanishes.
         However, in contrast with our $d$-mode,
         the above figure shows that for the analytic $d$-mode the
         isocurvature perturbation variable significantly changes its behavior at the
         bounce.
         Based on these behaviors we argue the following.
         Although the adiabatic
         condition for $\varphi_v$ of our $d$-mode initial condition
         is severely broken near the bounce, because the isocurvature
         perturbation shows the smooth behavior, the isocurvature
         perturbation does not give physical impact on the adiabatic nature
         of our $d$-mode throughout the bounce.
         Thus, although our $d$-mode initial condition shows
         apparently non-vanishing isocurvature mode (see Table
         \ref{Table-initiall2}), we argue that our $d$-mode initial
         condition keeps the adiabatic nature throughout the bounce.
         On the other hand, for the analytic $d$-mode both
         the adiabatic conditions are severely broken near the bounce
         and the isocurvature perturbation shows sharp change at the
         bounce.
         Thus, we argue that the isocurvature perturbation
         gives the physical impact on the adiabatic nature
         of the analytic $d$-mode;
         this analytic $d$-mode is based on an initial condition with
         `precise'
         adiabatic condition (see Table \ref{Table-initiall1}).
         That is, for the analytic $d$-mode, although the initial
         condition is imposed assuming the adiabatic condition is exactly met,
         as the evolution proceeds the isocurvature perturbation
         is also excited, especially near the bounce.
         Thus, the numerical evolution shows that after the bounce the analytic
         initial $d$-mode in fact becomes a substantial mixture of the adiabatic and
         isocurvature modes.
         }
         \label{Fig-isocurvature-d-mode}
\end{figure*}

The adiabatic conditions for $\varphi_v$ in \ref{Fig-anlc}(c) and
\ref{Fig-anld}(c) are severely broken near the bounce. Evolutions of
$\varphi_v$ in Figs.\ \ref{Fig-anlc}(a) and \ref{Fig-anld}(a) also
show switching of the $d$-mode before the bounce into the $C$-mode
after the bounce. Furthermore, Figures \ref{Fig-isocurvature-C-mode}
and \ref{Fig-isocurvature-d-mode} show that, for the analytic $C$-
and $d$-modes, the isocurvature perturbation is significantly
excited at the bounce. Based on these observations we conclude that
we {\it cannot} trace the $C$-modes after the bounce to the
$d$-modes before the bounce in Figs.\ \ref{Fig-anlc}(a) and
\ref{Fig-anld}(a). This conclusion should be compared with the one
made in our $d$-mode case. Although, the adiabatic condition for our
$d$-mode is also severely broken near the bounce [see Fig.\
\ref{Fig-d}(c)], there we have argued how this could happen for the
$d$-mode of $\varphi_v$ without affecting the $d$-mode nature before
and after the bounce. In our $d$-mode case, Fig.\ \ref{Fig-d}(a)
shows that the $d$-mode nature is preserved before and after the
bounce, and Fig.\ \ref{Fig-isocurvature-d-mode} shows that the
isocurvature perturbation is not excited near the bounce.

The evolutions show that each mode is soon dominated by the
relatively growing mode (the $d$-mode in a collapsing phase, and the
$C$-mode in an expanding phase are the relatively growing). In fact,
these analytic initial conditions correspond to mixtures of our
precise $C$- and $d$-mode initial conditions. Based on these
observations we conclude that we {\it cannot} trace the final
$C$-modes after the bounce in all these figures to the ones provided
initially.

\begin{table}
\caption{The analytic initial conditions in Eq.\
         (\ref{eq-connectioncd}).
         This initial condition
         is valid for an ideal situation
         assuming exact validity of the adiabatic condition, and
         taking the leading order solution in large-scale expansion
         presented in Eqs.\ (\ref{eq-varphiv}) and (\ref{eq-varphichi}).
         We call these `analytic $C$- and $d$-mode initial conditions',
         or simply `analytic $C$- and $d$-modes'.
         We take
         $k=0.0005$ and the initial $\eta=-500$. The numerical results based
         on these initial conditions are presented in Figs.\
         \ref{Fig-anlc} and \ref{Fig-anld}.
         }
         \label{Table-initiall1}
\begin{center}
\begin{tabular}{r r r} \hline\hline
       variable &  C-mode  & d-mode  \\ \hline
       $\delta\phi_{\delta\sigma}$ & 0. & 0. \\
       $\delta\phi_{\delta\sigma}^\prime$ & 0. & 0. \\
       $\varphi_{\delta\sigma}$ & 1.0E$-$05 & 1.0E$-$05 \\
       $\varphi_{\delta\sigma}^\prime$ & 0.
       & $-$3./$\eta \times \varphi_{\delta\sigma}$ \\
       $\varphi_v$  &1.0E$-$05  & 1.0E$-$05 \\
       $\varphi_v^\prime$  &0.  &  $-$3./$\eta \times \varphi_{v}$\\
                                \hline\hline
\end{tabular}
\end{center}
\end{table}


\begin{figure*}
\centering%
\includegraphics[width=8cm]{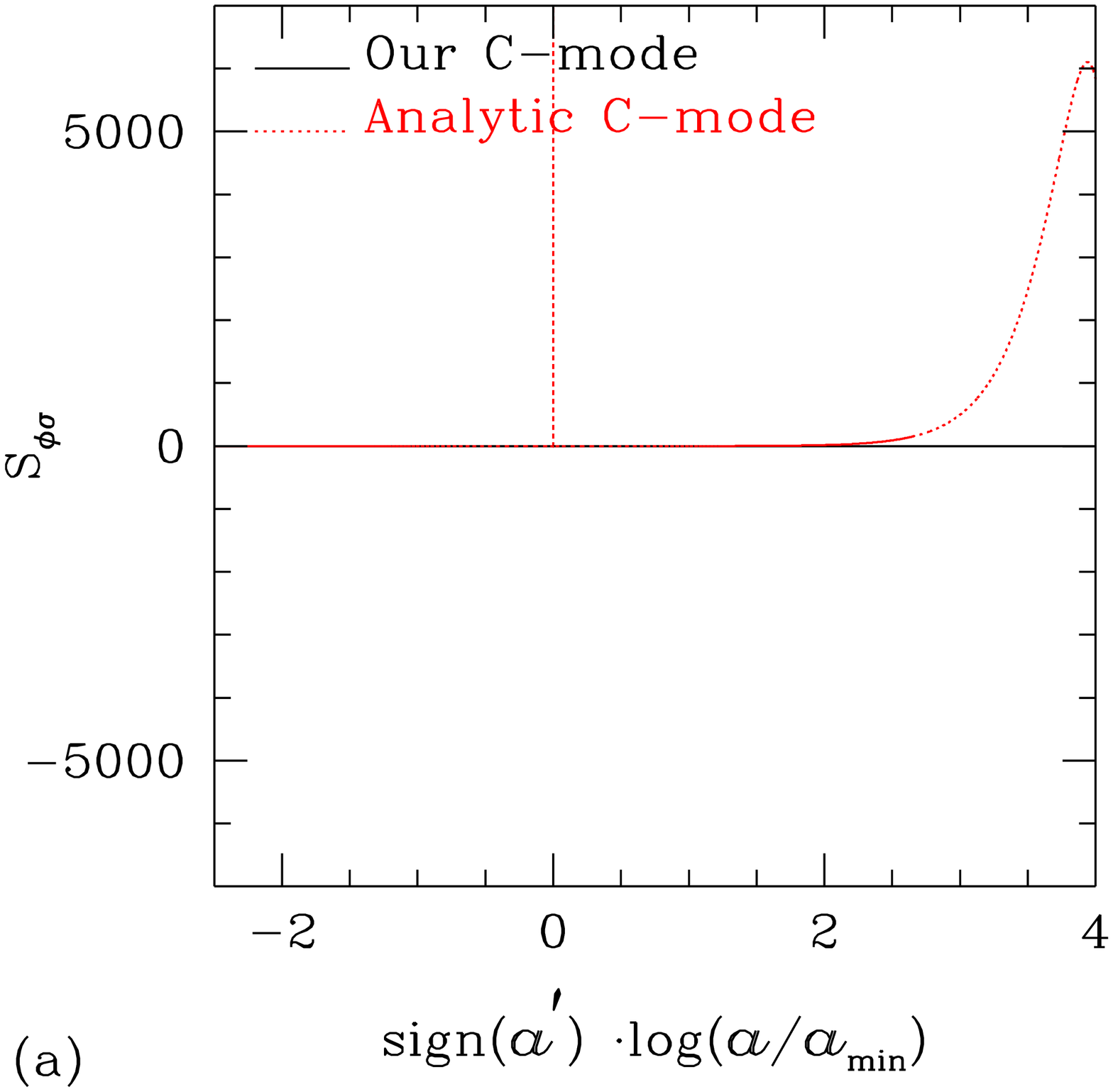}\hfill
\includegraphics[width=8cm]{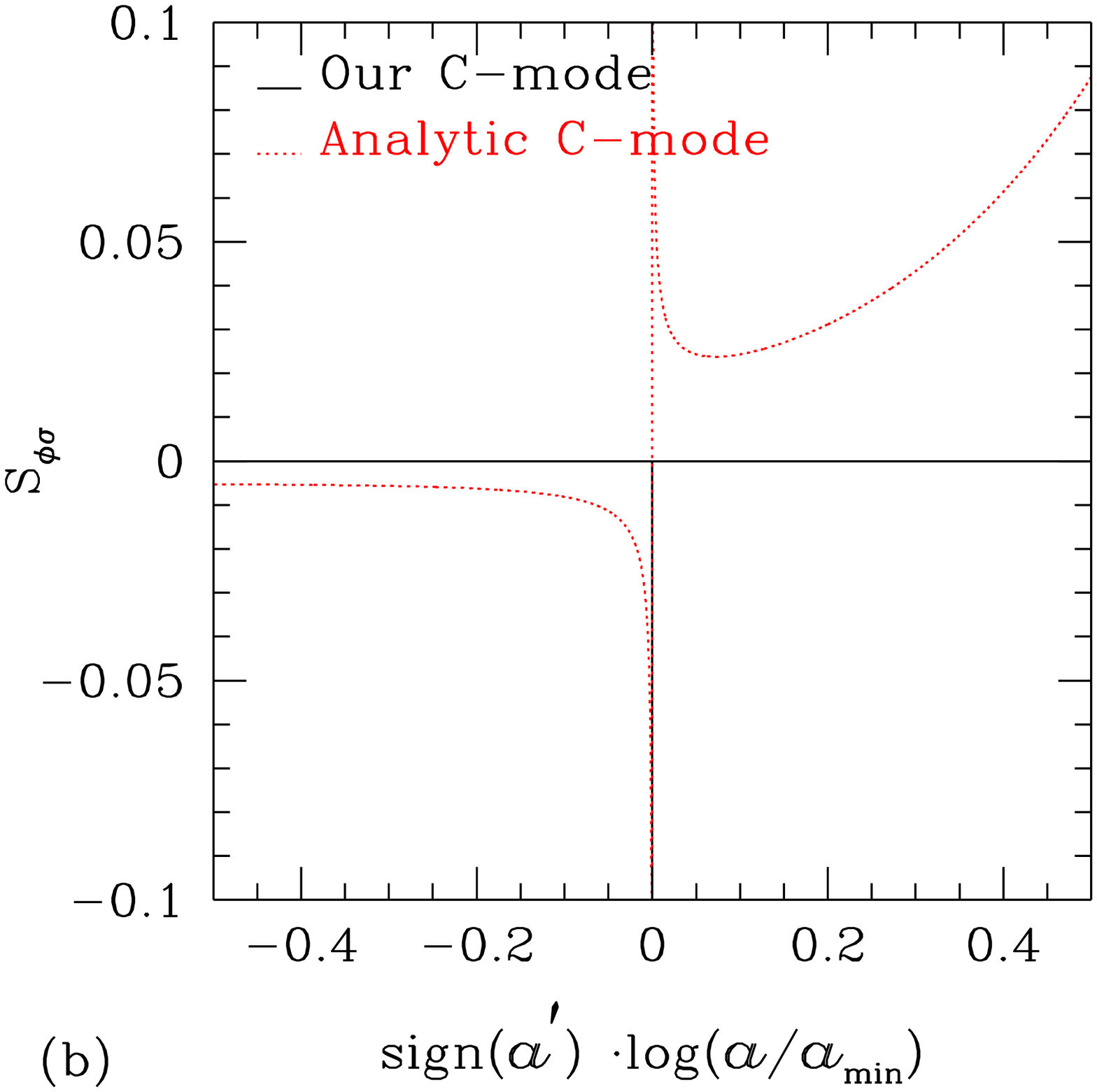}\\
\end{figure*}
\begin{figure*}
\centering%
\includegraphics[width=8cm]{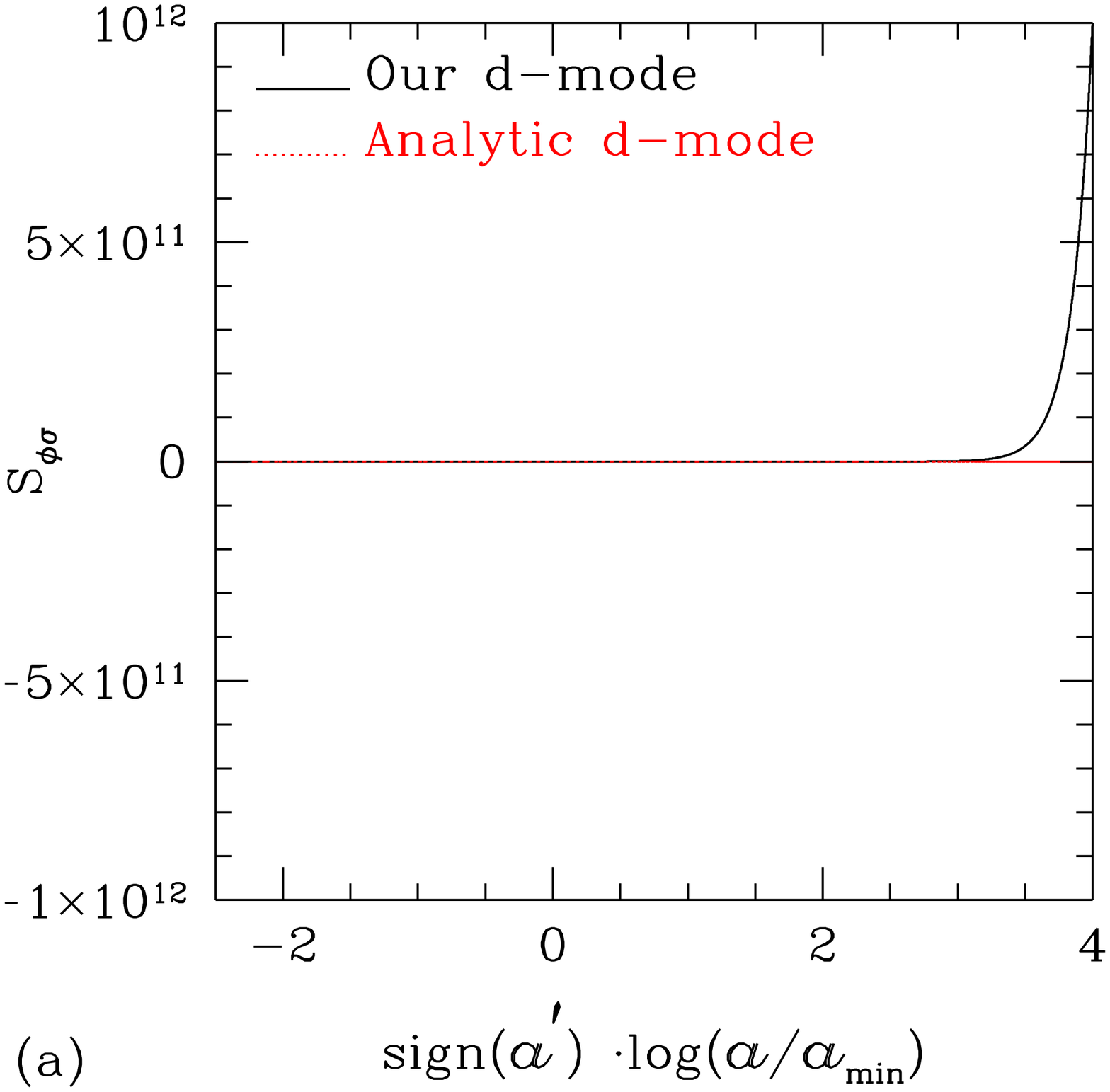}\hfill
\includegraphics[width=8cm]{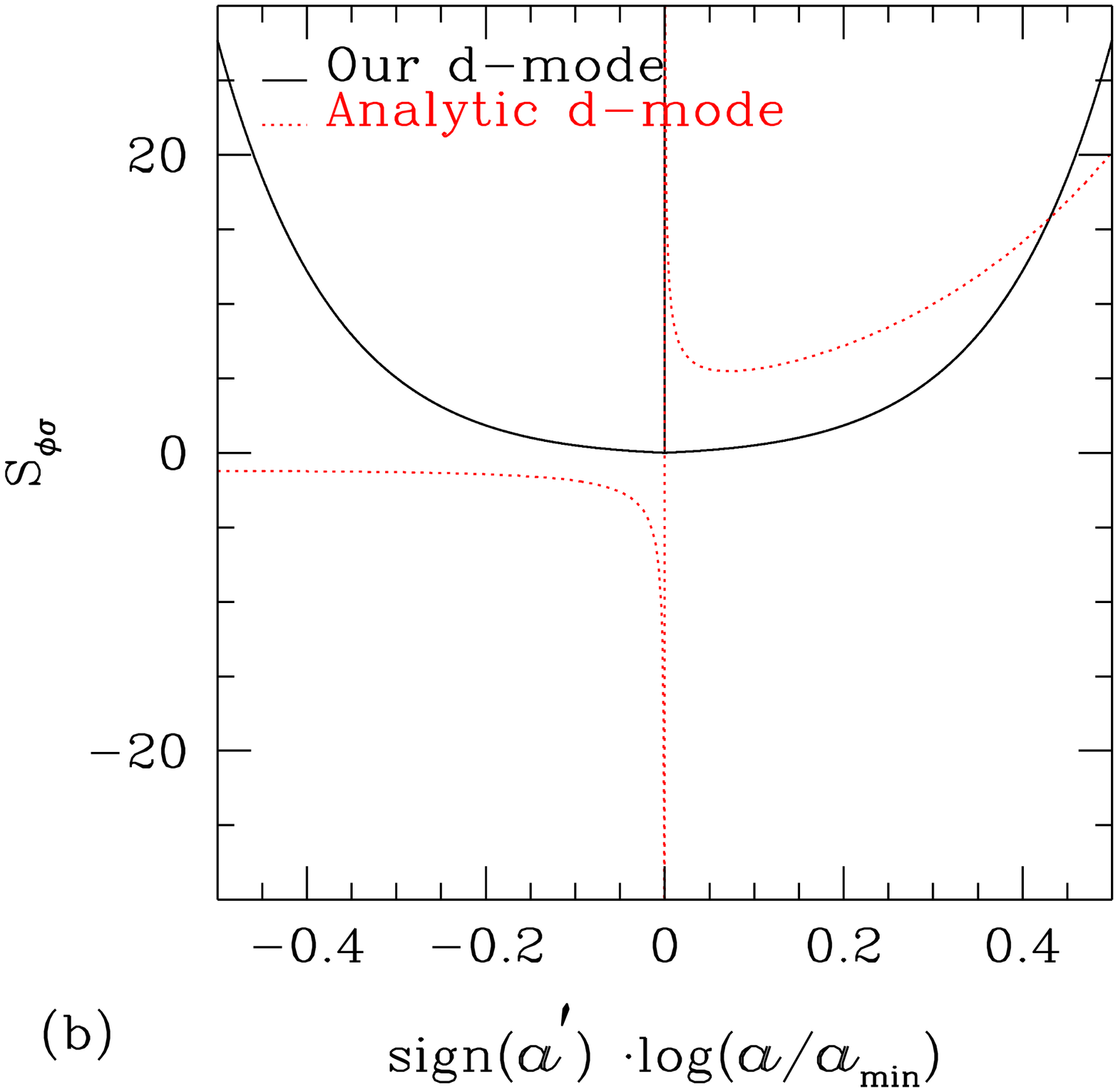}\\
\end{figure*}

\subsubsection{Method of finding our $C$- and $d$-modes}
                                            \label{scalar-numerical-IC}

Here, we explain how we found `our $C$- and $d$-mode initial
conditions' which lead to correct behaviors of the $C$- and $d$-mode
throughout the bounce. Since the $d$-mode is rapidly growing in a
collapsing phase, thus dominating, we can easily find $d$-mode in
that phase; similarly it is easy to find $C$-mode in an expanding
phase. However, unless we impose a quite precise initial condition,
it is very likely that, the relatively growing mode in that phase
soon dominates the evolution. Thus, after the bounce a $d$-mode
before the bounce is likely to switch to a dominating mode in an
expanding phase which is the $C$-mode. We have shown that our $C$-
and $d$-modes can be identified as the proper $C$- and $d$-mode
which preserve their nature before and after the bounce. An
imprecise initial condition, which tends to reproduce only the
dominating solutions, can be regarded as a mixture of our (precise)
$C$- and $d$-mode initial conditions. In Sec.\
\ref{sec:analytic-IC}, we showed that such a mixed initial condition
leads to the break down of the adiabatic condition near the bounce
and excitation of the iscurvature perturbation. Without a precise
initial condition, it is hard to maintain $d$-mode nature after the
bounce. Similarly, it is quite difficult to find $C$-mode initial
condition which maintains its $C$-mode nature during the collapsing
phase.

In order to find the precise initial condition for the $d$-mode we
begin with the analytic initial condition imposed close to the
bounce, and integrate forward and backward in time; the analytic
initial condition is presented in Eq.\ (\ref{eq-connectioncd}) which
is based on Eqs.\ (\ref{eq-varphiv}) and (\ref{eq-varphichi}). Our
$d$-mode initial condition found in this way is presented in Table
\ref{Table-initiall2}. As we have shown this $d$-mode initial
condition leads to ``the $d$-mode'' which maintains its $d$-mode
natures even after the bounce; see Fig.\ \ref{Fig-d}(a) and
\ref{Fig-d}(b). Since the analytic initial condition in Eq.\
(\ref{eq-connectioncd}) assumes the exact adiabatic condition, the
isocurvature perturbation of our $d$-mode naturally vanishes near
the bounce, i.e., $S_{\phi\sigma} = 0$; see Fig.\
\ref{Fig-isocurvature-d-mode}.

\begin{figure*}
\centering%
\includegraphics[width=8cm]{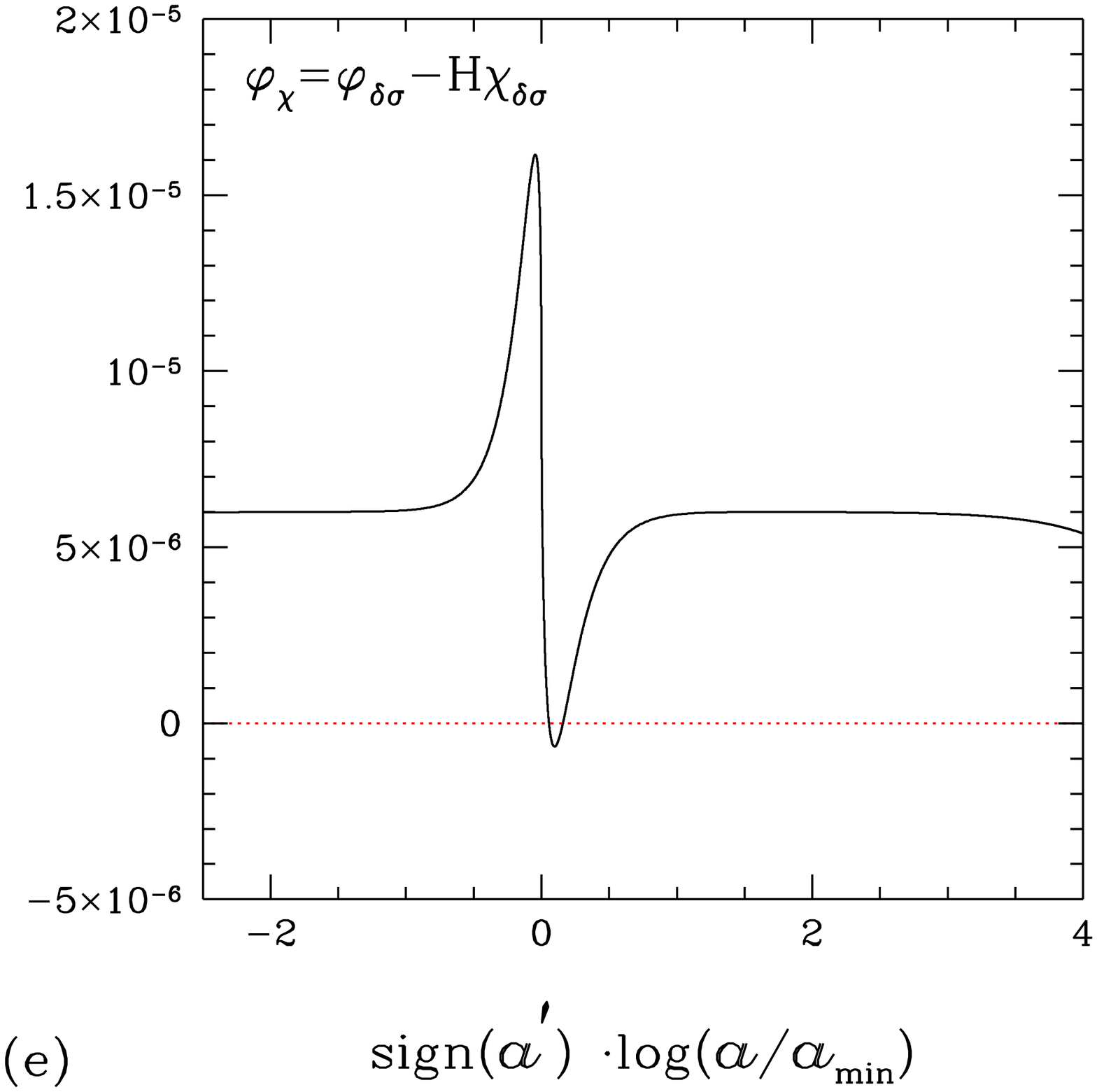}\hfill
\includegraphics[width=8cm]{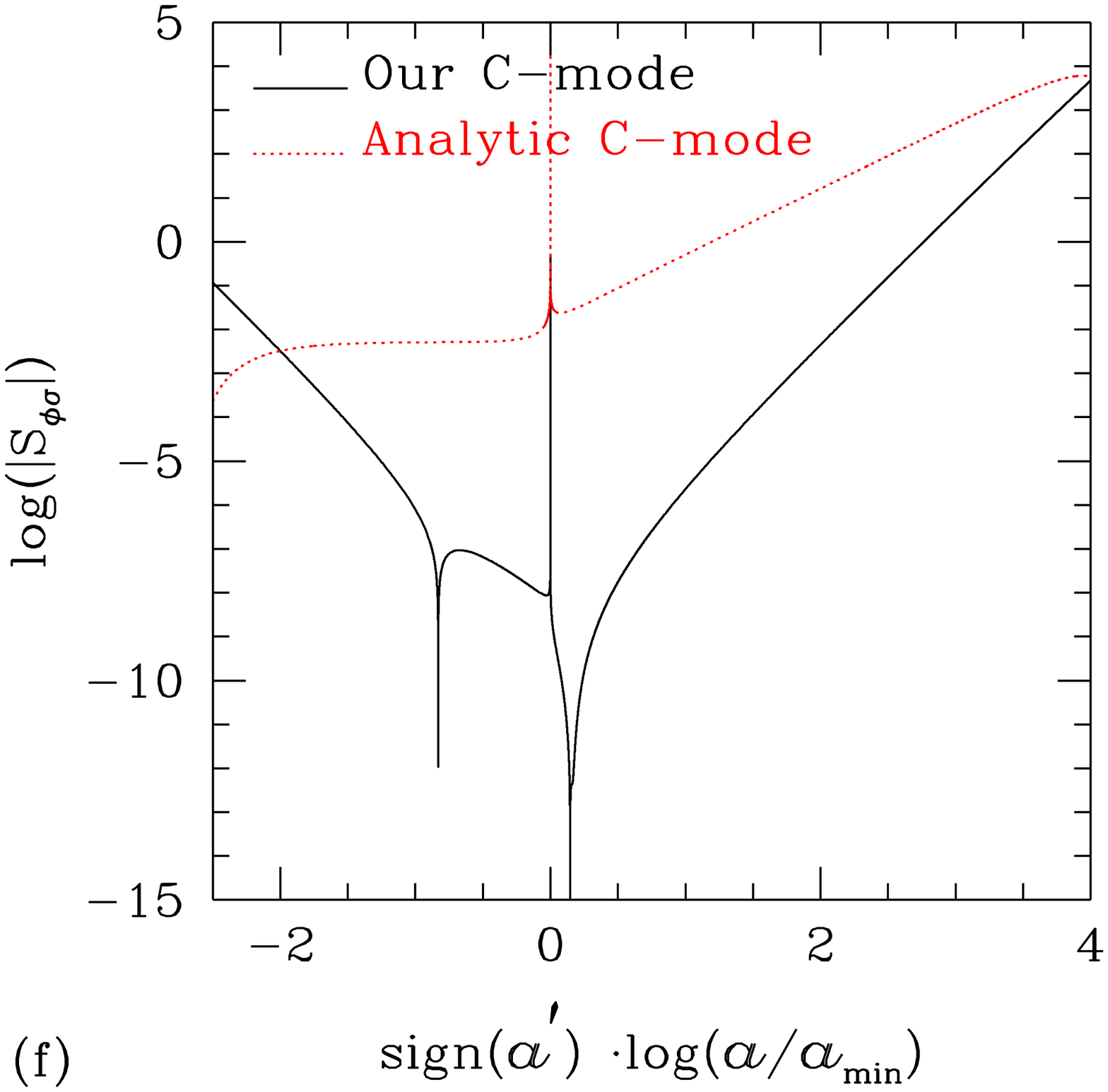}\\
\caption{The same $C$-mode as in Figs.\
         \ref{Fig-c} and \ref{Fig-usecalc} now using slightly different
         way of finding our $C$-mode initial condition explained in
         the text.
         Panels (a)-(d) correspond to Figs.\ \ref{Fig-c}(a)-(d).
         Panel (e) corresponds to Fig.\ \ref{Fig-usecalc}(f).
         Panel (f) corresponds to Fig. \ref{Fig-isocurvature-C-mode}(c).
         Near the bounce these figures show reflection symmetry at the
         bounce compared with
         Figs.\ \ref{Fig-c}, \ref{Fig-usecalc}(f), and \ref{Fig-isocurvature-C-mode}(c).
         Figs.\ \ref{Fig-usecalc}(a), \ref{Fig-usecalc}(b), and \ref{Fig-usecalc}(e) almost remain the same.
         }
\label{Fig-cafter}
\end{figure*}

\begin{figure*}
\centering%
\includegraphics[width=8cm]{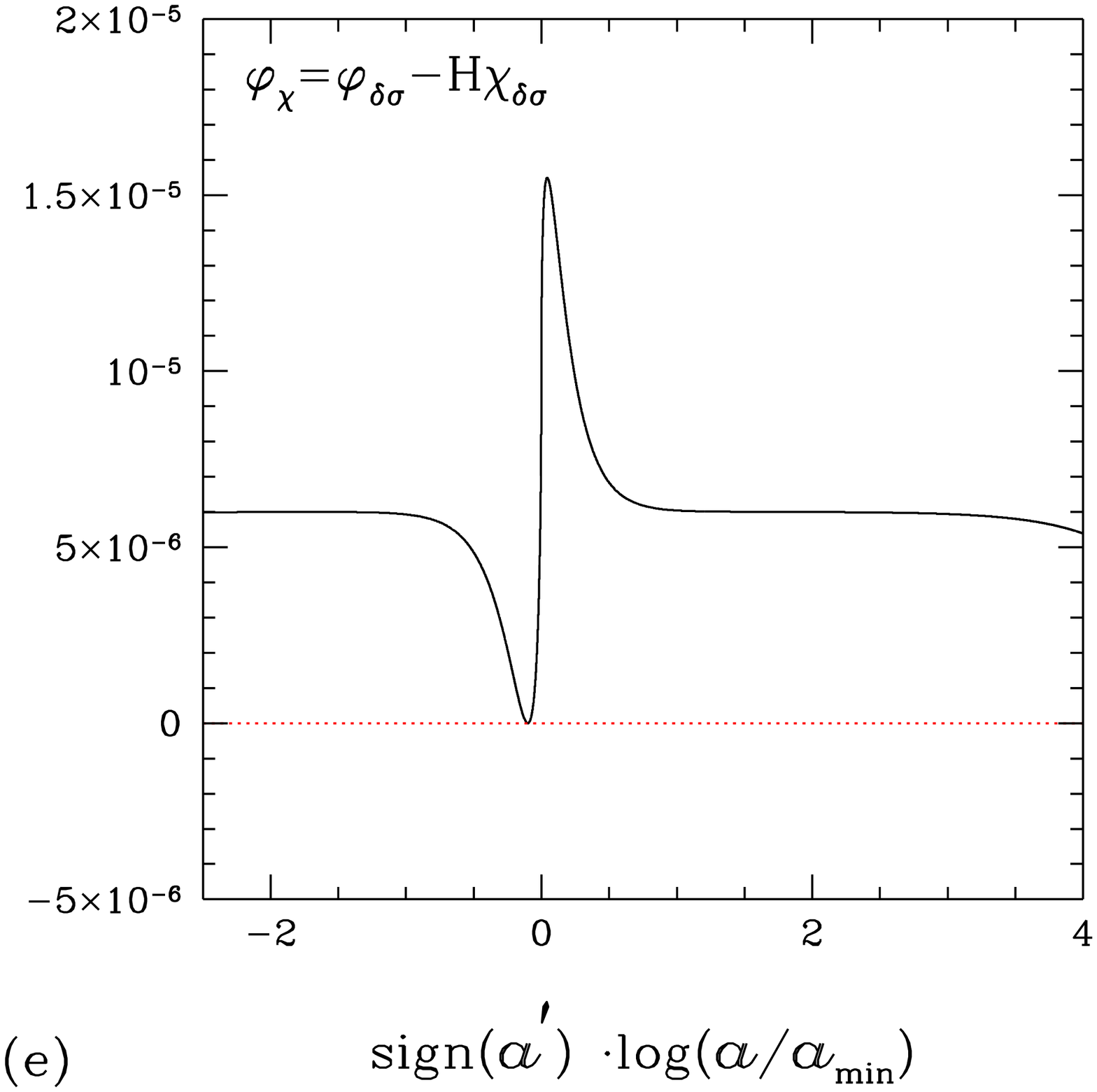}\hfill
\includegraphics[width=8cm]{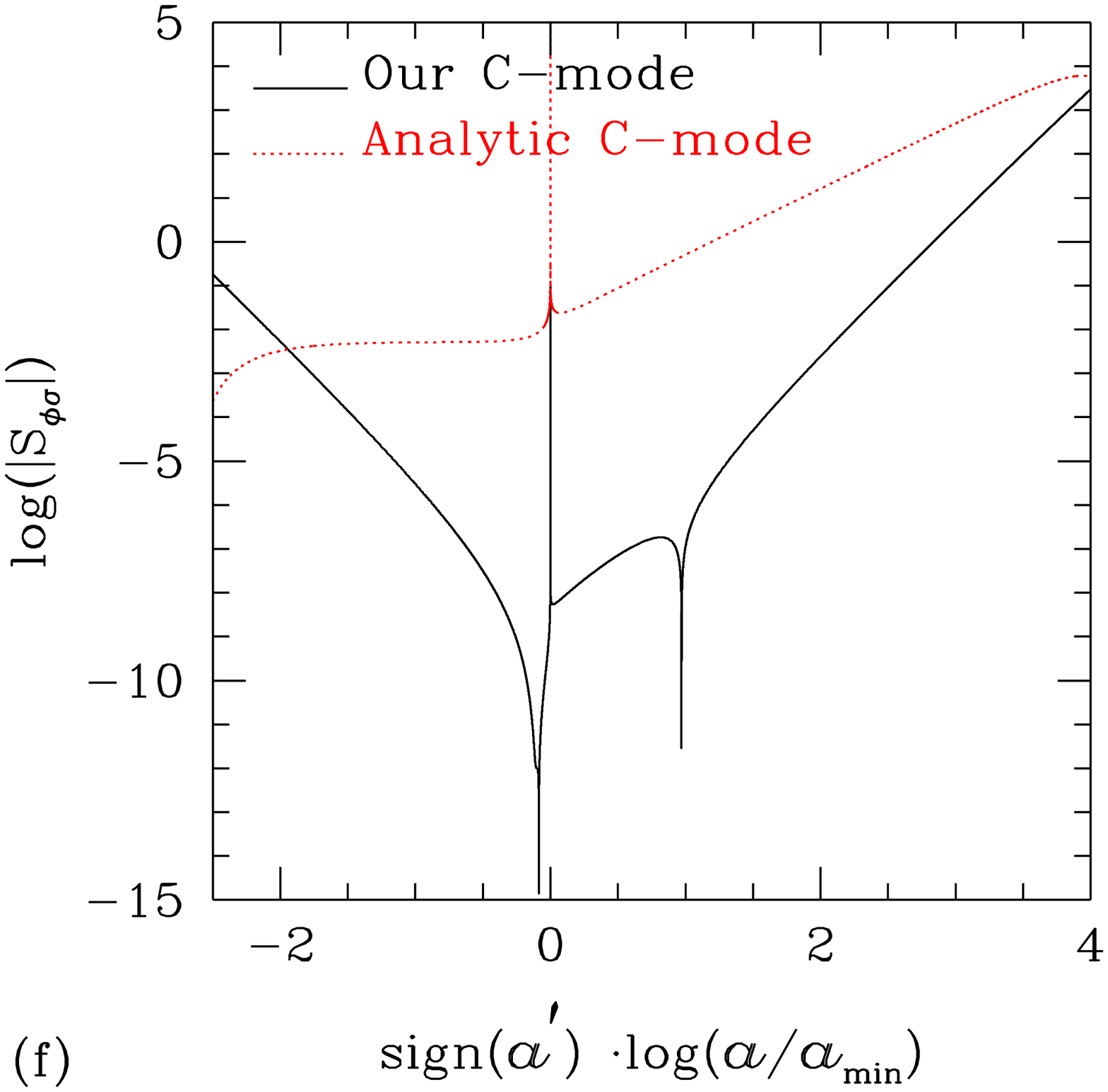}\\
\caption{The same $C$-mode as in Figs.\
         \ref{Fig-c}, \ref{Fig-usecalc}(f), and \ref{Fig-isocurvature-C-mode}(c)
         or Fig. \ref{Fig-cafter}, now
         for a particular
         way of finding our $C$-mode initial condition explained in
         the text.
         We imposed the analytic initial condition at one of the two $\mu +p
         =0$ epochs.
         Thus, Panels (a) and (c) have one less peak compared with Figs.\
         \ref{Fig-c}(a) and (c); in Panel (a) we do not have a spike at the location
         of first arrow, whereas we do have such a spike in Fig.\ \ref{Fig-c}(a)
         though not visible.
         Evolution of $\varphi_\chi$ in Panel (b) shows only one vanishing point, see Panel (e).
         In Panel (d), the adiabatic condition of $\varphi_\chi$ also has one peak
         which is caused by vanishing $\varphi_\chi$.
         Panel (f) corresponds to Fig.\
         \ref{Fig-isocurvature-C-mode}(c), and shows slightly
         different location of minimum $S_{\phi\sigma}$.
         }
\label{Fig-c-atzero}
\end{figure*}

Our $C$-mode initial condition is also found similarly. The $C$-mode
behavior of $\varphi_v$ can be found if we begin with the analytic
initial condition imposed close to the bounce, and integrate forward
and backward in time. The behavior of $\varphi_\chi$ found in this
way, however, shows $d$-mode behavior shortly before and shortly
after the bounce. In order to find the $C$-mode behavior of
$\varphi_\chi$ without appearance of the $d$-mode behavior near the
bounce as in Fig.\ \ref{Fig-c}(b), we should begin with the analytic
initial condition imposed away from bounce, and integrate forward
and backward in time. In order to have the $C$-mode behavior in
Fig.\ \ref{Fig-c}, we used the initial condition found by imposing
the analytic initial condition at a point where $S_{\phi\sigma}$
vanishes (we call it an $S_{\phi\sigma} = 0$ condition) in Fig.\
\ref{Fig-isocurvature-C-mode}(c). In this way, behaviors of
$C$-modes in Fig.\ \ref{Fig-c}, however, depend on whether we found
our initial condition by imposing $S_{\phi\sigma} = 0$ before or
after the bounce as in Figs.\ \ref{Fig-isocurvature-C-mode}(c) and
\ref{Fig-cafter}(f). The $C$-mode behaviors in Figs.\ \ref{Fig-c}
and \ref{Fig-usecalc} are related to Fig.\
\ref{Fig-isocurvature-C-mode}(c). In Fig.\ \ref{Fig-cafter} we
present the counterparts of Figs.\ \ref{Fig-c} and \ref{Fig-usecalc}
now related to Fig.\ \ref{Fig-cafter}(f). Comparison of these
figures will show that near the bounce Fig.\ \ref{Fig-cafter} is the
reflection of Figs.\ \ref{Fig-c} and \ref{Fig-usecalc} at the
bounce. In Fig.\ \ref{Fig-c-atzero} we present an additional case
where we begin by imposing the analytic initial condition at one of
the epochs of vanishing $\mu + p$. In this way, $\varphi_\chi = 0$
at only one point, see Fig.\ \ref{Fig-c-atzero}(b) and
\ref{Fig-c-atzero}(e). If we begin by imposing $S_{\phi\sigma} = 0$
condition closer to or more away from the bounce we will have
$d$-mode behavior of $\varphi_\chi$ near the bounce region. Thus,
the $C$-mode behaviors in Figs.\ \ref{Fig-c}(b),
\ref{Fig-cafter}(b), and \ref{Fig-c-atzero}(b) are closest to the
analytic behavior of $C$-mode of $\varphi_\chi$ in Eq.\
(\ref{eq-varphichi}).

\begin{figure*}
\centering%
\includegraphics[width=8cm]{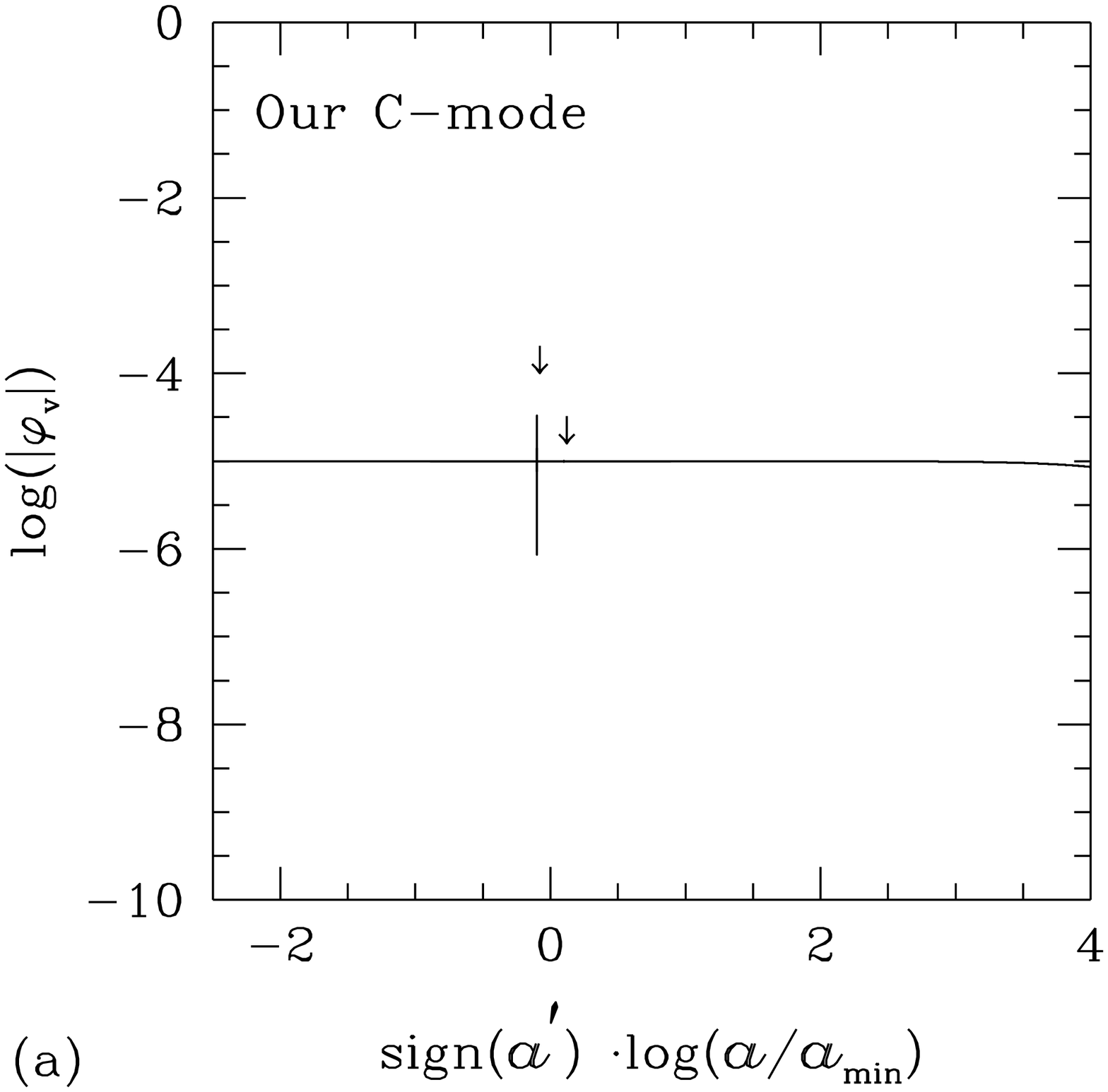}\hfill
\includegraphics[width=8cm]{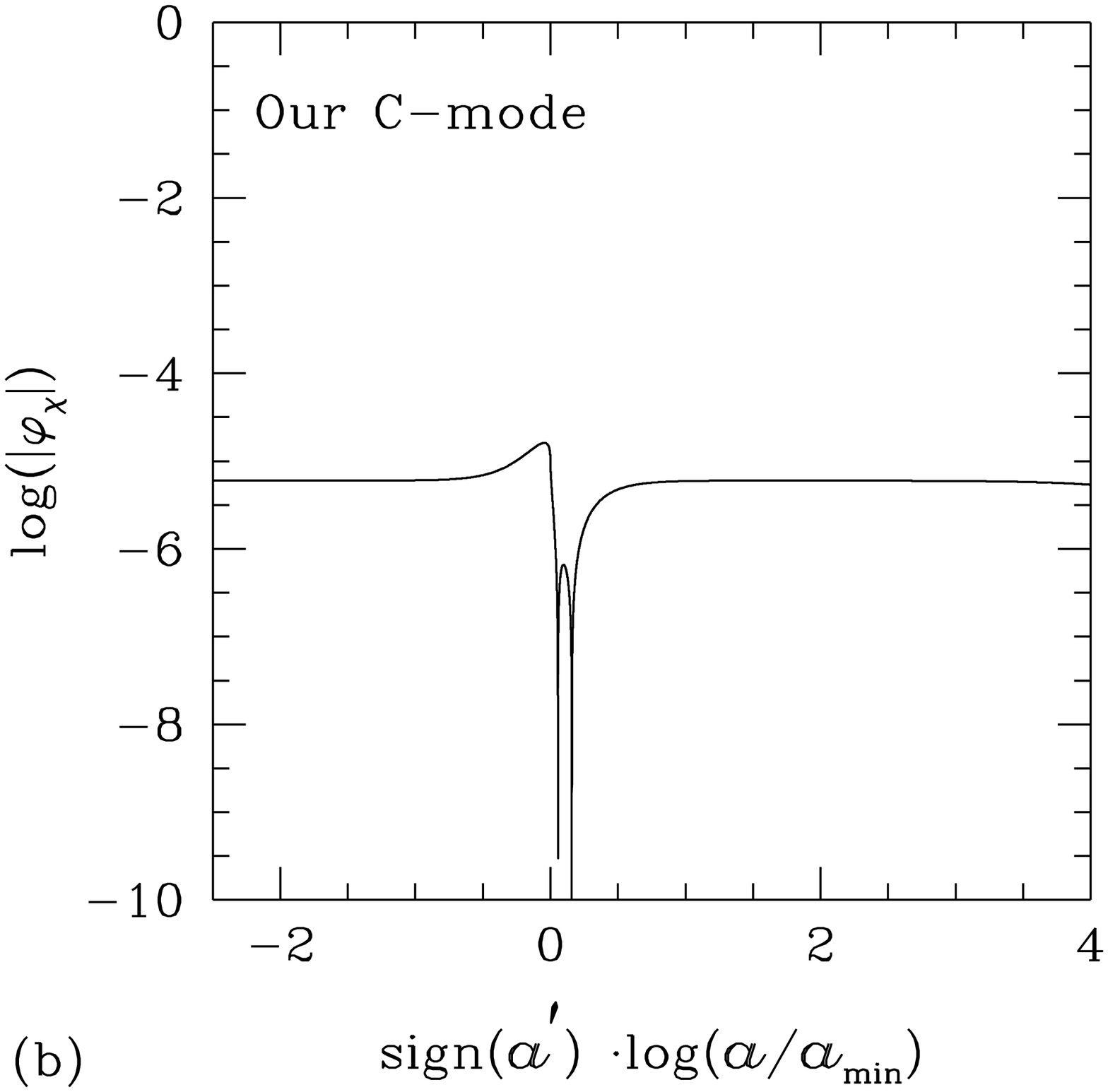}\\
\includegraphics[width=8cm]{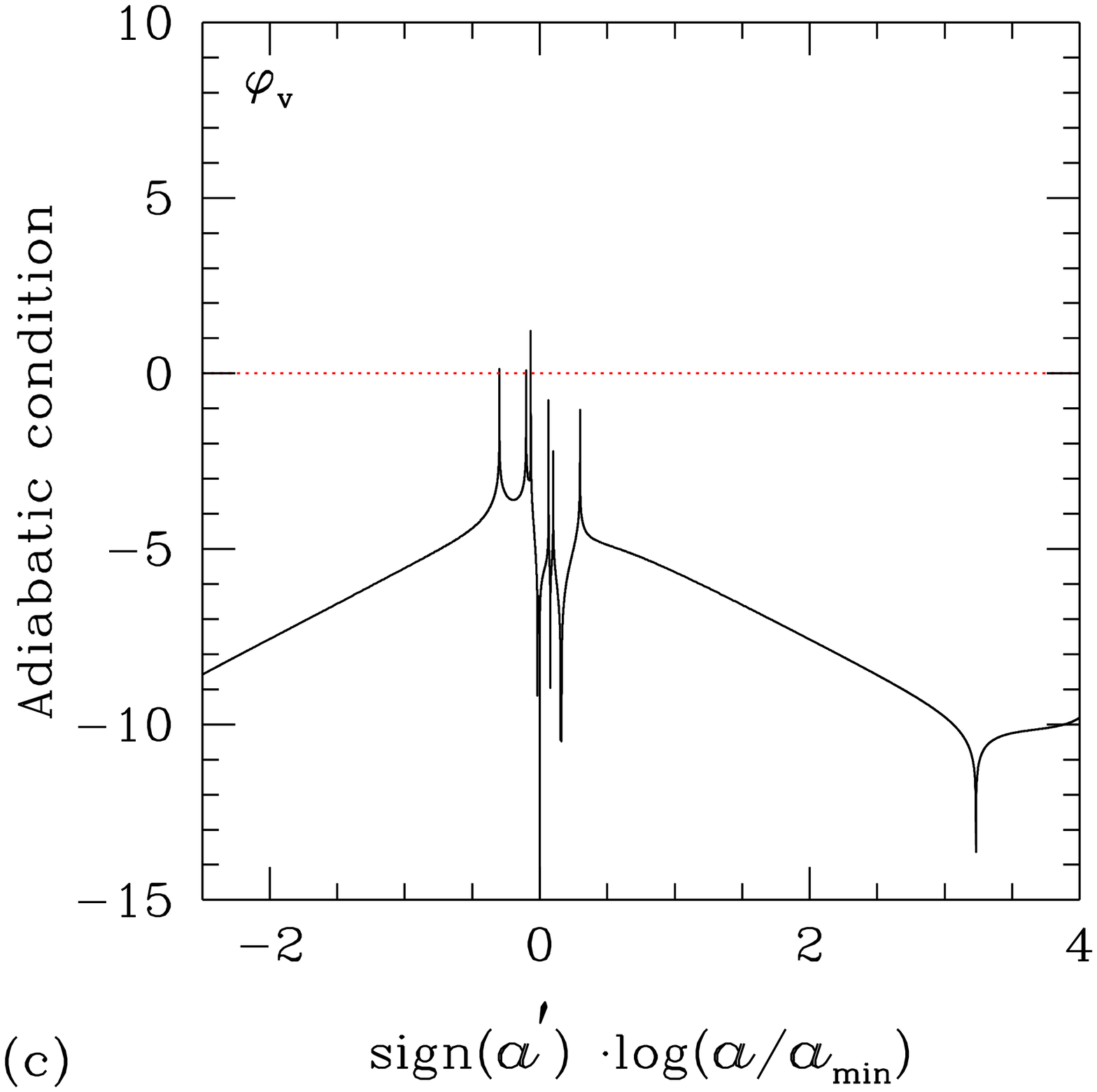}\hfill
\includegraphics[width=8cm]{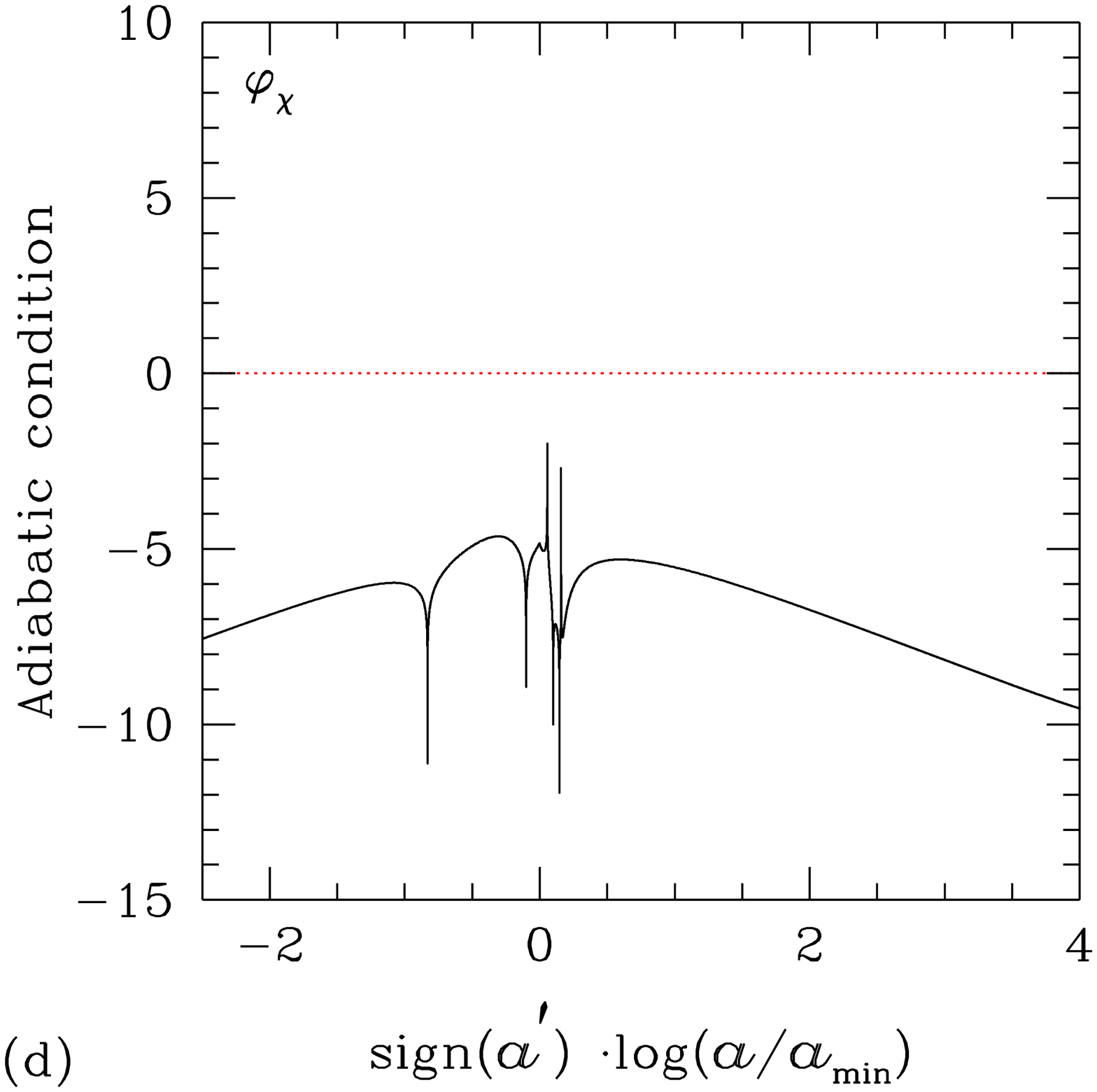}\\
\end{figure*}

\begin{figure*}
\centering%
\includegraphics[width=8cm]{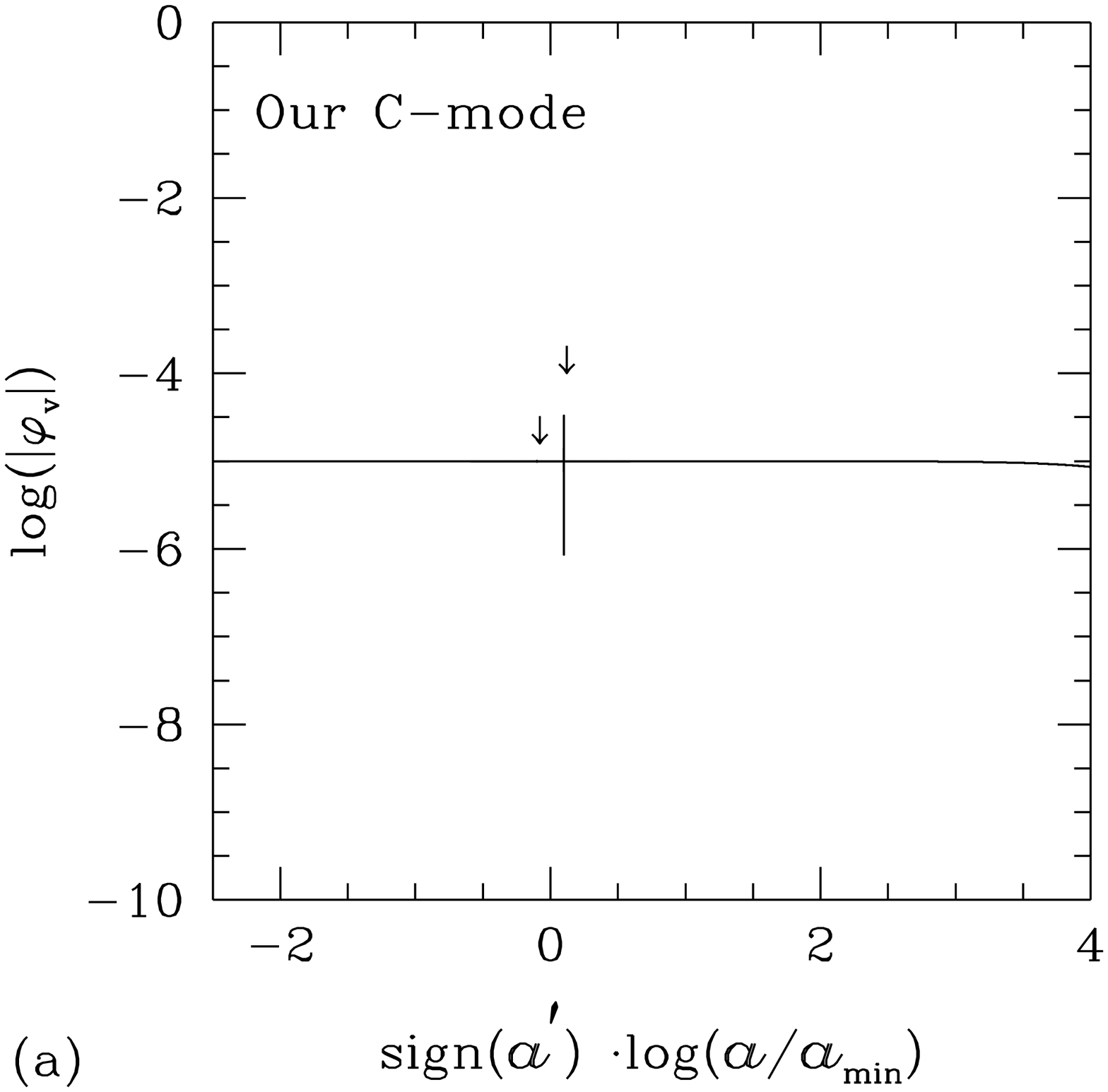}\hfill
\includegraphics[width=8cm]{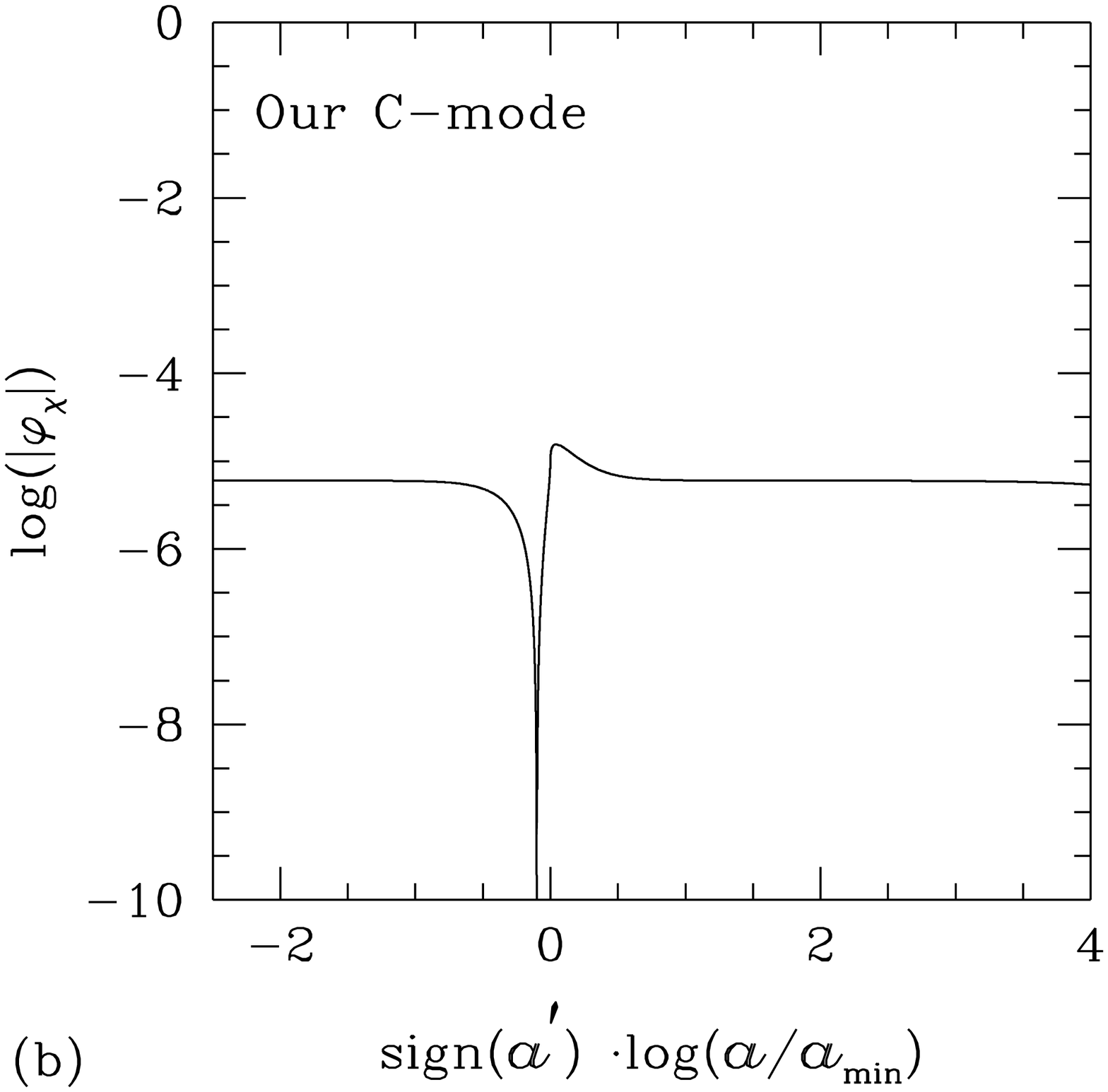}\\
\includegraphics[width=8cm]{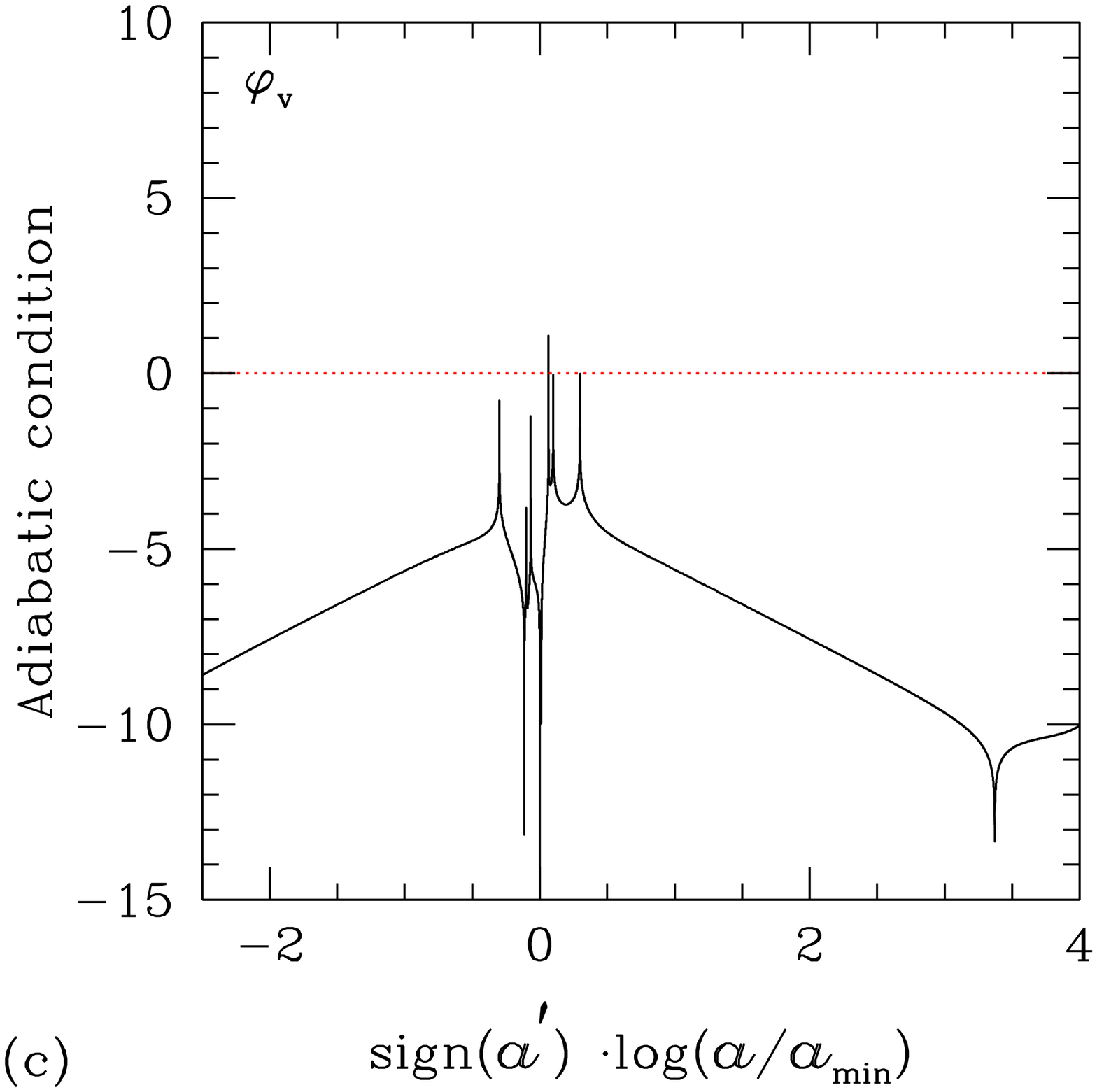}\hfill
\includegraphics[width=8cm]{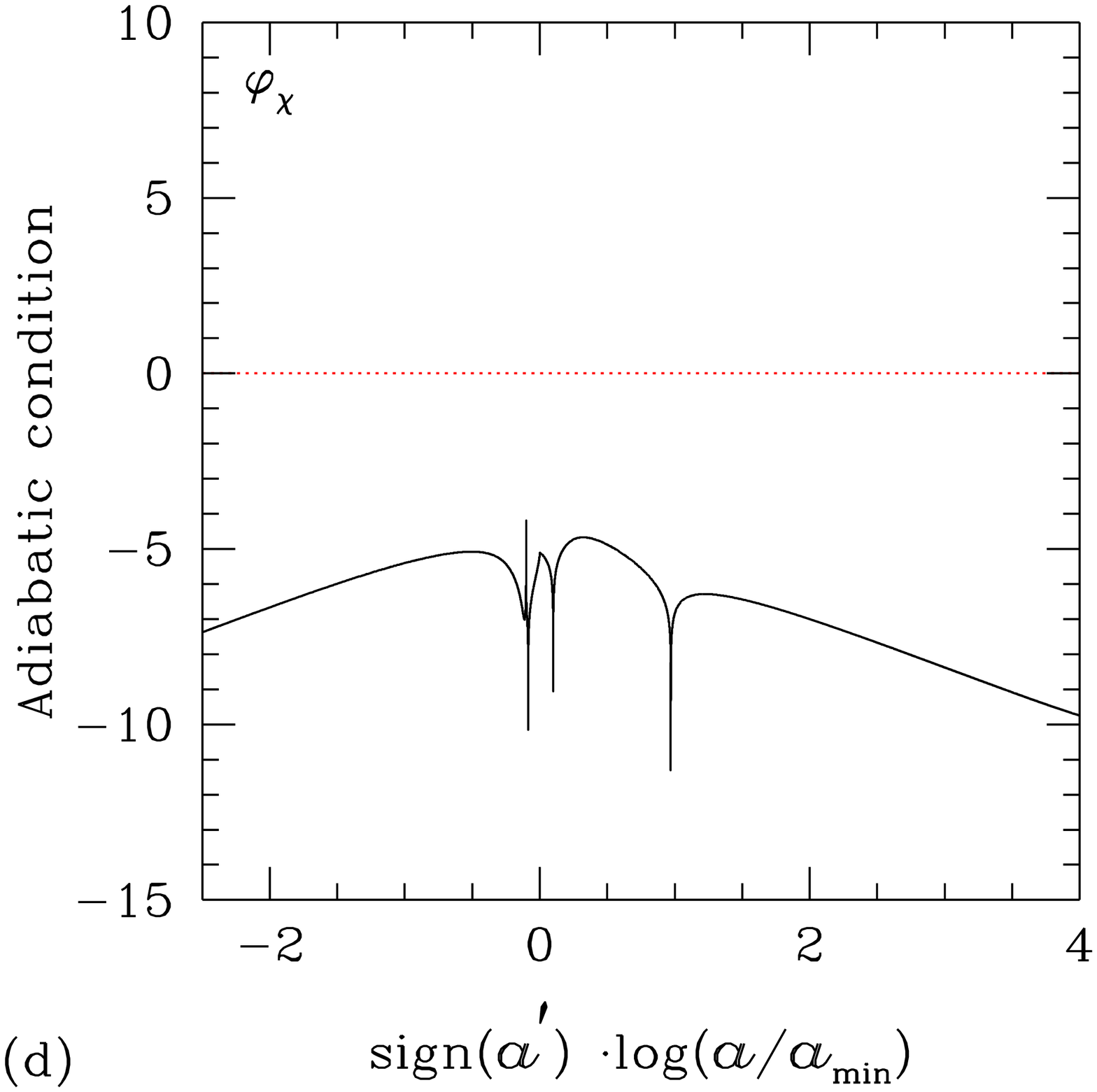}\\
\end{figure*}
\subsection{Tensor-type perturbations}
                                              \label{GW-numerical}

\begin{figure}
\centering%
\includegraphics[width=8cm]{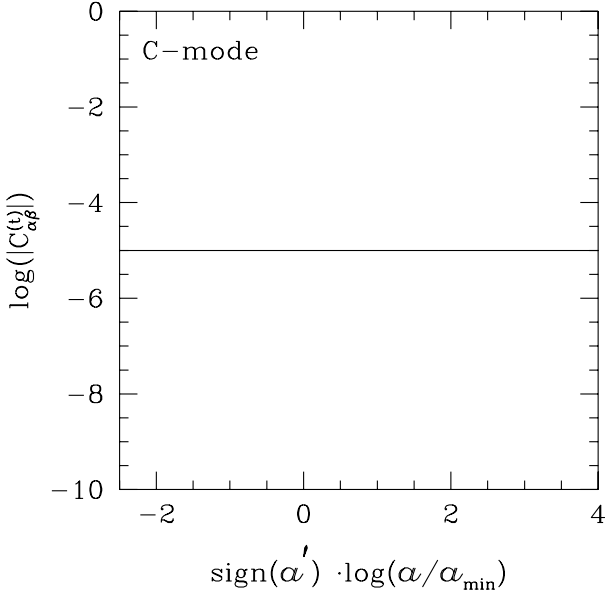}
\caption{
         Evolution of $\bar{C}_{\alpha\beta}$-mode tensor-type perturbation.
         The behavior of $\bar{C}_{\alpha\beta}$-mode is similar to the
         C-mode of $\varphi_v$. The $\bar{C}_{\alpha\beta}$-mode nature of
         this mode is preserved throughout the bounce;
         i.e., the time derivative of $C^{(t)}_{\alpha\beta}$
         vanishes throughout the evolution.
        }
        \label{Fig-GW-C-mode}
\end{figure}

\begin{figure*}
\centering%
\includegraphics[width=8cm]{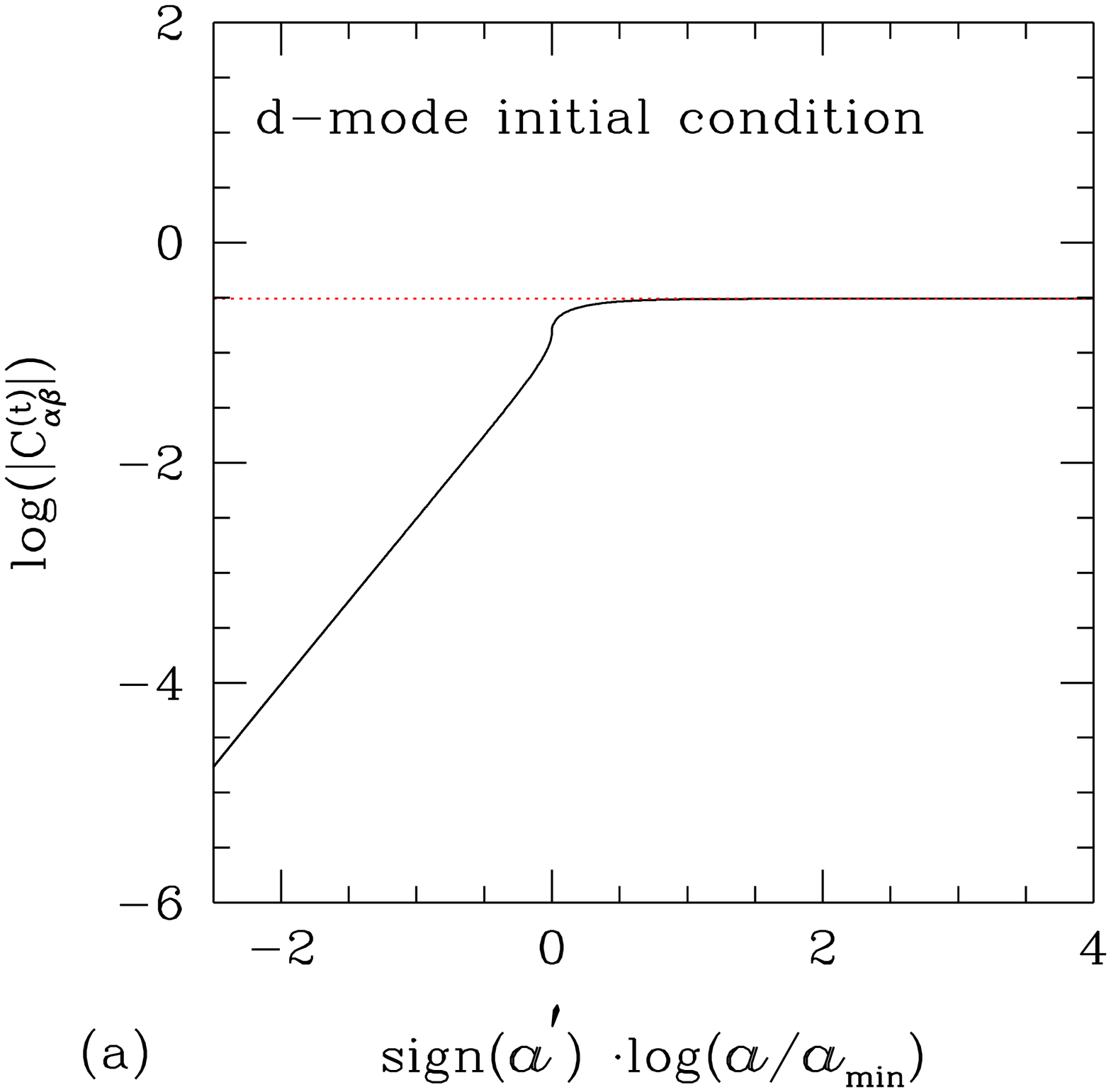}\hfill
\includegraphics[width=8cm]{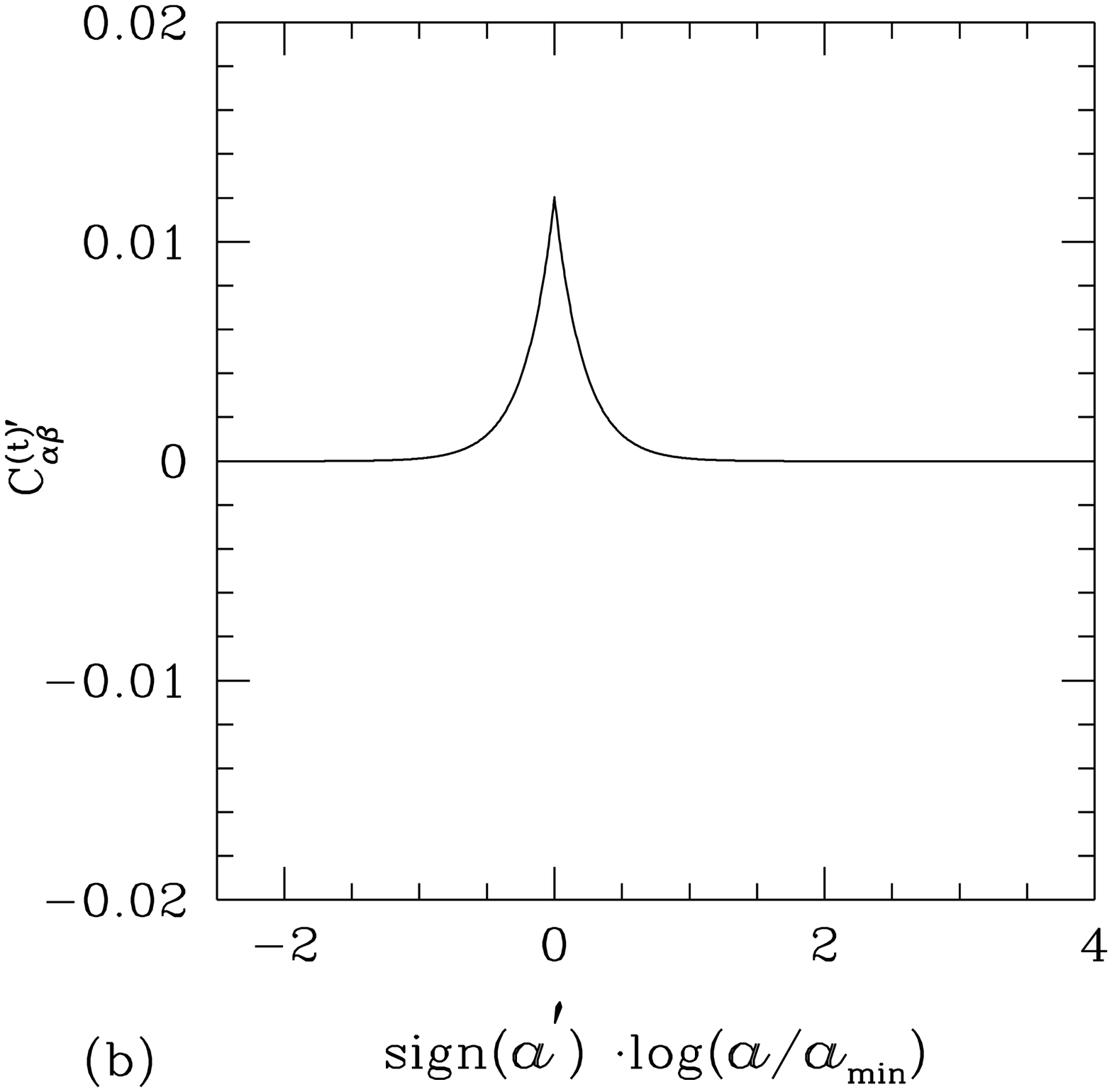}\\
\caption{
         Panel (a) shows the behavior of initial $\bar{d}_{\alpha\beta}$-mode
         which apparently transits to the $\bar{C}_{\alpha\beta}$-mode after the bounce.
         It is not possible to find the initial
         condition which maintains the $\bar{d}_{\alpha\beta}$-mode
         behavior throughout the bounce:
         see the main text.
         Figure (b) presents the time derivative of $C^{(t)}_{\alpha\beta}$
         which clearly shows that the initially provided
         $\bar{d}_{\alpha\beta}$-mode is present even after the
         bounce, this coincides with the analytic solution
         in Eq.\ (\ref{eq-tensorprime-1}).
         Meanwhile, Panel (a) shows that the initially introduced
         $\bar{d}_{\alpha\beta}$-mode is dominated by a newly
         generated $\bar{C}_{\alpha\beta}$-mode after the bounce.
         We take the initial condition in Table
         \ref{Table-tensorinitial}, thus $k = 0$.
         Even in the case of nonvanishing
         $k$, the large-scale condition which compares $k^2$ with
         $a^{\prime\prime} /a$ term in Eq.\ (\ref{eq-tensoru}) is well
         satisfied near the bounce.
         }
         \label{Fig-GW-d-mode}
\end{figure*}

\begin{figure*}
\centering%
\includegraphics[width=8cm]{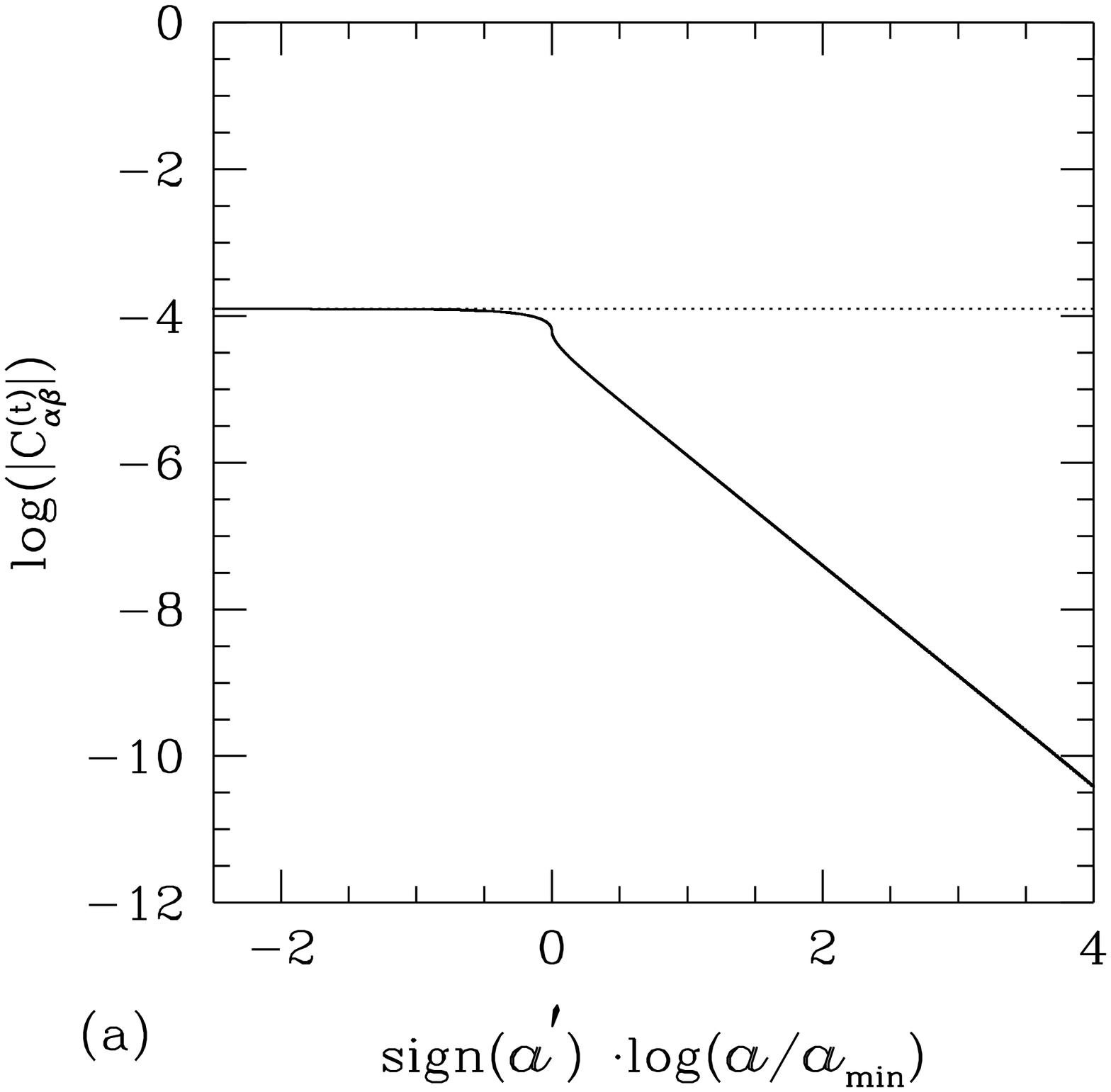}\hfill
\includegraphics[width=8cm]{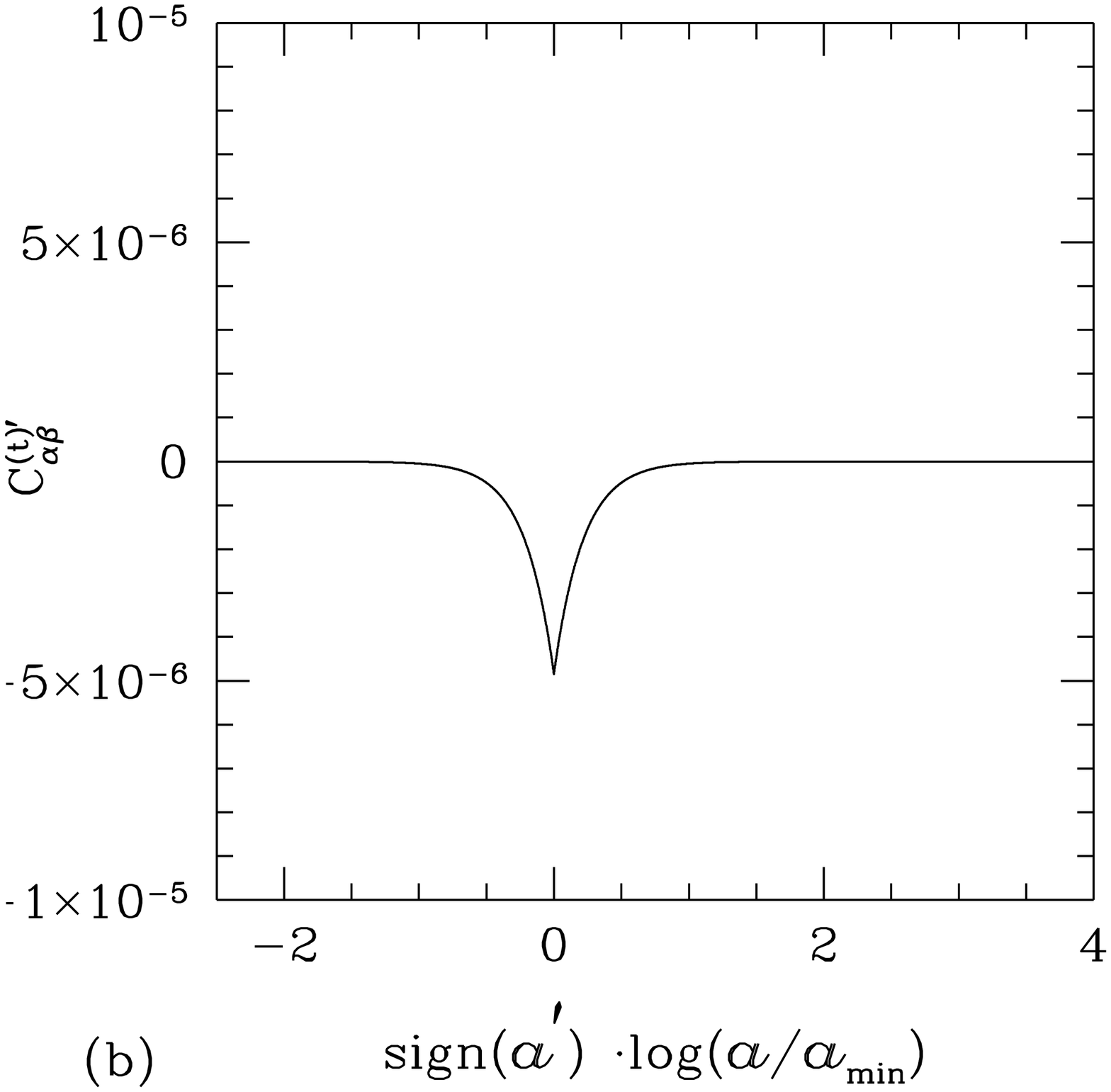}\\
\caption{
         Panel (a) shows the behavior of initially near $\bar{C}_{\alpha\beta}$-mode
         which transits to the $\bar{d}_{\alpha\beta}$-mode after the bounce.
         As we are considering $k=0$ limit,
         the precise $\bar{C}_{\alpha\beta}$-mode in Fig.\
         \ref{Fig-GW-C-mode} sets $C^{(t)\prime}_{\alpha\beta} = 0$.
         In this Figure we introduced a small negative value of
         $C^{(t)\prime}_{\alpha\beta} $= $-$1.9463109018142E-07$\times {C}^{(t)}_{\alpha\beta}
         $. This means that the present initial condition is a mixture
         of the dominating $\bar{C}_{\alpha\beta}$-mode plus a
         initially negligible but non-vanishing amount of the
         $\bar{d}_{\alpha\beta}$-mode.
         Panel (b) presents the time derivative of $C^{(t)}_{\alpha\beta}$
         which shows that the apparent $\bar{d}_{\alpha\beta}$-mode in
         the expanding phase is actually the one provided initially.
        }
        \label{Fig-GW-Cd-mode}
\end{figure*}

The evolution of tensor-type perturbation is described by Eq.\
(\ref{eq-tensoru}). Figures \ref{Fig-GW-C-mode}-\ref{Fig-GW-Cd-mode}
show diverse behaviors of the tensor-type perturbation depending on
initial conditions. Using the $\bar{C}_{\alpha\beta}$-mode initial
condition in Eq.\ (\ref{eq-analten}), Fig.\ \ref{Fig-GW-C-mode}
shows a conserved evolution of $C^{(t)}_{\alpha\beta}$ throughout
the bounce: thus, we can clearly identify this as the
$\bar{C}_{\alpha\beta}$-mode in Eq.\ (\ref{eq-analten}). However,
although we naively expected to be able to find a precise ${\bar
d}_{\alpha\beta}$-mode initial condition in the collapsing phase
which will be preserved as the same ${\bar d}_{\alpha\beta}$-mode
even in the expanding phase, the numerical result in Fig.\
\ref{Fig-GW-d-mode} shows that the ${\bar d}_{\alpha\beta}$-mode
initial condition switches to the $\bar{C}_{\alpha\beta}$-mode.

This result differs from our following naive anticipation. The
equation for tensor-type perturbation is quite similar to the one
for $\varphi_v$ in an adiabatic limit: compare Eq.\
(\ref{eq-tensoru}) with Eq.\ (\ref{eq-varphiz}). In large-scale
limits we have analytic solutions in Eqs.\ (\ref{eq-analten}) and
(\ref{eq-varphiv}). In case of $d$-mode for the scalar-type
perturbation in Fig.\ \ref{Fig-d}(a) we presented a proper $d$-mode
evolution before and after the bounce. Thus, previously we
anticipated that even for the tensor-type perturbation similarly we
can find the proper initial condition which produces the ${\bar
d}_{\alpha\beta}$-mode before the bounce continuing as the ${\bar
d}_{\alpha\beta}$-mode after the bounce. In the following paragraph
we can show that it is not possible to have such a behavior. Thus,
any $\bar d_{\alpha\beta}$-mode initial condition in a collapsing
phase is later dominated by an apparent $\bar C_{\alpha\beta}$-mode
solution in an expanding era. This implies that we do {\it not} have
the proper $\bar{d}_{\alpha\beta}$-mode which preserves its nature
throughout the bounce. Figure \ref{Fig-GW-d-mode}(b) shows that the
initially supplied $\bar d_{\alpha\beta}$-mode still survives as the
same $\bar d_{\alpha\beta}$-mode even in the expanding era. Figure
\ref{Fig-GW-d-mode}(a) shows that, after the bounce,
$C^{(t)}_{\alpha\beta}$ is simply dominated by an additional $\bar
C_{\alpha\beta}$-mode appearing after the bounce; this apparently
comes from the lower-bound of integration of Eq.\
(\ref{eq-analten}).

Here we can show why it is {\it not} possible to find the initial
condition which remains as the $\bar d_{\alpha\beta}$-mode after the
bounce. {\it If} the $\bar{d}_{\alpha\beta}$-mode in the collapsing
phase could somehow switch to the same mode in the expanding phase,
at the bounce $C^{(t)\prime}_{\alpha\beta}$ should either change its
sign or cease to be differentiable. However, from Eq.\
(\ref{eq-analten}), for $k=0$, we have \bea
    & & C^{{(t)}\prime}_{\alpha\beta}={\bar d_{\alpha\beta}(k) \over a^2},
    \label{eq-tensorprime-1}
\eea where $\bar d_{\alpha\beta}$ is a temporal constant. Thus, once
the sign of $\bar d_{\alpha\beta}$ is decided, then its sign is
fixed throughout the evolution, and $C^{(t)\prime}_{\alpha\beta}$ is
always differentiable. In case of the large-scale solution of
$\varphi_v$ in Eq.\ (\ref{eq-varphiv}), changing sign of $z^2$ plays
an important role in explaining the different behaviors between
$d$-mode of $\varphi_v$ and $\bar{d}_{\alpha\beta}$-mode of
$C^{(t)}_{\alpha\beta}$; Fig.\ \ref{Fig-usecald}(d) shows that
$\varphi_v$ does not only change its sign but also is not
differentiable in a couple of places. For a non-vanishing $k$, from
Eq.\ (\ref{eq-tensore}) we have \bea
        C^{{(t)}\prime}_{\alpha\beta}=
        {\bar d_{\alpha\beta} \over a^2}-{k^2 \over a^2}\left[\int^\eta
        {a^2}C^{(t)}_{\alpha\beta}(k,\eta)d\eta\right].
    \label{eq-tensorprime-2}
\eea Since we are considering the ${\bar d}_{\alpha\beta}$-mode
solution in the large-scale limit, the large-scale second-order
correction term (the second term in RHS) in Eq.\
(\ref{eq-tensorprime-2}) should be negligible compared with the
leading order term in Eq.\ (\ref{eq-tensorprime-1}); thus, the sign
of $C^{{(t)}\prime}_{\alpha\beta}$ cannot be changed in that limit.
This confirms that our numerical result for the ${\bar
d}_{\alpha\beta}$-mode presented in Fig.\ \ref{Fig-GW-d-mode}(a) is
unavoidable.

In Fig.\ \ref{Fig-GW-Cd-mode}, we present an evolution of initially
near $\bar{C}_{\alpha\beta}$-mode which transits to the
$\bar{d}_{\alpha\beta}$-mode after the bounce. Compared with the
precise $\bar{C}_{\alpha\beta}$-mode in Fig.\ \ref{Fig-GW-C-mode},
in this Figure we introduced a small amount of initially negligible
but non-vanishing $\bar{d}_{\alpha\beta}$-mode. The time derivative
of $C^{(t)}_{\alpha\beta}$ in Fig.\ \ref{Fig-GW-Cd-mode}(b) shows
that the apparent $\bar{d}_{\alpha\beta}$-mode in the expanding
phase is exactly the one provided initially. After the bounce, the
initially dominating $\bar C_{\alpha\beta}$-mode has been canceled
out by a contribution from the lower-bound of integration of Eq.\
(\ref{eq-analten}).

\begin{table}

\caption{The initial conditions for tensor-type perturbation with
         $k=0$. For $k=0$ the large-scale condition is met exactly. We take
         initial $\eta=-500$. Results based on these initial conditions are
         presented in Figs.\ \ref{Fig-GW-C-mode}-\ref{Fig-GW-d-mode}.
         }
\begin{center}
\begin{tabular}{ r r r r} \hline\hline
  variable   & C-mode  & d-mode\\ \hline
     $C_{\alpha\beta}^{(t)}$ & 1.0E$-$05
 & 1.0E$-$05\\
     ${C^{(t)}_{\alpha\beta}}^\prime$ & 0. & $-3./\eta \times
     C_{\alpha\beta}^{(t)}$\\
                                \hline\hline
\end{tabular}\label{Table-tensorinitial}
\end{center}
\end{table}


\section{Power spectra}
                                                   \label{sec:Implication}

\subsection{Scalar-type perturbations}
                                                \label{scalar-implication}

As the origin of seed fluctuations for the large-scale structure we
consider quantum fluctuations of the fields and the space-time
metric ever present in the collapsing phase. Assuming certain
dynamics of the scale factor we have the perturbation equations
described by Bessel equations with the solutions in Hankel functions
as in Eq.\ (\ref{eq-Hankel.}). The canonical quantization process
well known in the literature leads to the same equations and
solutions applicable to the quantum fluctuations. In order to
correspond the large-scale solutions with the $C$- and $d$-mode
decomposition in Eqs.\ (\ref{eq-varphiv}) and (\ref{eq-varphichi})
to the exact solutions with Hankel functions in Eq.\
(\ref{eq-Hankel.}), we follow \cite{hw2}. The power spectrum and
spectral index are defined as\bea
   & & \mathcal{P}_x={k^3 \over 2
       \pi^2}|x|^2, \quad
       n_S-1 \equiv {d \ln{\mathcal{P}_x} \over d \ln{k} },
   \label{eq76.}
\eea where $x$ can be $\varphi_v (k,\eta)$ or $\varphi_\chi
(k,\eta)$ in the Fourier space.

In the power-law case away from the bounce where the ghost field is
negligible, Eqs.\ (\ref{eq-varphiz}) and (\ref{eq-varphizb}) lead to
Bessel equations. Using the quantization based on the action
formulation, we have the exact mode function solutions \bea
    && \varphi_v(k,\eta)
    \nonumber \\
    & & \qquad
       =\left|{H \over \dot\phi}\right|{\sqrt{\pi|\eta|} \over 2a}
       [c_1(k)H_{\nu_v}^{(1)}(k|\eta|)+c_2(k)H_{\nu_v}^{(2)}(k|\eta|)],
    \nonumber \\
    && \varphi_\chi(k,\eta)
    \nonumber \\
    & & \qquad
       ={H\sqrt{\pi |\eta|} \over 2k
       \sqrt{2q}}[c_1(k)H_{\nu_u}^{(1)}(k|\eta|)+c_2(k)H_{\nu_u}^{(2)}(k|\eta|)],
       \nonumber\\
    && \qquad
       \nu_v \equiv {3q-1 \over 2(q-1)},\quad \nu_u \equiv {q+1 \over 2(q-1)}.
    \label{eq-power}
\eea The Hankel functions can be expanded as \cite{HM}\bea
    H_{\nu}^{(1,2)}(k|\eta|)&=&\sum_{n=0}^\infty {1\over n!}
    \Bigg[-{(k|\eta|)^2 \over 4}\Bigg]^n{1 \over
    \sin\nu\pi}\times\Bigg[\Bigg({k|\eta| \over 2}\Bigg)^\nu\nonumber\\
    &&\times{\pm ie^{\mp i \nu\pi} \over
    \Gamma(\nu +n+1)}+\Bigg({k|\eta| \over 2}\Bigg)^{-\nu}\nonumber\\
    &&\times{\mp i \over \Gamma(-\nu+n+1)}\Bigg].\label{eq-hankelex}
    \eea
In large scale limit, the first (second) term in the parenthesis
dominates for $\nu < 0$ $(\nu > 0)$. In Eqs.\ (\ref{eq-varphiv}) and
(\ref{eq-varphichi}) the leading orders of the $C$-modes are time
independent whereas the leading orders of the $d$-modes behave as
$\varphi_v \propto |\eta|^{2\nu_v}$ and $\varphi_\chi \propto
|\eta|^{2\nu_u}$. Since \bea
    & & \varphi_{v}(k,\eta) \propto |\eta|^{\nu_v}H_{\nu_v}^{(1,2)}(k|\eta|),
        \nonumber\\
     &&   \varphi_\chi(k,\eta) \propto
        k^{-1}|\eta|^{\nu_u}H_{\nu_u}^{(1,2)}(k|\eta|),
    \label{eq-vpvc}
\eea we can easily identify the first and the second terms in the
parenthesis of Eq.\ (\ref{eq-hankelex}) as the $d$-mode and the
$C$-mode, respectively.

After taking the simple vacuum choice $c_1 = 1$ and $c_2 = 0$, the
spectral indices for the $C$-mode and $d$-mode of $\varphi_v$ and
$\varphi_\chi$ can be read as \cite{hw2} \bea
    && (n_S-1)_{\varphi_v,C}=(n_S-1)_{\varphi_\chi,C}={2 \over 1-q},
    \nonumber \\
    && (n_S-1)_{\varphi_v,d} = {4-6q \over 1-q}, \quad
    (n_S-1)_{\varphi_\chi,d} = -{2q \over 1-q}.\nonumber\\
    \label{eq-power2}
\eea Notice that for the $C$-mode, the spectral index of $\varphi_v$
and $\varphi_\chi$ coincide.

The near Harrison-Zel'dovich scale-invariant spectrum corresponds to
$n_S - 1 \simeq 0$ which is consistent with current observations of
the cosmic microwave background radiation \cite{CMB} and the
large-scale structures \cite{LSS}. The ekpyrotic or the cyclic
scenario has $0<q \ll 1$, thus from Eq.\ (\ref{eq-power2}) these
scenarios have a scale-invariant spectrum for $d$-mode of
$\varphi_\chi$, but these have a quite blue spectrum with $n_S - 1
\simeq 2$ for the $C$-mode \cite{hw2}; incidentally, we have $n_S-1
\simeq 4$ for the $d$-mode of $\varphi_v$. Although the $d$-mode of
$\varphi_\chi$ shows a scale-invariant spectrum, we have shown that
the $d$-mode eventually decays away in the expanding phase. Thus, we
cannot see this spectrum in the current expanding phase of the
universe. The observationally relevant one is the $C$-mode spectrum.
Thus, for the ekpyrotic or the cyclic scenarios we have $n_S - 1
\simeq 2$ which is too blue compared with observations. Here, we
assume a smooth nonsingular bounce with valid linear perturbation
through the bounce; this differs from the original ekpyrotic/cyclic
scenarios where the effective four dimensional spacetime goes
through a singularity \cite{Ek}. In the pre-Big Bang scenario we
have $q=1/3$, thus $n_S - 1 = 3$ for the $C$-mode \cite{pbb-pert}.

\begin{figure*}
\centering%
\includegraphics[width=8cm]{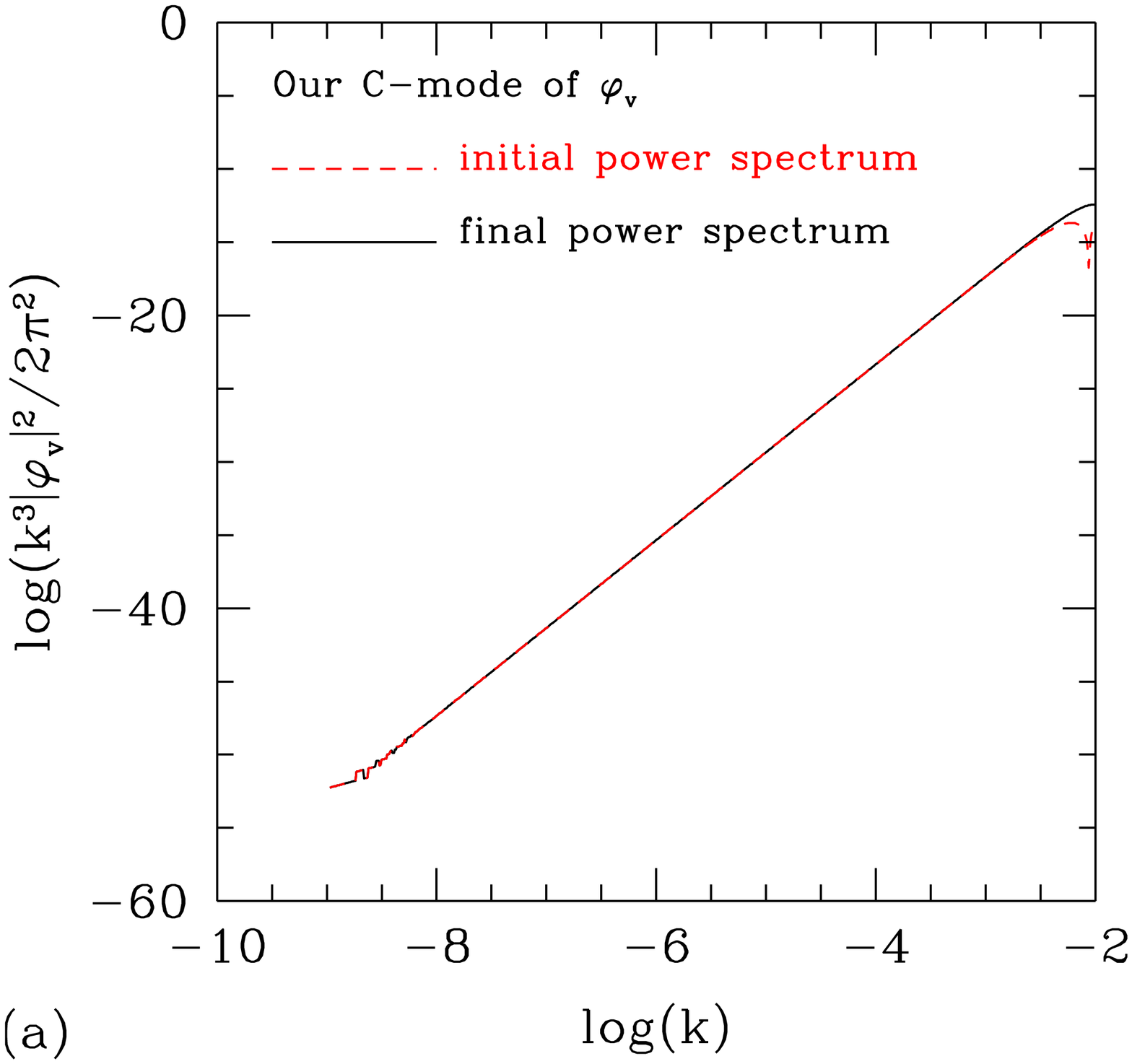}\hfill
\includegraphics[width=8cm]{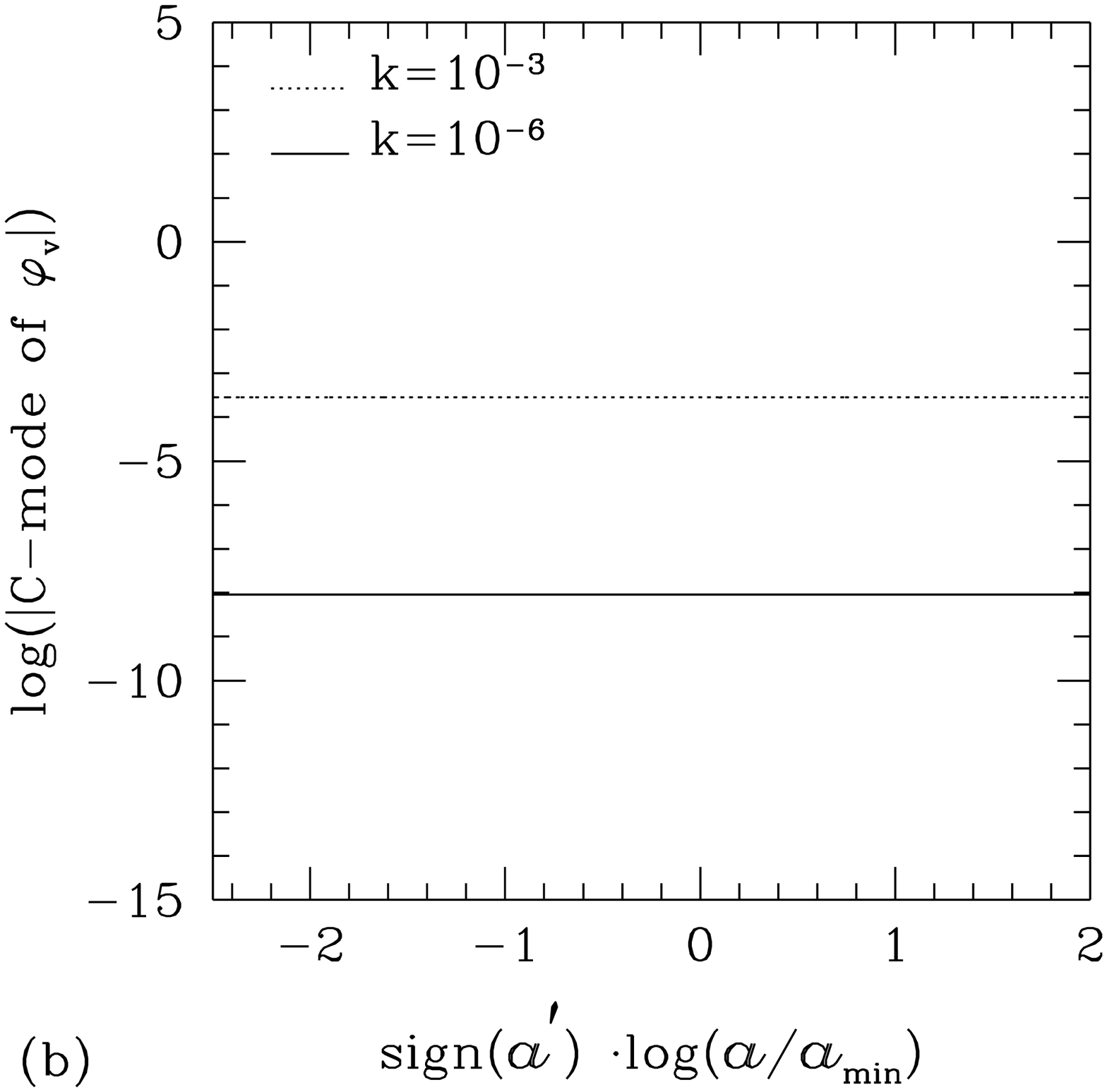}\\
\includegraphics[width=8cm]{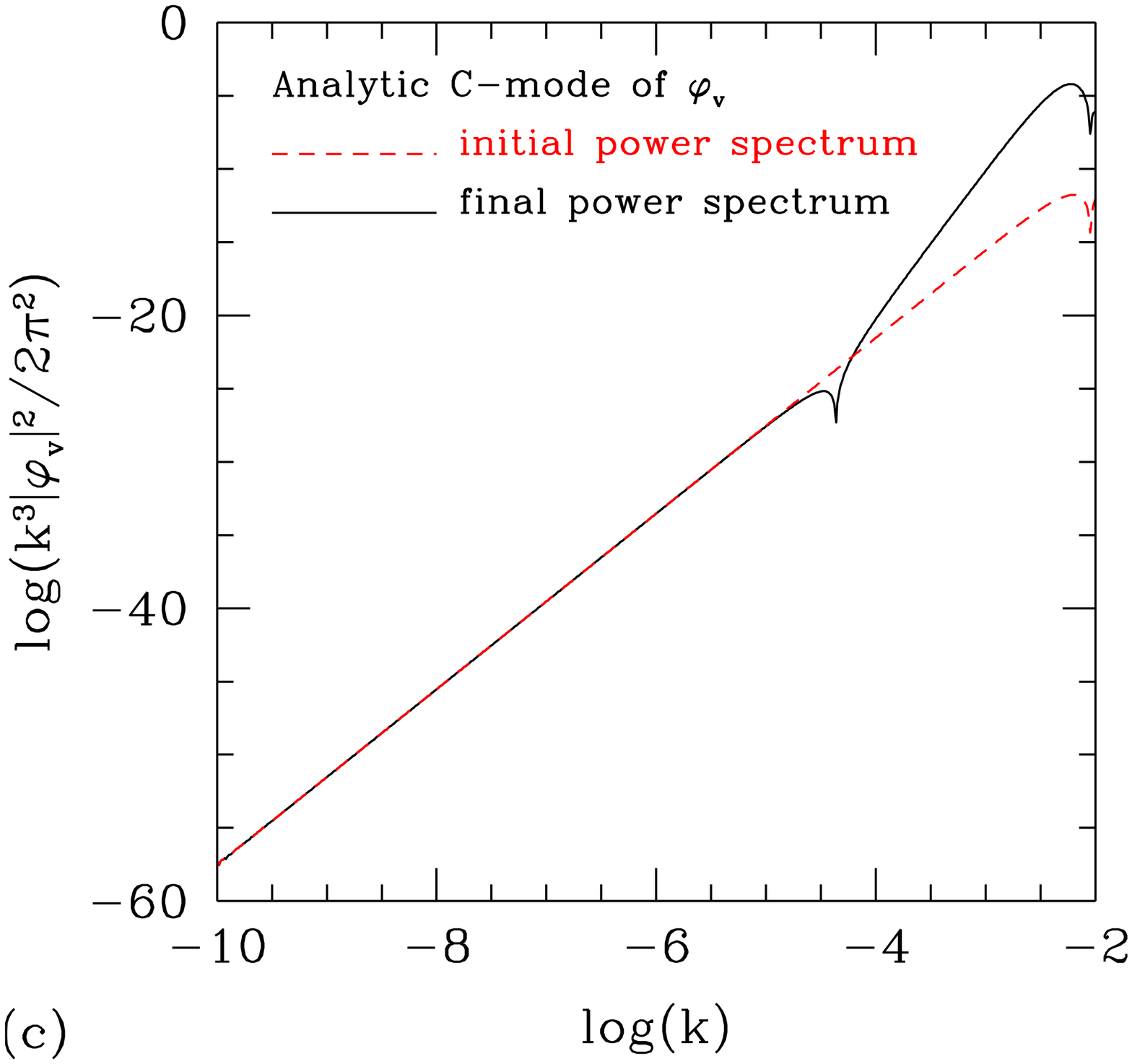}\hfill
\includegraphics[width=8cm]{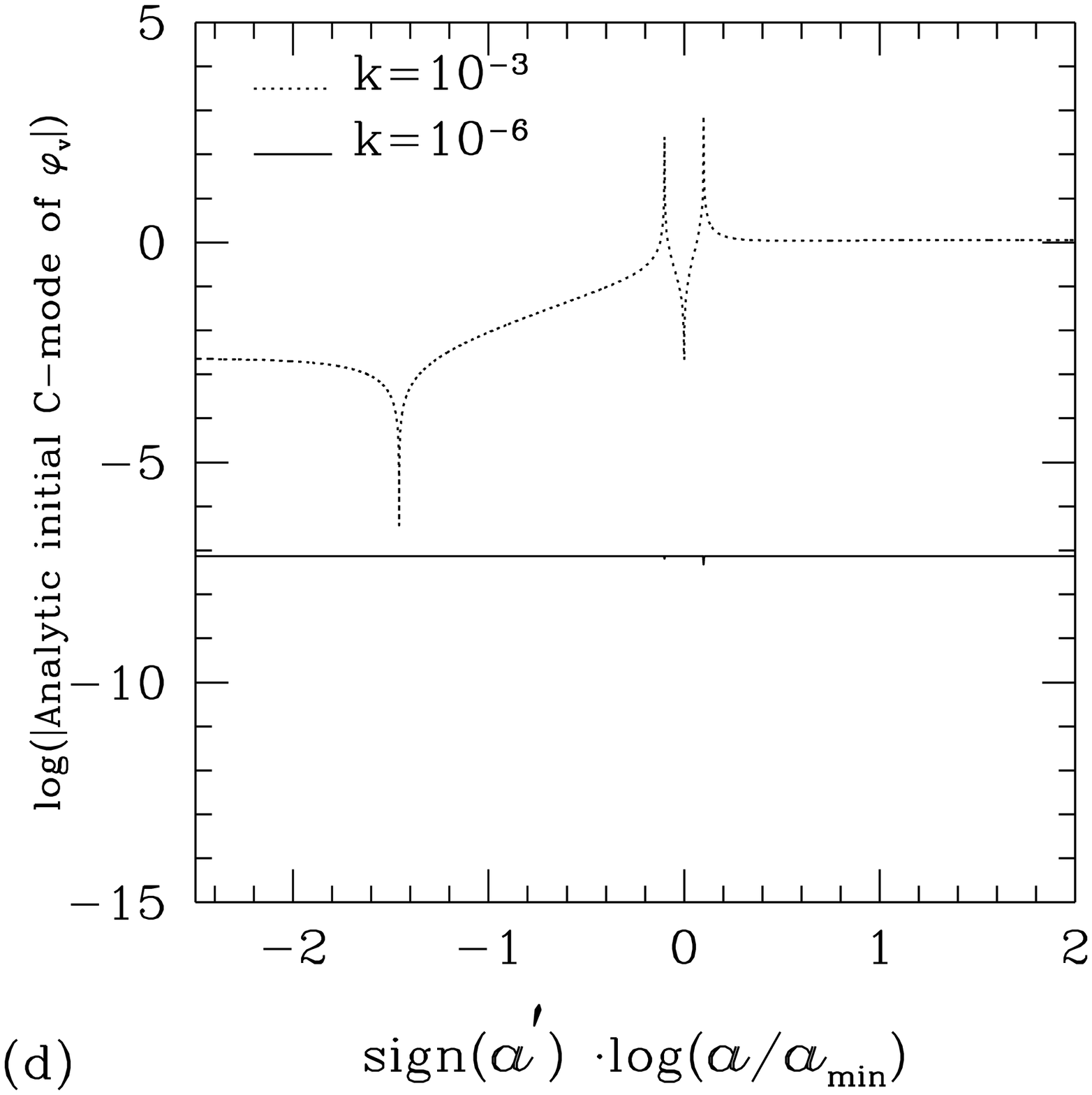}
\caption{Power spectra and typical evolutions for the $C$-mode of
         $\varphi_v$.
         Panel (a) shows the initial and the final power
         spectra for our $C$-mode of $\varphi_v$.
         Panel (a) shows that the spectral index is $n_S-1 = 6$
         which agrees with
         our analytic result in Eq.\ (\ref{eq-power2}).
         Panel (c) shows the initial and the final power
         spectra for the analytic $C$-mode of $\varphi_v$.
         In this case, only for large-scale ($\log{k} < -4.5$) perturbation
         the spectral index coincides with our $C$-mode case.
         Panels (b) and (d) show typical
         evolutions of $\varphi_v$ corresponding to Panels (a) and (c), respectively,
         for two different wavenumbers.
         Panel (d) shows why the power spectrum at
         large wavenumber is enhanced compared with the one at small
         wavenumber.
         In Figs.\ \ref{Fig-comparepowervpc}-\ref{Fig-comparepowervpchid}
         we take initial epoch at $\eta=-500$
         [sign$(a^\prime)\cdot$log$({a_i /a_{min}}) \simeq -2.65$]
         and final epoch at
         $\eta=300$ [sign$(a^\prime)\cdot$log$({a_f / a_{min}}) \simeq 2.25$].
         In order to get these power spectra, as the initial
         conditions both for our and analytic $C$- and $d$-modes,
         we use the Hankel function solutions in Eq.\
         (\ref{eq-power}) in the large scale limit.
         In order to find our $C$- and $d$-mode initial
         conditions for different wavenumbers we use the same method
         explained in Sec.\ \ref{scalar-numerical-IC}.
         }
         \label{Fig-comparepowervpc}
\end{figure*}

\begin{figure*}
\centering%
\includegraphics[width=8cm]{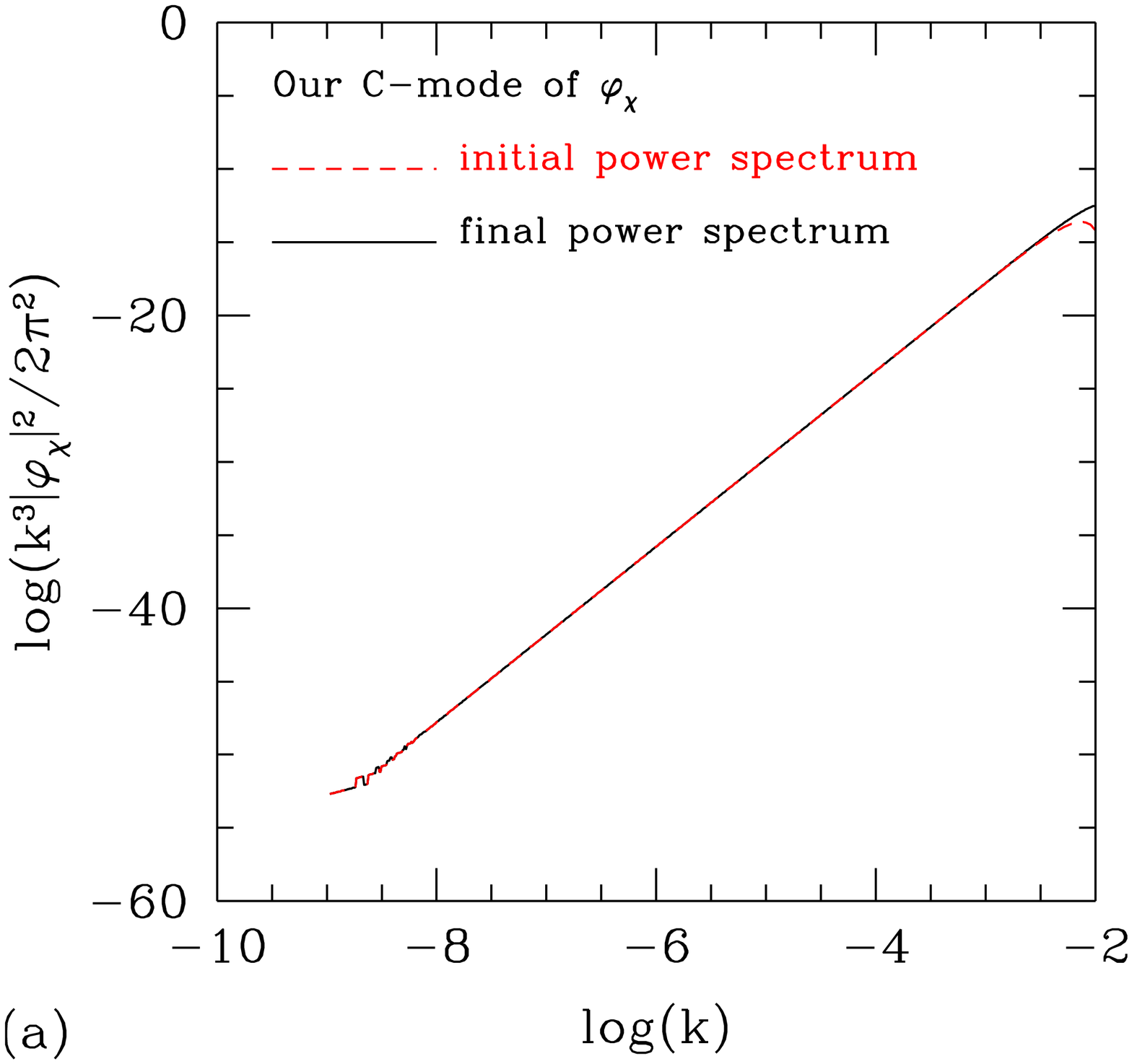}\hfill
\includegraphics[width=8cm]{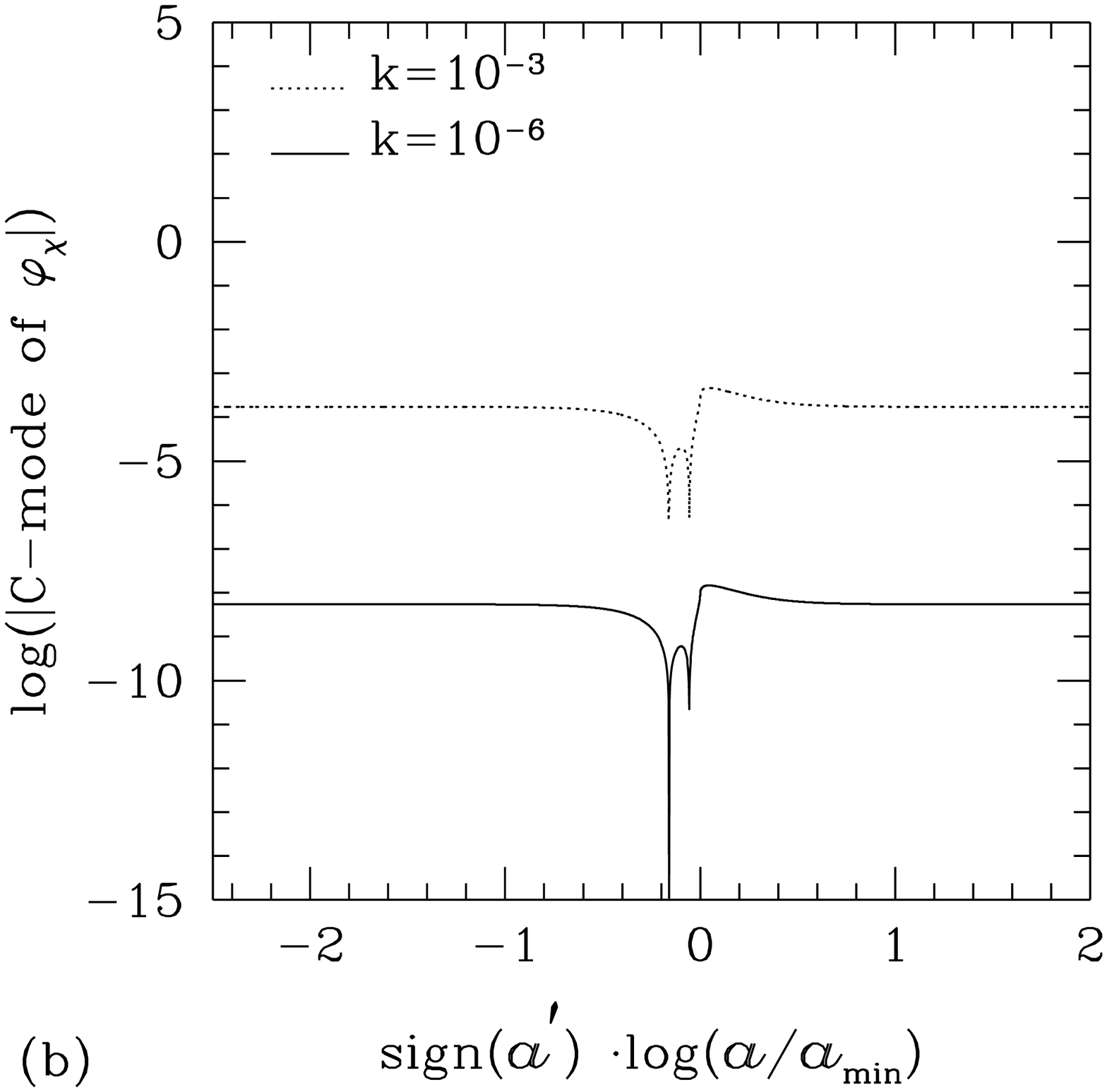}\\
\includegraphics[width=8cm]{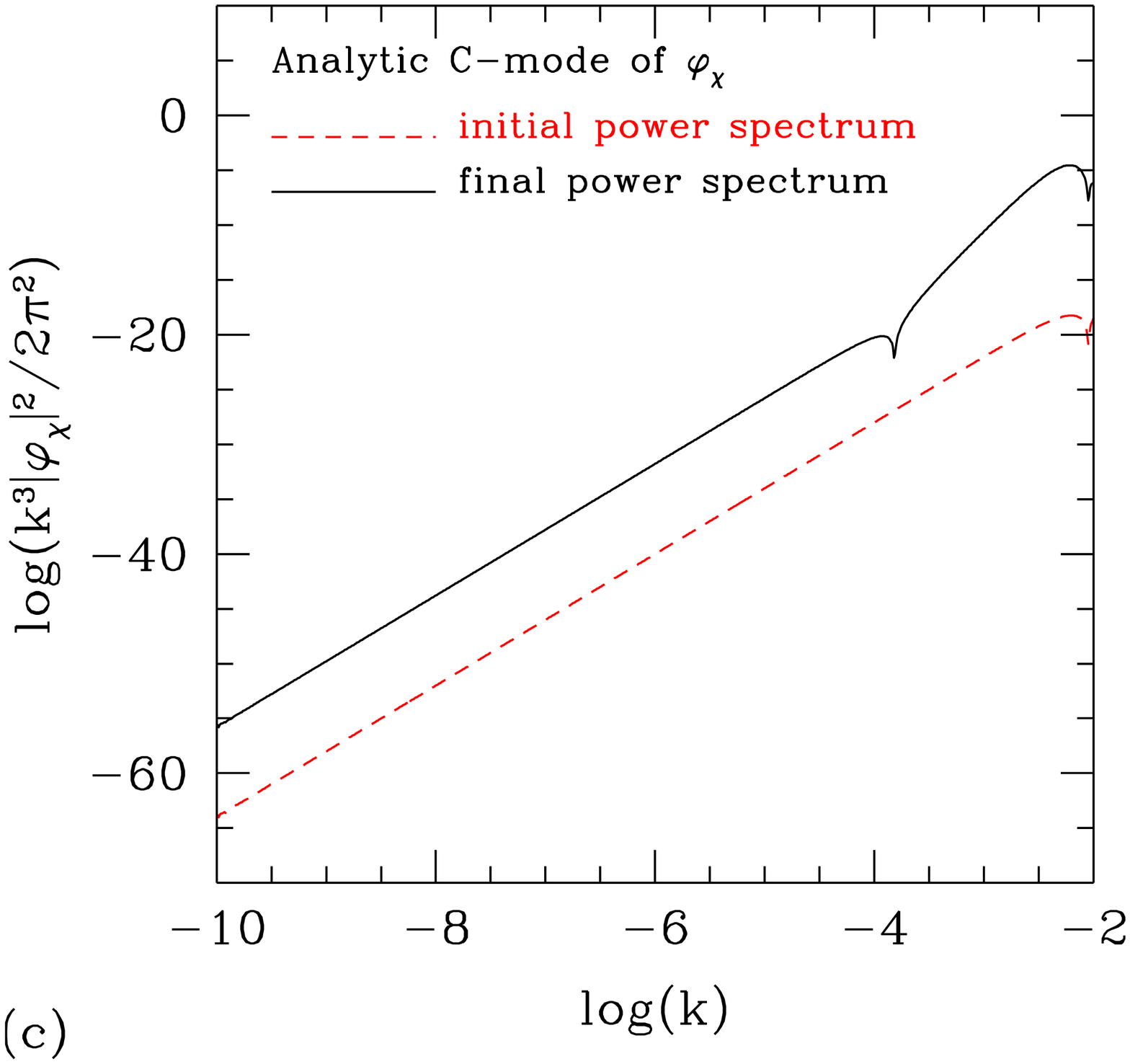}\hfill
\includegraphics[width=8cm]{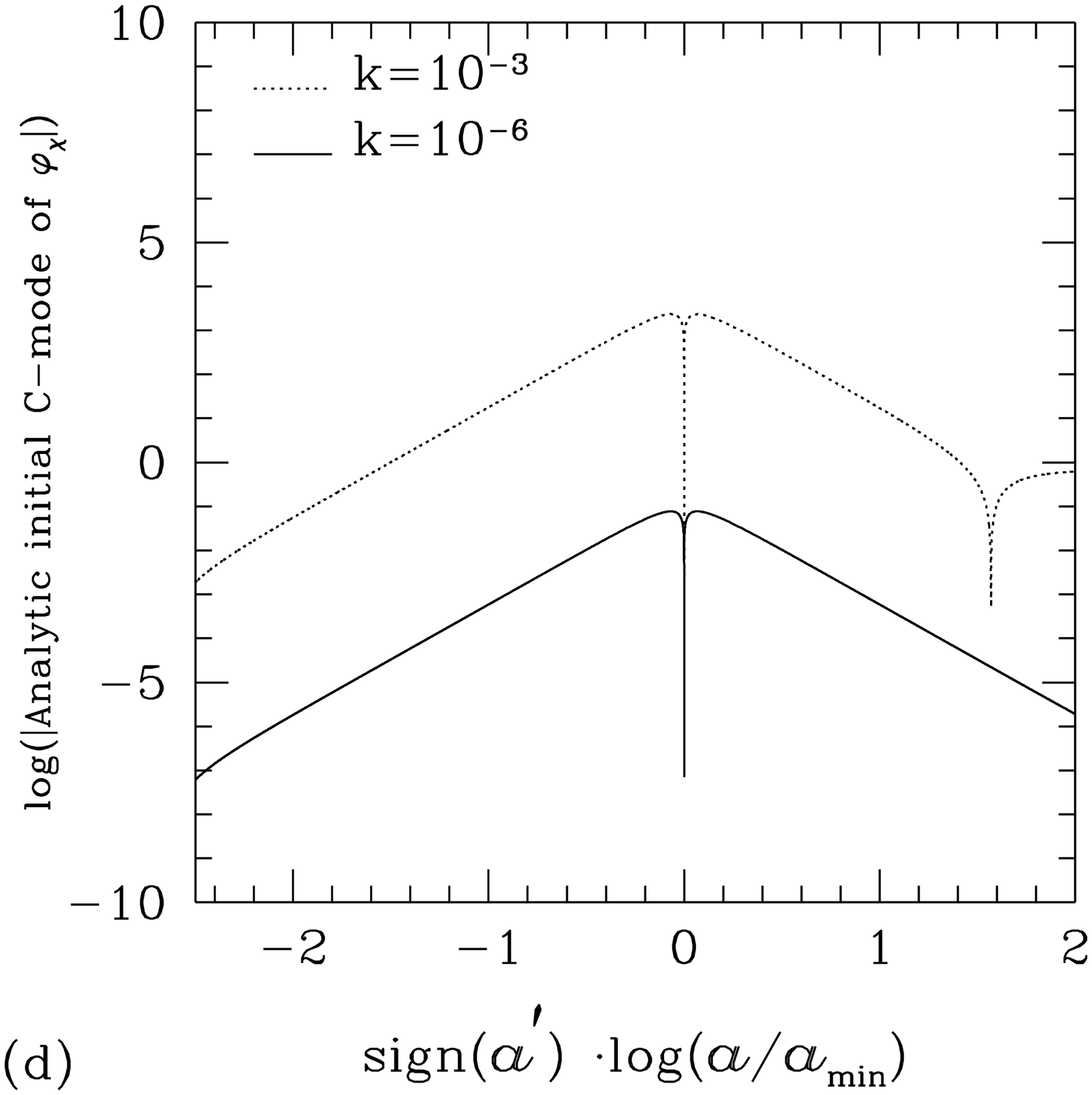}
\caption{The same as Fig.\ \ref{Fig-comparepowervpc}, now for
         $C$-mode of $\varphi_\chi$.
         Panel (a) shows that the spectral index for our $C$-mode
         is also $n_S-1 = 6$ which agrees with
         our analytic result in Eq.\ (\ref{eq-power2}).
         In case of analytic $C$-mode,
         only for large-scale ($\log{k} < -4$) perturbations
         the spectral index coincides with our $C$-mode case.
         In this case, however, due to time dependent evolution
         of $\varphi_\chi$, the amplitude of power spectrum
         simply depends on our choice of initial and final epochs.
         Panels (b) and (d) show typical
         evolution of $\varphi_v$ corresponding to Panels (a) and (c), respectively,
         for two different wavenumbers.
         Panel (d) shows why the power spectrum at
         large wavenumber is enhanced compared with the one at small
         wavenumber.
         }
         \label{Fig-comparepowervpchic}
\end{figure*}

\begin{figure*}
\centering%
\includegraphics[width=8cm]{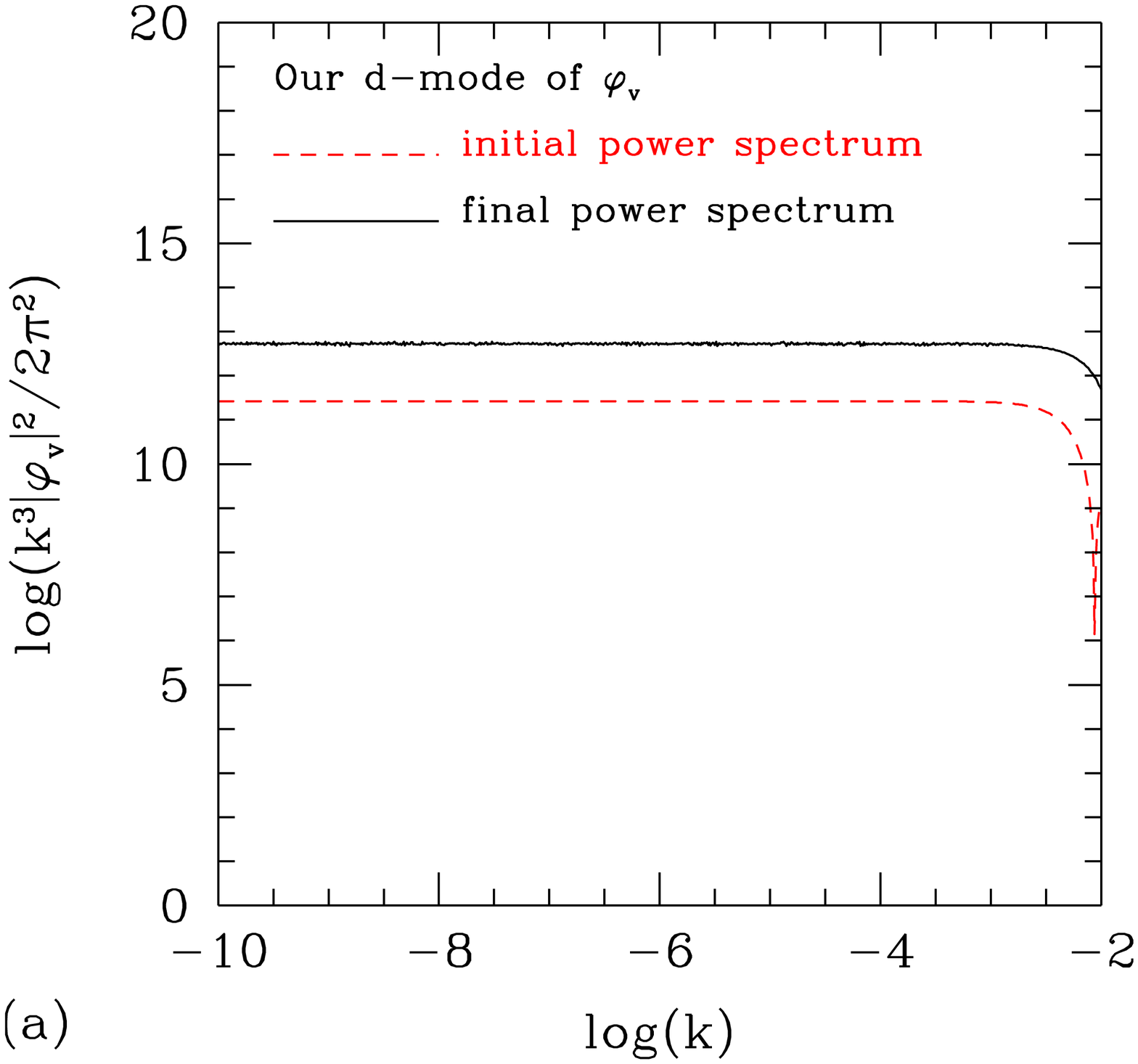}\hfill
\includegraphics[width=8cm]{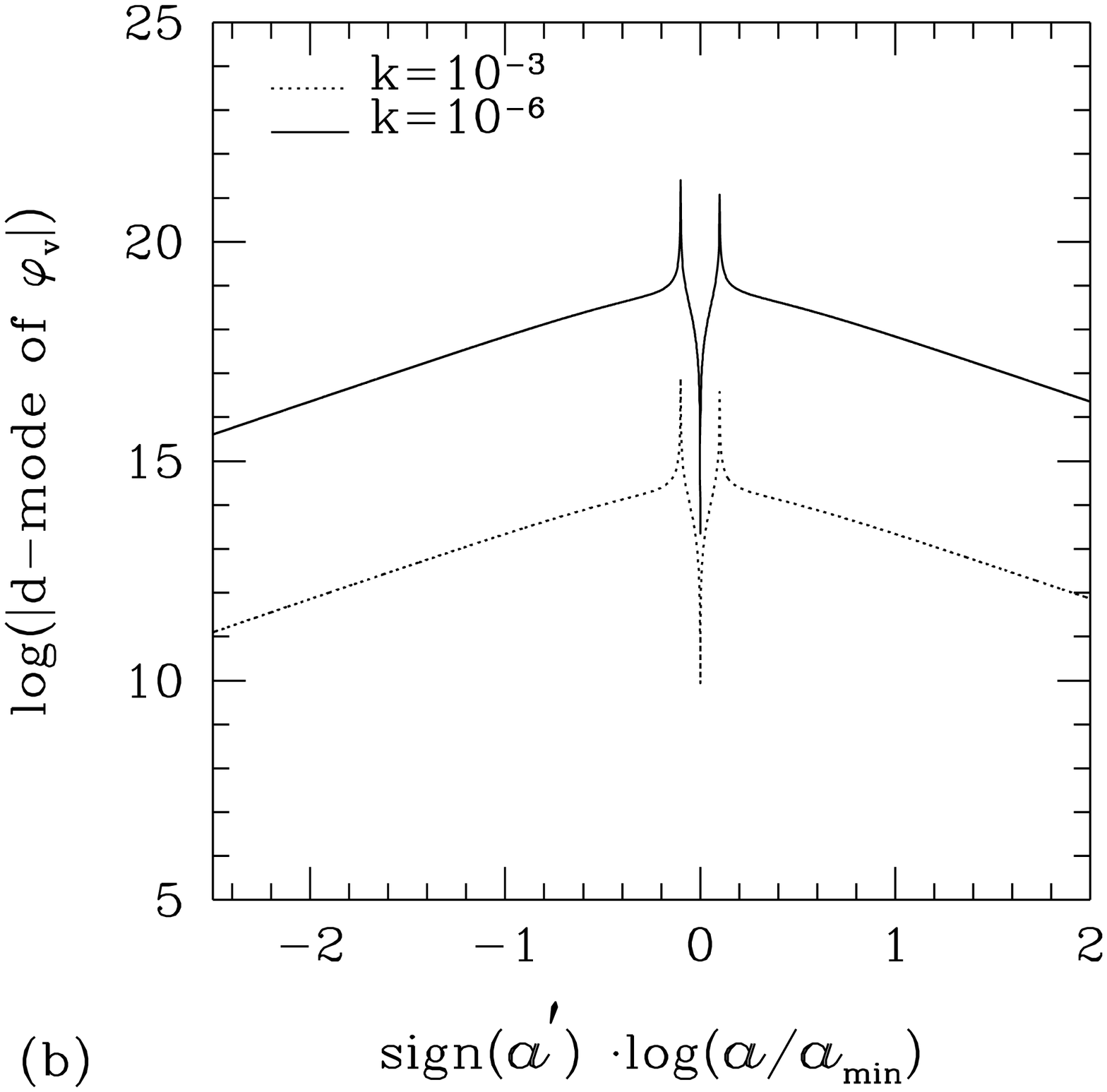}\\
\includegraphics[width=8cm]{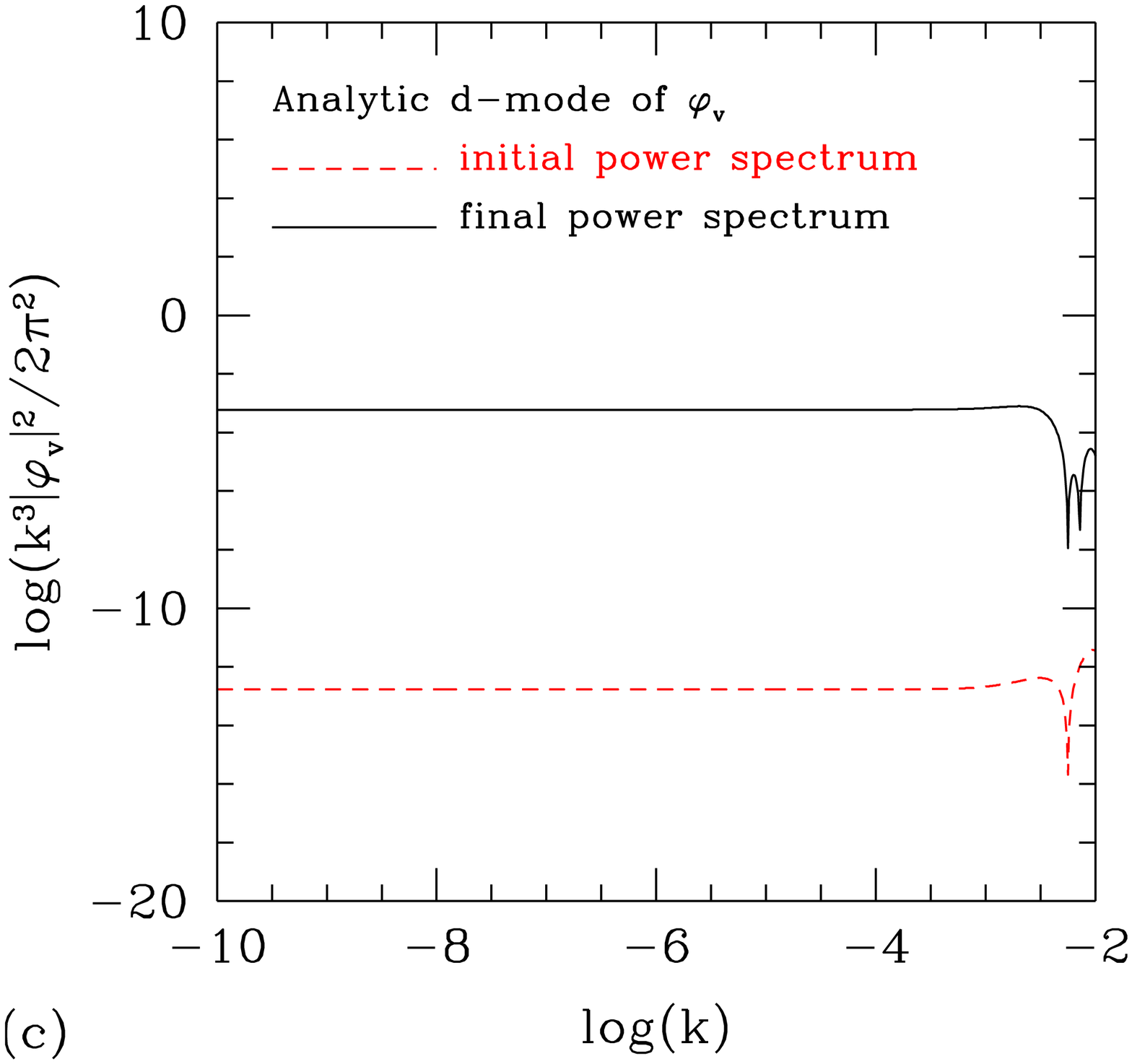}\hfill
\includegraphics[width=8cm]{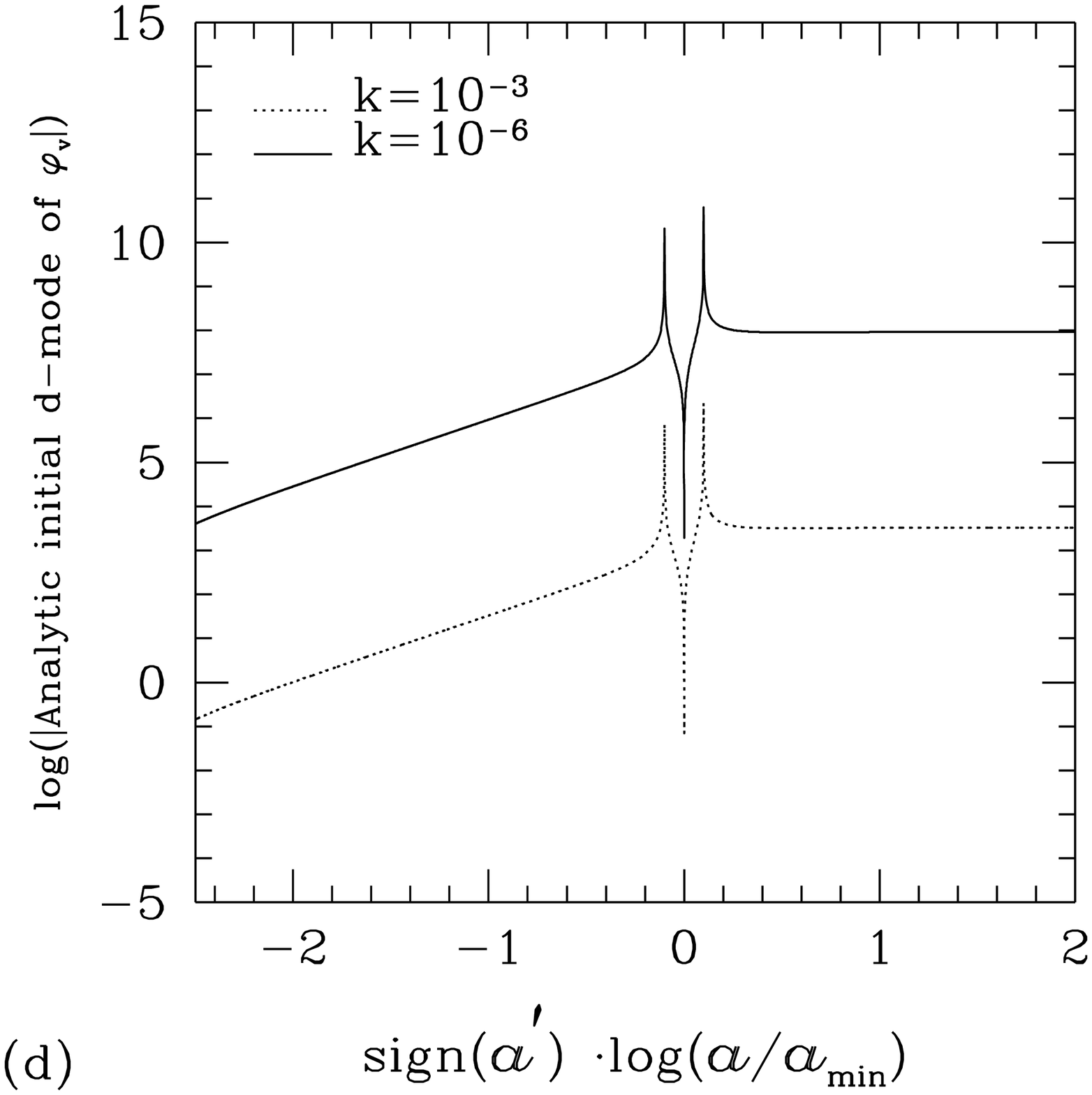}
\caption{The same as Fig.\ \ref{Fig-comparepowervpc}, now for
         $d$-mode of $\varphi_v$.
         Panel (a) shows that the spectral index for our $d$-mode gives
         near Harrison-Zel'dovich scale-invariant spectrum ($n_S-1 =
         0$) which agrees with
         our analytic result in Eq.\ (\ref{eq-power2}).
         Panel (c) shows that the spectral index of
         the analytic $d$-mode also gives $n_S-1=0$.
         Panels (b) and (d) show typical
         evolution of $\varphi_v$ corresponding to Panels (a) and (c), respectively,
         for two different wavenumbers;
         these evolutions show why the final spectral indices
         resemble the initial ones.
         In both Panels (a) and (c), due to time dependent evolution
         of $\varphi_\chi$, the amplitudes of power spectrum
         depend on our choice of initial and final epochs.
         For $\log(k) > -2.5$ the scale comes inside the horizon when the
         final spectra are measured.
         Although the figure is not presented in \cite{AW},
         based on the power spectrum in Panel (c) the authors of
         \cite{AW} reported that the bouncing world model can generate
         scale-invariant spectrum.
         }
         \label{Fig-comparepowervpd}
\end{figure*}

\begin{figure*}
\centering%
\includegraphics[width=8cm]{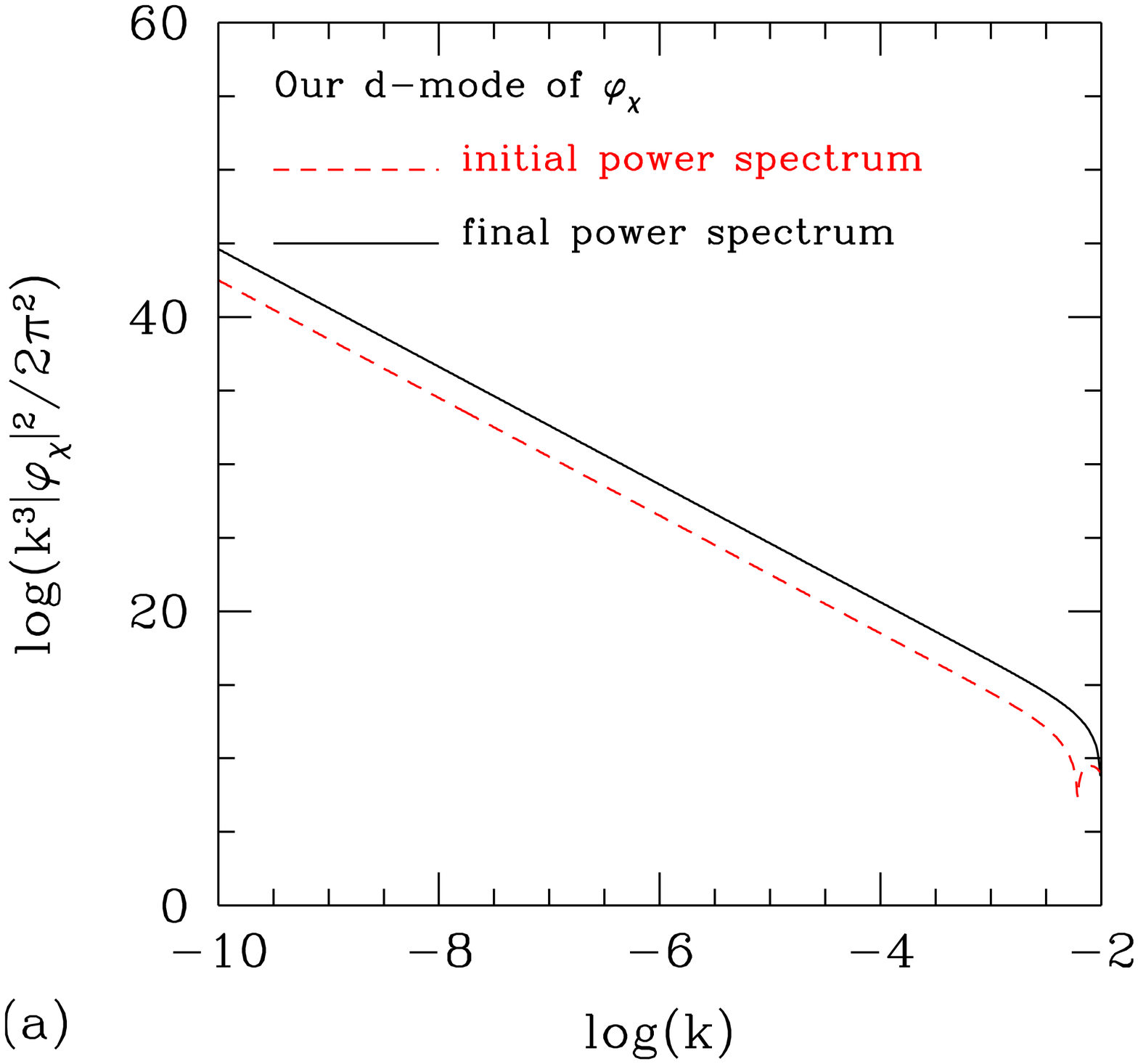}\hfill
\includegraphics[width=8cm]{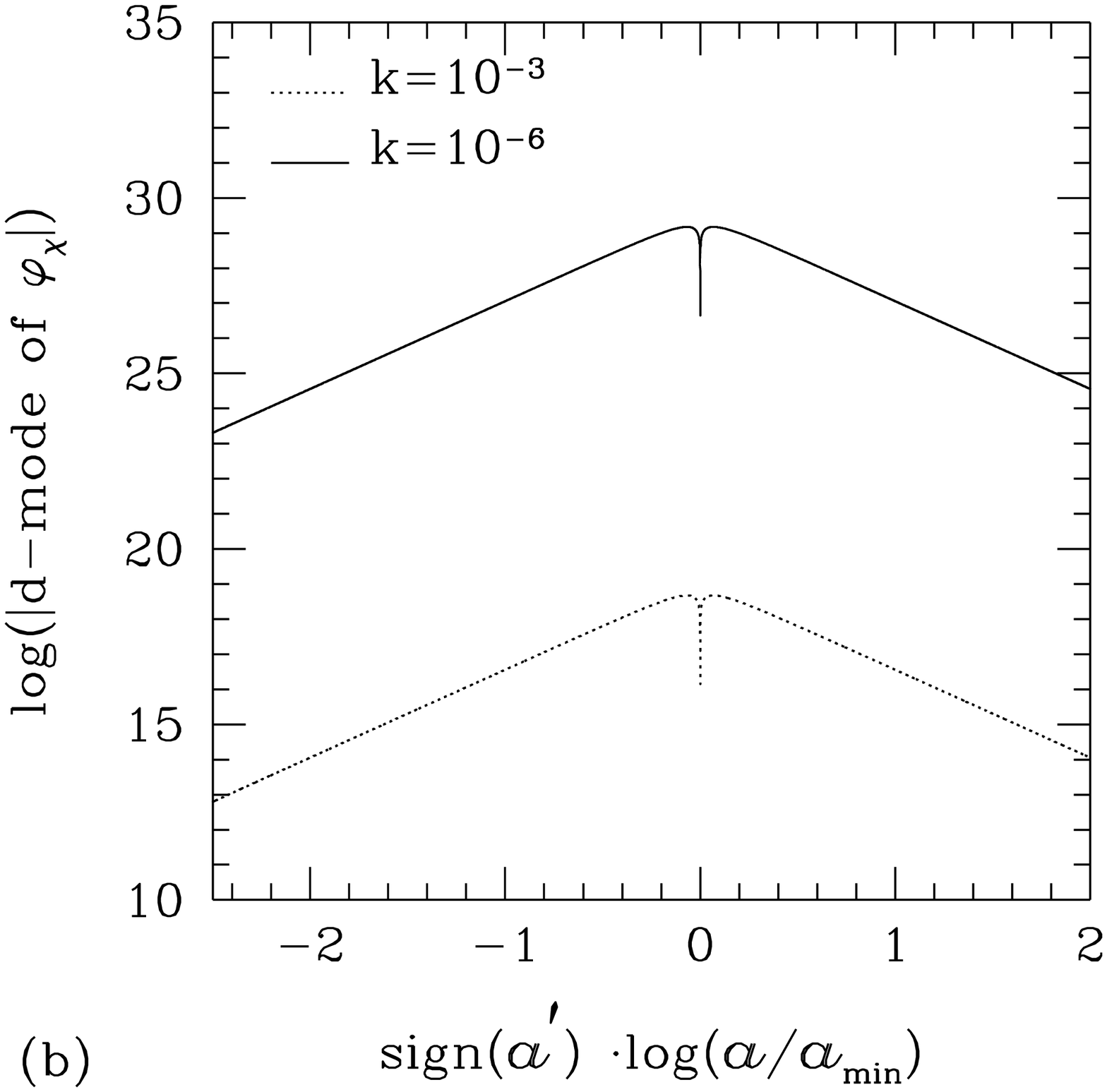}\\
\includegraphics[width=8cm]{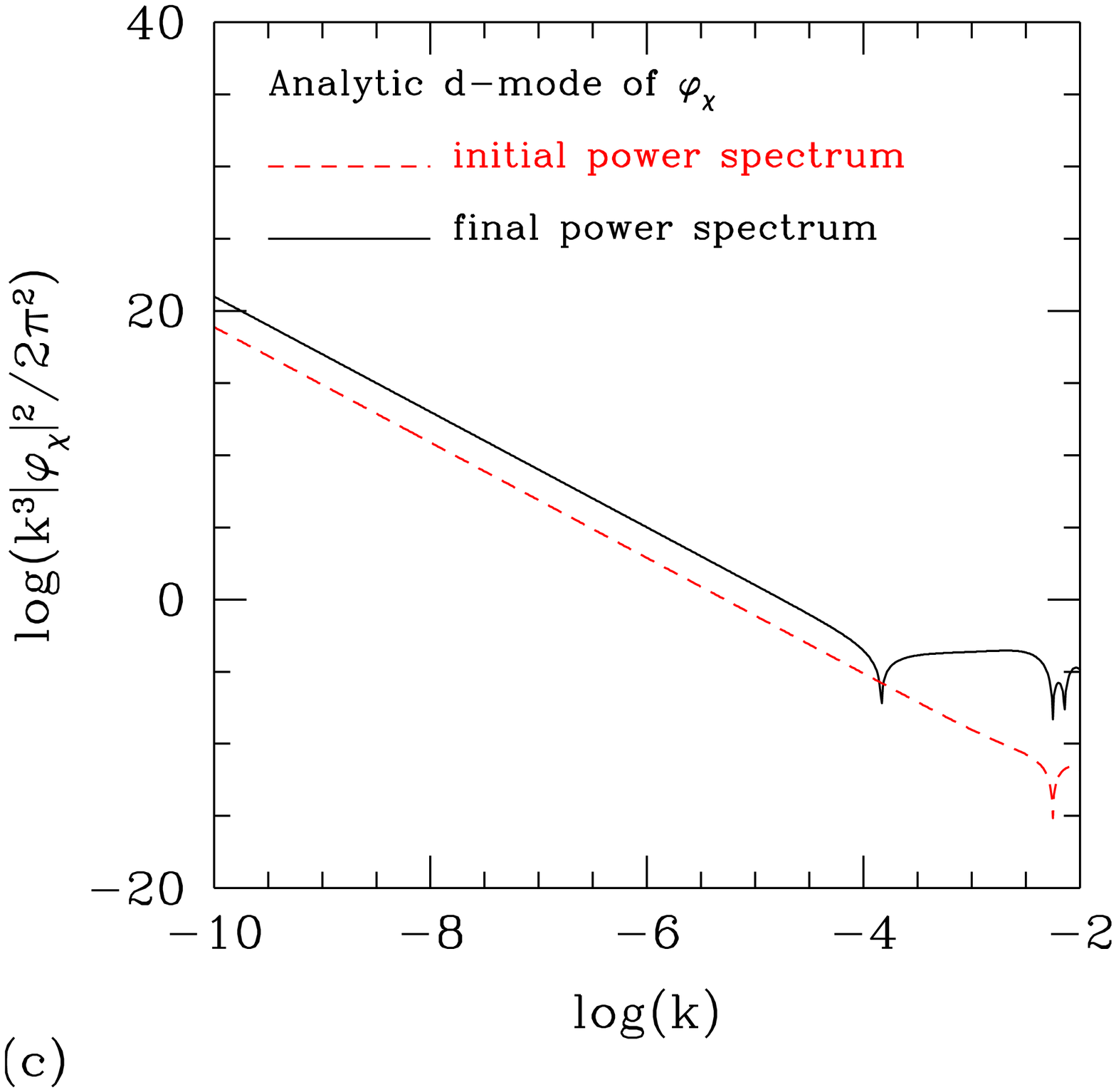}\hfill
\includegraphics[width=8cm]{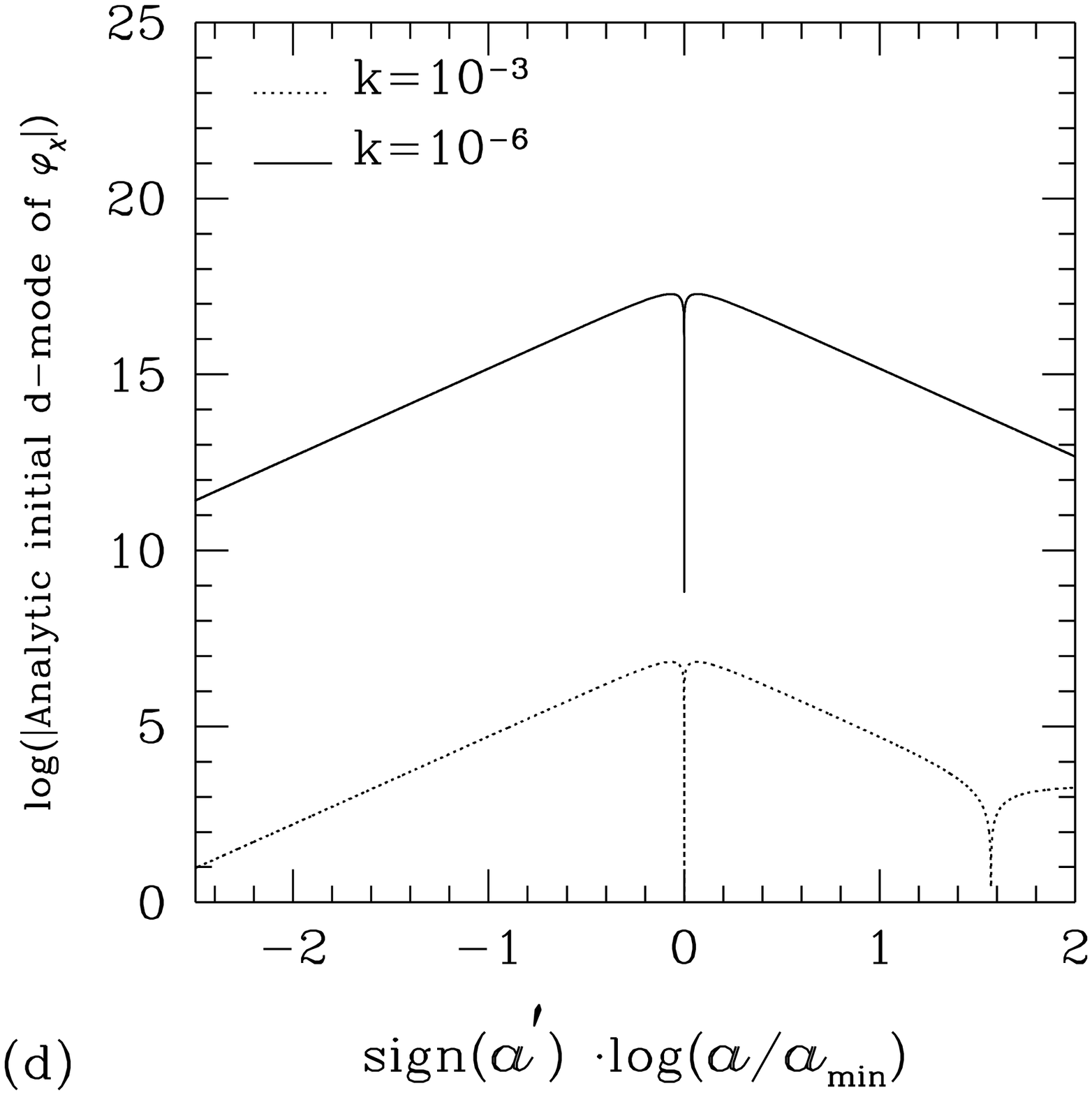}
\caption{The same as Fig.\ \ref{Fig-comparepowervpc}, now for
         $d$-mode of $\varphi_\chi$.
         Panel (a) shows that the spectral index for our $d$-mode
         is $n_S-1 = -4$ which agrees with
         our analytic result in Eq.\ (\ref{eq-power2}).
         In case of analytic $d$-mode,
         only for large-scale ($\log{k} < -4$) perturbation
         the spectral index coincides with our $d$-mode case.
         In both Panels (a) and (c), due to time dependent evolution
         of $\varphi_\chi$, the amplitudes of power spectrum
         depend on our choice of initial and final epochs.
         Panels (b) and (d) show typical
         evolution of $\varphi_v$ corresponding to Panels (a) and (c), respectively,
         for two different wavenumbers.
         Panel (d) shows why the power spectrum at
         large wavenumber is enhanced compared with the one at small
         wavenumber.
         }
         \label{Fig-comparepowervpchid}
\end{figure*}

In our simple bounce model case with $q={2/3}$, we have a
scale-invariant spectrum for $d$-mode of $\varphi_v$, but the
$d$-mode is meaningless because it decays away in an expanding
phase. Our model shows a quite blue spectra for $C$-mode with
$n_S-1=6$. In Figs.\
\ref{Fig-comparepowervpc}-\ref{Fig-comparepowervpchid} we present
the power spectra of $C$- and $d$-modes of $\varphi_v$ and
$\varphi_\chi$ based on numerical integrations; we also present
characteristic evolutions of $\varphi_v$ and $\varphi_\chi$. The
initial spectra are based on the large-scale limit of Hankel
function solutions in Eq.\ (\ref{eq-power}) which have their origins
in quantum fluctuations in the collapsing phase. The spectral
indices of final power spectra coincide with our analytic
estimations made in Eq.\ (\ref{eq-power2}). In Figs.\
\ref{Fig-comparepowervpc}-\ref{Fig-comparepowervpchid} we present
the initial and final power spectra from the analytic $C$- and
$d$-modes together, and the characteristic evolutions of $\varphi_v$
and $\varphi_\chi$.

Based on numerical evolutions of certain initial conditions which
are similar to the `analytic $C$- and $d$-modes' of ours, the
authors of \cite{AW} reported that the $d$-mode in the collapsing
phase survives as the $C$-mode in the expanding phase, see Fig.\
\ref{Fig-anld}(a); similar results were also reported in
\cite{Wands-1999,BF,FB}. This result leads these authors to claim
that for $q=2/3$ the quantum fluctuations in the collapsing phase
give rise to a scale-invariant spectrum in an expanding phase. Our
numerical result in Fig.\ \ref{Fig-comparepowervpd}(c) apparently
confirms this result. However, based on the behaviors of our $C$-
and $d$-modes, the initial conditions used in these analytic $C$-
and $d$-modes are mixture of the pure (our) $C$- and $d$-modes.
Under these initial conditions, as the evolution proceeds we have
shown that the isocurvature perturbation is also excited near the
bounce, see Fig.\ \ref{Fig-isocurvature-d-mode}. Thus, it is {\it
not correct} to interpret that the final $C$-mode spectrum in the
expanding phase is generated from the initial $d$-mode spectrum in
the collapsing phase. In our interpretation, although the evolution
begins with near $d$-mode (which is actually a mixture of our, thus
precise, $C$- and $d$-modes), as we approach the bounce not only the
$C$- and $d$-modes but also additional two isocurvature-type
perturbations are all excited; this eventually leaves only the
dominating mode (which is the $C$-mode in an expanding phase) after
the bounce. In any case, our numerical result in Fig.\
\ref{Fig-comparepowervpd}(c) shows that near Harrison-Zel'dovich
spectrum can be generated from the analytic $d$-mode initial
condition with the help of the isocurvature perturbation
simultaneously excited near the bounce. In case of our $C$- and
$d$-modes where the $C$- and $d$-mode natures are preserved before
and after the bounce, and the isocurvature perturbation is not
excited. In that case, near Harrison-Zel'dovich spectrum in the
$d$-mode of $\varphi_v$ is observationally not relevant.

A scale-invariant spectrum for the $C$-mode can be obtained for $q
\gg 1$ which is the power-law inflation in an expanding phase. In
the collapsing phase, however, $q>1$ implies that the relevant
scales come inside the horizon near the bounce, thus invalidating
the large-scale condition we used to get the above result. In the
sub-horizon scale the modes oscillate, and lose their previous
large-scale identity of the $C$- and the $d$-modes \cite{hw2}. From
Eq.\ (\ref{eq-power2}), for any model with $q<1$ during the quantum
generation stage, we have $n_S-1>0$, thus having a blue spectrum.

Therefore, we conclude that while the large-scale condition is
satisfied and the adiabatic condition is met (i.e., the isocurvature
perturbation is not excited) during the bounce, it is {\it not
possible} to obtain near Harrison-Zel'dovich scale-invariant density
spectrum through a bouncing world model as long as the seed
fluctuations were generated from quantum fluctuations of the
curvature perturbation in the collapsing phase. This conclusion {\it
assumes} that Einstein gravity is dominating throughout the bounce
including the quantum generation stage.


\subsection{Tensor-type perturbations}
                                                \label{tensor-implication}

The quantum fluctuations of the space-time metric detached from the
fields provide seed fluctuations for the tensor-type perturbation
(gravitational waves). For tensor-type perturbation, the power
spectrum and spectral index are defined as \bea
   & & \mathcal{P}_{C_{\alpha\beta}^{(t)}}({\bf k},\eta)
       \equiv 2 \sum_l{\mathcal{P}_{h_l}({\bf k}, \eta)}
       \equiv 2 \sum_l{k^3 \over 2 \pi^2}|h_{lk}|^2,
   \nonumber \\
   & & n_T \equiv
       {d\ln{\mathcal{P}_{C^{(t)}_{\alpha\beta}}} \over d\ln{k}},
   \label{eq-tensorpol}
\eea where we introduced a decomposition based on the two
polarization states \cite{gw-quantum}\bea
    C^{(t)}_{\alpha\beta}({\bf x}, t)
        &\equiv& \sqrt{Vol}\int{d^3k \over (2\pi)^3}C^{(t)}_{\alpha\beta}({\bf x},t ; {\bf
    k})\nonumber\\
    &\equiv& \sqrt{Vol}\int{d^3k \over (2\pi)^3}\sum_l e^{i{\bf k
    \cdot x}}h_{l{\bf k}}(t)e^{(l)}_{\alpha\beta}({\bf
    k}),\nonumber\\
    h_l({\bf x}, t) &\equiv& {1 \over 2}\sqrt{Vol}\int{d^3k \over
    (2\pi)^3}C^{(t)}_{\alpha\beta}({\bf x},t ;{\bf
    k})e^{(l)\alpha\beta}({\bf k})\nonumber\\
    &=& \sqrt{Vol}\int{d^3k \over (2\pi)^3}e^{ik \cdot x}h_{lk}(t),
\eea with $l=+, \times$; the polarization tensors
$e^{(+)}_{\alpha\beta}$ and $e^{(\times)}_{\alpha\beta}$ indicate
the plus (+) and cross ($\times$) polarization states with
$e^{(l)}_{\alpha\beta}({\bf k})e^{(l^{\prime})\alpha\beta}({\bf
k})=2\delta_{ll^\prime}$.

In order to correspond our large-scale solution in Eq.\
(\ref{eq-analten}) to the exact solution in Eq.\
(\ref{eq-tensorhankel}), we follow a similar reasoning used for the
$\varphi_v$. In Eq.\ (\ref{eq-analten}) the leading order of the
$\bar{C}_{\alpha\beta}$-mode is time independent whereas the leading
order of $\bar{d}_{\alpha\beta}$-mode behaves as
$C_{\alpha\beta}^{(t)} \propto |\eta|^{-2\nu_g}$. Since \bea
    & & C_{\alpha\beta}^{(t)}(k,\eta) \propto
        |\eta|^{-\nu_g}H_{\nu_g}^{(1,2)}(k|\eta|),
    \label{eq-connecttensorhankel}
\eea we can easily identify the first term in the parenthesis of
Eq.\ (\ref{eq-hankelex}) as the $\bar{C}_{\alpha\beta}$-mode and
second term as the $\bar{d}_{\alpha\beta}$-mode. We can read the
spectral indices for $\bar C_{\alpha\beta}$ and $\bar
d_{\alpha\beta}$ as \bea
    & & n_{T,\bar C_{\alpha\beta}}={2 \over 1-q}, \quad
        n_{T,\bar d_{\alpha\beta}}={4-6q \over 1-q}.
    \label{tensor-spectrum}
\eea These spectral indices of tensor-type perturbation {\it
coincide} with the spectral indices of the $\varphi_v$ in Eq.\
(\ref{eq-power2}).

{}From the classical evolution of the tensor-type perturbation
studied in Sec.\ \ref{GW-numerical} we notice that both the initial
$\bar C_{\alpha\beta}$- and $\bar d_{\alpha\beta}$-modes survive as
the $\bar C_{\alpha\beta}$-mode after the bounce.  If both modes are
generated from the quantum fluctuations with comparable amplitude,
as the collapsing phase proceeds the initial $\bar
C_{\alpha\beta}$-mode becomes negligible compared with the initial
$\bar d_{\alpha\beta}$-mode. Thus, after the bounce the dominant
$\bar C_{\alpha\beta}$-mode should have origin from the initial
$\bar d_{\alpha\beta}$-mode; as the $\bar d_{\alpha\beta}$-mode
grows rapidly in the collapsing phase, its final amplitude surely
depends on the duration of the collapse after its quantum origin.
This is in contrast to the situation of the scalar-type perturbation
where the initial $d$-mode remains as the $d$-mode even after the
bounce, thus decays away in the expanding phase.

\begin{figure}
\centering%
\includegraphics[width=8cm]{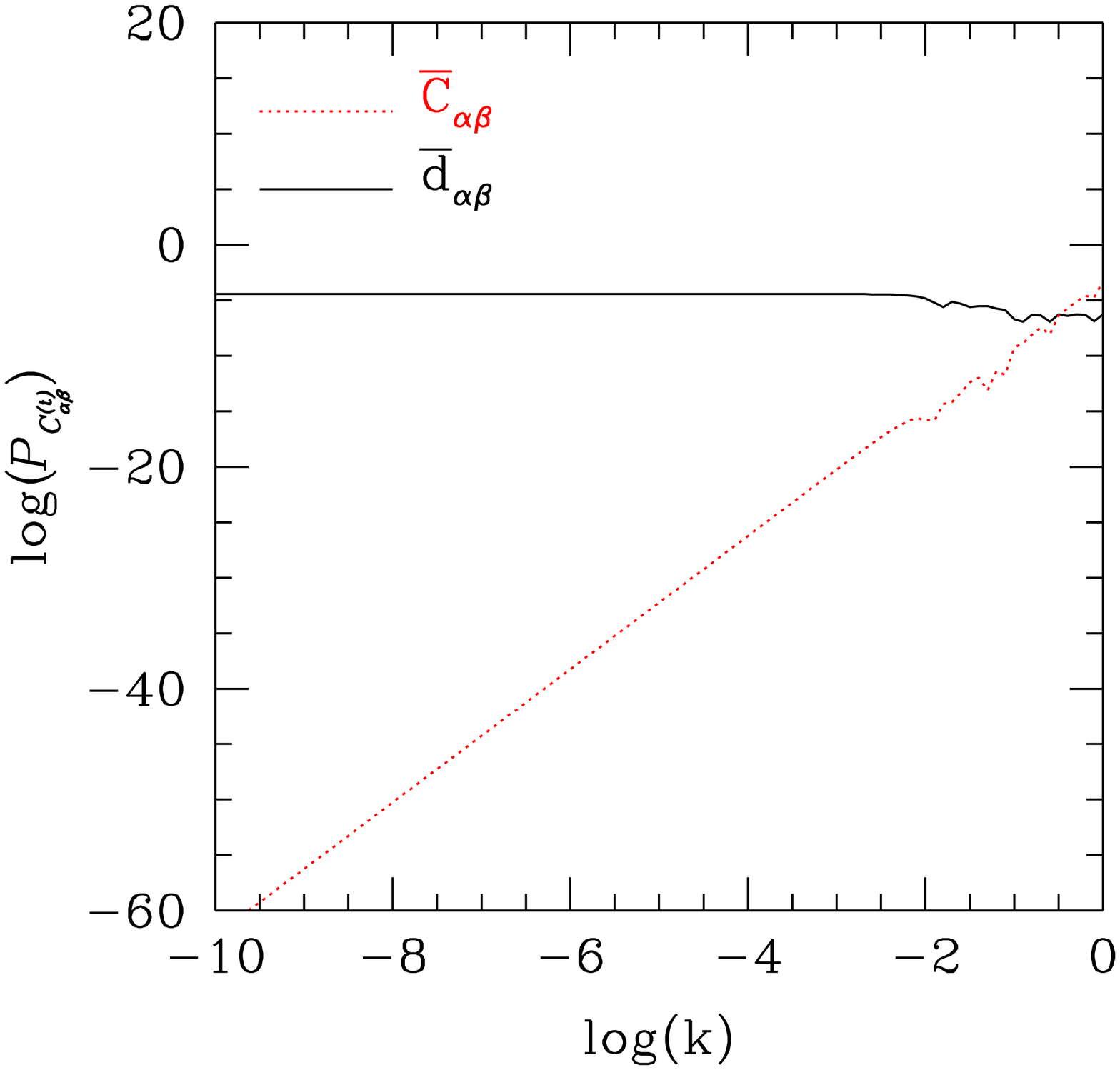}\hfill
\caption{This figure shows the final power spectra for initial
         ${\bar C}_{\alpha\beta}$- and ${\bar d}_{\alpha\beta}$-modes.
         The initial ${\bar d}_{\alpha\beta}$-mode has a
         scale-invariant spectrum  with $n_{T,\bar d_{\alpha\beta}} = 0$,
         and ${\bar C}_{\alpha\beta}$-mode has very
         blue spectrum with $n_{T,\bar C_{\alpha\beta}} = 6$;
         these agree with our analytic result in Eq.\ (\ref{tensor-spectrum}).
         The amplitudes of power spectra depend on background properties.
         For $\log(k) > -2$ the scale comes inside the horizon when the
         final spectra are measured.
         }
         \label{Fig-GW-Power}
\end{figure}

In the pre-big bang scenario with $q = 1/3$, we have $n_{T,{\bar
C_{\alpha\beta}}} = 3$ and $n_{T,{\bar d_{\alpha\beta}}} = 3$. In
the ekpyrotic or cyclic models with $q \simeq 0$, we have
$n_{T,{\bar C_{\alpha\beta}}} = 2$ and $n_{T,{\bar d_{\alpha\beta}}}
= 4$. Notice that it is the $\bar d_{\alpha\beta}$-mode initial
condition in the collapsing phase which survives as the dominating
$\bar C_{\alpha\beta}$-mode after the bounce in case of the
tensor-type perturbation. Thus, the dominating $\bar
C_{\alpha\beta}$-mode perturbation in expanding phase produced in
the ekpyrotic and cyclic scenarios has $n_T = 4$ spectrum; this also
differs from $n_T = 2$ which is claimed by the authors of these
scenarios \cite{Ek}. Meanwhile, in our simple bounce model with $q =
2/3$, we have $n_{T,{\bar C_{\alpha\beta}}}=6$ and $n_{T,{\bar
d_{\alpha\beta}}}=0$. From Eq.\ (\ref{tensor-spectrum}) we notice
that only for $q=2/3$, thus $w=0$, we have a scale-invariant near
Harrison-Zel'dovich type spectrum $n_{T,{\bar d_{\alpha\beta}}}=0$
for the tensor-type perturbation. In Fig.\ \ref{Fig-GW-Power} we
present the final spectra of the initial $\bar C_{\alpha\beta}$- and
$\bar d_{\alpha\beta}$-modes in case of $w=0$; these numerical
results coincide with our analytic estimation made above.

\section{Discussion}
                                                     \label{sec:Discussion}

In order to explain the origin of large-scale structure in the
universe, several models of bouncing scenarios \cite{pbb,Ek} were
suggested as possible alternatives to the inflation. The pre-big
bang scenario in \cite{pbb} failed to produce the observationally
required scale-invariant spectrum for scalar-type perturbation; this
model has $q=1/3$ thus leads to $(n_S-1)_{\varphi_v,C} \simeq 3$ and
$n_{T,\bar d_{\alpha\beta}} \simeq 3$. Whereas, the authors of
ekpyrotic or cyclic scenarios \cite{Ek} claimed that their scenarios
successfully produce the required spectrum. The latter authors find
that the quantum fluctuations of $d$-mode of $\varphi_\chi$ show
near Harrison-Zel'dovich scale-invariant spectrum and claimed that
this mode gives rise to the $C$-mode after the bounce. As long as
the adiabatic condition and the large-scale condition are met during
the bounce (and, of course, assuming the linear perturbation theory
survives the bounce) from the analytic point of view we can see that
$d$-mode before the bounce remains as the same $d$-mode after the
bounce and similarly for the $C$-mode. Since the $d$-mode is
transient in an expanding phase, we have to consider the $C$-mode as
the proper seed for the large-scale structure. For the ekpyrotic or
cyclic scenarios with $0<q \ll 1$ the quantum fluctuations of the
$C$-mode of both $\varphi_\chi$ and $\varphi_v$ show very blue
spectra with $n_S - 1 \simeq 2$, thus fail to explain the
observation.

In this work we have confirmed the above expectation based on
analytic arguments made in \cite{HN-bounce} by following the $C$-
and $d$-modes numerically in a simple nonsingular bouncing world
model. We found the $C$- and $d$-mode initial conditions in which
the nature of the $C$- and $d$-modes are preserved through our
nonsingular bounce world model. Thus, we have shown that the
scale-invariant spectrum claimed by the proponents of ekpyrotic or
cyclic models is not possible if the bounce occurs non-singularly as
in our toy bounce model. In the original ekpyrotic/cyclic scenarios
based on five dimension the effective four dimensional spacetime
goes through a singularity \cite{Ek}, and the situation cannot be
handled using perturbation theory in Einstein gravity. Our results
also imply that it is {\it not possible} to generate near
Harrison-Zel'dovich scale-invariant spectrum from the quantum
fluctuations in the collapsing phase as long as the isocurvature
perturbation is not excited significantly near the bounce. We state
again the assumptions we made to get such a conclusion: (i)
Einstein's gravity and linear perturbation theory work throughout
the evolution, (ii) quantum fluctuations of a minimally coupled
scalar field in a collapsing phase provide seed curvature
(adiabatic) mode fluctuations for the large-scale structure, (iii)
the large-scale conditions are met during the bounce, (iv) the
adiabatic conditions are met during the bounce. We have carefully
examined that the latter two conditions are often violated subtly
near the bounce without affecting curvature perturbation natures in
the large-scale limit.

In our simple bounce model we introduced a ghost field $\sigma$ only
to have a natural bounce in the background. The presence of
additional ghost field in addition to the ordinary minimally coupled
scalar field $\phi$ (which is introduced to have curvature-type seed
fluctuations), however, forces us to handle a fourth-order
differential equation for the scalar-type perturbation. In order to
have a nonsingular bounce we need $\sigma$ to be important only near
the bounce. However, the presence of ghost field could potentially
generates the isocurvature-type perturbation $S_{\phi\sigma}$ due to
the perturbed ghost field near the bounce. Our above conclusion
forbidding near Harrison-Zel'dovich spectrum from the bouncing model
is based on assuming no isocurvature perturbation being excited near
the bounce (i.e., the adiabatic conditions are well met during the
bounce). Under this condition we have found the $C$- and $d$-modes
which preserve their nature before and after the bounce.

By using the analytic $C$- and $d$-mode initial conditions we have
shown that it is actually possible to have near Harrison-Zel'dovich
spectrum from the analytic $d$-mode initial condition of $\varphi_v$
for $q = 2/3$, see Figs.\ \ref{Fig-comparepowervpd}(c) and
\ref{Fig-comparepowervpd}(d); this result was reported in
\cite{AW,Wands-1999,BF,FB}. In this case, although Fig.\
\ref{Fig-comparepowervpd}(d) shows that the initial $d$-mode before
the bounce is switched into an apparently $C$-mode after the bounce,
such a transition occurs because isocurvature perturbation is
significantly excited near the bounce; i.e., under such an initial
condition all four modes (two $C$- and $d$- curvature modes and two
isocurvature modes) are excited near the bounce, thus mixing modes
before and after the bounce. Our numerical results show that, unless
we impose a rather precise initial conditions, it is very likely
that we can easily have only relatively dominating mode during the
evolution: these are the $d$-mode in a collapsing phase, and the
$C$-mode in an expanding phase. Figures\ \ref{Fig-anlc}(a),
\ref{Fig-anlc}(b), \ref{Fig-anld}(a), and \ref{Fig-anld}(b) based on
the analytic $C$- and $d$-modes clearly show such mixed behaviors
between the $C$- and $d$-modes. We have shown that these are because
the analytic $C$- and $d$-mode initial conditions can be viewed as
mixtures of our (precise) $C$- and $d$-modes, i.e., imprecise. In
addition, even the isocurvature perturbation caused by the presence
of the ghost field is excited near the bounce.

We have shown that for the tensor-type perturbation, however, it is
the $\bar d_{\alpha\beta}$-mode initial condition which survives as
the dominating $\bar C_{\alpha\beta}$-mode in the expanding phase.
Thus, for the ekpyrotic or cyclic scenarios with $0 < q \ll 1$, we
have $n_{T,\bar d_{\alpha\beta}} \simeq 4$; this also {\it differs}
from the predictions made by the original authors of these scenarios
who claimed $n_{T,\bar C_{\alpha\beta}} \simeq 2$ as the dominating
mode.

In conclusion, as long as the large-scale condition and the
adiabatic condition are satisfied, we numerically show that the
$C$-mode in a collapsing phase leads to the observationally relevant
quantity in the expanding phase. The $d$-mode, in contrast, remains
as the same $d$-mode even after the bounce. Thus, under these
conditions, it is not possible to obtain a scale-invariant spectrum
in the bounce cosmological model based on the quantum fluctuations
in the collapsing phase. Equation (\ref{eq-power2}) shows that only
for $q \gg 1$ during quantum generation stage the bounce model can
produce near Harrison-Zel'dovich scale-invariant density spectrum.
However, since this leads to a violation of the large-scale
condition near the bounce we cannot use this result; the modes are
mixed up and begin to oscillate, thus we cannot trace the $C$- and
$d$-mode nature through the bounce. That is, as the scales come
inside the horizon, perturbations start oscillations and lose their
large-scale memory of the original $C$- and $d$-modes. For the
tensor-type perturbation, it is the initial $\bar
d_{\alpha\beta}$-mode which survives as the dominating $\bar
C_{\alpha\beta}$-mode in the expanding phase.

We may summarize the final spectral indices for the scalar- and
tensor-type perturbations as \bea
    & & ( n_{s} - 1 )_{C} ={2 \over 1-q}, \quad
        n_{T,{\bar d_{\alpha\beta}}}={4-6q \over 1-q},
\eea where we have $a \propto |t|^q$ during the quantum generation
stage {\it before} bounce. Meanwhile, if the initial seed
fluctuations were generated from quantum fluctuations during
expanding phase (without or after bounce), the observationally
relevant perturbations have \bea
    & & ( n_{s} - 1 )_{C} ={2 \over 1-q}
        = n_{T,{\bar C_{\alpha\beta}}},
\eea which is the well known result in the literature
\cite{Stewart-Lyth-1993,hw1}. Notice the difference in the
gravitational wave power spectrum generated between the bounce and
the inflation scenarios.

Our analysis in this work is based on a specific nonsingular bounce
model with two scalar fields. Using this specific model we were able
to follow numerically the evolutions of scalar- and tensor-type
perturbations through the bounce. We have shown that the nature of
large-scale decomposition of the adiabatic solutions far away from
the bounce, i.e., the $C$- and $d$-modes, is preserved throughout
the bounce as long as the two conditions (the large-scale and
adiabatic ones) are met. Throughout the bounce we have numerically
monitored that these conditions are physically met as long as we
take the correct and precise initial conditions which assure the
$C$- and $d$-mode natures; we also have monitored the evolution of
isocurvature perturbation. In order to help readers to reproduce our
numerical results, in Table \ref{Table-initiall2} we have provided
the precise initial conditions we used. Because the above argument
was originally based on the analytic solutions with general equation
of state or field potential, our numerical analysis based on
specific bounce model may have more generic implications which go
beyond the specific model we used. Our concrete numerical study
assures that our previous conclusions based on analytic reasons in
\cite{hw1,hw2,HN-bounce} are valid.

In this work, we have {\it assumed} Einstein's gravity and spatially
homogeneous-isotropic background world model with linear
perturbations are valid throughout the bounce. In realistic
situation, however, it is likely that in the collapsing phase as the
model approaches the bounce the perturbations of all three types
grow rapidly and eventually become nonlinear, thus invalidating the
spatially homogeneous-isotropic Friedmann background world model.
Furthermore, as the scale factor shrinks we reach the high energy
scale where quantum corrections of Einstein's gravity likely lead to
serious correction terms in the gravity itself. Although we have
used a ghost field as the bounce agent in Einstein's gravity, it is
also likely that the quantum correction terms can cause nonsingular
bounce more naturally \cite{bounce-pert-others}. In this work we
have assumed that the linear perturbation theory survives throughout
the bounce which certainly restricts the allowed bounce model.

$ $ \vskip .5cm
\centerline{\bf Acknowledgments}

We thank Drs.\ Hyerim Noh and Ewan Stewart for useful discussions.
This work was supported by the Korea Research Foundation Grant No.\
2003-015-C00253 and KNURT (Kyungpook National University Research
Team) Fund 2002.


\end{document}